\documentclass[11pt]{article}

\usepackage[english]{babel}
\usepackage[normalem]{ulem}

\usepackage{orcidlink}

\usepackage{amsmath,amsthm,amssymb,amscd}
\usepackage{amsfonts,mathrsfs}
\usepackage[smalltableaux,centertableaux]{ytableau}
\usepackage{mathtools}
\usepackage{tikz}
\usepackage{tikz-cd}
\usepackage{braket}
\usepackage{tensor}
\usepackage{slashed}
\usepackage{cancel}
\usepackage{nicefrac}
\usepackage{bm}
\usepackage{diagbox}
\usepackage[version=4]{mhchem}
\usepackage{bbm}
\usepackage{centernot}
\usepackage{upgreek}
\usepackage{fontawesome}

\usepackage{graphicx}
\usepackage[export]{adjustbox}
\usepackage{xcolor}
\usepackage{float}

\usepackage{cite}
\usepackage{hyperref}
\usepackage[nottoc]{tocbibind}

\usepackage{booktabs}
\usepackage{multirow}
\usepackage{adjustbox}
\usepackage{caption}
\usepackage{subcaption}
\usepackage{arydshln}
\usepackage{nicematrix}

\usepackage[titletoc,title]{appendix}

\usepackage{relsize}
\usepackage{csquotes}
\usepackage{enumitem}
\usepackage{multicol}

\numberwithin{equation}{section}
\numberwithin{figure}{section}
\numberwithin{table}{section}

\hypersetup{
    colorlinks=true,      
    linktoc=all,           
    linktocpage=true,      
    citecolor=jazzberryjam,  
    urlcolor=jazzberryjam,   
    linkcolor=jazzberryjam    
}

\definecolor{blue(ncs)}{rgb}{0.0, 0.53, 0.74}
\definecolor{myblue}{cmyk}{0.65, 0.37, 0.0, 0.19}
\definecolor{darkblue}{rgb}{0.0, 0.0, 0.55}
\definecolor{carnelian}{rgb}{0.7, 0.11, 0.11}
\definecolor{cerise}{rgb}{0.87, 0.19, 0.39}
\definecolor{fulvous}{rgb}{0.86, 0.52, 0.0}
\definecolor{jazzberryjam}{rgb}{0.65, 0.04, 0.37}
\definecolor{gamboge}{rgb}{0.89, 0.61, 0.06}

\title{
\bf \LARGE
Exploring Dimuon Higgs Decay in an Extended Scotogenic Model
\vspace{.2cm}}

\author{
\begin{tabular}[t]{c@{\hskip 10mm}c}
Pablo Escribano~\orcidlink{0000-0003-3701-0969}\footnote{pablo.escribano@ific.uv.es}\hspace{2mm}$^{a}$, &
Víctor Martín Lozano~\orcidlink{0000-0002-9601-0347}\footnote{victor.lozano@ific.uv.es}\hspace{2mm}$^{a,b}$, \\
[2mm]
Sebastián Norero~\orcidlink{0000-0002-2064-9494}\footnote{sanorero@uc.cl}\hspace{2mm}$^{c,d}$, &
Avelino Vicente~\orcidlink{0000-0002-1137-4695}\footnote{avelino.vicente@ific.uv.es}\hspace{2mm}$^{a,b}$. 
\end{tabular}
\\[10mm]
\normalsize\itshape $^a$ Instituto de Física Corpuscular,\\
\normalsize\itshape CSIC-Universitat de València, 46980 Paterna, Spain.\\
\normalsize\itshape $^b$ Departament de F\'{i}sica Te\`orica, Universitat de Val\`encia, 46100 Burjassot, Spain\\
\normalsize\itshape $^c$ Instituto de Física, Pontificia Universidad Católica de Chile,\\
\normalsize\itshape Avenida Vicuña Mackenna 4860, Santiago, Chile.\\
\normalsize\itshape $^d$ Millennium Institute for Subatomic Physics at the High Energy Frontier (SAPHIR),\\
\normalsize\itshape Fernández Concha 700, Santiago, Chile.\\
}

\allowdisplaybreaks

\begin{document}

\maketitle


\begin{abstract}
We investigate the dimuon Higgs decay $h \to \mu^+\mu^-$ in the context of an extended Scotogenic model. The model introduces a singlet complex scalar in addition to the standard Scotogenic scalar doublet and singlet fermions, charged under a dark $\mathbb{Z}_2$ symmetry. By exploring the one-loop contributions, we show that the model allows for sizable deviations in the Higgs dimuon decay rate, quantified by the quotient $R_{\mu\mu} = \text{Br}(h \to \mu^+\mu^-)/\text{Br}(h \to \mu^+\mu^-)_{\text{SM}}$. Crucially, these deviations comply with experimental limits, including those on $\text{Br}(\mu^{+} \to e^{+}\gamma)$ and $\text{Br}(h \to \gamma\gamma)$. Such deviations can be tested and constrained by future precision measurements at the LHC.
\end{abstract}




\section{Introduction}
\label{section:introduction}

The smallness of neutrino masses and the mechanism behind their generation, along with a deeper understanding of the scalar sector of the Standard Model (SM) and the Higgs mechanism, remain some of the most important and interesting open questions in particle physics nowadays. In particular, upcoming precision measurements at colliders offer a promising avenue to shed light on potential extensions of the SM. A crucial test of the Higgs mechanism in the near future will be the precise measurement of the 125 GeV Higgs boson decaying into second- and third-generation leptons, such as the dimuon decay channel $h \to \mu^+\mu^-$. The high luminosity LHC (HL-LHC) is expected to measure the muon Yukawa coupling with about $5\%$ precision \cite{ATLAS:2022hsp, ATLAS:2019mfr}, while future colliders could improve this to sub-percent levels \cite{FCC:2018evy}. Any deviation from the SM prediction in this decay channel would signal the presence of new physics, potentially linked to modified Yukawa couplings or additional fields in the scalar sectors.

On the other hand, the search for lepton flavor violation (LFV) processes, particularly the radiative decay $\mu^{+} \to e^{+} \gamma$, imposes some of the strictest constraints on a wide range of beyond the Standard Model (BSM) scenarios. The MEG II experiment has placed stringent bounds on this process, with the current limit being $\text{Br}(\mu^{+} \to e^{+} \gamma) < 3.1 \times 10^{-13}$~\cite{MEGII:2023ltw}. Moreover, near-future experiments are expected to substantially improve their sensitivity to other flavor-violating processes, such as $\mu^+ \to e^+ e^+ e^-$ or $\mu^- - e^-$ conversion in nuclei~\cite{Renga:2019mpg}. These precision experiments are crucial for testing BSM scenarios where LFV effects play an important role, such as those in which neutrino masses are generated radiatively~\cite{Cai:2017jrq}.

The Scotogenic Model (ScM)~\cite{Tao:1996vb,Ma:2006km} is one of the simplest BSM models capable of explaining the generation of neutrino masses and dark matter at the TeV scale. The model introduces a new scalar doublet and three additional singlet fermions, all odd under a discrete $\mathbb{Z}_2$ symmetry. Neutrino masses are generated at the one-loop level, as the $\mathbb{Z}_2$ symmetry forbids tree-level contributions to Dirac neutrino masses. Simultaneously, this symmetry stabilizes the dark matter candidate, which can be either the lightest singlet fermion or the neutral component of the scalar doublet. However, achieving the observed dark matter relic abundance in the original ScM may pose some challenges. For instance, for a fermion dark matter candidate, saturating the relic density typically requires enhancing lepton flavor violating (LFV) couplings~\cite{Vicente:2014wga}, and coannihilation processes may become necessary~\cite{Hagedorn:2018spx}. To address these tensions and expand the viable parameter space of the ScM, various extensions have been explored. Variants with new Scotogenic states and/or symmetries usually lead to additional viable regions~\cite{DeRomeri:2021yjo,DeRomeri:2022cem}. Alternatively, Ref.~\cite{Beniwal:2020hjc} proposes the addition of a real singlet scalar field to alleviate these constraints. A real scalar singlet was also added to the Scotogenic model in Ref.~\cite{Escribano:2023hxj} to accommodate several collider hints of a new scalar state with a mass of about $95$ GeV.

In this work, we take a next-to-minimal approach and introduce a singlet complex scalar to the ScM. Inspired by the analogous model with a real singlet scalar studied in Ref.~\cite{Beniwal:2020hjc} --referred to as the ``ScotoSinglet Model'' (ScSM)-- we will refer to our model as the ``Complex ScotoSinglet Model'' (CxScSM). This new charged scalar singlet, charged under a discrete $\mathbb{Z}_2$ symmetry, mediates LFV processes like $\mu^{+} \to e^{+} \gamma$. Among many other experimental constraints, we pay particular attention to this branching ratio, which has a relatively strong impact on the prospects of observing a deviation in $h \to \mu^{+} \mu^{-}$. Similarly, the new Scotogenic particles also induce large one-loop contributions to $h \to \gamma \gamma$, which turns out to be another process that restricts the allowed parameter space of the model significantly.

The paper is organized as follows. In Section~\ref{section:the_complex_scotosinglet_model}, we introduce the model in a general setup and then focus on the specific case of a single complex singlet scalar ($n_\phi = 1$). We also discuss some of its salient features, such as neutrino mass generation. Several observables of interest are discussed in Section~\ref{section:observables_of_interest}. In particular, we analytically derive an expression for $\text{Br}(h \to \ell^+\ell^-)$ at the one-loop level. We also briefly discuss the evaluation of the LFV process $\mu^{+} \to e^{+}\gamma$, the anomalous magnetic moment of the muon, and the Higgs diphoton decay $h \to \gamma \gamma$. In Section~\ref{section:constraints}, we outline the most relevant experimental constraints in our model, while the results of our numerical analysis are presented in Section~\ref{section:numerical_analysis}, with a focus on the ratio $\text{Br}(h \to \mu^+ \mu^-)/\text{Br}(h \to \mu^+ \mu^-)_{\text{SM}}$. We also analyze in this Section some limiting cases for the analytical expressions of this observable (as well as of $\mu^+ \to e^+ \gamma$). Finally, we conclude in Section~\ref{section:conclusions} and include additional details of our work in several Appendices.

\section{The Complex ScotoSinglet Model}
\label{section:the_complex_scotosinglet_model}

The Complex ScotoSinglet Model extends the SM by introducing $n_N \geq 2$ heavy singlet fermions, $N_i$, one $\text{SU}(2)_L$ scalar doublet $\eta$, and $n_\phi \geq 1$ complex scalars $\phi_a$.~\footnote{Variants of the Scotogenic model with more doublets are possible~\cite{Escribano:2020iqq}. Still, we stick to the usual choice of a single $\eta$ field.} Additionally, we impose a dark $\mathbb{Z}_2$ parity, under which the new states are odd, while the SM fields remain even. As with the ScM, this symmetry is integral to the model's architecture in two aspects: (1) it prevents tree-level Higgs-neutrino couplings, ensuring that the neutrinos remain massless at the tree level, and (2) it ensures that the lightest odd particle under this symmetry is stable and thus a viable dark matter candidate. We assume that the dark parity remains unbroken after electroweak symmetry breaking, which requires the vacuum expectation value of $\eta$ to be zero, i.e., $\braket{\eta} = 0$. The scalar and fermion particle content of the model, along with their representations under $\text{SU}(3)_c \times \text{SU}(2)_L \times \text{U}(2)_Y$ and the $\mathbb{Z}_2$ parity, are summarized in Table~\ref{tab:matter_content}. The relevant Yukawa and bare mass terms for our discussion are
\begin{equation}
\begin{aligned}
    -\mathcal{L} &\supset \overline{\ell^0_{L}} \, \hat{y} \, \tilde{\eta} \, N_{R} + \overline{N^{c}_{R}} \, \hat{\kappa}^a \, e_{R} \, \phi^\ast_a + \frac{1}{2}\overline{N^{c}_{R}} \, \hat{M} \, N_{R} + \text{h.c.} \\
    &= y_{ij} \, \overline{\ell^0_{iL}} \, \tilde{\eta} \, N_{jR} + \kappa^a_{ji} \, \overline{N^{c}_{jR}} \, e^0_{iR}\phi^\ast_a + \frac{1}{2} \, M_i \, \overline{N^{c}_{iR}} \, N_{iR} + \text{h.c.} \, , \label{eq:yukawa_sector}
\end{aligned}
\end{equation}
where $\tilde{\eta} := i\sigma^2\eta^\ast$, h.c.~stands for ``Hermitian conjugate'' and repeated indices (two or three) are summed over. We assume that singlet fermion fields, $N_i$, are in their mass eigenbases, as this involves no loss of generality in doing this. Indeed, the most general mass term for the singlet fermions would be of the form $\sim M_{ij} \, \overline{N^{0c}_{iR}} \, N^{0}_{jR}$ with $M^{0}_{ij}$ a non-diagonal symmetric matrix with mass dimensions. This term can always be diagonalized using a unitary transformation $U^N_R$, such that $M^0 \equiv \text{diag}(M_1, \ldots, M_{n_N}) = U^{N\dagger}_R \, M^0 \, U^N_R$, with $N_{iR} = (U^{N\dagger}_R)_{ij}N^{0}_{jR}$. The matrix elements $(U^N_R)_{ij}$ would typically appear in the vertex factors but can be absorbed by redefining the couplings $y_{ij}$ and $\kappa^a_{ij}$.

A standard Yukawa coupling between left-handed neutrinos, the singlet fermions, and the SM Higgs doublet is forbidden by the $\mathbb{Z}_2$ symmetry. The complex Yukawa couplings $y_{ij}$ will be reconstructed using a Casas-Ibarra parameterization~\cite{Casas:2001sr}, adequately adapted to the Scotogenic neutrino mass mechanism~\cite{Toma:2013zsa, Cordero-Carrion:2018xre, Cordero-Carrion:2019qtu}, as will be shown in Section~\ref{subsection:neutrino_mass_generation}, and their structure is closely related to neutrino physics. In contrast, the Yukawa couplings $\kappa^a_{ji}$, which are also complex, remain entirely arbitrary.

{
\setlength{\tabcolsep}{12pt}
\renewcommand\arraystretch{1.4}
\begin{table}[t!]
\centering
\begin{NiceTabular}{ccccc}
\CodeBefore
\rowcolors{1}{white}{gray!15}
\Body
\toprule
Field       & Generations & $\text{SU}(2)_L$ & $\text{U}(1)_Y$ & $\mathbb{Z}_2$ \\ \midrule\midrule
$\ell^0_{iL}$ & $3$          & $\bm{2}$         & $-1/2$          & $+1$            \\
$e^0_{iR}$    & $3$          & $\bm{1}$         & $-1$            & $+1$            \\
$H$         & $1$          & $\bm{2}$         & $+1/2$           & $+1$            \\ \midrule
$\phi_a$      & $n_\phi$     & $\bm{1}$         & $-1$             & $-1$            \\
$\eta$      & $1$          & $\bm{2}$         & $+1/2$           & $-1$            \\
$N_{iR}$    & $n_N$        & $\bm{1}$         & $0$             & $-1$            \\ \bottomrule
\end{NiceTabular}
\caption{Particle content and corresponding quantum numbers of the model. The electric charge is defined as $Q = T^3 + Y$. Fermionic fields with a superscript $0$ denote weak eigenstates. Latin indices $i, j, k$ label fermion generations. On the other hand, Latin indices $a, b, c$, and $d$ are reserved for labeling scalar weak eigenstates, while Greek letters $\alpha, \beta$, and $\gamma$ label scalar mass eigenstates. All fields listed are singlets under $\text{SU}(3)_c$.
\label{tab:matter_content}}
\end{table}
}

\subsection{Scalar sector}
\label{subsection:scalar_sector}

The most general $\mathbb{Z}_2$-symmetric, CP-conserving, renormalizable, and gauge-invariant potential involving the scalar fields under consideration:
\begin{align}\label{eq:potential}
    &V(H, \eta, \{\phi_a\}) \nonumber \\
    &= m^2_H(H^\dagger H) + \frac{\lambda_1}{2}(H^\dagger H)^2 + m^2_\eta(\eta^\dagger \eta) + \frac{\lambda_2}{2}(\eta^\dagger \eta)^2 + (m^2_\phi)_{ab}(\phi^\ast_a \phi_b) + (\lambda_\phi)_{ab,cd}(\phi^\ast_a \phi_b)(\phi^\ast_c \phi_d) \nonumber \\
        &\quad + \lambda_3(H^\dagger H)(\eta^\dagger \eta) + \lambda_4(H^\dagger \eta)(\eta^\dagger H) + \frac{1}{2}\Bigl[\lambda_5(H^\dagger\eta)^2 + \text{h.c.}\Bigr] \nonumber \\
        &\quad + (\lambda_{\eta\phi})_{ab}(\eta^\dagger \eta)(\phi^\ast_a \phi_b) + (\lambda_{H\phi})_{ab}(H^\dagger H)(\phi^\ast_a \phi_b) + \Bigl[\mu_a(\eta^\dagger\tilde{H}\phi^\ast_a) + \text{h.c.}\Bigr] \, ,
\end{align}
where $\tilde{H} := i\sigma_2H^\ast$. A few key observations follow. First, the parameters $\lambda_5$ and $\mu_a$ (with $a,b = 1, \ldots, n_\phi$) can be both made real through complex field rotations of $\eta$ and $\phi_a$, respectively; hence, we assume these parameters are real from here on. Second, the matrices $(m^2_\phi)_{ab}$, $(\lambda_{\eta\phi})_{ab}$, and $(\lambda_{H\phi})_{ab}$ are Hermitian.

Thus, each of these matrices contributes with $n^2_\phi$ real free parameters. Similarly, the 4-dimensional array $(\lambda_\phi)_{ab,cd}$ is Hermitian in the sense that $(\lambda_\phi)_{ab,cd} = (\lambda_\phi)^\ast_{ba,dc}$. It also satisfies the symmetry $(\lambda_\phi)_{ab,cd} = (\lambda_\phi)_{cd,ab}$. Overall, this array introduces $n^2_\phi(n^2_\phi + 1)/2$ real free parameters. In particular, for $n_\phi = 1, 2$ and $3$, there are $1, 10$ and $45$ new independent real parameters, respectively, from this term alone. However, these coefficients do not play a role in the phenomenology considered in this work.\\

The neutral components $H^0$ and $\eta^0$ can be split into their CP-even and CP-odd parts as follows:
\begin{equation}
    H = \frac{1}{\sqrt{2}}\begin{pmatrix}\sqrt{2} \, H^{+} \\ h + v + iA\end{pmatrix},\quad \eta = \frac{1}{\sqrt{2}}\begin{pmatrix} \sqrt{2} \, \eta^+ \\ \eta_R + i\eta_I \end{pmatrix} \, ,
\end{equation}
where $v \equiv \sqrt{2}\braket{H^0} \simeq 246\ \text{GeV}$ is the Higgs vacuum expectation value. The tadpole condition is given by
\begin{equation}
    \frac{\partial V}{\partial h}\Bigr|_{\text{fields} \to 0} = 0 \quad\Rightarrow\quad m^2_H = -\frac{1}{2}v^2\lambda_1 \, .
\end{equation}
In this basis, the charged component of the scalar doublet $\eta$ and the charged singlet scalars $\phi_\beta$ mix at tree level, with the relevant terms being
\begin{equation}\label{eq:non_diagonal_elements}
    V \supset
    \begin{pmatrix}
        \phi^+_1 & \ldots & \phi^+_{n_\phi} & \eta^+
    \end{pmatrix}
    \begin{pmatrix}
        (\tilde{m}^2_\phi)_{ab} & \frac{1}{\sqrt{2}}v\mu_a \\
	\frac{1}{\sqrt{2}}v\mu_b & \tilde{m}^2_\eta
    \end{pmatrix}
    \begin{pmatrix}
        \phi^-_1 \\ \ldots \\ \phi^-_{n_\phi} \\ \eta^-
    \end{pmatrix}
        \equiv
    \Xi^\dagger_\text{w} \, \mathcal{M}^2 \, \Xi_\text{w} \, ,
\end{equation}
where we have defined $\Xi^\dagger_\text{w(eak)} := (\phi^+_1, \ldots, \phi^+_{n_\phi}, \eta^+)$ and used the notation $\phi^{-} \equiv \phi$ and $\phi^{+} \equiv \phi^\ast$. Additionally, we have introduced the auxiliary mass variables
\begin{equation}
    (\tilde{m}_\phi)_{ab} := \Bigl(m^2_\phi + \frac{1}{2}v^2\lambda_{H\phi}\Bigr)_{ab},\quad \tilde{m}_\eta := m^2_\eta + \frac{1}{2}v^2\lambda_3 \, .
\end{equation}
This leads to a corresponding non-diagonal square mass matrix of dimension $(n_\phi+1)\times(n_\phi+1)$:
\begin{equation}\label{eq:mass_matrix}
    \mathcal{M}^2
        :=
    \begin{pmatrix}
        (\tilde{m}^2_\phi)_{ab} & \frac{1}{\sqrt{2}}v\mu_a \\
        \frac{1}{\sqrt{2}}v\mu_b & \tilde{m}^2_\eta
    \end{pmatrix} \, .
\end{equation}
Since the mass matrix of Eq.~\eqref{eq:mass_matrix} is Hermitian, the $n_\phi+1$ mass eigenstates, $S^{-}_\alpha$, are obtained by diagonalizing Eq.~\eqref{eq:mass_matrix} via an unitary rotation $R \in \text{SU}(n_\phi+1)$. Details for the special case $n_\phi = 1$, in which case the mass matrix is real and $R \in \text{SO}(2)$, are provided in Appx.~\ref{appendix:mass_eigenstate_basis}. This rotation introduces $(n_\phi+1)^2-1$ additional free real parameters. Consequently, the $n_\phi+1$ mass eigenstates, $S^{-}_{1}, \ldots, S^{-}_{n_\phi+1}$, are related to the weak states, $\phi^{-}_1, \ldots, \phi^{-}_{n_\phi}, \eta^{-}$, via the relation $\Xi_\text{w} = R \, \Xi_\text{mass}$ or, more explicitly,
\begin{equation}
    \begin{aligned}
        \phi^{-}_{a} &= R_{a\beta}S^{-}_{\beta} \\
        \eta^{-} &= R_{n_\phi+1,\beta}S^{-}_{\beta}
    \end{aligned}
    \quad\Leftrightarrow\quad
    \begin{pmatrix}
        \phi_1 \\
        \vdots \\
        \phi_{n_\phi} \\
        \eta^-
    \end{pmatrix}
    =
    \begin{pmatrix}
        R_{11} & \cdots & R_{1,n_\phi} & R_{1,n_\phi+1} \\
	\vdots & \ddots & \cdots & \vdots \\
        R_{n_\phi,1} & \cdots & R_{n_\phi,n_\phi} & R_{n_\phi,n_\phi+1} \\
        R_{n_\phi+1,1} & \cdots & R_{n_\phi+1,n_\phi} & R_{n_\phi+1,n_\phi+1} \\
    \end{pmatrix}
    \begin{pmatrix}
        S^-_1 \\
        \vdots \\
        S^-_{n_\phi} \\
        S^-_{n_\phi+1}
    \end{pmatrix} \, ,
\end{equation}
where $\alpha, \beta = 1, \ldots, n_\phi+1$ and we make the identification $\phi^{-}_{n_\phi+1} \equiv \eta^{-}$.\\

Henceforth, we proceed with the case $n_\phi=1$. In this case, there is a single mixing angular parameter, $\theta$, associated with the diagonalization process, which is determined by:
\begin{equation}\label{eq:tan2theta}
    \tan(2\theta) = \frac{\sqrt{2}v\mu}{\tilde{m}^2_\eta - \tilde{m}^2_\phi},\quad \sin(2\theta) = \frac{\sqrt{2}v\mu}{m^2_{S^\pm_2} - m^2_{S^\pm_1}},\quad \cos(2\theta) = \frac{\tilde{m}^2_\eta - \tilde{m}^2_\phi}{m^2_{S^\pm_2} - m^2_{S^\pm_1}} \, .
\end{equation}
We restrict the mixing angle to the range $\theta \in [-\pi/4, \pi/4]$ using the first identity to account for all possible cases. In this scenario, the charged mass eigenstates are
\begin{equation}
    S^{\pm}_1 = \phi^{\pm}\cos\theta - \eta^{\pm}\sin\theta,\quad S^{\pm}_2 = \phi^{\pm}\sin\theta + \eta^{\pm}\cos\theta \, .
\end{equation}
The masses of these mass eigenstates are given by
\begin{subequations}
\begin{align}
    m^2_{S^\pm_1} &:= \frac{1}{2}\Biggl[\tilde{m}^2_\eta + \tilde{m}^2_\phi - \sqrt{(\tilde{m}^2_\eta - \tilde{m}^2_\phi)^2 + 2v^2\mu^2}\Biggr] = \frac{1}{2}\Biggl[\tilde{m}^2_\eta + \tilde{m}^2_\phi - |\tilde{m}^2_\eta - \tilde{m}^2_\phi|\sec(2\theta)\Biggr]  \, , \\
    m^2_{S^\pm_2} &:= \frac{1}{2}\Biggl[\tilde{m}^2_\eta + \tilde{m}^2_\phi + \sqrt{(\tilde{m}^2_\eta - \tilde{m}^2_\phi)^2 + 2v^2\mu^2}\Biggr] = \frac{1}{2}\Biggl[\tilde{m}^2_\eta + \tilde{m}^2_\phi + |\tilde{m}^2_\eta - \tilde{m}^2_\phi|\sec(2\theta)\Biggr] \, .
\end{align}
\end{subequations}
Without loss of generality, we set $m^2_{S^\pm_1} \leq m^2_{S^\pm_2}$. After electroweak symmetry breaking (EWSB) takes place, we are left with two CP-even scalars, $\eta_R$ and $h$, one CP-odd scalar, $\eta_I$, and two pairs of charged scalars, $S^{\pm}_1$ and $S^{\pm}_2$. The masses of the neutral scalars are
\begin{subequations}
\begin{align}
    m^2_{R} &= m^2_\eta + \frac{1}{2}v^2(\lambda_3 + \lambda_4 + \lambda_5) \, , \\
    m^2_{I} &= m^2_\eta + \frac{1}{2}v^2(\lambda_3 + \lambda_4 - \lambda_5) \, , \\
    m^2_{h} &= v^2\lambda_1 \, .
\end{align}
\end{subequations}
As usual, the mass splitting between $\eta_R$ and $\eta_I$ is governed by $\lambda_5$ as $m^2_R - m^2_I = v^2\lambda_5$. Perturbative unitarity bounds on dimensionless couplings. Particularly, $|\lambda_5| < 4\pi$, makes the square mass difference to be constrained to $|m^2_R - m^2_I| \lesssim (872\ \text{GeV})^2$.

Experimental bounds arising from dark matter phenomenology will also have an impact on the parameter $\lambda_5$, regardless of the specific nature of the DM candidate considered in our model. If the DM candidate is chosen from the inert neutral scalars, bounds from direct detection experiments \cite{LZ:2022lsv} must be considered. In this case, the DM phenomenology of our model closely resembles that of the ScM (or the Inert Doublet Model~\cite{Deshpande:1977rw}) \cite{Barbieri:2006dq,LopezHonorez:2006gr,Diaz:2015pyv,LopezHonorez:2010eeh}. Additionally, note that the choice between the two scalar candidates determines the sign of $\lambda_5$. Alternatively, if the DM candidate is the lightest singlet fermion, the phenomenology becomes similar to that of the ScM, but with strong constraints on the Yukawa couplings --and, consequently, on $\lambda_5$-- from LFV processes~\cite{Vicente:2014wga}. Nevertheless, achieving the observed DM relic density remains feasible in this scenario \cite{Kubo:2006yx,AristizabalSierra:2008cnr,Suematsu:2009ww,Adulpravitchai:2009gi,Vicente:2014wga}.

Assuming $n_N = 3$ (i.e., three singlet fermions $N_i$ with respective masses $M_i$), the model contains 14 independent parameters, chosen as follows:
\begin{equation}
\begin{aligned}
    \text{8 dimensionless couplings} :&\quad \{\lambda_2, \lambda_3, \lambda_4, \lambda_5, \lambda_\phi, \lambda_{\eta\phi}, \lambda_{H\phi}, \kappa_{ij}\} \\
    \text{4 mass parameters} :&\quad \{\mu, M_1, M_2, M_3\} \\
    \text{2 mass squared parameters} :&\quad \{m^2_\phi, m^2_\eta\}
\end{aligned}
\end{equation}

\subsection{Neutrino Mass Generation}
\label{subsection:neutrino_mass_generation}

\begin{figure}[tb!]
\centering
    \includegraphics[width=0.5\linewidth]{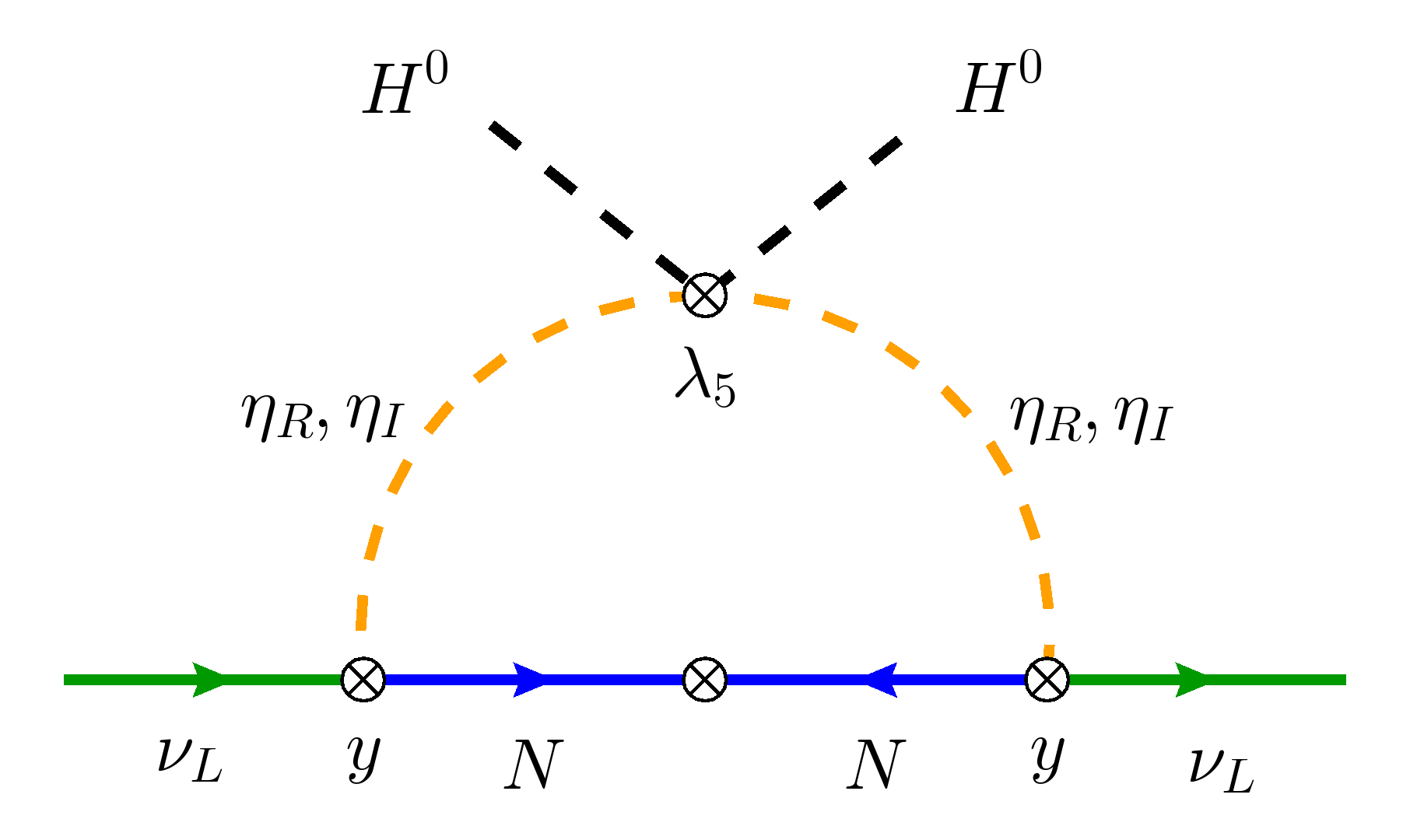}
\caption{One-loop Feynman diagrams contributing to neutrino masses in the model under consideration.}
\label{fig:neutrino_mass_loop}
\end{figure}

One of the most interesting features of the ScM is its ability to generate neutrino masses. This process occurs at the loop level, with the Scotogenic particles $\eta$ and $N$ circulating in the loop, and it is weighted by the couplings $\lambda_5$ and $y$ as shown in Fig.~\ref{fig:neutrino_mass_loop}. The Majorana neutrino mass matrix resulting from the loop amplitude is expressed as
\begin{align}
\label{eq:non_diagonal_neutrino_mas_matrix_to_one_loop}
    (m^0_\nu)_{ij} &= \sum^3_{k=1} \frac{y_{ki}y_{kj}}{(4\pi)^2}M_{k}\left[\mathcal{F}\left(\frac{m^2_R}{M^2_k}\right) - \mathcal{F}\left(\frac{m^2_I}{M^2_k}\right)\right] \equiv (y^T \Lambda y)_{ij} \, ,
\end{align}
where $\mathcal{F}(x) = x\ln(x)/(x-1)$, and the diagonal matrix $\Lambda$ is defined as
\begin{equation}
\label{eq:capital_lambda_matrix}
    \Lambda = 
    \begin{pmatrix}
        \Lambda_1 & 0 & 0 \\
        0 & \Lambda_2 & 0 \\
        0 & 0 & \Lambda_3
    \end{pmatrix},\quad
    \Lambda_j := \frac{M_{j}}{(4\pi)^2}\left[\mathcal{F}\left(\frac{m^2_R}{M^2_j}\right) - \mathcal{F}\left(\frac{m^2_I}{M^2_j}\right)\right] \, .
\end{equation}
We note that in the limit $\lambda_5 \to 0$, $m_R = m_I$ and neutrino masses vanish. This is because this limit allows for a definition of a conserved lepton number.
Assuming a diagonal charged lepton sector (i.e., no mixing among charged leptons), we can the neutrino diagonalize the mass matrix as
\begin{equation}
    U^T \, m^0_\nu \, U
        =
    \begin{pmatrix}
        m_1 & 0 & 0 \\
        0 & m_2 & 0 \\
        0 & 0 & m_3
    \end{pmatrix}
        \equiv
    m_\nu\, ,
\end{equation}
where $m_1, m_2$, and $m_3$ are the masses of the light left-handed neutrinos, and $U$ is the leptonic mixing matrix, also known as Pontecorvo-Maki-Nakagawa-Sakata (PMNS) matrix. The PMNS matrix can be parameterized as
\begin{equation}\label{eq:PMNS_matrix}
    U = \begin{pmatrix}
		c_{12}c_{13} & s_{12}c_{13} & s_{13}e^{-i\delta_{\text{CP}}} \\
		-s_{12}c_{23} - c_{12}s_{13}s_{23}e^{i\delta_{\text{CP}}} & c_{12}c_{23} - s_{12}s_{13}s_{23}e^{i\delta_{\text{CP}}} & c_{13}s_{23} \\
		s_{12}s_{23} - c_{12}s_{13}c_{23}e^{i\delta_{\text{CP}}} & -c_{12}s_{23} - s_{12}s_{13}c_{23}e^{i\delta_{\text{CP}}} & c_{13}c_{23}
	\end{pmatrix}
	\begin{pmatrix}
		1 & 0 & 0 \\
		0 & e^{i\frac{\rho}{2}} & 0 \\
		0 & 0 & e^{i\frac{\sigma}{2}}
	\end{pmatrix} \, .
\end{equation}
Here, $c_{ij} = \cos\theta_{ij}$, $s_{ij} = \sin\theta_{ij}$, and the angles $\theta_{ij}$ are taken, without loss of generality, to lie in the first quadrant, $\theta_{ij} \in [0, \pi/2]$. The parameter $\delta_{\text{CP}}$ corresponds to the Dirac CP-violating phase, while $\rho$ and $\sigma$ denote two CP-violating Majorana phases. We can take all three phases to lie within the range $[0, 2\pi]$.

One can always express the Yukawa matrix $\hat{y}$ in terms of the left-handed neutrino masses $m_j$ and the parameters measured in neutrino oscillation experiments using a modified version of the Casas-Ibarra parametrization~\cite{Casas:2001sr, Toma:2013zsa, Cordero-Carrion:2018xre, Cordero-Carrion:2019qtu} as 
\begin{equation}
\label{eq:yukawa_matrix}
   y = \Lambda^{-1/2} \, R \, m^{1/2}_\nu \, U^\dagger \, ,
\end{equation}
where $\Lambda^{-1/2} = \text{diag}(\Lambda^{-1/2}_1, \Lambda^{-1/2}_2, \Lambda^{-1/2}_3)$, $m^{1/2}_\nu = \text{diag}(m^{1/2}_1, m^{1/2}_2, m^{1/2}_3)$, and $R$ is a $3\times3$ complex orthogonal matrix, which introduces three additional free complex parameters. Finally, in our numerical analysis below, the mass of the lightest neutrino, $m_1$, will be randomly chosen, assuming a normal hierarchy.~\footnote{The mass of the lightest neutrino will be taken randomly within the range $[0, 0.028]$ eV, where we discuss the upper limit in Section~\ref{subsection:cosmological_observations}.} The masses of the remaining two left-handed neutrinos are then determined as
\begin{equation}
     m_2 = \sqrt{m^2_1 + \Delta m^2_\text{sol}},\quad m_3 = \sqrt{m^2_1 + \Delta m^2_\text{atm}} \, ,
\end{equation}
where the neutrino mass-squared differences, $\Delta m^2_\text{sol} := m^2_2 - m^2_1$ and $\Delta m^2_\text{atm} := m^2_3 - m^2_1$, are parameters determined experimentally. Similarly, the neutrino mixing parameters, $\sin^2\theta_{ij}$, are also input parameters randomly chosen within their experimental ranges. These mixing parameters are used to construct the PMNS matrix of Eq.~\eqref{eq:PMNS_matrix}, which is then employed to reconstruct the corresponding neutrino Yukawa matrix using the Casas-Ibarra parametrization, as detailed in this Section. We sample all values from the $3\sigma$ experimental ranges listed in Table~\ref{tab:neutrino_oscillation_parameters} obtained from Ref.~\cite{deSalas2021}.
{
\setlength{\tabcolsep}{12pt}
\renewcommand\arraystretch{1.4}
\begin{table}[]
\centering
\begin{NiceTabular}{cccc}
\CodeBefore
\rowcolors{1}{white}{gray!14}
\Body
\toprule
Parameter       & Best fit $\pm1\sigma$ & $3\sigma$ range & Units \\ \midrule\midrule
$\Delta m^2_{21} / 10^{-5}$    & $7.50^{+0.22}_{-0.20}$ & $6.94$ - $8.14$ & $\text{eV}^2$ \\
$\Delta m^2_{31} / 10^{-3}$    & $2.55^{+0.02}_{-0.03}$ & $2.47$ - $2.63$ & $\text{eV}^2$ \\ \midrule
$\sin^2\theta_{12} / 10^{-1}$    & $3.18\pm 0.16$ & $2.71$ - $3.69$ & - \\
$\sin^2\theta_{23} / 10^{-1}$    & $5.74\pm 0.14$ & $4.34$ - $6.10$ & - \\
$\sin^2\theta_{13} / 10^{-2}$    & $2.20^{+0.069}_{-0.062}$ & $2.00$ - $2.405$ & - \\ \midrule
$\delta_{CP} / \pi$    & $1.08^{+0.13}_{-0.12}$ & $0.71$ - $1.99$ & rad \\ \bottomrule
\end{NiceTabular}
\caption{Experimental ranges for neutrino oscillation parameters assuming normal ordering from Ref.~\cite{deSalas2021}.}
\label{tab:neutrino_oscillation_parameters}
\end{table}
}

\section{Observables of Interest}
\label{section:observables_of_interest}

In this Section, we review and provide analytical expressions for some of the observables studied in our analysis.

\subsection{Dimuon Higgs decay}
\label{subsection:dimuon_higgs_decay}

The experimental study of the decay $h \to \mu^{+}\mu^{-}$ holds particular significance in the coming years as the LHC is starting to measure the Yukawa coupling of the second generation leptons~\cite{CMS:2020xwi, ATLAS:2022vkf, CMS:2022dwd}. With the advent of the High-Luminosity Large Hadron Collider (HL-LHC) program at CERN, expected to be operational by mid-2030, a significant improvement in the precision of Higgs boson coupling measurements is anticipated \cite{Apollinari:2015bam,Arduini:2016xsb,BejarAlonso:2015ldp,Bruning:2023tzn}. The program aims to increase the integrated luminosity by a factor of 10 compared to the LHC's original design value ($100\ \text{fb}^{-1}$ per year during its initial operation phase). For instance, the HL-LHC is projected to measure the muon Yukawa coupling with a precision of approximately $5\%$ \cite{ATLAS:2022hsp, ATLAS:2019mfr}, while future colliders could push this precision to sub-percent levels \cite{FCC:2018evy}. In this context, it turns out to be crucial to explore the impact of BSM scenarios in these processes, such as the one investigated in this work. Such studies will help to determine whether these models can account for potential new physics, providing a framework to explain any deviations that might emerge from upcoming experimental data.

\begin{figure}[t!]
\centering
\begin{subfigure}{.30\textwidth}
    \centering
    \includegraphics[width=1.0\linewidth]{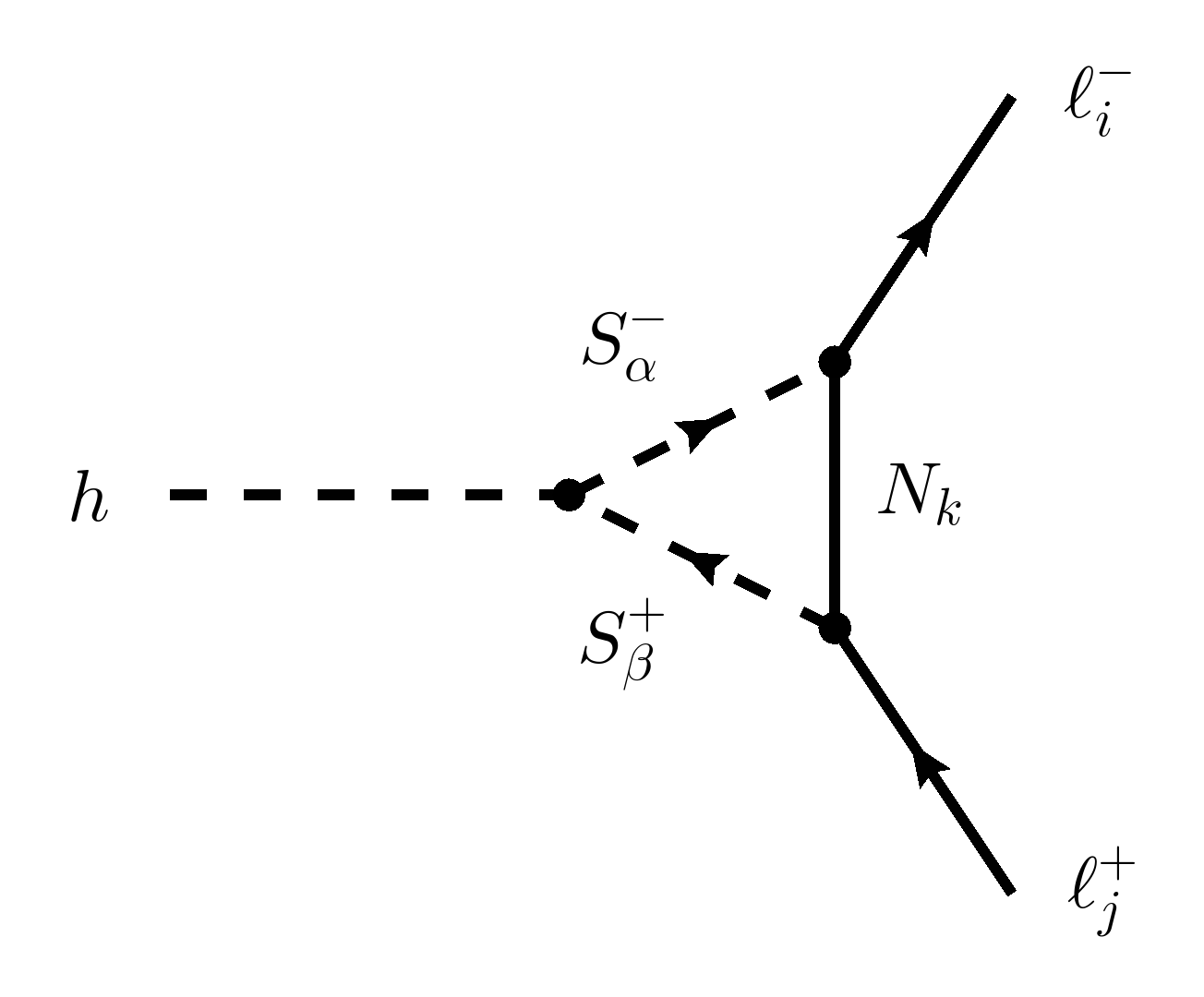}
    \caption{}
    \label{fig:hTOll_loop_triangle}
\end{subfigure}%
    \hfill
\begin{subfigure}{.30\textwidth}
    \centering
    \includegraphics[width=1.0\linewidth]{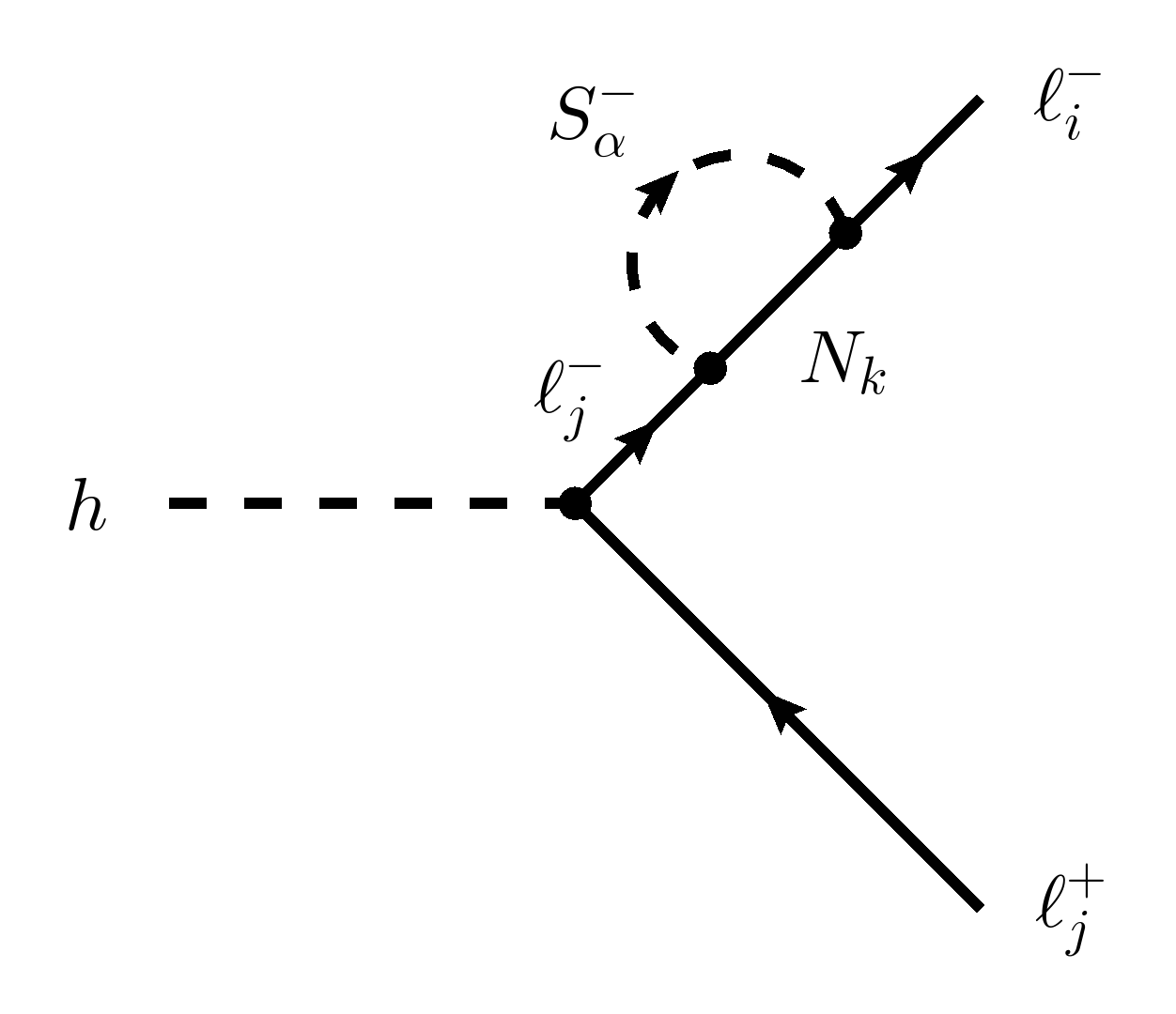}
    \caption{}
    \label{fig:hTOll_loop_lminus}
\end{subfigure}%
    \hfill
\begin{subfigure}{.30\textwidth}
    \centering
    \includegraphics[width=1.\linewidth]{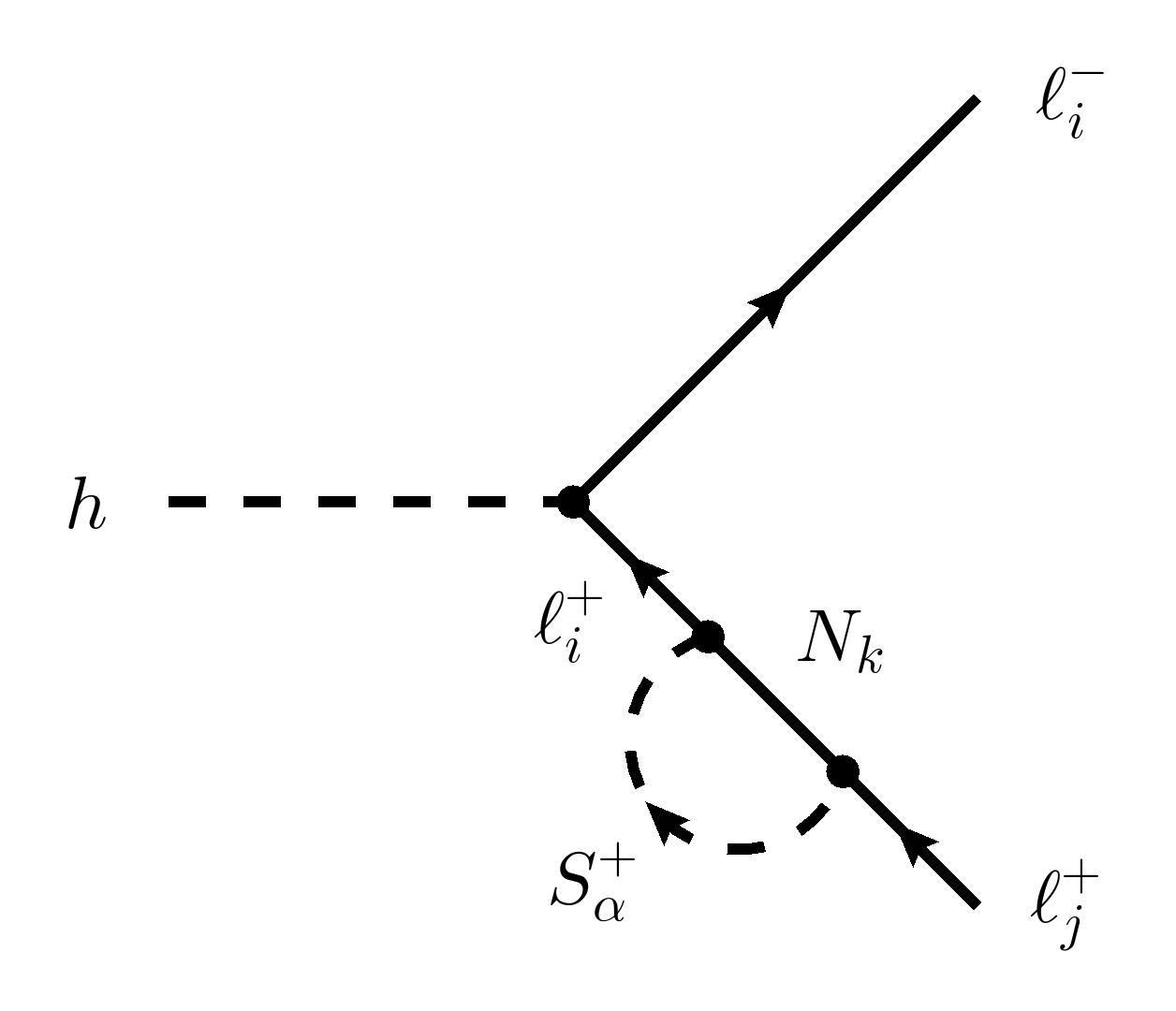}
    \caption{}
    \label{fig:hTOll_loop_lplus}
\end{subfigure}
\caption{One-loop Feynman diagrams contributing to $h \to \ell^{+}_i\ell^{-}_j$. The leg-correction diagrams cancel out.}
\label{fig:hTOll_diagrams}
\end{figure}

In our model, new Higgs couplings to pairs of charged leptons are generated at one-loop level, as shown in the Feynman diagram in Figure~\ref{fig:hTOll_diagrams}. However, the leg-corrections~\eqref{fig:hTOll_loop_lminus} and \eqref{fig:hTOll_loop_lplus} cancel each other out, so we only consider the vertex-correction in the diagram~\eqref{fig:hTOll_loop_triangle}. The vertex correction must be combined with the tree-level contribution from the SM. The resulting amplitude for the process $h \to \ell^{-}_i\ell^{+}_j$ can then be interpreted algebraically as a ``tree-level'' process with an effective vertex:
\begin{equation}
\label{eq:hTOmumu_equation}
    \underbrace{\includegraphics[height=4.0cm, valign=c]{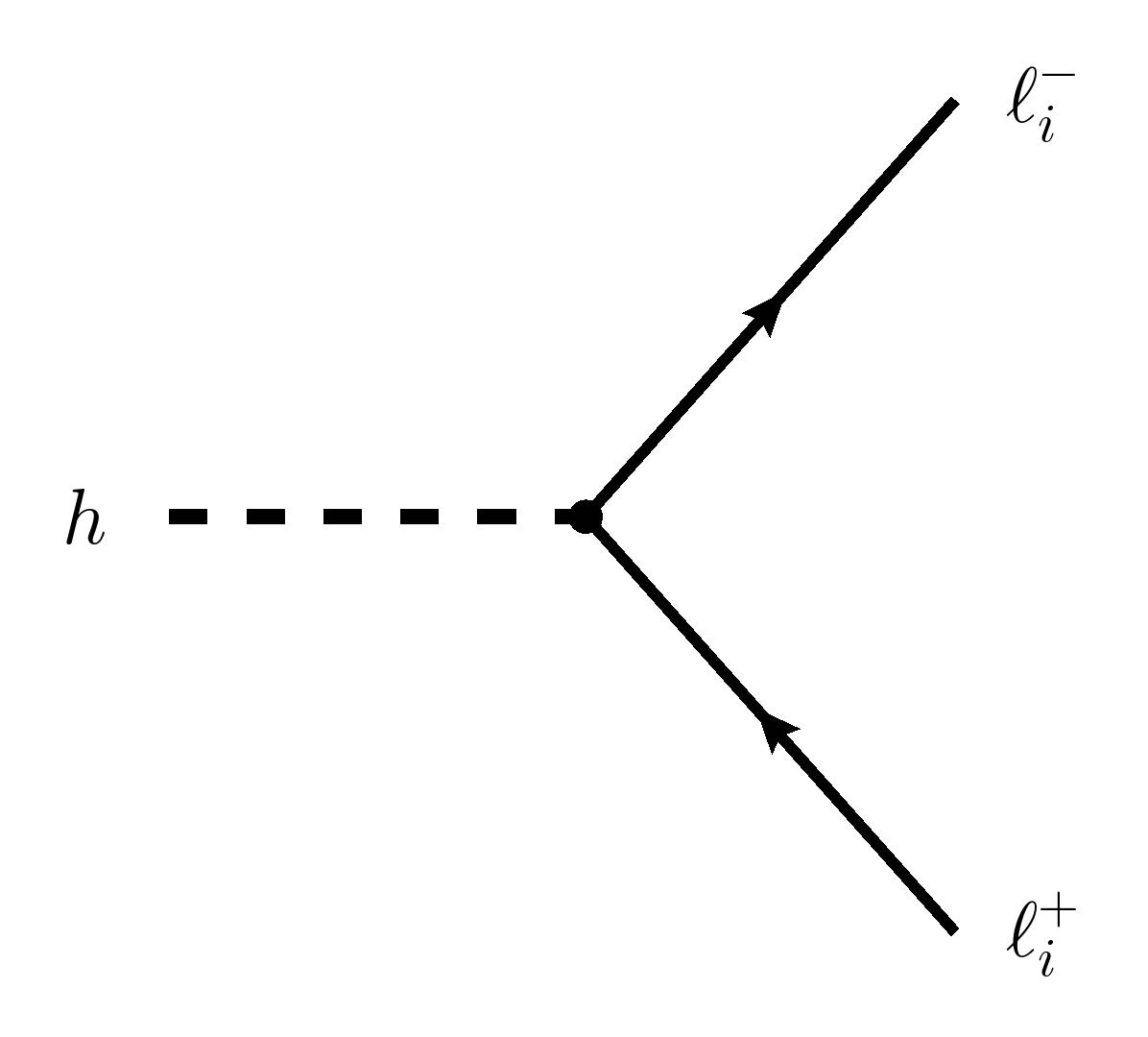}}_{-i(\vartheta^\text{SM}_{ii}\delta_{ij}P_L + \vartheta^\text{SM}_{ii}\delta_{ij}P_R)}
     \; +\; \underbrace{\includegraphics[height=4.0cm, valign=c]{figures/hTOll/hTOll_1loop_black.png}}_{-i(\vartheta^L_{ij}P_L + \vartheta^R_{ij}P_R)}\; =\; \underbrace{\includegraphics[height=4.0cm, valign=c]{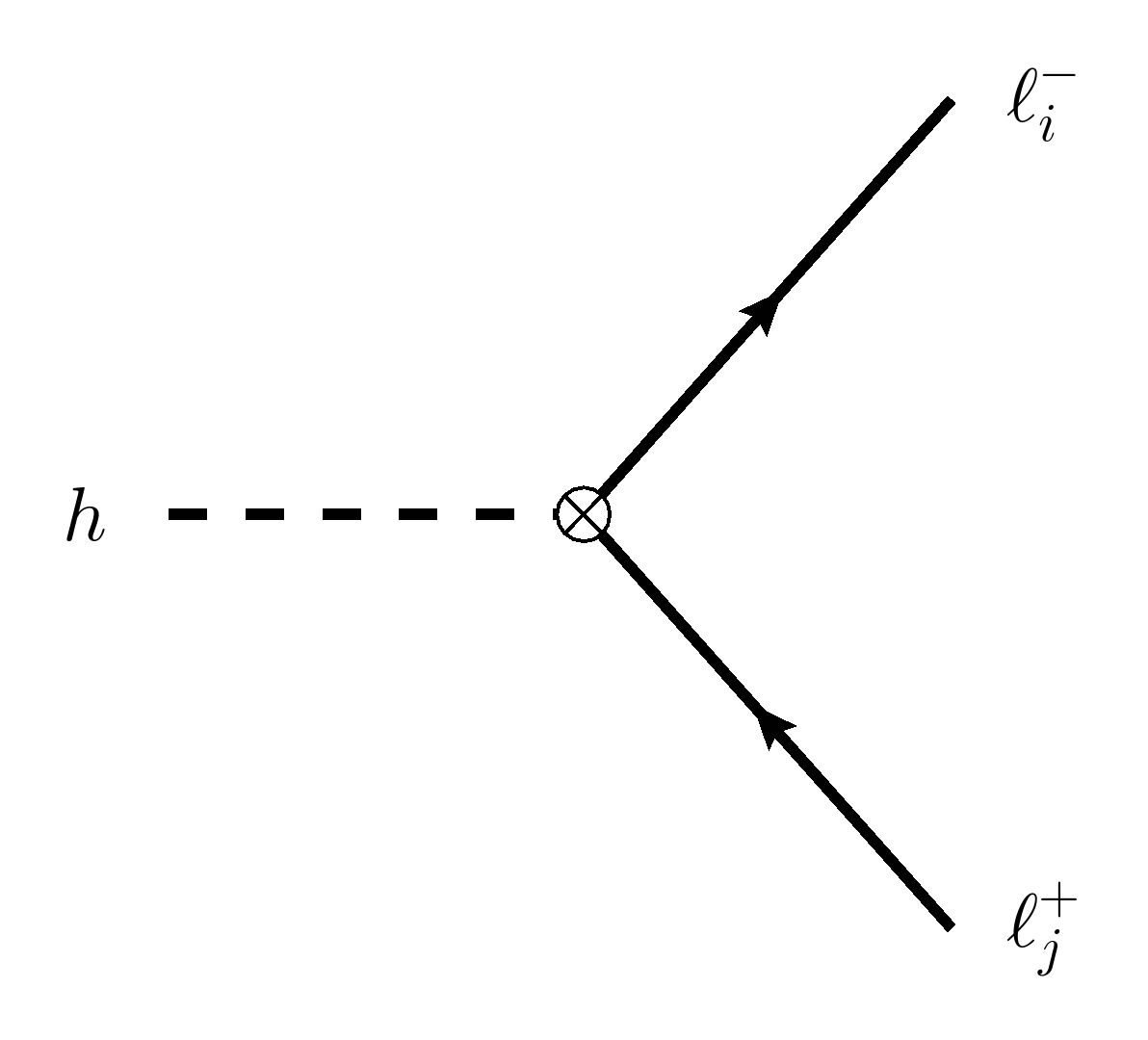}}_{-i(F^L_{ij}P_L + F^R_{ij}P_R)}
\end{equation}
Consequently, the Higgs decay branching ratio takes the form
\begin{equation}\label{eq:BR_hTOlili}
    \text{Br}(h \to \ell^{+}_i\ell^{-}_j) = \frac{\sqrt{\lambda(m^2_h, m^2_i, m^2_j)}}{16\pi m^3_h \Gamma_h}\Biggl\{(|F^L_{ij}|^2 + |F^R_{ij}|^2)(m^2_h - m^2_i - m^2_j) - 4m_im_j\mathfrak{Re}[F^L_{ij}F^{R\ast}_{ij}]\Biggr\} \, ,
\end{equation}
where $\Gamma_h$ is the total decay width of the $125$-GeV Higgs boson, $\lambda$ is the Källén function, and $F^{L,R}_{ij} := \vartheta^\text{SM}_{ii}\delta_{ij} + \vartheta^{L,R}_{ij}$. The SM coefficients $\vartheta^\text{SM}_{ii}$ can be computed numerically, accounting for QED and EW corrections~\cite{Dabelstein:1991ky}. Analytical expressions for the loop contributions $\vartheta^{L,R}_{ij}$ are provided in Appendix~\ref{appendix:coefficients_hTOlilj}. Note that $\vartheta^{R}_{ij}$ is not independent, but related to $\vartheta^{L}_{ij}$ via Eq.~\eqref{eq:relation_between_varthetaLR_ijkalphabeta_coefficients}. Therefore, the loop contribution to the Higgs dimuon rate is governed by the single complex parameter $\vartheta^{L}_{ij}$. In the specific case where $i = j$, we define the observable
\begin{equation}
\label{eq:hTOmumu_branching_quotient}
    R_{ii} := \frac{\text{Br}(h \to \ell^{+}_i\ell^{-}_i)}{\text{Br}(h \to \ell^{+}_i\ell^{-}_i)_{\text{SM}}} = 1 + 2\left(\frac{\mathfrak{Re}[\vartheta^L_{ii}]}{\vartheta^{\text{SM}}_{ii}}\right) + \left(\frac{\mathfrak{Re}[\vartheta^L_{ii}]}{\vartheta^{\text{SM}}_{ii}}\right)^2 + \frac{1}{\beta^2_i}\left(\frac{\mathfrak{Im}[\vartheta^L_{ii}]}{\vartheta^{\text{SM}}_{ii}}\right)^2 \, ,
\end{equation}
where $\beta_i := \sqrt{1 - 4m^2_i/m^2_h}$ and $i = e, \mu, \tau$. When treated as a function of the complex variable $\vartheta^L_{ii}$, the quotient of Eq.~\eqref{eq:hTOmumu_branching_quotient} takes the form
\begin{equation}
\label{eq:f_ellipse}
    R_{ii}(\mathfrak{Re}[\vartheta^L_{ii}], \mathfrak{Im}[\vartheta^L_{ii}]) = \frac{\Bigl(\mathfrak{Re}[\vartheta^L_{ii}] + \vartheta^{\text{SM}}_{ii}\Bigr)^2}{(\vartheta^{\text{SM}}_{ii})^2} + \frac{\mathfrak{Im}[\vartheta^L_{ii}]^2}{(\beta_i\vartheta^{\text{SM}}_{ii})^2} \, .
\end{equation}
Thus, geometrically, the value of $R_{ii}$ will be less than, equal to, or greater than 1, depending on whether the point $(\mathfrak{Re}[\vartheta^L_{ii}], \mathfrak{Im}[\vartheta^L_{ii}])$ lies inside, on the boundary, or outside the ellipse defined by this equation, respectively. Note that any one of these conditions can always be fulfilled, as this ellipse never degenerates to a point. Moreover, the minimum of Eq.~\eqref{eq:hTOmumu_branching_quotient} is always reached on the inside of the ellipse, at $(\mathfrak{Re}[\vartheta^L_{ii}], \mathfrak{Im}[\vartheta^L_{ii}]) = (-\vartheta^{\text{SM}}_{ii}, 0)$, and it corresponds to $R_{ii} = 0$. This result is independent of the number of complex scalar singlets or singlet fermions in the model, as it depends solely on the structure of the branching ratio of Eq.~\eqref{eq:BR_hTOlili} and the relation $(\vartheta^R_{ii})^\ast = \vartheta^L_{ii}$. Additionally, note that $R_{ii} = 1$ (i.e., no deviations from the SM prediction) does not necessarily imply $\vartheta^{L}_{ii} = 0$; new physics can be hidden such that Eq.~\eqref{eq:f_ellipse} is close to 1 without vanishing contributions from $\vartheta^{L}_{ii}$.

\subsection{Lepton flavor violating processes}
\label{subsection:lepton_flavour_violation_processes}

\begin{figure}[t!]
\centering
\begin{subfigure}{.30\textwidth}
    \centering
    \includegraphics[width=1.0\linewidth]{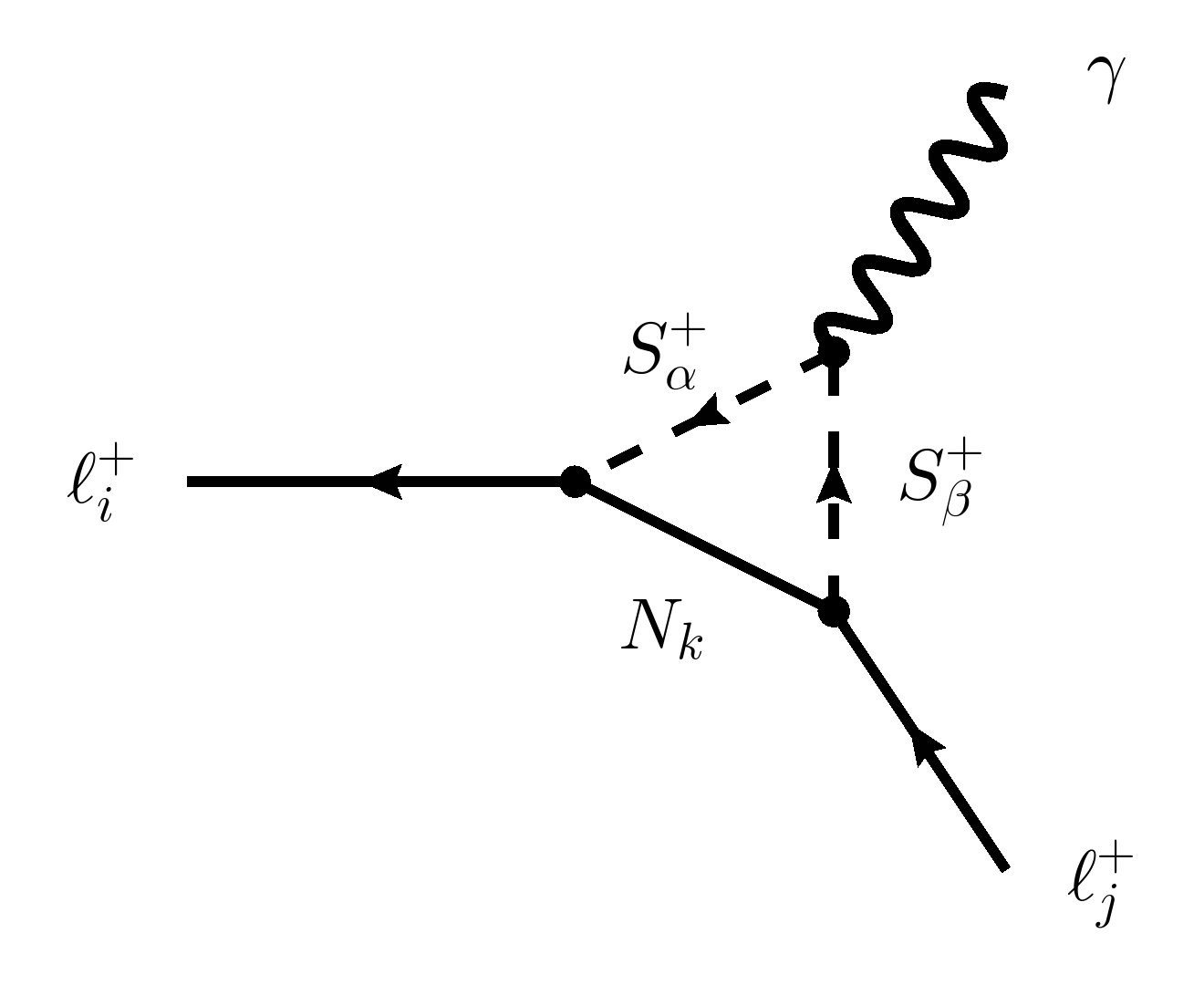}
    \caption{}
    \label{fig:liTOljgamma_loop_triangle}
\end{subfigure}%
    \hfill
\begin{subfigure}{.30\textwidth}
    \centering
    \includegraphics[width=1.0\linewidth]{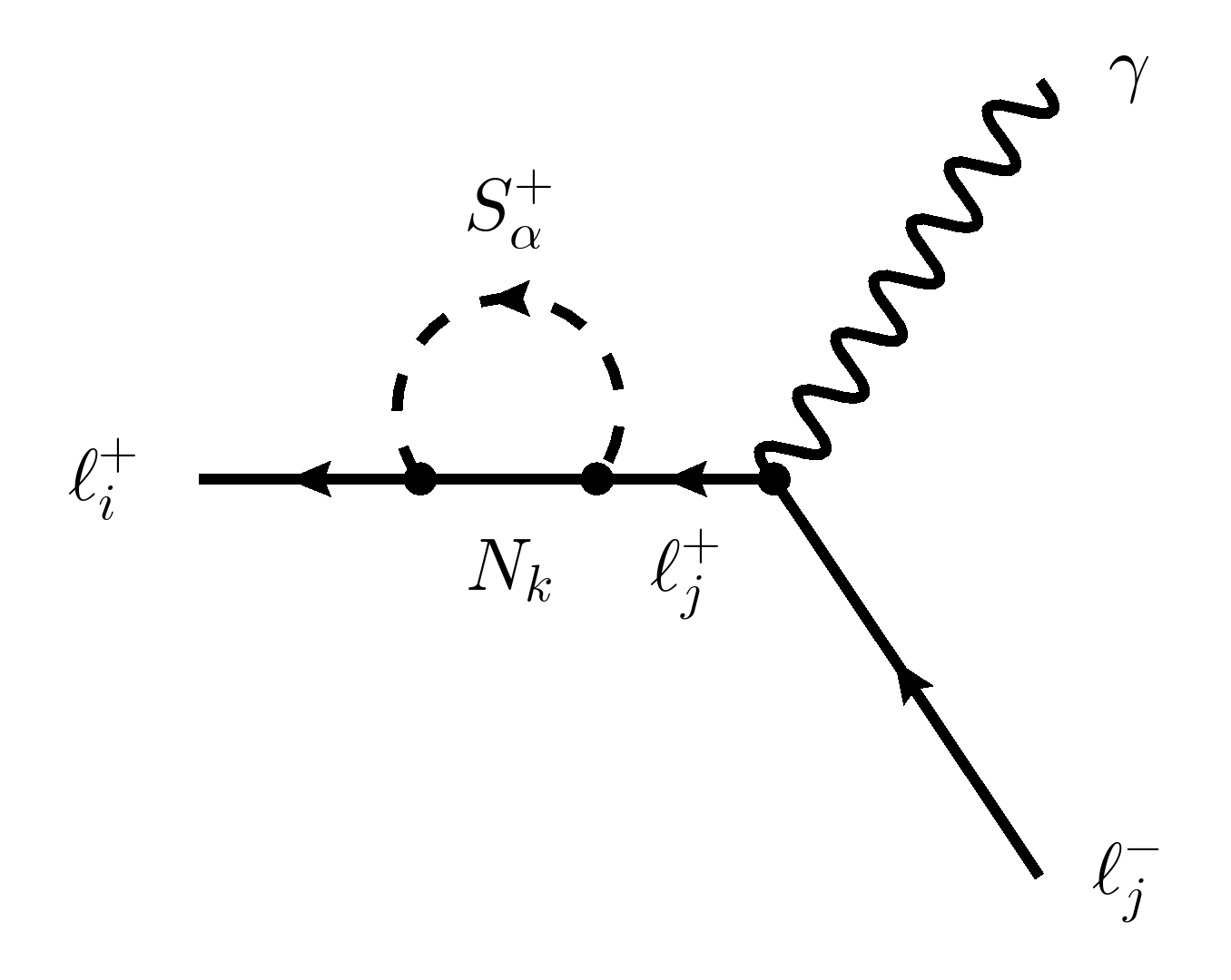}
    \caption{}
    \label{fig:liTOljgamma_loop_initial_leg}
\end{subfigure}%
    \hfill
\begin{subfigure}{.30\textwidth}
    \centering
    \includegraphics[width=1.0\linewidth]{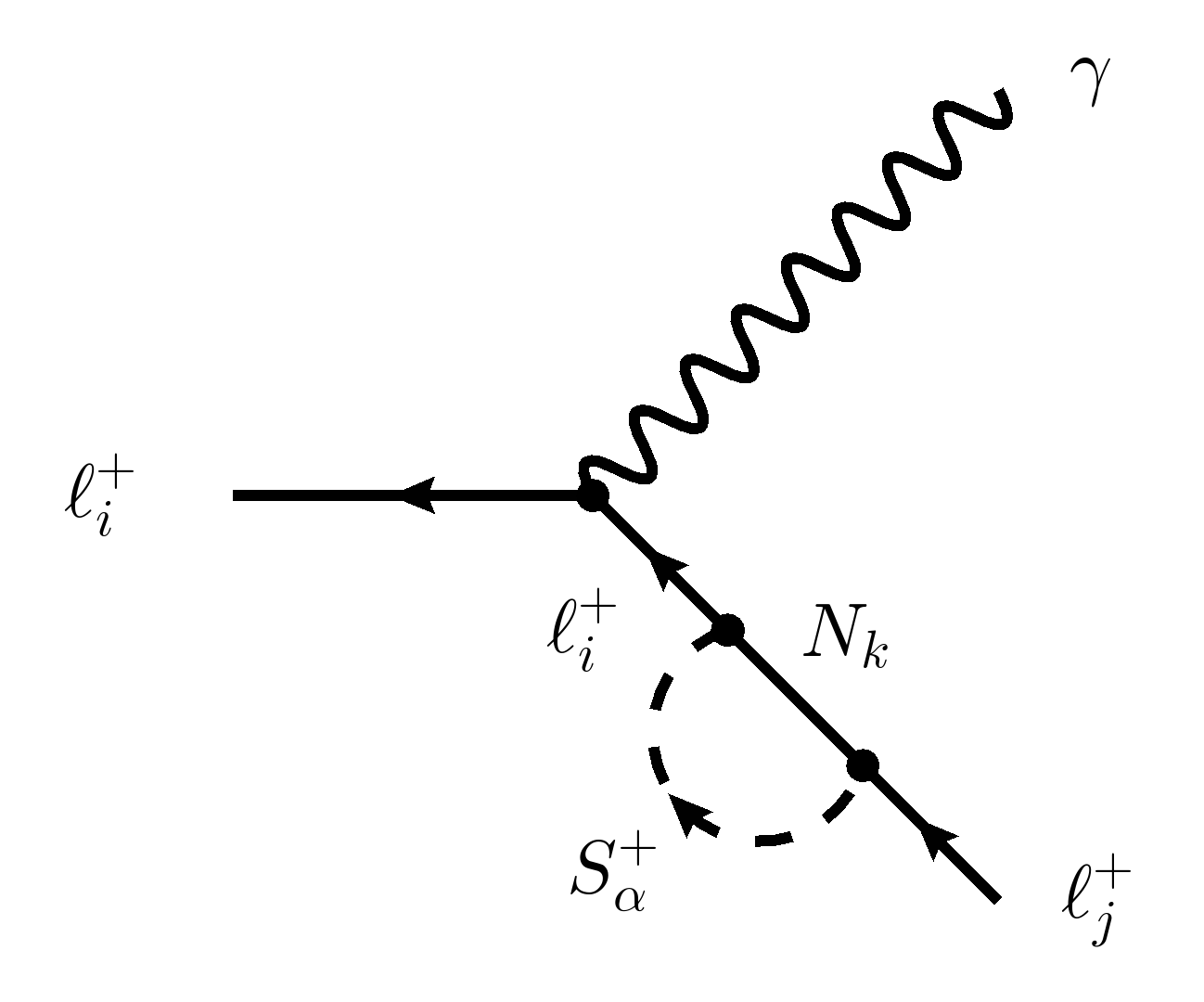}
    \caption{}
    \label{fig:liTOljgamma_loop_final_leg}
\end{subfigure}
\caption{One-loop Feynman diagrams leading to $\ell^{+}_i \to \ell^{+}_j\gamma$}
\label{fig:lTOlgamma_diagrams}
\end{figure}

In the present work, we only focus on the LFV process $\ell^{+}_i \to \ell^{+}_j\gamma$ since it typically leads to the most stringent bounds compared to other LFV processes such as $\ell^{+}_i \to 3\ell^{+}_j$ and $\mu^--e^-$ conversion in nuclei.~\footnote{A detailed study of the LFV processes $\ell^{+}_i \to \ell^{+}_j\gamma$, $\ell^{+}_i \to 3\ell^{+}_j$ and $\mu^--e^-$ conversion in nuclei, in the ScM can be found in Ref.~\cite{Toma:2013zsa}.}

We compute the one-loop decay $\mu^{+} \to e^{+}\gamma$ branching ratio from the expression given in Ref.~\cite{Crivellin:2018qmi},
\begin{equation}\label{eq:br_muTOegama}
    \text{Br}(\mu^{+} \to e^{+}\gamma) = \frac{m^3_\mu}{4\pi\Gamma_\mu}\Bigl(|c^{e\mu}_R|^2 + |c^{\mu e}_R|^2\Bigr) \, ,
\end{equation}
where $\Gamma_\mu \simeq 2.99 \times 10^{-19}\ \text{GeV}$ is the total decay width of the muon. We proceed to analyze some specific scenarios of Eq.~\eqref{eq:br_muTOegama}. If we assume that $\kappa_{ij}$ is a real, diagonal matrix, i.e. $\kappa_{ij} = \kappa_i\delta_{ij} \in \mathbb{R}$, then the coefficient $c^{e\mu}_R$ is given by
\begin{subequations}
\begin{equation}
    c^{e\mu}_R = \frac{e}{16\pi^2} \Biggl[\kappa_{11} \, y_{21} \sum^2_{\alpha=1} R_{1\alpha}R_{2\alpha} \frac{M_1}{m^2_{S^\pm_\alpha}} f_\Phi\left(\frac{M^2_1}{m^2_{S^\pm_\alpha}}\right) + m_e \sum^3_{k=1}y^\ast_{1k}y_{2k} \sum^2_{\alpha=1} R^2_{2\alpha} \frac{1}{m^2_{S^\pm_\alpha}}\tilde{f}_\Phi\left(\frac{M^2_k}{m^2_{S^\pm_\alpha}}\right)\Biggr] \, .
\end{equation}
Similarly, the expression for $c^{\mu e}_R$ takes the form
\begin{equation}
    c^{\mu e}_R = \frac{e}{16\pi^2} \Biggl[\kappa_{22} \, y_{12} \sum^2_{\alpha=1} R_{1\alpha}R_{2\alpha} \frac{M_2}{m^2_{S^\pm_\alpha}} f_\Phi\left(\frac{M^2_2}{m^2_{S^\pm_\alpha}}\right) + m_\mu \sum^3_{k=1} y^\ast_{2k}y_{1k} \sum^2_{\alpha=1} R^2_{2\alpha} \frac{1}{m^2_{S^\pm_\alpha}}\tilde{f}_\Phi\left(\frac{M^2_k}{m^2_{S^\pm_\alpha}}\right)\Biggr] \, .
\end{equation}
\end{subequations}
As long as the singlet fermion masses are not significantly smaller than the masses of the charged scalars, i.e., $M_1, M_2, M_3 \centernot\ll m_{S^\pm_{1}},m_{S^\pm_{2}}$, the second term in each expression becomes negligible compared to the first term. Consequently, the coefficients can be approximated as
\begin{subequations}
\begin{align}
    c^{e\mu}_R &\simeq \frac{e}{16\pi^2} \kappa_{11} \, y_{21} \sum^2_{\alpha=1} R_{1\alpha}R_{2\alpha} \frac{M_1}{m^2_{S^\pm_\alpha}} f_\Phi\left(\frac{M^2_1}{m^2_{S^\pm_\alpha}}\right) \, , \\
    c^{\mu e}_R &\simeq \frac{e}{16\pi^2} \kappa_{22} \, y_{12} \sum^2_{\alpha=1} R_{1\alpha}R_{2\alpha} \frac{M_2}{m^2_{S^\pm_\alpha}} f_\Phi\left(\frac{M^2_2}{m^2_{S^\pm_\alpha}}\right) \, .
\end{align}
\end{subequations}
If we further assume that $M_1 \ll M_2, M_3$, the coefficient $c^{e\mu}_R$ dominates numerically, and the branching ratio is primarily determined by the lightest singlet fermion mass and the masses of the charged scalars:
\begin{equation}
    \text{Br}(\mu^{+} \to e^{+}\gamma) \simeq \frac{e^2}{2^{12}\pi^5}\frac{m^3_\mu M^2_1}{\Gamma_\mu}|\kappa_{11}y_{21}|^2\sin^2(2\theta) \left|\frac{1}{m^2_{S_2}} f_\Phi\left(\frac{M^2_1}{m^2_{S_2}}\right) - \frac{1}{m^2_{S_1}} f_\Phi\left(\frac{M^2_1}{m^2_{S_1}}\right)\right|^2 \, .
\end{equation}
Therefore, in the case of degenerate charged scalars, $\mu^+ \to e^+ \gamma$ is highly suppressed.

\subsection{Anomalous magnetic moment of the muon}
\label{subsection:anomalous_magnetic_moment_of_the_muon}

\begin{figure}[t!]
\centering
    \includegraphics[width=0.5\linewidth]{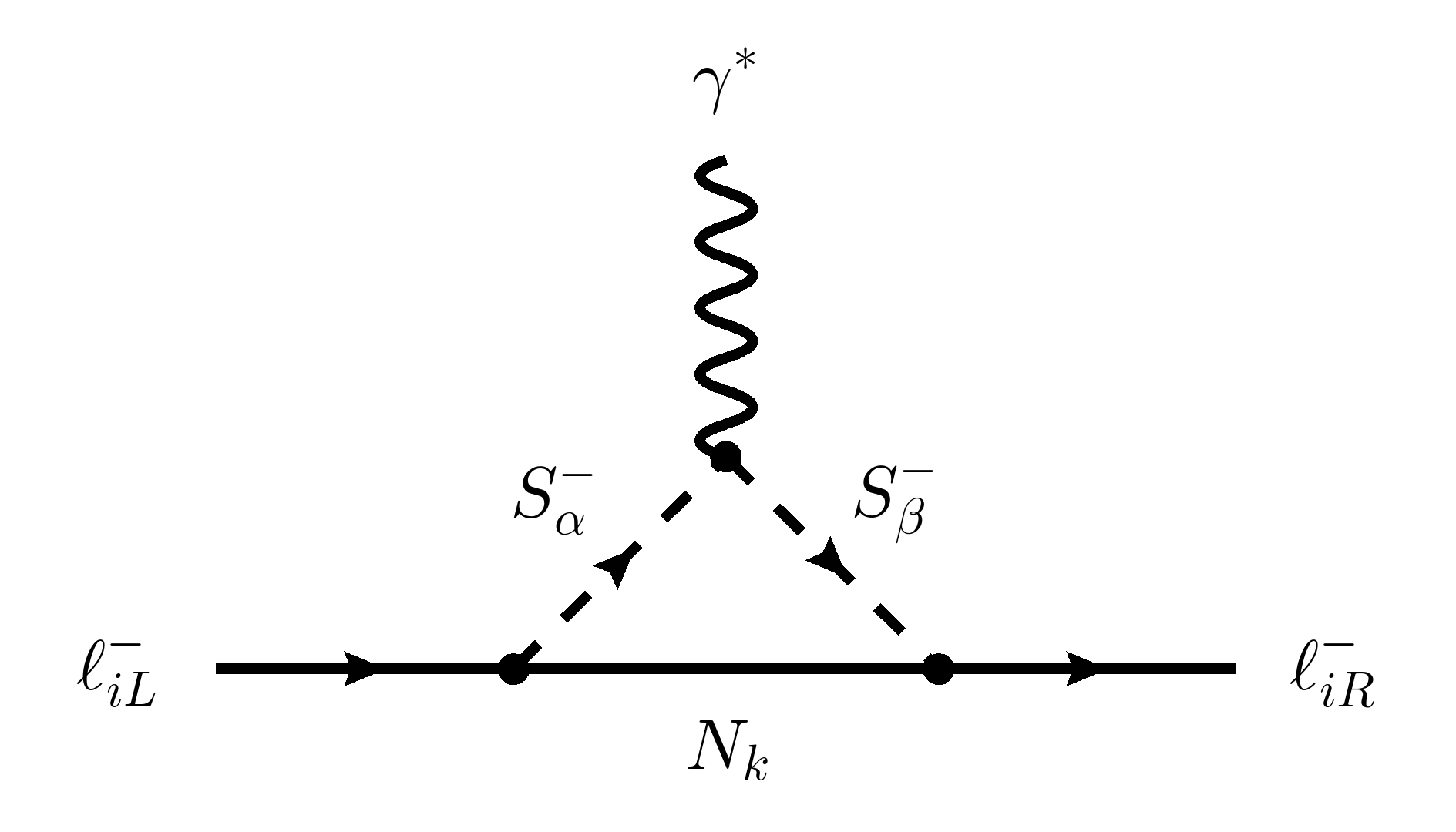}
\caption{One-loop Feynman diagrams contributing to the muon's anomalous magnetic moment. Here, $k = 1,2,3$ and $\alpha, \beta = 1,2$.}
\label{fig:g2_HHN}
\end{figure}

At the one-loop level, new contributions to the muon's anomalous magnetic moment follow from the diagrams shown in Fig.~\ref{fig:g2_HHN}. Addressing a possible tension between the SM prediction and an experimental determination would require a chiral enhancement, which is usually crucial when heavy new particle contributions generate the deviation. A significant enhancement like this demands at least three different fields in the loop. In the model under consideration, the chirality flip arises from large couplings between the muon, singlet fermions, and charged scalars.

We calculate the contribution to the muon's anomalous magnetic moment as~\cite{Crivellin:2018qmi}:
\begin{equation}
    a_{\ell_i} = -\frac{4m_{\ell_i}}{e}\mathfrak{Re}[c^{\ell_i\ell_i}_R] \, ,
\end{equation}
where the coefficient $c^{\ell_i\ell_i}_R$ is provided in Appendix \ref{appendix:coefficients_a_and_muTOegamma}.~\footnote{The correspondence between the notation used in Ref.~\cite{Crivellin:2018qmi} and that adopted in this work is given by $\Gamma^{\ell_i(L,R)}_{N_kH_{n}} \equiv g^{L,R}_{nki}$.}\\

For many years, experimental measurements of the muon magnetic anomaly conducted at Brookhaven National Laboratory (BNL) and Fermi National Accelerator Laboratory (FNAL) have appeared to be in tension with the SM predictions. In 2023, the Fermilab $g-2$ Collaboration presented its latest results~\cite{Muong-2:2023cdq}, combining their data with the 2006 final report from the E821 experiment at BNL \cite{Muong-2:2006rrc}. When comparing this combined result with the Muon $g-2$ Theory Initiative in 2020~\cite{Aoyama:2020ynm}, they find a discrepancy close to 5 standard deviations with the SM prediction. Recent developments, however, challenge the significance of this discrepancy. Updated measurements of the $e^+e^- \to \pi^+\pi^-$ cross-section from the CMD-3 experiment~\cite{CMD-3:2023rfe, CMD-3:2023alj} have provided an updated value for the lowest-order hadronic contribution, $a^\text{had;LO}_\mu$, to the SM predicted value of $a_\mu$.~\footnote{The SM prediction for the muon anomalous magnetic moment is calculated as the sum of three contributions: $a^\text{SM}_\mu = a^\text{QED}_\mu + a^\text{weak}_\mu + a^\text{had}_\mu$, where the hadronic contribution, $a^\text{had}_\mu$, is typically divided into the lowest-order term, $a^\text{had;LO}_\mu$ (hadronic vacuum polarization), and higher-order terms. The CMD-3 Collaboration calculates this contribution as the dispersion integral over the cross-section of the hadronic production in $e^+e^-$ annihilation.} Their resulting SM prediction agrees with the most recent experimental value of Fermilab. Additionally, recent lattice QCD results for the hadronic vacuum polarization (HVP) from the BMW Collaboration \cite{Borsanyi:2020mff} (see also Ref.~\cite{Djukanovic:2024cmq} for another but more recent analysis) have also estimated the hadronic contribution, yielding an $a^\text{SM}_\mu$ value that is just within $1.7$ standard deviations of the experimental result.
    
These latest theoretical results suggest that the tension may be less significant than previously thought. Motivated by these findings, we adopt a more conservative perspective and do not prioritize an explanation for the muon's anomalous magnetic moment.

\subsection{Diphoton Higgs Decay}
\label{subsection:diphoton_higgs_decay}

In the SM, the diphoton Higgs decay channel $h \to \gamma\gamma$ is dominated by the $W$ loop diagram, which is roughly four times larger than the subdominant contribution from the top quark loop. The channel receives additional contributions from loops with the new charged scalars in our model. We define the SM-normalized branching ratio as~\cite{Carena:2012xa}
\begin{equation}\label{eq:R_gammagamma}
    R_{\gamma\gamma} := \frac{\Gamma(h \to \gamma\gamma)}{\Gamma(h \to \gamma\gamma)_{\text{SM}}} = \left|1 + \sum^2_{i=1}\frac{g_{hS_iS_i}v}{2m^2_{S_i}}\frac{A_0(\tau_{S_i})}{A_1(\tau_W) + \frac{4}{3}A_{1/2}(\tau_t)}\right|^2 \, ,
\end{equation}
where $\tau_i \equiv 4m^2_i/m^2_h$. The only relevant couplings are those of the form $hS^{\pm}_\alpha S^{\mp}_\alpha$. While the coupling $hS^{\pm}_1S^{\mp}_2$ does exist, it does not contribute to the diphoton width due to its vector-like nature. The relevant couplings (see Appendix.~\ref{appendix:vertex_factors_scalar_sector}) are
\begin{subequations}
\begin{align}
    g_{hS_1S_1} &= v\Bigl(\lambda_3\sin^2\theta + \lambda_{H\phi}\cos^2\theta\Bigr) - \frac{\mu}{\sqrt{2}}\sin(2\theta) \, , \\
    g_{hS_2S_2} &= v\Bigr(\lambda_3\cos^2\theta + \lambda_{H\phi}\sin^2\theta\Bigr) + \frac{\mu}{\sqrt{2}}\sin(2\theta) \, .
\end{align}
\end{subequations}
The loop functions are defined as
\begin{equation}
\begin{aligned}
    A_0(x) &= -x^2[x^{-1} - f(x^{-1})] \, , \\
    A_{1/2}(x) &= 2x^2[x^{-1} + (x^{-1} - 1)f(x^{-1})] \, , \\
    A_1(x) &= -x^2[2x^{-2} + 3x^{-1} + 3(2x^{-1} - 1)f(x^{-1})] \, ,
\end{aligned}
\end{equation}
where
\begin{equation}
    f(x) = 
\begin{cases}
    \arcsin^2(\sqrt{x}) & \text{when $x \leq 1$} \, , \\
    -\frac{1}{4}\left[\ln\left(\frac{1 + \sqrt{1 - x^{-1}}}{1 - \sqrt{1 - x^{-1}}}\right) - i\pi\right]^2 & \text{when $x > 1$} \, .
\end{cases}
\end{equation}

\section{Experimental Constraints}
\label{section:constraints}

This Section delves into all phenomenological constraints and corresponding experimental bounds. Additionally, we discuss essential theoretical aspects, including perturbativity and vacuum stability, which also impose restrictions on the parameter space of our model. We will incorporate these factors into the numerical analysis presented in Section~\ref{section:numerical_analysis}.

\subsection{Perturbative unitarity and naturalness}
\label{subsection:perturbativity_unitarity_and_naturalness}

We demand all the quartic couplings in the potential to be below,
\begin{align}
    |\lambda_i| < 4\pi,\quad (i=1,2,3,4,5,\phi,\eta\phi,H\phi) \, ,
\end{align}
to ensure perturbative unitarity of scalar field scattering amplitudes. 
Similarly, for the Yukawa coupling, we demand
\begin{equation}
    |y_{ij}|^2 < 4\pi,\quad (i,j=1,2,3) \, .
\end{equation}
Moreover, the Higgs boson mass is sensitive to radiative corrections induced by the trilinear interaction term that is proportional to~\cite{Beniwal:2020hjc}
\begin{equation}
    \delta m^2_h \simeq \frac{\mu^2}{16\pi^2} \, .
\end{equation}
Therefore, if we require a small amount of fine-tuning, $\delta m_h/m_h < \varepsilon$, where $\varepsilon$ is the tolerance level, it follows that
\begin{equation}
    |\mu| \lesssim 4\pi\varepsilon \, m_h \, .
\end{equation}
For instance, for a value of $\epsilon = 1$, we obtain the upper limit $\mu \lesssim 1.5\ \text{TeV} $. 

\subsection{Stability of the vacuum}
\label{subsection:stability_of_the_vacuum}

The scalar potential must be bounded from below to ensure a stable vacuum. This is guaranteed by the inequalities (see Appendix.~\ref{appendix:vacuum_stability} for their derivation)
\begin{equation}
    0 \leq \lambda_1, \lambda_2, \lambda_\phi \quad\text{and}\quad 0 \leq c_1, c_2, c_3, c_4 \, ,
\end{equation}
where
\begin{subequations}\label{eq:system_inequalities_vacuum_stability}
\begin{align}
    c_1 &:=
    \begin{cases}
        \lambda_3 + \sqrt{\lambda_1\lambda_2} \, , & \text{if $\lambda_4 - |\lambda_5| \geq 0$} \, , \\
        \lambda_3 + \lambda_4 - |\lambda_5| + \sqrt{\lambda_1\lambda_2} \, , & \text{if $\lambda_4 - |\lambda_5| < 0$} \, ,
    \end{cases}
    \\
    c_2 &:= \lambda_{H\phi} + \sqrt{\lambda_1\lambda_\phi} \, , \\
    c_3 &:= \lambda_{\eta\phi} + \sqrt{\lambda_2\lambda_\phi} \, , \\
    c_4 &:=
    \begin{cases}
        \sqrt{\lambda_1\lambda_2\lambda_\phi} + \lambda_3\sqrt{\lambda_\phi} + \lambda_{H\phi}\sqrt{\lambda_2} + \lambda_{\eta\phi}\sqrt{\lambda_1} + \sqrt{2 c_1 \, c_2 \, c_3} \, , & \text{if $\lambda_4 - |\lambda_5| \geq 0$} \, , \\
        \sqrt{\lambda_1\lambda_2\lambda_\phi} + (\lambda_3 + \lambda_4 - |\lambda_5|)\sqrt{\lambda_\phi} + \lambda_{H\phi}\sqrt{\lambda_2} \\
        + \lambda_{\eta\phi}\sqrt{\lambda_1} + \sqrt{2 c_1 \, c_2 \, c_3} \, , & \text{if $\lambda_4 - |\lambda_5| < 0$} \, .
    \end{cases}
\end{align}
\end{subequations}

\subsection{Cosmological observations}
\label{subsection:cosmological_observations}

Observations of the cosmic microwave background and large-scale structures provide a stringent limit on the sum of three neutrino masses
\begin{equation}\label{eq:sigma_m1_m2_m3}
    \Sigma := m_1 + m_2 + m_3 \, .
\end{equation}
The latest result from Planck Collaboration \cite{eBOSS:2020yzd, Planck:2018vyg} and Ly-$\alpha$ measurements \cite{Palanque-Delabrouille:2015pga} show that:
\begin{equation}\label{eq:sigma_m1m2m3}
    \Sigma < 0.115\ \text{eV}\; \text{at $95\%$ \text{C.L.}} \, ,
\end{equation}
with a slight preference for a normal hierarchy over an inverted one. When interpreted as a function of $m_1$, and using the central values of the neutrino oscillation parameters, the bound of Eq.~\eqref{eq:sigma_m1m2m3} imposes an upper limit of $m_1 \lesssim 0.028\ \text{eV}$. This is the most stringent constraint for the mass of the lightest active neutrino.

\subsection{Neutrinoless double-beta decay}
\label{subsection:neutrinoless_double_beta_decay}

To date, there is no evidence for neutrinoless double-beta decay ($0\nu\beta\beta$), a lepton-number-violating process in which two neutrons are converted into two protons with the emission of two electrons (in contrast to the standard two-neutrino double-beta decay, which emits two electrons and two antineutrinos). The simplest mechanism for $0\nu\beta\beta$ decay involves the virtual exchange of a light Majorana neutrino between two nucleons. In this scenario, the half-life of the process is proportional to the square of the effective Majorana mass, defined as
\begin{align}\label{eq:m_betabeta}
    m_{\beta\beta} &:= \Biggl|\sum^3_{j=1}m_j \, U_{ej}^2\Biggr|= \Bigl|m_1c^2_{12}c^2_{13} + m_2s^2_{12}c^2_{13}e^{i\rho} + m_3s^2_{13}e^{i\sigma - 2i\delta_{\text{CP}}}\Bigr| \, .
\end{align}
After its Phase-II operation, the KamLAND-Zen Collaboration has set the most stringent limit on the half-life of $\ce{^{136}_{54}Xe}$ in the $0\nu\beta\beta$ decay mode, corresponding to $T^{1/2}_{0\nu} > 1.07 \times 10^{26}\ \text{yr}$ at $90\%$ CL~\cite{PhysRevLett.117.082503}. This result translates into a restrictive upper bound on the effective Majorana mass:
\begin{equation}\label{eq:m_betabeta_constrain}
    m_{\beta\beta} < 0.165\ \text{eV}\; \text{at $90\%$ C.L.} \, .
\end{equation}
Using the central values of the neutrino oscillation parameters, the effective Majorana mass bound directly translates into an upper limit for the mass of the lightest active neutrino. Moreover, for larger values of $m_1$, the approximation $m_{\beta\beta} \simeq m_1$ becomes increasingly accurate, the constraint of Eq.~\eqref{eq:m_betabeta_constrain} yields $m_1 \lesssim 0.165\ \text{eV}$. However, since the numerical scan in Section~\ref{section:numerical_analysis} already takes into account the constraint derived in Section~\ref{subsection:neutrinoless_double_beta_decay}, which turns out to be more stringent, this bound will not have any practical effect.

\subsection{Beta decay}
\label{subsection:beta_decay}

Unlike the cosmological bound of Eq.~\eqref{eq:sigma_m1_m2_m3}, which heavily depends on the assumption of a $\nu\Lambda$CDM Universe, and the neutrinoless double-beta decay limit of Eq.~\eqref{eq:m_betabeta}, which is valid only under the assumption that light neutrinos are Majorana particles mediating the decay, the kinematics of single beta decay offers a method that is free from these assumptions. As such, it provides one of the most direct and model-independent approaches to determine neutrino masses. In this context, the Karlsruhe Tritium Neutrino (KATRIN) experiment has used the single beta decay of molecular tritium, $\text{T}_2 \to {}^3\text{HeT}^+ + e^- + \bar{\nu}_e$, to measure the so-called effective electron antineutrino mass, defined as
\begin{equation}\label{eq:effective_electron_anti-neutrino_mass}
    m^\text{eff}_{\nu_e} := \sqrt{\sum^3_{j=1}m^2_j \, |U_{ej}|^2} = \sqrt{m^2_{1}c^2_{12}c^2_{13} + m^2_{2}s^2_{12}c^2_{13} + m^2_{3}s^2_{13}} \, .
\end{equation}
Through direct measurement, KATRIN has set the upper limit on this observable~\cite{KATRIN:2021uub}:
\begin{equation}
    m^\text{eff}_{\nu_e} < 0.8\ \text{eV}\; \text{at $90\%$ C.L.} \, .
\end{equation}
The effective electron antineutrino mass, as defined in Eq.~\eqref{eq:effective_electron_anti-neutrino_mass}, also provides an upper limit on the lightest active neutrino mass. For larger $m_1$, we find that $m^\text{eff}_{\nu_e} \simeq m_1$ holds to a good approximation, leading to $m_1 \lesssim 0.8\ \text{eV}$. Again, this bound has no practical effect on the results of our analysis in Section~\ref{section:numerical_analysis}.

\subsection{LFV \texorpdfstring{$\mu^{+} \to e^{+}\gamma$}{mu+ -> e+ gamma} decay}

As mentioned above, in this work, we focus on the radiative decay $\mu^{+} \to e^{+}\gamma$. This process, which has not been observed to date, usually provides the most stringent constraint from LFV processes. In our numerical analysis, we make use of the results provided by the MEG II experiment~\cite{MEGII:2023ltw}:
\begin{equation}\label{eq:MEG_bound}
    \text{Br}(\mu^{+} \to e^{+}\gamma) < 3.1 \times 10^{-13}\; \text{at $90\%$ C.L.} \, .
\end{equation}

\subsection{Lower bounds on charged scalars masses}
\label{subsection:lower_bounds_on_charged_scalar_masses}

The masses of charged scalars, assuming that the decays $S^{+}_{1,2} \to c\bar{s}$ and $S^{+}_{1,2} \to \tau^{+}\nu$ (and their charge conjugates) exhaust their decay widths, and the decays $S^{+}_{1,2} \to W^{+}h$ are absent~\cite{ALEPH:2013htx}, are bounded from below by the LEP experiment~\cite{ALEPH:2001oot,DELPHI:2003uqw,L3:2003fyi,OPAL:2003nhx,ParticleDataGroup:2024cfk}:
\begin{equation}\label{eq:lower_bound_charged_scalar_masses}
    m_{S^\pm_1}\, , \, m_{S^\pm_2} > 80\ \text{GeV}\; \text{at $90\%$ C.L.} \, .
\end{equation}
Additionally, because the decay widths of the $W$ and $Z$ bosons are measured with high precision~\cite{ParticleDataGroup:2024cfk}, we require that no new decay channels be kinematically allowed. These considerations lead to the following conditions:
\begin{equation}
\begin{aligned}
    2m_{S^\pm_1} &> m_Z \, , & 2m_{S^\pm_2} &> m_Z \, , & m_{S^\pm_1} + m_{S^\pm_2} &> m_Z \, , & m_{\eta_R} + m_{\eta_I} &> m_Z \, , \\
    m_{S^\pm_1} + m_{\eta_R} &> m_W \, , & m_{S^\pm_2} + m_{\eta_R} &> m_W \, , & m_{S^\pm_1} + m_{\eta_I} &> m_W \, , & m_{S^\pm_2} + m_{\eta_I} &> m_W \, .
\end{aligned}
\end{equation}
However, note that some of these conditions, such as the first two, are automatically satisfied by the more stringent LEP bounds in Eq.~\eqref{eq:lower_bound_charged_scalar_masses}. 

\subsection{Dimuon Higgs channel}

The ratio $R_{\mu\mu}$ given by Eq.~\eqref{eq:hTOmumu_branching_quotient}, can be bounded from the dimuon Higgs signal strength, $\hat{\mu}_{\mu\mu}$, defined by both the ATLAS and CSM collaborations as
\begin{equation}
    \hat{\mu}_{\mu\mu} := \frac{\sigma(pp \to h)}{\sigma(pp \to h)_\text{SM}} \times \frac{\text{Br}(h \to \mu^{+}\mu^{-})}{\text{Br}(h \to \mu^{+}\mu^{-})_\text{SM}} \, .
\end{equation}
Assuming that the $pp$ production cross-section of the Higgs boson in our model is approximately equal to the SM one, then we have the approximation $R_{\mu\mu} \simeq \hat{\mu}_{\mu\mu}$. We can then use the experimental bounds from ATLAS~\cite{ATLAS:2020fzp} and CMS~\cite{CMS:2020xwi}
\begin{equation}
    \hat{\mu}^{\text{ATLAS}}_{\mu\mu} \in [0.6, 1.8]\; \text{at $68\%$ C.L.} \, ,\quad \hat{\mu}^{\text{CMS}}_{\mu\mu} \in [0.79, 1.6] \; \text{at $68\%$ C.L.} \, ,
\end{equation}
and naively combine them to get the average bound:
\begin{equation}\label{eq:naive_bound_on_Rmumu}
    R_{\mu\mu} \in [0.69, 1.7] \; \text{at $68\%$ C.L.} \, .
\end{equation}

\subsection{Diphoton Higgs channel}
\label{sec:diphoton}

\begin{figure}[!t]
\centering
\begin{subfigure}[b]{0.33\textwidth}
    \includegraphics[width=\textwidth]{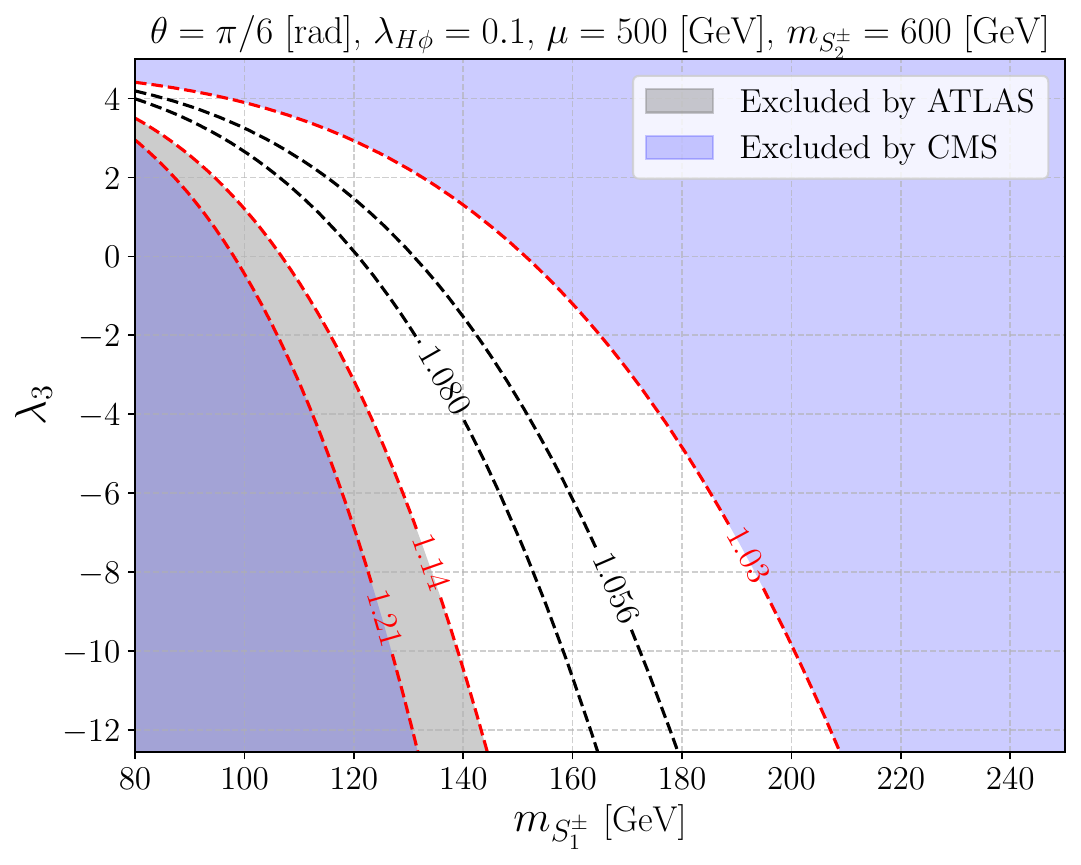}
    \caption{}
    \label{fig:plot_hTOgammagamma_contour_lines_mS1_vs_lam3}
\end{subfigure}%
\hfill
\begin{subfigure}[b]{0.33\textwidth}
    \includegraphics[width=\textwidth]{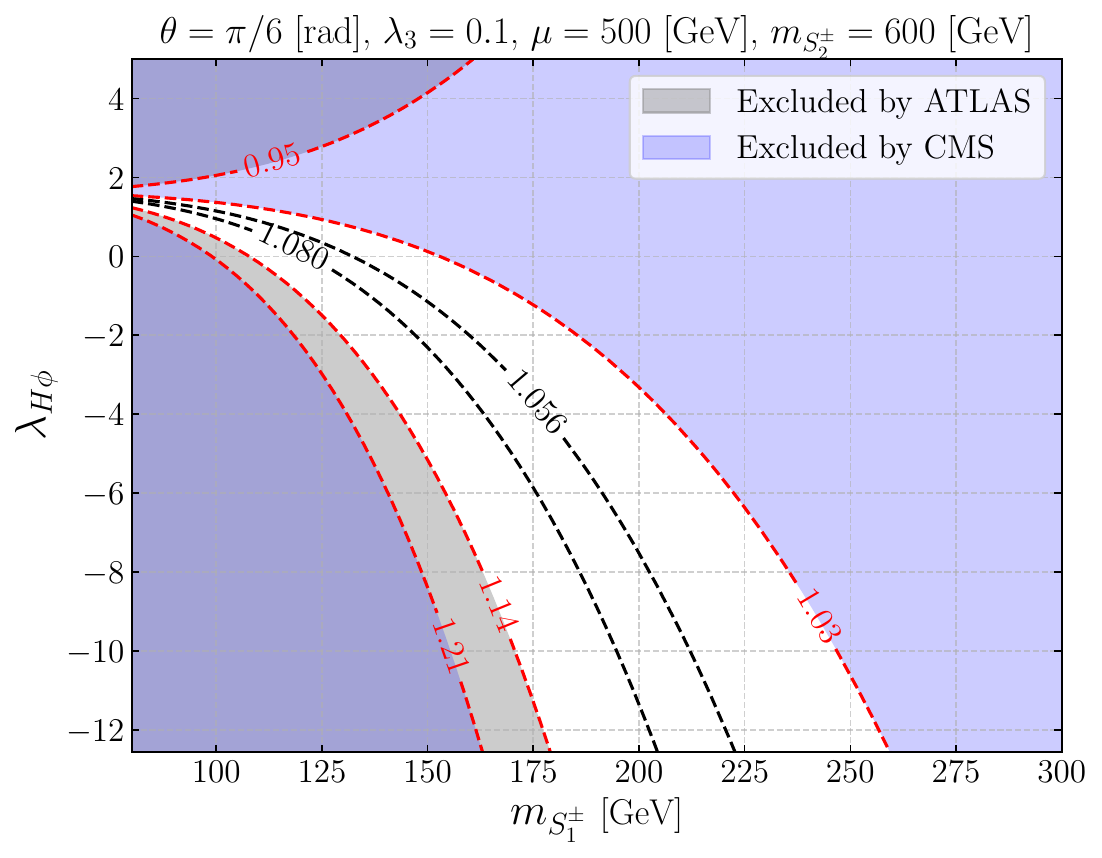}
    \caption{}
    \label{fig:plot_hTOgammagamma_contour_lines_mS1_vs_lamHphi}
\end{subfigure}%
\hfill
\begin{subfigure}[b]{0.33\textwidth}
    \includegraphics[width=\textwidth]{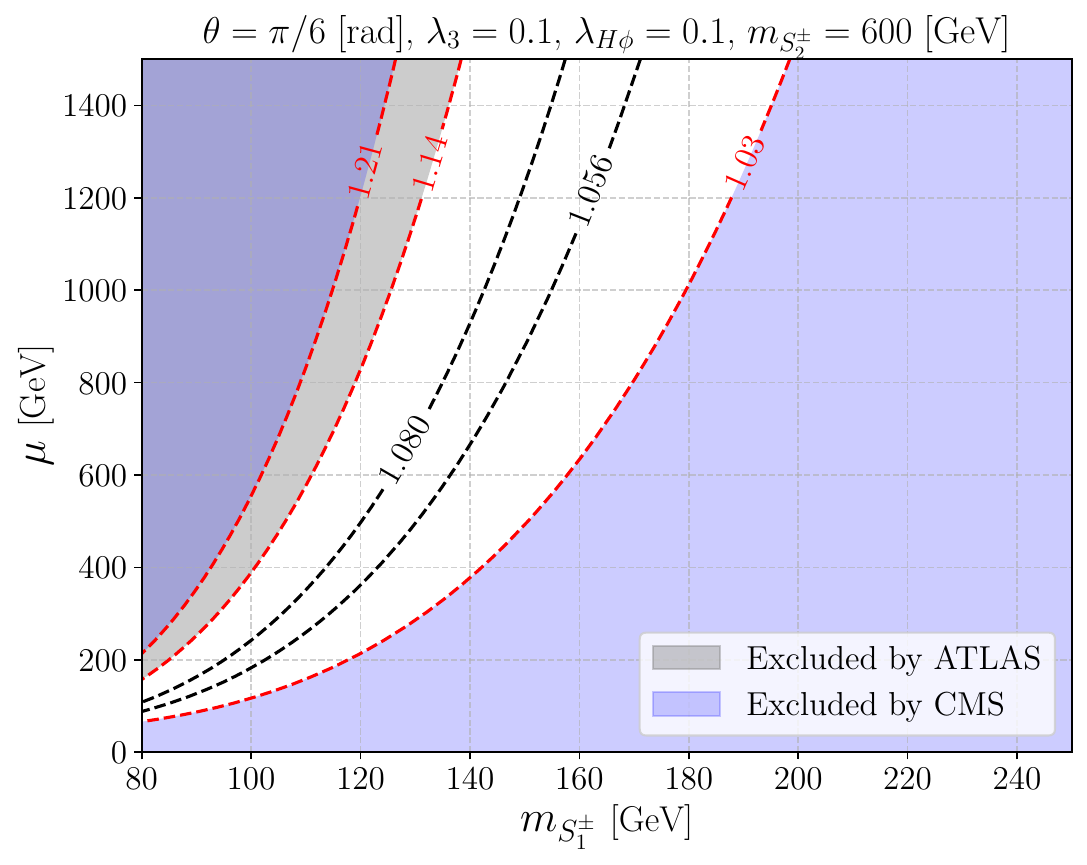}
    \caption{}
    \label{fig:plot_hTOgammagamma_contour_lines_mS1_vs_mu}
\end{subfigure}
\caption{Contour plots to illustrate the effect of the $R_{\gamma\gamma}$ constrain \eqref{eq:R_gammagamma_experimental_bounds} in the planes of the variables relevant for this observable. These plots highlight the regions of parameter space excluded by the ATLAS and CMS experimental constraints. Fixed parameter values were chosen as representative values from the random scan to illustrate the characteristic behavior of $R_{\gamma\gamma}$.
\label{fig:plot_hTOgammagamma_contour_lines}}
\end{figure}

Similar to what happened with the dimuon Higgs channel, the ratio $R_{\gamma\gamma}$ given by Eq.~\eqref{eq:R_gammagamma}, can be bounded this time from the diphoton Higgs signal strength, $\hat{\mu}_{\gamma\gamma}$, defined by the ATLAS and CSM collaborations as:
\begin{equation}
    \hat{\mu}_{\gamma\gamma} := \frac{\sigma(pp \to h)}{\sigma(pp \to h)_\text{SM}} \times \frac{\text{Br}(h \to \gamma\gamma)}{\text{Br}(h \to \gamma\gamma)_\text{SM}} \, .
\end{equation}
Again, if we assume that the $pp$ production cross-section of the Higgs boson in our model is approximately given by that of the SM, then we may use the approximation $R_{\gamma\gamma} \simeq \hat{\mu}_{\gamma\gamma}$. The experimental bounds from ATLAS~\cite{ATLAS:2022tnm} (see also Ref.~\cite{Cepeda:2019klc} for future prospects) and CMS~\cite{CMS:2022dwd, CMS:2021kom} are
\begin{equation}\label{eq:R_gammagamma_experimental_bounds}
    \hat{\mu}^{\text{ATLAS}}_{\gamma\gamma} \in [0.95, 1.14]\; \text{at $95\%$ C.L.} \, ,\quad \hat{\mu}^{\text{CMS}}_{\gamma\gamma} \in [1.03, 1.21]\; \text{at $95\%$ C.L.}
\end{equation}
We illustrate the effect of this constraint in Fig.~\ref{fig:plot_hTOgammagamma_contour_lines}, although for the scan of Section \ref{section:numerical_analysis}, we use the averaged bound
\begin{equation}\label{eq:naive_bound_on_Rgammagamma}
    R_{\gamma\gamma} \in [0.99, 1.17]\; \text{at $95\%$ C.L.}
\end{equation}
On the left panel, we can see the dependence of the Higgs diphoton rate as a function of the lightest charged scalar mass, $m_{S_{1}^{\pm}}$ and the coupling $\lambda_3$. We can see the constraints imposed by the ATLAS and CMS measurements as gray and purple areas. The allowed region for a given value of the coupling $\lambda_3$ is restricted into a range of values of the mass of the scalar that becomes wider when the coupling value gets smaller. The central panel shows the behavior of the diphoton channel as a function of $m_{S_{1}^{\pm}}$ and the coupling $\lambda_{H\phi}$. We find a behavior similar to the previous case. On the right panel, we can see the dependence of the diphoton channel as a function of $m_{S_{1}^{\pm}}$ and $\mu$. In this case, for greater values of the parameter $\mu$, the allowed region for $m_{S_{1}^{\pm}}$ gets wider.

Although Fig.~\ref{fig:plot_hTOgammagamma_contour_lines} provides only a rough visualization, it effectively illustrates the general impact that ATLAS and CMS bounds will have when scanning the entire parameter space of the model. The ATLAS bound is slightly stronger at ruling out lower masses, while CMS sets an upper bound on these. The latter is easy to understand since CMS disfavors the SM prediction (it is excluded at $95\%$ C.L.). Due to the loop nature of the charged scalars' contributions to the processes $h \to \gamma\gamma$ and $h \to \mu^+\mu^-$, lighter charged scalars would result in more significant contributions to both decays. However, due to the stringent nature of the ATLAS and CMS lower bounds on $h \to \gamma\gamma$ for light-charged scalars, we can expect this constraint to restrict the parameter space for significant deviations significantly. Consequently, while deviations on $R_{\mu\mu}$ are anticipated to be small, they are not entirely excluded and will be explored in Section~\ref{section:numerical_analysis}.

\subsection{Higgs invisible decays}
\label{subsection:invisible_decays_of_the_higgs_boson}

The invisible decay channel of the Higgs boson is currently constrained by the ATLAS collaboration to the upper limit \cite{ATLAS:2023tkt}:
\begin{equation}\label{eq:higgs_to_invisible_ATLAS_bound}
    \text{Br}(h \to \text{inv}) < 0.107\; \text{at $95\%$ C.L.} \, .
\end{equation}
In our numerical analysis, the singlet fermions $N_i$ will always be assumed to be heavier than the Higgs boson. However, in some parameter points, the inert neutral scalars, $\eta_R$ and $\eta_I$, turn out to be light, thus opening invisible decay channels for the Higgs boson. In this case, the constraint in Eq.~\eqref{eq:higgs_to_invisible_ATLAS_bound} depends on which of them, $\eta_R$ or $\eta_I$, is lighter, something determined by the sign of $\lambda_5$. In this context, we must also consider the corresponding bound with:
\begin{equation}
    \Gamma(h \to \eta_0 \, \eta_0) = \frac{g^2}{32\pi^2}\frac{1}{m_h}\Biggl[1 - \frac{4m^2_{\eta_0}}{m^2_h}\Biggr]^{1/2} \, , \quad \text{where }\; g =
    \begin{cases}
        v(\lambda_3 + \lambda_4 + \lambda_5) \, , & \text{if $\eta_0 = \eta_R$} \, , \\
        v(\lambda_3 + \lambda_4 - \lambda_5) \, , & \text{if $\eta_0 = \eta_I$} \, .
    \end{cases}
\end{equation}
The bound of Eq.~\eqref{eq:higgs_to_invisible_ATLAS_bound} translates then into
\begin{equation}
    \Gamma(h \to \eta_0 \, \eta_0) < \frac{0.107}{1 - 0.107} \times \Gamma^{\text{SM}}_h \simeq 0.120 \times \Gamma^{\text{SM}}_h \, ,
\end{equation}
where $\Gamma^{\text{SM}}_h$ is the total decay width of the Higgs boson according to the SM.

\section{Numerical Analysis}
\label{section:numerical_analysis}

{
\setlength{\tabcolsep}{12pt}
\renewcommand\arraystretch{1.4}
\begin{table}[!t]
\centering
\begin{NiceTabular}{ccc}
\CodeBefore
\rowcolors{1}{white}{gray!14}
\Body
\toprule
Parameter                                   & Range                 & Units \\ \midrule\midrule
$\lambda_i,\ (i=2,\phi)$ & $[0, 4\pi]$ & -       \\
$\lambda_i,\ (i=3,4,5,\eta\phi,H\phi)$ & $[-4\pi, 4\pi]$ & -       \\
$|\kappa_{ij}|,\ (i,j = 1,2,3)$                               & $[0, 4\pi]$ & -         \\
$\theta_{ij},\ (i,j = 1,2,3)$                               & $[0, 2\pi]$ & rad         \\
$M_i,\ (i=1,2,3)$                             & $[100, 1500]$ & GeV   \\
$m^2_\phi, m^2_\eta$                                    & $[-10^6, 10^6]$ & GeV${}^2$ \\
$\mu$                                       & $[-1500, 1500]$ & GeV \\ \bottomrule
\end{NiceTabular}
\caption{Free parameters and their ranges used in the full random scan for the $n_\phi = 1$ scenario. The $\kappa_{ij}$ values are complex, written as $\kappa_{ij} = |\kappa_{ij}|e^{i\theta_{ij}}$, with modulus and phase selected independently. Each scan done in Section \ref{section:numerical_analysis} includes $N=10^5$ points. Every parameter's value is picked randomly, assuming a flat distribution. Note that any interval containing zero is effectively cut off at $10^{-12}$ due to numerical precision limitations during the scan.}
\label{tab:free_parameters_and_ranges}
\end{table}
}

In this Section, we present the results of a numerical analysis performed across the full parameter space of the model, focusing on deviations in $\text{Br}(h \to \mu^{+}\mu^{-})$ with respect to the SM value. We also take particular emphasis on the processes $\mu^{+} \to e^{+}\gamma$ and $h \to \gamma\gamma$, as they impose significant constraints on the magnitude of $|R_{\mu\mu} - 1|$.

To identify viable regions of the model, we performed different random linear samplings over the ranges of the input parameters listed in Tab.~\ref{tab:free_parameters_and_ranges} with each scan containing $N=10^5$ points. The neutrino oscillation parameters, $\Delta m^2_{\text{sol}}$, $\Delta m^2_{\text{atm}}$, and $\sin\theta_{ij}$, were treated as free parameters within their respective $3\sigma$ ranges, as detailed in Tab.~\ref{tab:neutrino_oscillation_parameters}. For simplicity, the two Majorana CP-violating phases, $\rho$ and $\sigma$ from Eq.~\eqref{eq:PMNS_matrix}, were fixed to zero. Additionally, the three complex angles of the orthogonal matrix in the Casas-Ibarra parametrization, Eq.~\eqref{eq:yukawa_matrix}, were assumed to be real and varied freely within the interval $[0, 2\pi]$. Finally, we fix $m_h = 125.18\ \text{GeV}$~\cite{ParticleDataGroup:2024cfk} in our numerical analysis.

After getting the value of every parameter, each point is then tested against all the constraints described in Section~\ref{section:constraints}, with the exception of the $\mu^{+} \to e^{+}\gamma$ bound. This last constraint was applied \textit{a posteriori} in the plots to clearly illustrate its impact on the allowed parameter space.

\subsection{Impact of \texorpdfstring{$h \to \gamma\gamma$}{} and \texorpdfstring{$\mu^+ \to e^+ \gamma$}{}}
\label{subsection:impact_of_hTOgammagamma}

Let us now analyze the relationship between the processes $h \to \gamma\gamma$, $\mu^+ \to e^+ \gamma$ and $h \to \mu^+\mu^-$. Specifically, we aim to understand how experimental constraints from the former two processes impact potential deviations in the dimuon Higgs decay rate.\\

\begin{figure}[!t]
\centering
\begin{subfigure}[b]{0.49\textwidth}
    \includegraphics[width=\textwidth]{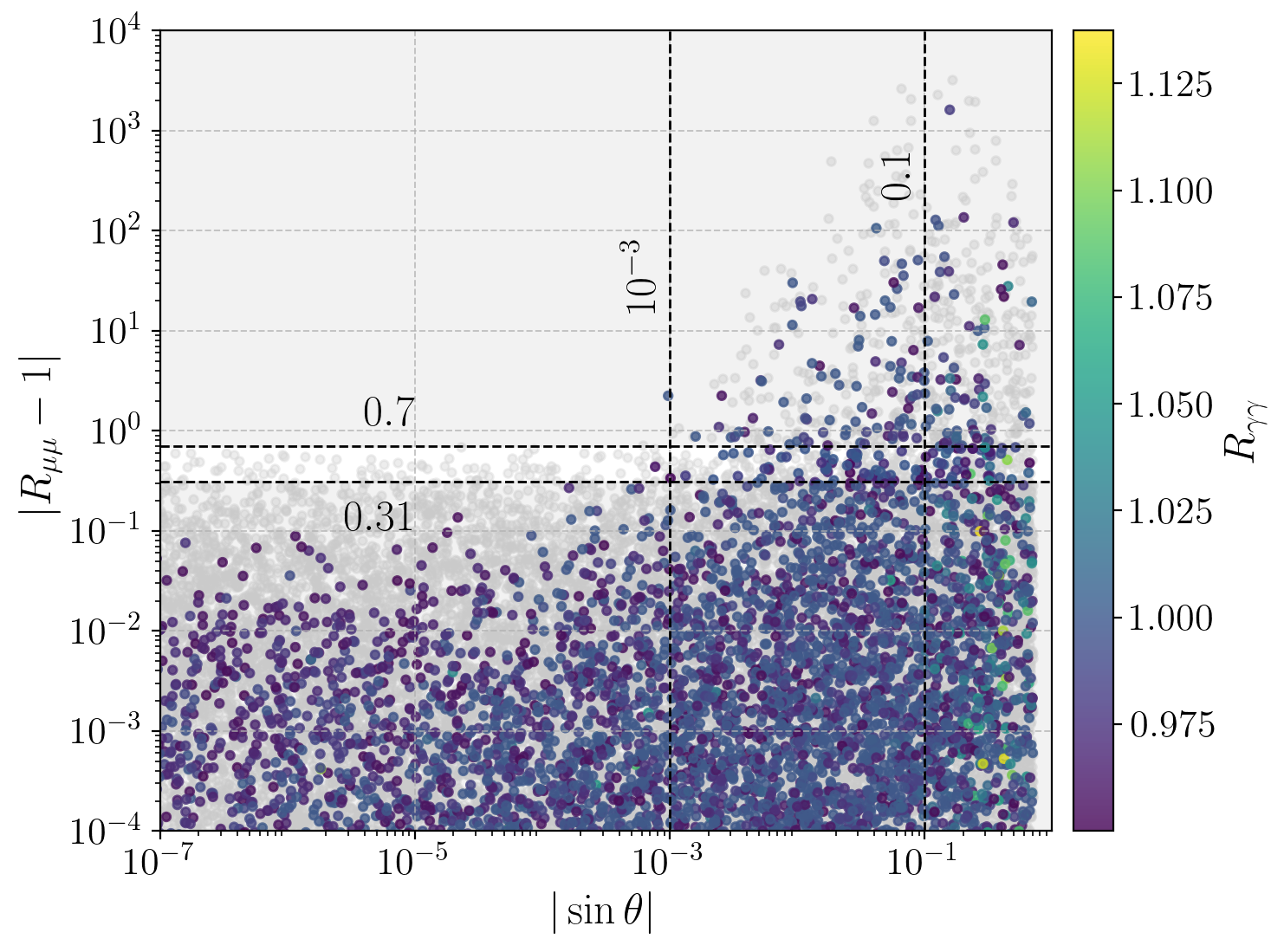}
    \caption{}
    \label{fig:plot_abs_sintheta_vs_abs_1mRmumu_vs_Rgammagamma}
\end{subfigure}%
\begin{subfigure}[b]{0.5\textwidth}
    \includegraphics[width=\textwidth]{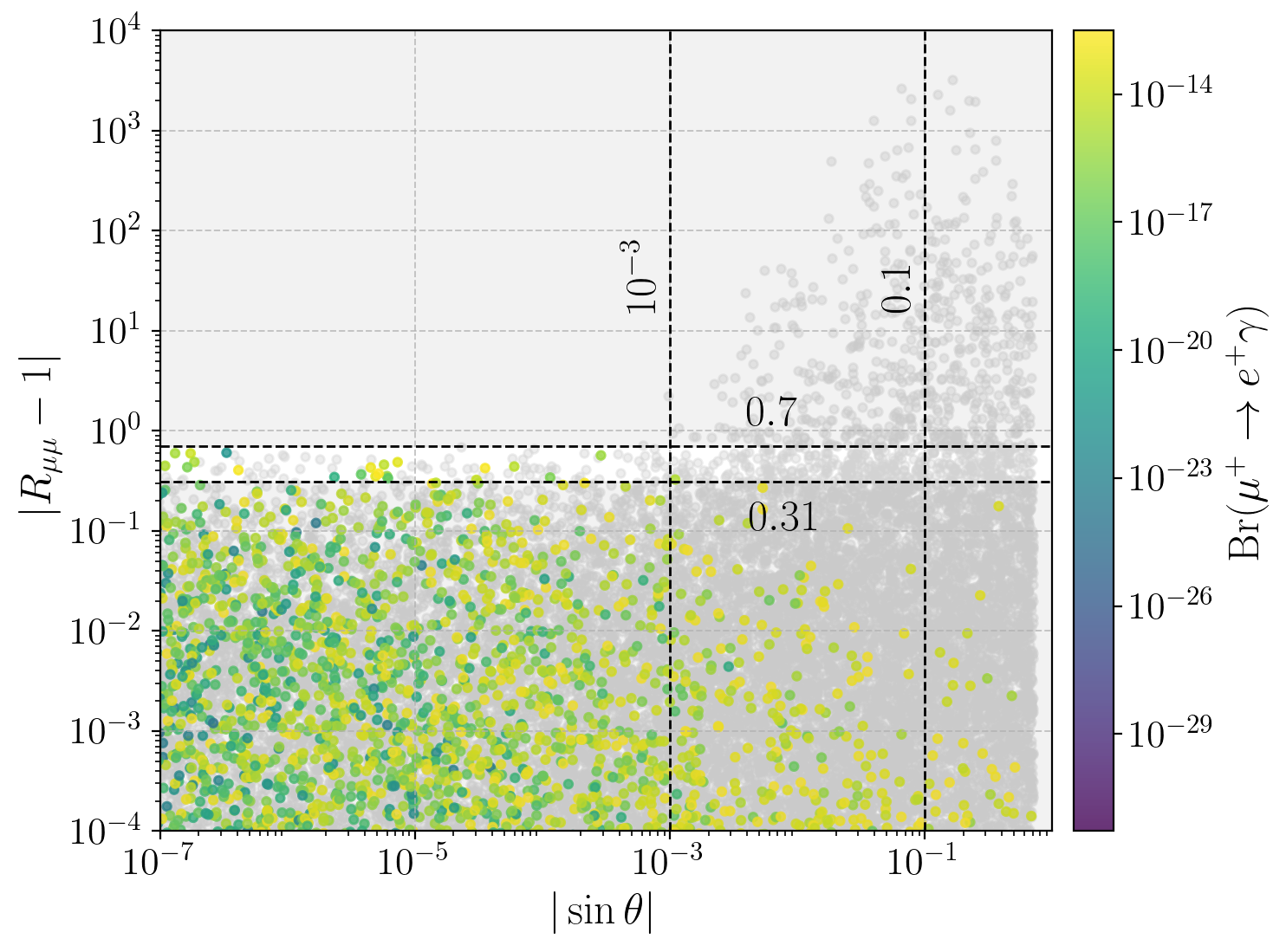}
    \caption{}
    \label{fig:plot_abs_sintheta_vs_abs_1mRmumu_vs_muTOegamma}
\end{subfigure}
\caption{These panels show the impact of $h \to \gamma\gamma$ and $\mu^+ \to e^+\gamma$ bound on the possible values that $|R_{\mu\mu} - 1|$ can reach as a function of the mixing angle. Both figures including all constraints except $h \to \gamma \gamma$ and $\mu^{+} \to e^{+}\gamma$, which are applied \textit{a posteriori} and visualized as color (satisfying) and gray (violating) points. The horizontal lines show the limits from Eq.~\eqref{eq:naive_bound_on_Rmumu}. On the left panel, we can see how bounds on $h \to \gamma\gamma$ bounds deviations in $R_{\mu\mu}$ at small angles. On the right panel, we see how the LFV process $\mu^{+} \to e^{+}\gamma$ rules out most of the points with large mixing angles and, consequently, large deviation on $R_{\mu\mu}$.}
\label{fig:effect_individual_constraints}
\end{figure}

We begin by examining the interplay between $h \to \gamma\gamma$ and $h \to \mu^+\mu^-$. A preliminary discussion of the $h \to \gamma\gamma$ process is provided in Section \ref{sec:diphoton}. Both panels in Fig.~\ref{fig:effect_individual_constraints} illustrate the plane $|\sin\theta|$ vs.~$|R_{\mu\mu} - 1|$, based on a scan that includes all constraints outlined in Section \ref{section:constraints}, except for those associated with $h \to \gamma\gamma$ and $\mu^+ \to e^+ \gamma$.

Figure \ref{fig:plot_abs_sintheta_vs_abs_1mRmumu_vs_Rgammagamma} highlights the impact of the $h \to \gamma\gamma$ constraint. Colored points in this figure denote parameter sets that satisfy the ATLAS and CMS combined average bound on $h \to \gamma\gamma$, as given in Eq.~\eqref{eq:naive_bound_on_Rgammagamma}. The $R_{\gamma\gamma}$ values are shown using a color gradient, with the corresponding scale provided in the chart. Gray points indicate parameter sets that fail to satisfy the $h \to \gamma\gamma$ constraint. As we can observe, the mixing angle $|\sin\theta|$ is a key driver of both large and small deviations in $R_{\mu\mu}$. For $|\sin\theta| > 10^{-3}$, deviations in $R_{\mu\mu}$ can become substantially large with respect to the SM value, peaking near $|\sin\theta| \simeq 0.1$. However, while the $h \to \gamma\gamma$ bounds from ATLAS and CMS allows for large deviations at sizable mixing angles, they strongly constrain deviations of $R_{\mu\mu}$ of the order of $\mathcal{O}(10^{-1})$ or smaller for very small angles.

On the other hand, the impact of the MEG II bound (see Eq.~\eqref{eq:MEG_bound}) on $R_{\mu\mu}$ deviations is shown in Fig.~\ref{fig:plot_abs_sintheta_vs_abs_1mRmumu_vs_muTOegamma}. Here, colored points correspond to parameter sets satisfying the MEG II constraint. Unlike $h \to \gamma\gamma$, this bound primarily excludes points with large mixing angles. Consequently, any significant deviation in $R_{\mu\mu}$ must align with small mixing angles.\\

\begin{figure}[!t]
\centering
\begin{subfigure}[b]{0.49\textwidth}
    \includegraphics[width=\textwidth]{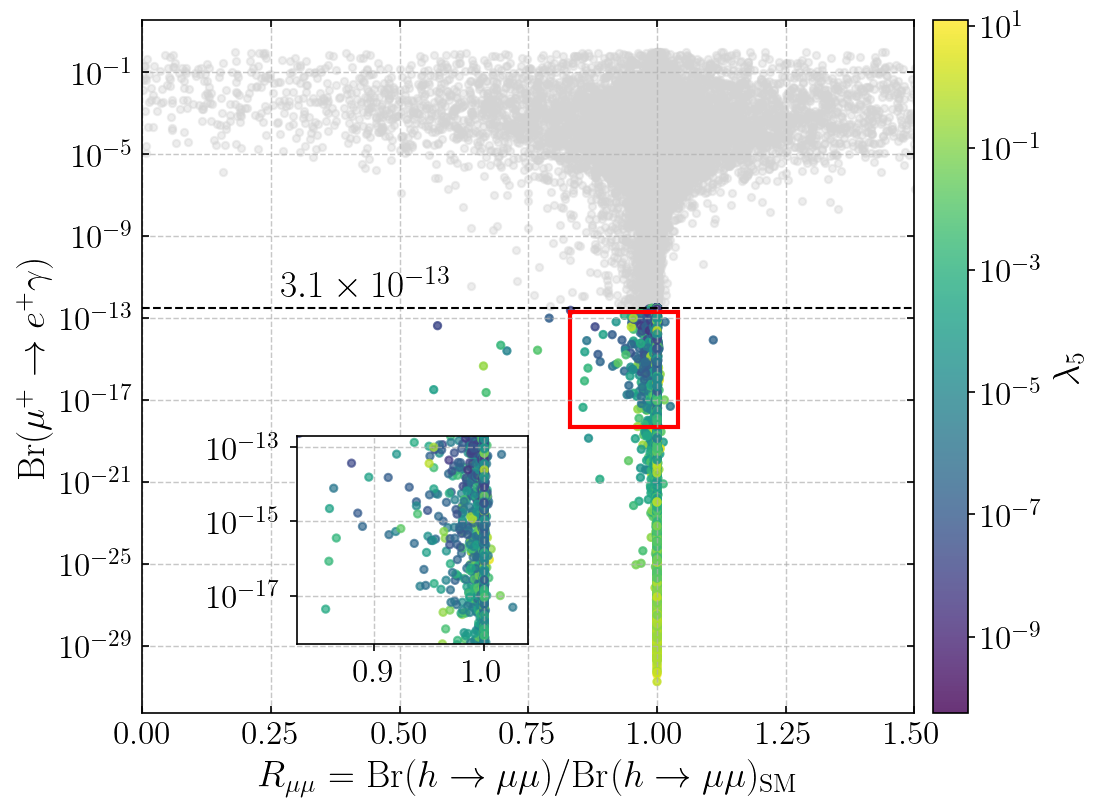}
    \caption{}
    \label{fig:plot_muTOegamma_hTOmumu_lambda5_discriminate_with_line_with_zoom}
\end{subfigure}%
\hfill
\begin{subfigure}[b]{0.49\textwidth}
    \includegraphics[width=\textwidth]{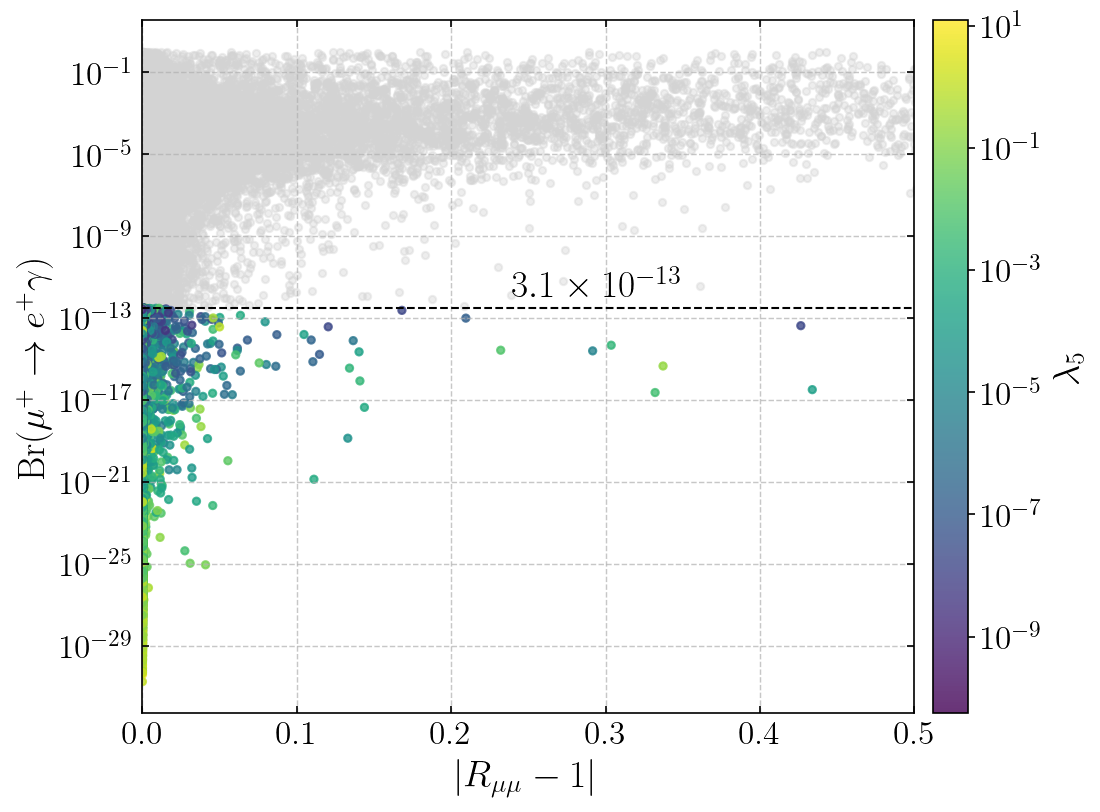}
    \caption{}
    \label{fig:plot_muTOegamma_hTOmumu_deviations_lambda5}
\end{subfigure}
\caption{The left panel shows the relation between the observables $h \to \mu^{+}\mu^{-}$ and $\mu^{+} \to e^{+}\gamma$. Points for which $\text{Br}(\mu^{+} \to e^{+}\gamma) > 3.1 \times 10^{-13}$ are gray and translucent. The minimum value reached by $\text{Br}(\mu^{+} \to e^{+}\gamma)$ is determined by the upper value of the masses $M_j$. While the right panel shows the possible values for variable $R_{\mu\mu}$, the right panel shows the corresponding deviations, i.e., $|R_{\mu\mu} - 1|$.}
\label{fig:plot_Rmumu_vs_muTOegamma}
\end{figure}

Figure \ref{fig:plot_Rmumu_vs_muTOegamma} provides another perspective on the parameter space. Here, the scan incorporates constraints from diphoton Higgs decay but excludes $\mu^+ \to e^+ \gamma$. The analysis is presented in the plane of $\text{BR}(\mu^+ \to e^+ \gamma)$ vs.~$R_{\mu\mu}$, with the parameter $\lambda_5$ highlighted for illustrative purposes. Points that violate the MEG II bound are shown in gray. Most allowed points exhibit $R_{\mu\mu}$ values slightly smaller than the SM prediction, with only a few exceeding unity. This behavior is more clearly illustrated in Fig.~\ref{fig:plot_muTOegamma_hTOmumu_deviations_lambda5}.

Despite these constraints, the model accommodates sizable $R_{\mu\mu}$ deviations in specific regions of the parameter space. As we will see in the next Section \ref{subsection:key_diagrams_and_key_parameters}, achieving these deviations, however, requires careful tuning of some parameters to simultaneously satisfy the constraints from $h \to \gamma \gamma$, $\mu^{+}\to e^{+}\gamma$ and $h \to \gamma \gamma$.

\subsection{Key diagrams and key parameters}
\label{subsection:key_diagrams_and_key_parameters}

\begin{figure}[!t]
\centering
\begin{subfigure}{.30\textwidth}
    \centering
    \includegraphics[width=1.0\linewidth]{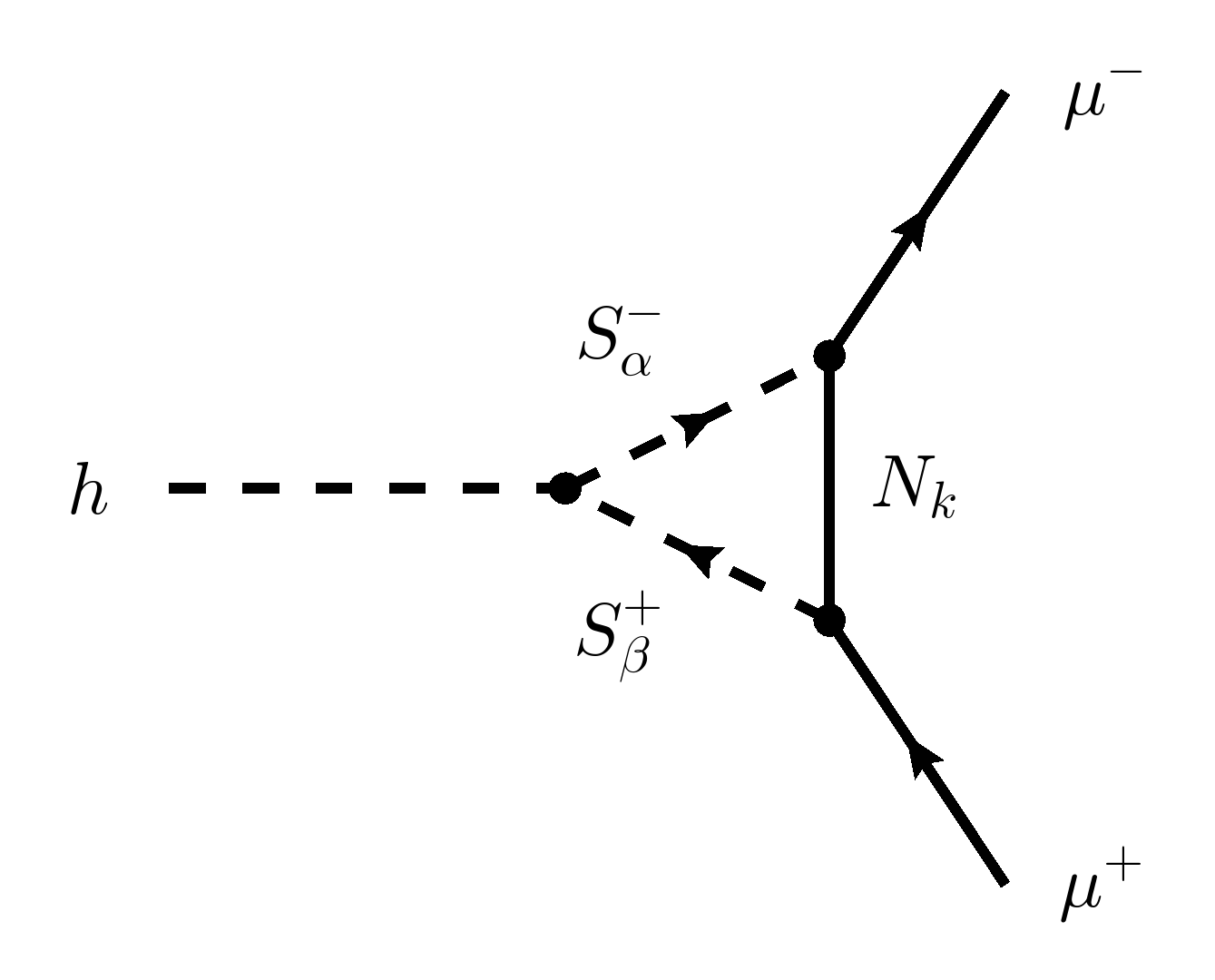}
    \caption{}
    \label{fig:key_process_hTOmumu}
\end{subfigure}%
    \hspace{2.5cm}
\begin{subfigure}{.30\textwidth}
    \centering
    \includegraphics[width=1.0\linewidth]{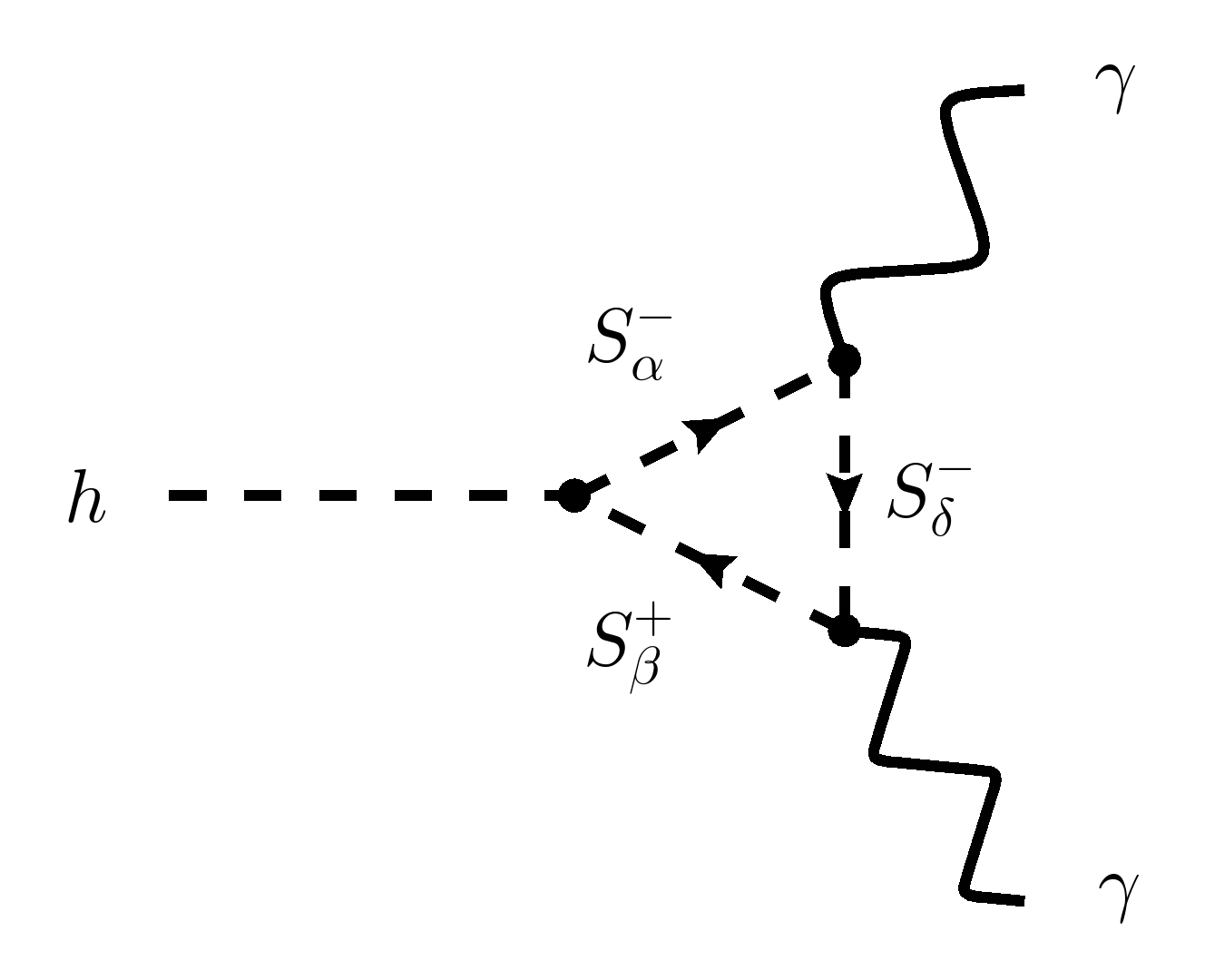}
    \caption{}
    \label{fig:key_process_hTOgammagamma}
\end{subfigure}\\
\begin{subfigure}{.30\textwidth}
    \centering
    \includegraphics[width=1.0\linewidth]{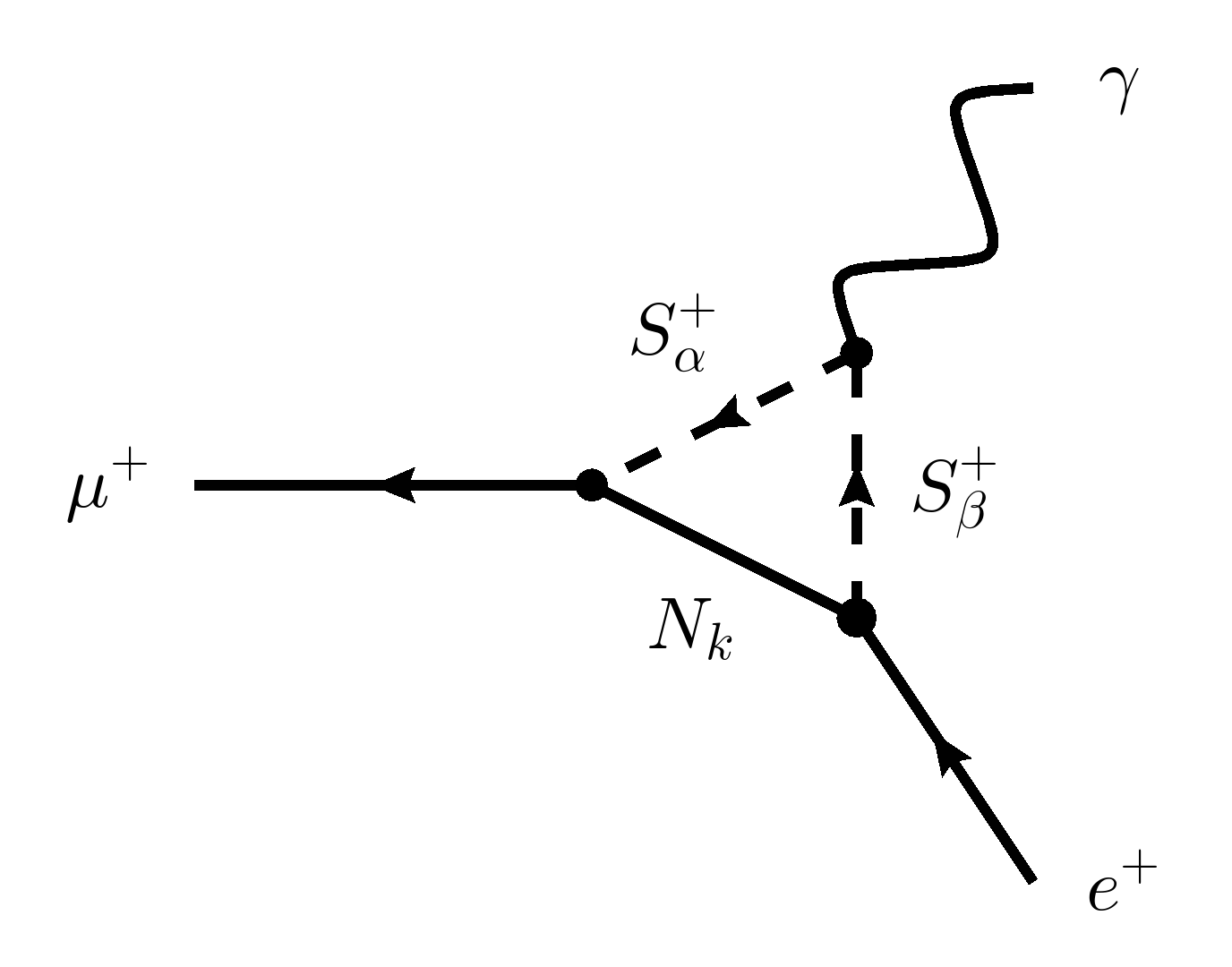}
    \caption{}
    \label{fig:key_process_muTOegamma}
\end{subfigure}%
    \hspace{2.5cm}
\begin{subfigure}{.30\textwidth}
    \centering
    \includegraphics[width=1.0\linewidth]{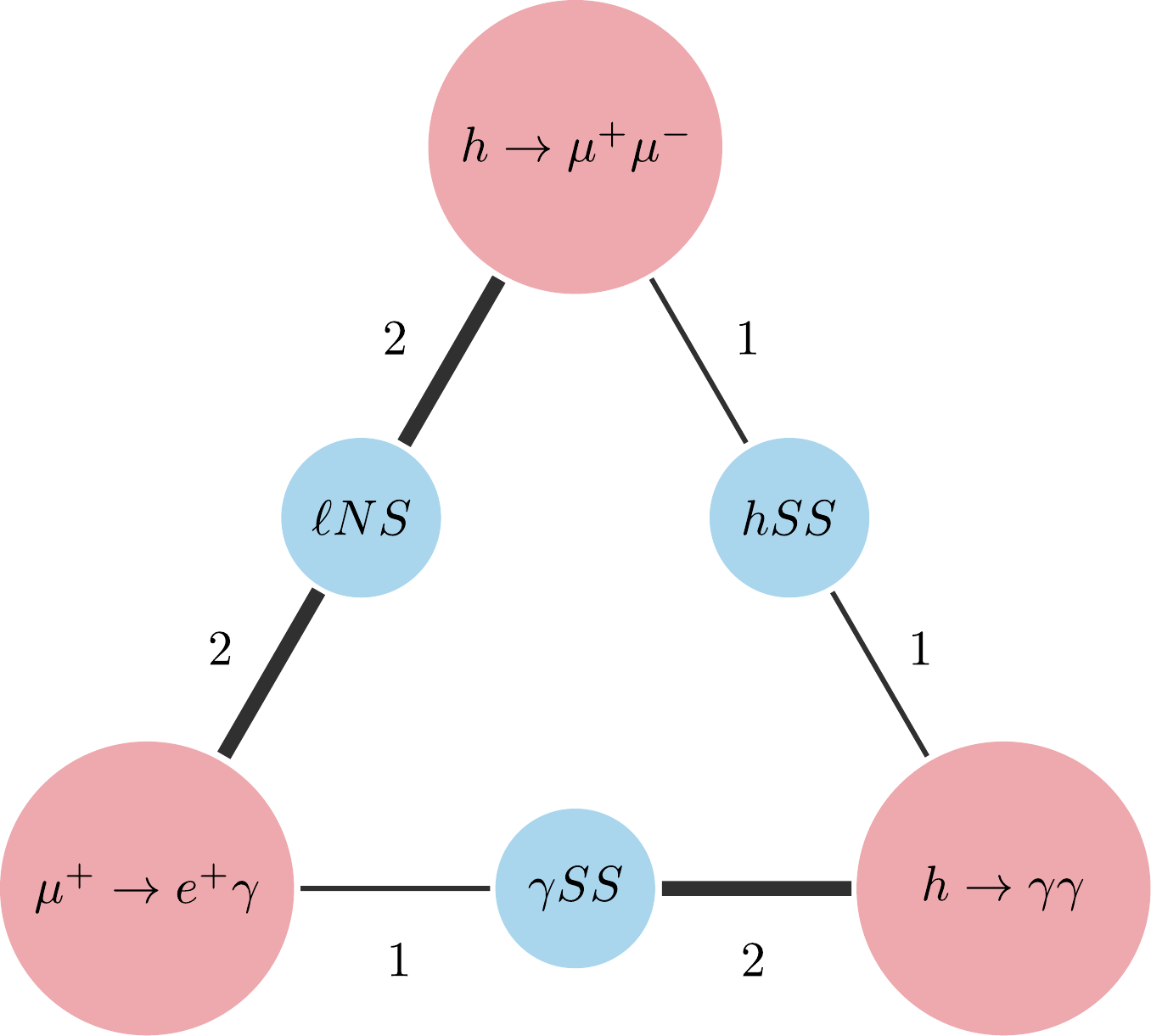}
    \caption{}
    \label{fig:graph_three_processes}
\end{subfigure}
\caption{The top left, top right, and bottom left panels depict the one-loop contributions to the processes $h \to \mu^+ \mu^-$, $h \to \gamma \gamma$, and $\mu^+ \to e^+ \gamma$, respectively. These processes are interconnected through three common vertices: $\ell N S$, $hSS$, and $\gamma SS$. The bottom right graph illustrates the connections between these vertices and the corresponding diagrams. The numbers along the edges indicate the number of times each vertex appears in the respective process.}
\label{fig:three_key_diagrams}
\end{figure}

The one-loop level processes $h \to \mu^+ \mu^-$ (Fig.~\ref{fig:key_process_hTOmumu}), $h \to \gamma \gamma$ (Fig.~\ref{fig:key_process_hTOgammagamma}), and $\mu^{+} \to e^+ \gamma$ (Fig.~\ref{fig:key_process_muTOegamma}) are strongly interconnected and are governed by three key vertices: $\ell N S$, $h\gamma\gamma$, and $hSS$ (see Appendix \ref{appendix:vertex_factors}). Specifically, the Higgs dimuon decay $h \to \mu^+ \mu^-$ is driven by $\ell N S$ and $hSS$; the LFV process $\mu^{+} \to e^+ \gamma$ depends on $\ell N S$ and $\gamma SS$; and the Higgs digamma decay $h \to \gamma\gamma$ is determined by $\gamma SS$ and $hSS$. Modifying one vertex to enhance or suppress a particular process will similarly affect other processes sharing the same vertex. These interconnections are summarized diagrammatically in Fig.~\ref{fig:graph_three_processes}, where the numbers along the edges indicate the frequency with which each vertex appears in the diagrams.

Deviations in the Higgs decay rate are driven by the one-loop diagram shown in Fig.~\eqref{fig:key_process_hTOmumu} and captured by the quotient of Eq.~\eqref{eq:hTOmumu_branching_quotient}. The significance of this loop contribution is controlled by a single complex parameter, $\vartheta^L_{\mu\mu}$ (with $\vartheta^R_{\mu\mu}$ being a dependent parameter through Eq.~\eqref{eq:relation_between_varthetaLR_ijkalphabeta_coefficients}), which can contribute comparably to the SM tree-level value, $\vartheta^\text{SM}_{\mu\mu} \simeq 4.45203 \times 10^{-4}$, in some areas of the parameter space. This sizable loop contribution occurs when the couplings $h S S$ and $N S \ell$ are enhanced (see Appendixes~\ref{appendix:vertex_factors_scalar_sector} and \ref{appendix:vertex_factors_yukawa_sector}, respectively).

The $hSS$ vertex is enhanced when $\lambda_3$, $\lambda_{H\phi}$ and $\mu$ take large values. However, these parameters also control the masses of the charged scalars $S^{\pm}_\alpha$. As these parameters increase, so do the scalar masses, which in turn amplify the contribution from the functions $xf(x^2)$ and $\tilde{f}(x^2)$ in Eq.~\eqref{eq:functions_f_and_ftilde} that leads to a larger value of $\text{Br}(\mu^{+} \to e^{+}\gamma)$. Consequently, maximizing $R_{\mu\mu}$ while keeping $\mu^{+} \to e^{+} \gamma$ within experimental bounds requires a careful balance of these parameters.

\begin{figure}[!t]
\centering
\begin{subfigure}[b]{0.33\textwidth}
    \includegraphics[width=\textwidth]{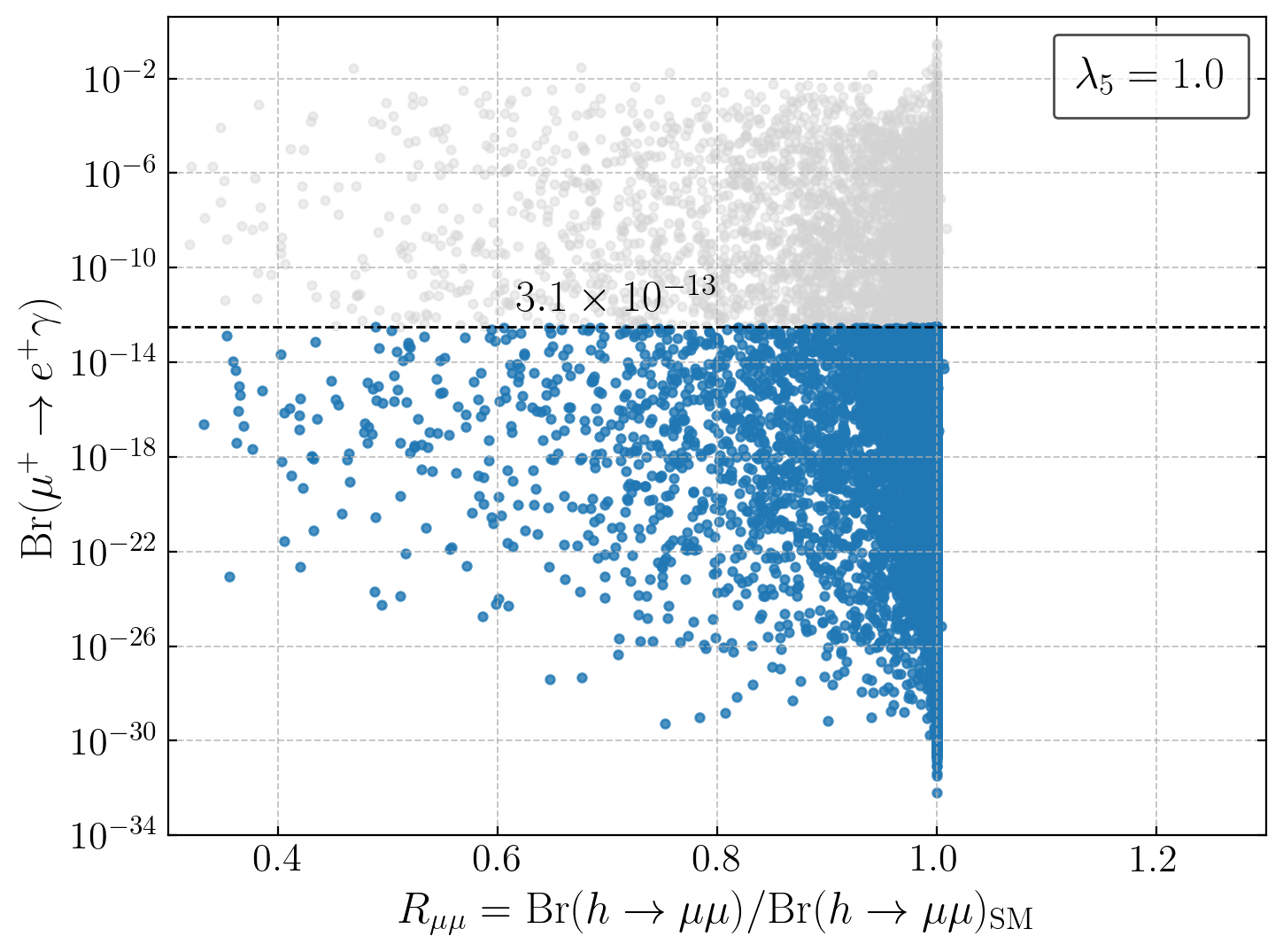}
    \caption{}
    \label{fig:plot_muTOegamma_hTOmumu_lambda5_1e0_discriminate_with_line_lambda5_kappa}
\end{subfigure}%
\hfill
\begin{subfigure}[b]{0.33\textwidth}
    \includegraphics[width=\textwidth]{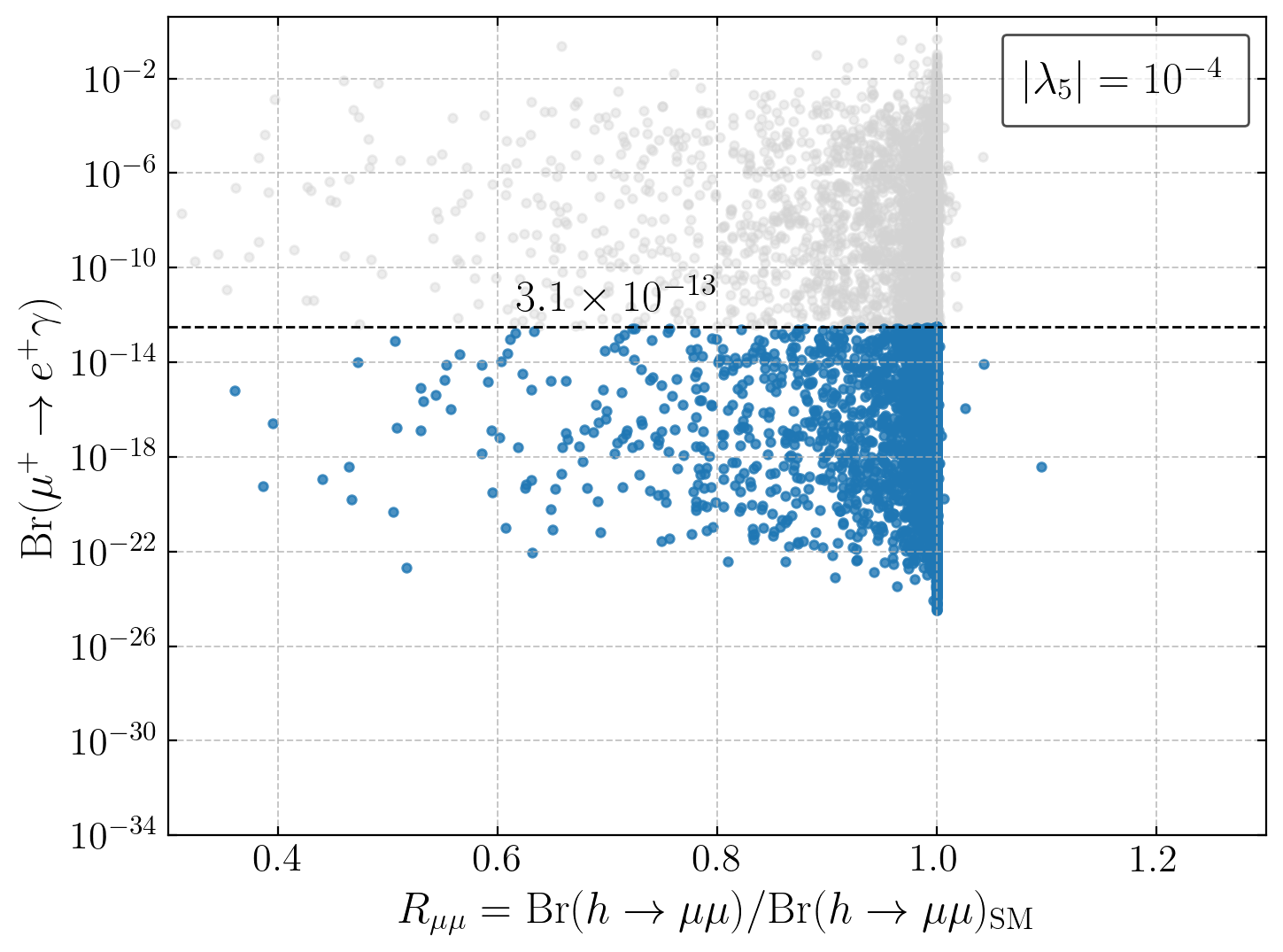}
    \caption{}
    \label{fig:plot_muTOegamma_hTOmumu_lambda5_1em4_discriminate_with_line}
\end{subfigure}%
\hfill
\begin{subfigure}[b]{0.33\textwidth}
    \includegraphics[width=\textwidth]{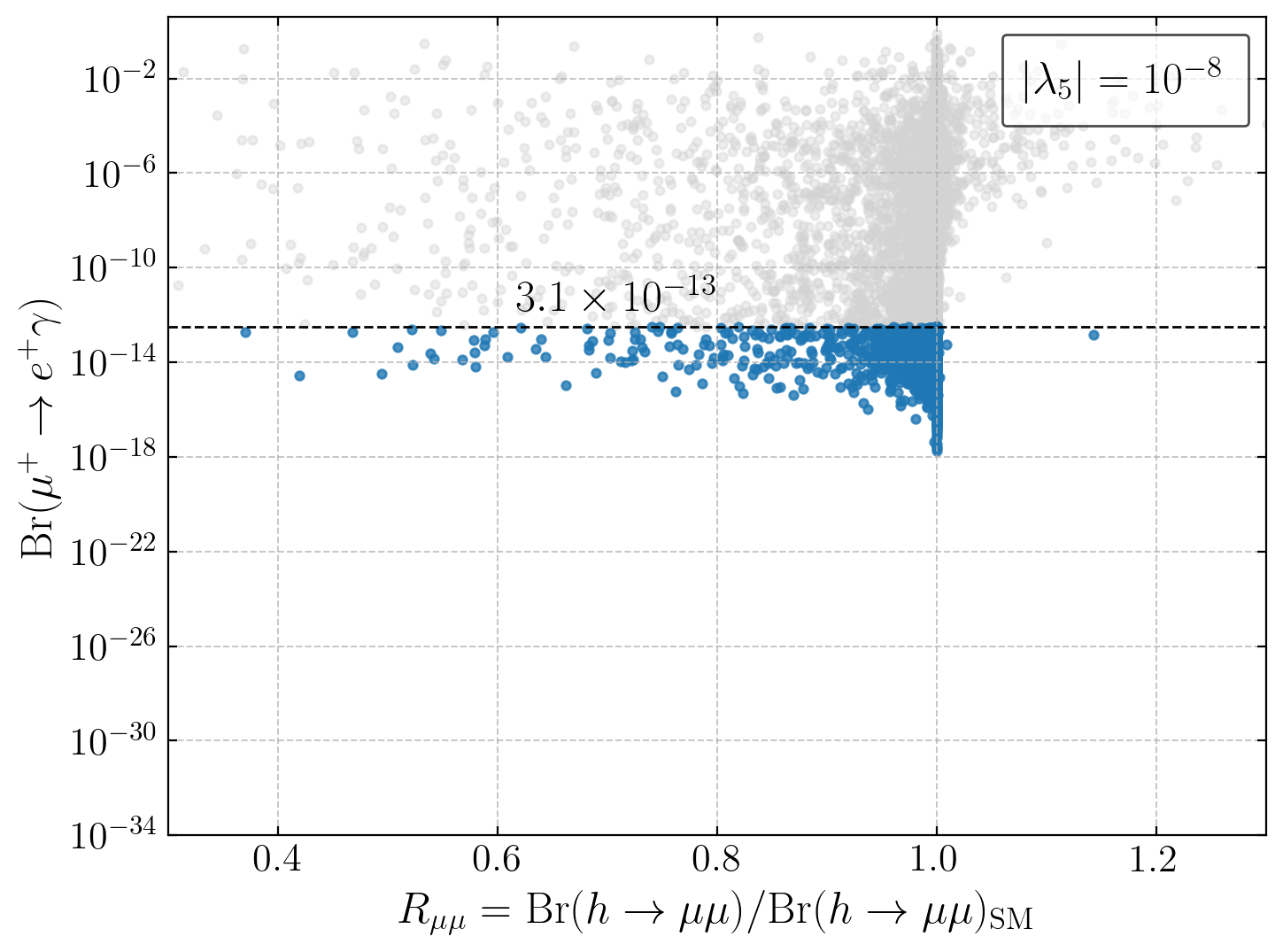}
    \caption{}
    \label{fig:plot_muTOegamma_hTOmumu_lambda5_1em8_discriminate_with_line}
\end{subfigure} \\
\begin{subfigure}[b]{0.33\textwidth}
    \includegraphics[width=\textwidth]{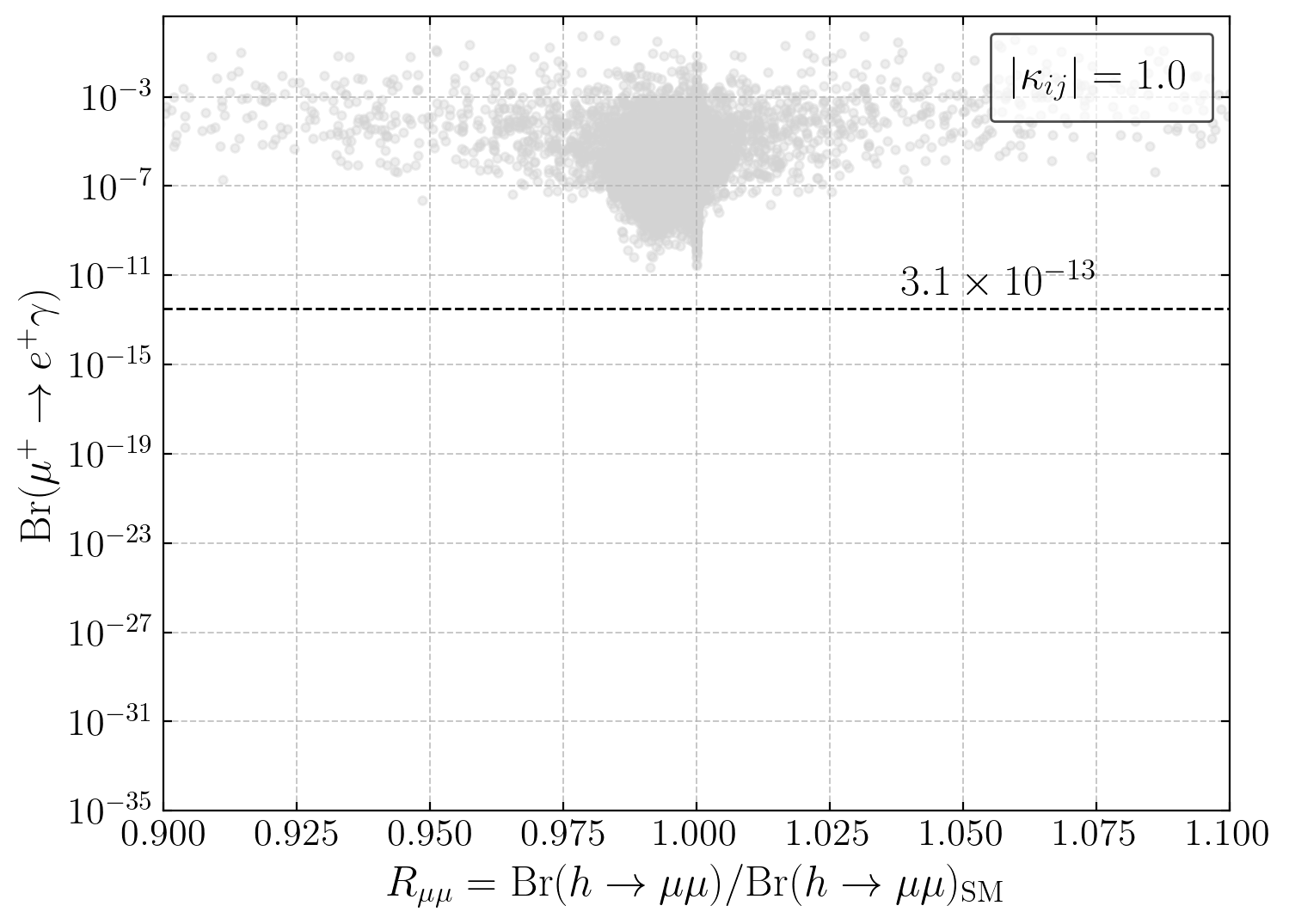}
    \caption{}
    \label{fig:plot_muTOegamma_hTOmumu_kappa_1e0_discriminate_with_line}
\end{subfigure}%
\hfill
\begin{subfigure}[b]{0.33\textwidth}
    \includegraphics[width=\textwidth]{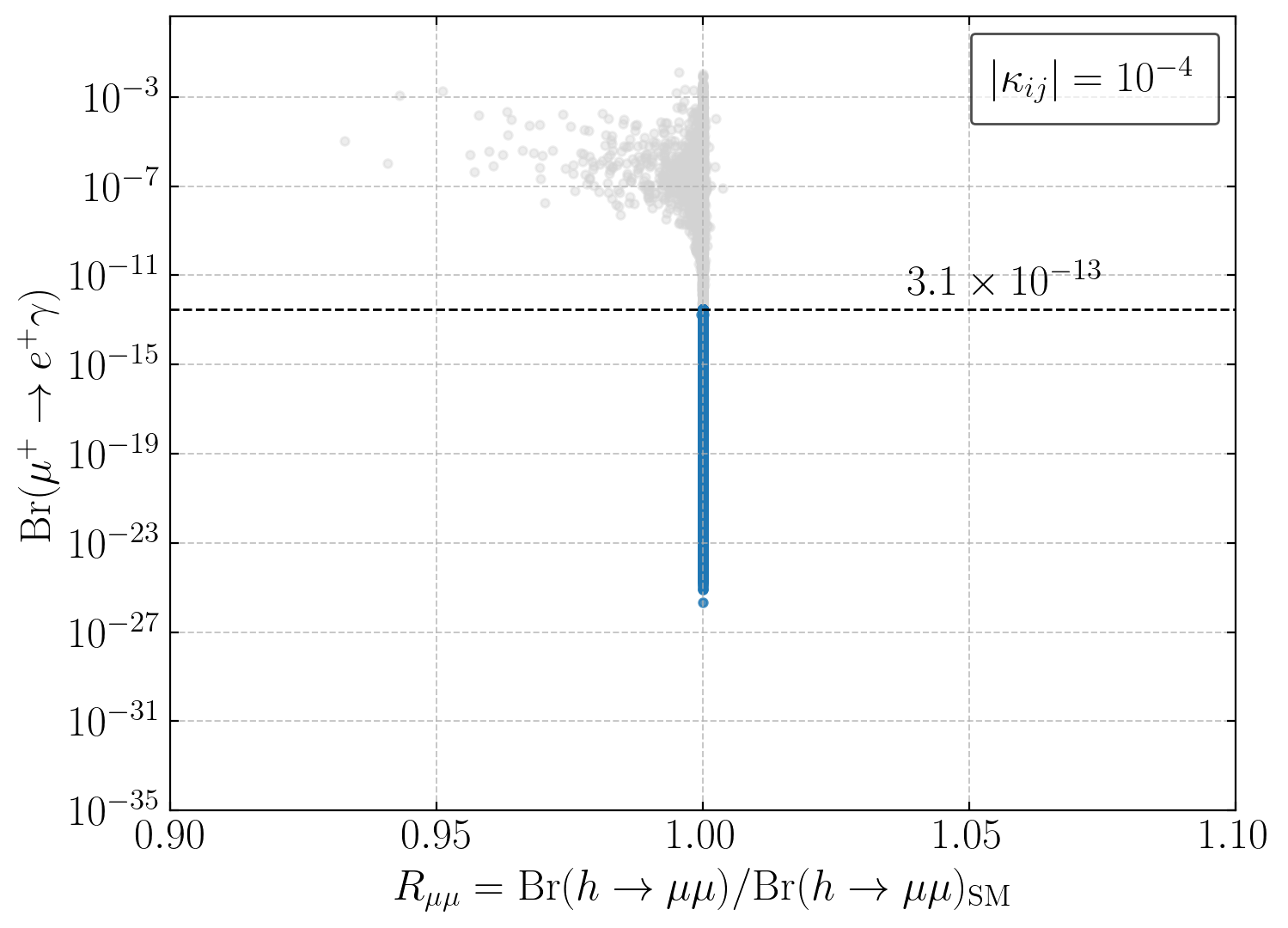}
    \caption{}
    \label{fig:plot_muTOegamma_hTOmumu_kappa_1em4_discriminate_with_line}
\end{subfigure}%
\hfill
\begin{subfigure}[b]{0.33\textwidth}
    \includegraphics[width=\textwidth]{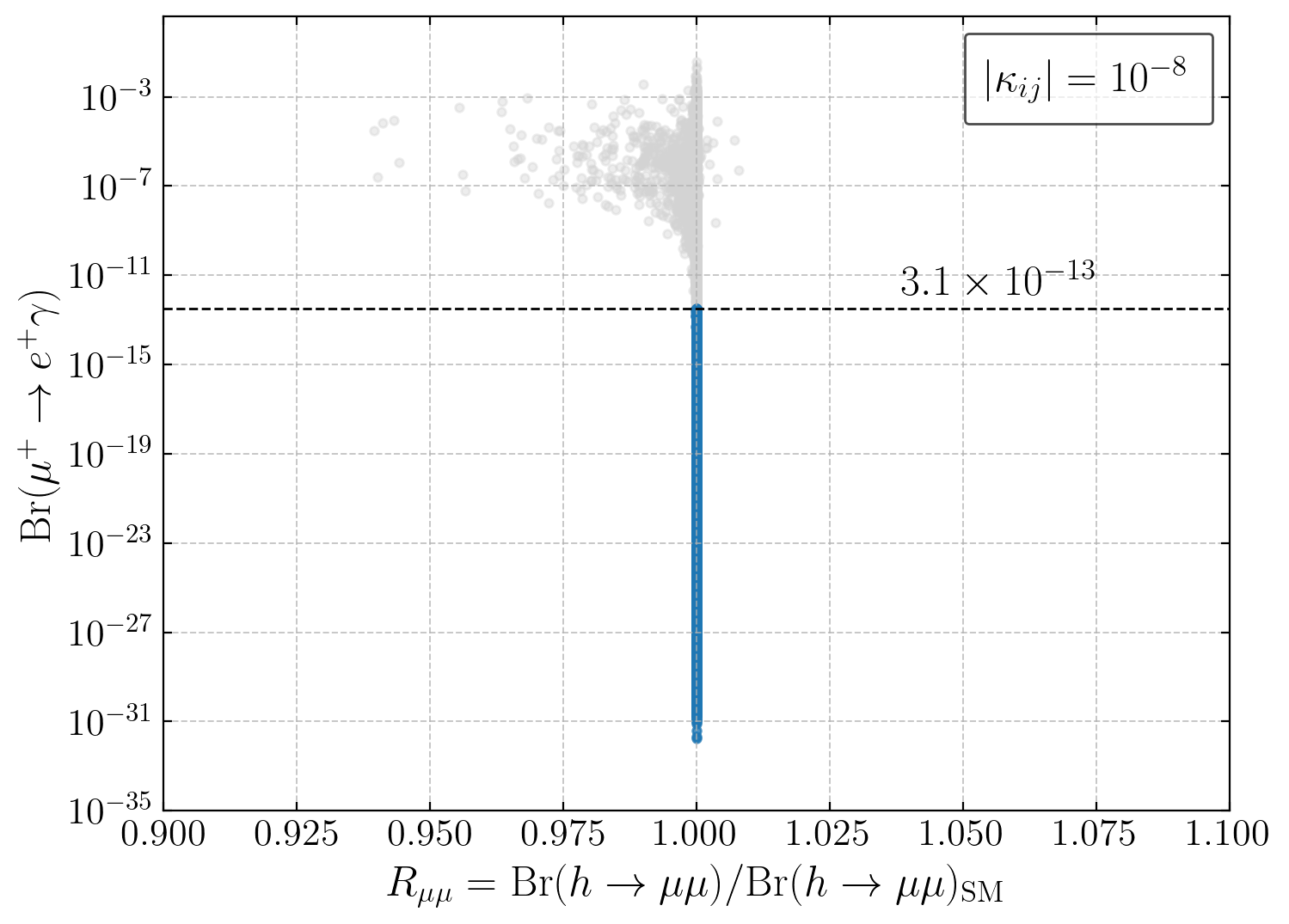}
    \caption{}
    \label{fig:plot_muTOegamma_hTOmumu_kappa_1em8_discriminate_with_line}
\end{subfigure}
\caption{Relation between the observables $h \to \mu^{+}\mu^{-}$ and $\mu^{+} \to e^{+}\gamma$ for three different fixed values of $\lambda_5$ (top row) and three different fixed values of $|\kappa_{ij}|$ (bottom row). Note that the scans shown in these six panels do not include the constraint from $h \to \gamma\gamma$. This exclusion is intentional, as the $h \to \gamma\gamma$ constraint, as discussed in the text, is very stringent. The focus of these panels is to highlight the effects described in the main discussion.}
\label{fig:plot_muTOegamma_hTOmumu_scans}
\end{figure}

On the other hand, the $NS\ell$ vertex depends on the couplings $\kappa_{ij}$ and $\lambda_5$ (the latter appears through the Yukawas $y_{ij}$, related by neutrino masses, see Sec.~\ref{subsection:neutrino_mass_generation}) via the coefficients $g^{L,R}_{k\alpha i}$ of Eqs.~\eqref{eq:auxiliary_coupling_definition}. Generally, increasing $\kappa_{ij}$ and decreasing $\lambda_5$ enhances this vertex, boosting the contribution to $h \to \mu^{+}\mu^{-}$, but also increases the branching ratio of $\mu^{+} \to e^{+}\gamma$ as it can be seen in Figs.~\ref{fig:plot_muTOegamma_hTOmumu_scans}. Again, the non-trivial dependencies prevent setting these parameters to extreme values without compromising control over $\mu^{+} \to e^{+}\gamma$.

Given these insights, the free parameters $|\kappa_{ij}|, |\lambda_5|, |\lambda_3|, |\lambda_{H\phi}|, \mu$ are key to assessing the model's viability in explaining $h \to \mu^{+} \mu^{-}$ while keeping $\mu^{+} \to e^{+} \gamma$ under control. In particular, some relations between the variables $|\kappa_{ij}|$ and $\lambda_5$ are shown in Figs.~\ref{fig:plot_muTOegamma_hTOmumu_scans}.

The three panels on the top row of this figure show the behavior of $\text{Br}(\mu^{+} \to e^{+}\gamma)$ and $R_{\mu\mu}$ for values of $\lambda_5=1.0,\, 10^{-4}$ and $10^{-8}$ respectively. In each panel, the value of the $\kappa_{11}$ parameter is shown as a color palette. It is manifest that increasing values of $\lambda_5$ drives $\text{Br}(\mu^{+} \to e^{+}\gamma)$ to lower values. Furthermore, smaller values of $\kappa_{11}$ tend to reduce the LFV process. This effect is clearly seen in the panels of the bottom row of Fig.~\ref{fig:plot_muTOegamma_hTOmumu_scans}. In this case we have fixed three values of $\kappa_{11}=1,10^{-4}, 10^{-8}$ while allowing $\lambda_5$ to expand in a range from $10^{-10}$ to 1. We can notice that the behavior is different from the previous case. For greater values of $\kappa_{11}$, the LFV process increases. We can also see that once $\kappa_{11}$ is fixed, smaller values of $\lambda_5$ increase substantially the value of the LFV process.

Finally, it is worth pointing out that the minimum value of $\mu^{+} \to e^{+}\gamma$ depends strongly on the maximum value set for the masses of the singlet fermions, $M_j$; the larger the value of these masses, the smaller the minimum value of BR($\mu^{+} \to e^{+}\gamma$). This can be seen in the coefficients of Eq.~\eqref{eq:coefficients_CRji} due to the behavior of the functions $xf(x^2)$ and $\tilde{f}(x^2)$ (with $x \sim M_j$). In both cases, both functions vanish as $x \to \infty$.

\subsection{Linearization of deviations}
\label{subsection:linearization_of_deviations}

For most points identified in the numerical analysis, the loop amplitude shown in Fig.~\ref{fig:hTOll_loop_triangle} is significantly smaller than its SM tree-level counterpart. This suppression arises naturally from the loop structure, as the amplitude scales with the fourth power of coupling factors such as $|\kappa|^4$, $|y|^4$, or $|\kappa|^2|y|^2$. Consequently, it is unsurprising that, for the majority of the scanned points, the parameters $|\mathfrak{Re}[\vartheta^L_{\mu\mu}]|$ and $|\mathfrak{Im}[\vartheta^L_{\mu\mu}]|$ are much smaller than the SM value, $\vartheta^\text{SM}_{\mu\mu} \simeq 4.45 \times 10^{-4}$.

\begin{figure}[!t]
\centering
\begin{subfigure}[t]{.5\textwidth}
  \centering
  \includegraphics[width=\linewidth,valign=t]{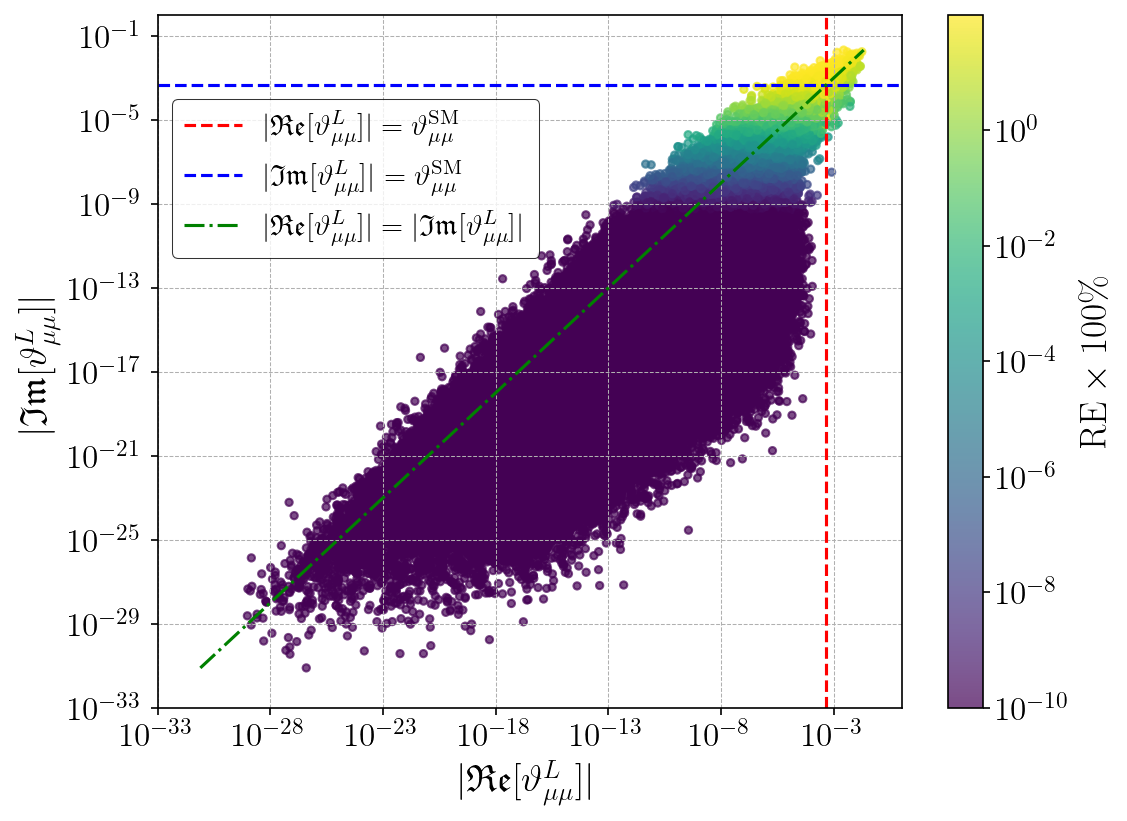}
  \caption{}
\end{subfigure}%
\begin{subfigure}[t]{.5\textwidth}
  \centering
  \includegraphics[width=0.93\linewidth,valign=t]{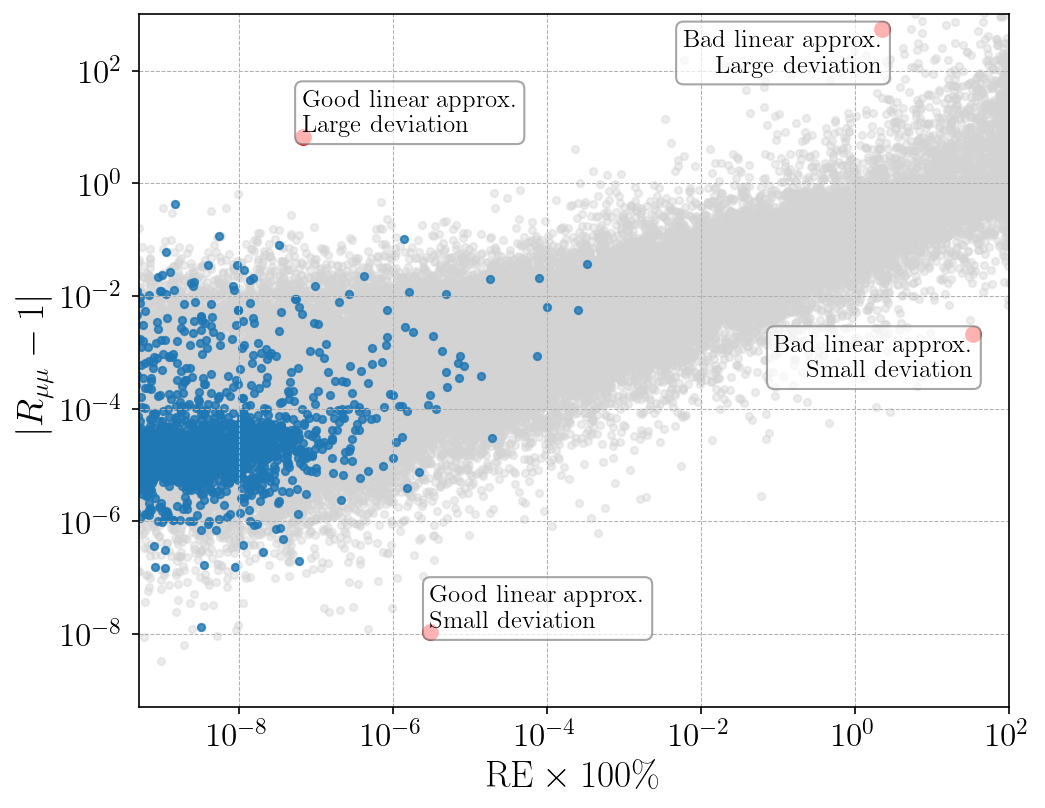}
  \caption{}
\end{subfigure}
\caption{The left panel shows $|\mathfrak{Re}[\vartheta^L_{\mu\mu}]|$ versus $|\mathfrak{Im}[\vartheta^L_{\mu\mu}]|$, with the percentage error of the approximation $R_{\mu\mu} \simeq R^\text{linear}_{\mu\mu}$ represented by color. For most points, the real and imaginary parts of the loop-generated parameters  $\vartheta^L_{\mu\mu}$ are several orders of magnitude smaller than the SM value $\vartheta^\text{SM}_{\mu\mu} \simeq 4.45 \times 10^{-4}$. For these points, the percentage error is negligible. Note that the LFV constraint from $\mu^+ \to e^+ \gamma$ has not been applied in this panel. The right panel shows the percentage error of the approximation $R_{\mu\mu} \simeq R^\text{linear}_{\mu\mu}$ as a function of the deviations in $R_{\mu\mu}$. The general trend, ``the smaller the deviation, the better the linear approximation,'' can be noted, although some points deviate from this rule. Points that satisfy the MEG II bound [Eq.~\eqref{eq:MEG_bound}] are highlighted in color.}
\label{fig:plot_support_approximations}
\end{figure}

For points where both the real and imaginary parts of $\vartheta^L_{\mu\mu}$ are sufficiently small, the quadratic terms in Eq.~\eqref{eq:hTOmumu_branching_quotient} can be safely neglected. Under these conditions, the deviations in $R_{\mu\mu}$ are well-approximated by:
\begin{equation}
\label{eq:Rmu_approximation_for_boring_points}
    R_{\mu\mu} \simeq R^{\text{linear}}_{\mu\mu} := 1 + 2\Biggl(\frac{\mathfrak{Re}[\vartheta^L_{\mu\mu}]}{\vartheta^{\text{SM}}_{\mu\mu}}\Biggr)
\end{equation}
As shown in the left panel of Fig.~\ref{fig:plot_support_approximations}, the majority of points in the scan satisfy the condition $|\mathfrak{Re}[\vartheta^L_{\mu\mu}]|, |\mathfrak{Im}[\vartheta^L_{\mu\mu}]| \ll \vartheta^\text{SM}_{\mu\mu}$. For these points, Eq.~\eqref{eq:Rmu_approximation_for_boring_points} provides a good approximation of $R_{\mu\mu}$. For example, we found that for such points, the relative error, $\mathrm{RE} = | (R_{\mu\mu} - R^{\mathrm{linear}}_{\mu\mu}) / R_{\mu\mu} |$, is less than $1\%$ for $98.99\%$ of cases, and the error drops below $0.001\%$ for $96.13\%$ of cases as it can be seen from the right panel of Fig.~\ref{fig:plot_support_approximations}. For these points, the constructive or destructive nature of the loop interference depends solely on the sign of $\mathfrak{Re}[\vartheta^L_{\mu\mu}]$.

Note that Eq.~\eqref{eq:Rmu_approximation_for_boring_points} can describe both small and large deviations in $R_{\mu\mu}$. As seen in the right panel of Fig.~\ref{fig:plot_support_approximations}, the general trend is that the smaller the deviation in $R_{\mu\mu}$, the better the linear approximation. However, exceptions to this rule of thumb do exist.

When restricting the analysis to points that satisfy the MEG II bound (see Eq.~\eqref{eq:MEG_bound}) highlighted in light blue, we observe that the linear approximation holds exceptionally well. For almost all of these points, the error is of order $\sim \mathcal{O}(10^{-4})$ or smaller, confirming the reliability of Eq.~\eqref{eq:Rmu_approximation_for_boring_points} in describing $R_{\mu\mu}$ deviations for these particular points.

\subsection{Analytical behavior of \texorpdfstring{$R_{\mu\mu}$}{Rmumu}}
\label{subsection:analytical_behaviour_of_R_mumu}

Let us analyze now the analytical behavior of $R_{\mu\mu}$. For simplicity, this section focuses exclusively on points that satisfy all the constraints outlined in Section \ref{section:constraints}, \textit{i.e.}, the colored points in Fig.~\ref{fig:plot_muTOegamma_hTOmumu_deviations_lambda5} and the right panel of Fig.~\ref{fig:plot_support_approximations}. These points adhere to the linear approximation in Eq.~\eqref{eq:Rmu_approximation_for_boring_points}.

Using the relations in Eq.~\eqref{eq:relations_between_C_coefficients}, the real parts of the coefficients $C^{(2)}_{\mu\mu}$, $C^{(3)}_{\mu\mu}$, $C^{(5)}_{\mu\mu}$, and $C^{(6)}_{\mu\mu}$. cancel exactly. This leaves only $C^{(1)}_{\mu\mu}$, $C^{(4)}_{\mu\mu}$ and $C^{(7)}_{\mu\mu}$ as relevant contributors. To better understand their behavior, we analyze their analytic forms in special cases, starting from their smallest constitutive block coefficients given in Appendix \ref{appendix:coefficients_hTOlilj}. These are given by
\begin{subequations}\label{eqs:C_coefficients_mumu}
\begin{align}
    C^{(1)}_{\mu\mu,k,\alpha\beta} &\simeq 2m^4_hm^3_\mu\Bigl(|\kappa_{k2}|^2R_{1\alpha}R_{1\beta} + |y_{2k}|^2R_{2\alpha}R_{2\beta}\Bigr)B_0(m^2_h, m_{S_\alpha}, m_{S_\beta})  \, , \label{eq:approximation_for_C1_non_degenerate_scenario} \\
    C^{(4)}_{\mu\mu,k,\alpha\beta} &\simeq m^2_hm^2_\mu\Biggl[-(m^2_h + m^2_{S_\alpha} - m^2_{S_\beta})|\kappa_{k2}|^2R_{1\alpha}R_{1\beta} \nonumber \\
    & + (m^2_h - m^2_{S_\alpha} + m^2_{S_\beta})|y_{2k}|^2R_{2\alpha}R_{2\beta}\Biggr]\ln\frac{m^2_{S_\alpha}}{m^2_{S_\beta}} \, , \label{eq:approximation_for_C4_non_degenerate_scenario} \\
    C^{(7)}_{\mu\mu,k,\alpha\beta} &\simeq -2m^2_hm^2_\mu\Biggl\{-m_\mu\Bigl[2m^2_\mu(-m^2_{S_\alpha} + m^2_{S_\beta}) + m^2_h(-m^2_{S_\beta} + M^2_k)\Bigr]|\kappa_{k2}|^2R_{1\alpha}R_{1\beta} \nonumber \\
        &\quad  - m_\mu\Bigl[2m^2_\mu(-m^2_{S_\beta} + m^2_{S_\alpha}) + m^2_h(-m^2_{S_\alpha} + M^2_k)\Bigr]|y_{2k}|^2R_{2\alpha}R_{2\beta} \nonumber \\
        &\quad + M_km^4_h\kappa^\ast_{k2}y^{\ast}_{2k}R_{1\beta}R_{2\alpha}\Biggr\}C_0(m^2_h, m^2_\mu, m^2_\mu|\, m_{S_\beta}, m_{S_\alpha}, M_k) \, , \label{eq:approximation_for_C7_non_degenerate_scenario}
\end{align}
\end{subequations}
where we have neglected terms proportional to $m^2_\ell/m^2_h$ with $\ell = e,\mu$.

To further simplify the behavior of the coefficients, note that $C^{(4)}_{\mu\mu,k,\alpha\beta}$ is highly sensitive to the mass splitting between the charged scalars. In the nearly degenerate case where $m_{S^\pm_1} \simeq m_{S^\pm_2} \equiv m_S \gg m_h$ and $m_{S_2} - m_{S_1} < 10^{-1}\ \text{GeV}$, the real part of $C^{(4)}_{\mu\mu}$ becomes negligible compared to $\mathfrak{Re}[C^{(1)}_{\mu\mu}]$ and $\mathfrak{Re}[C^{(7)}_{\mu\mu}]$. Thus, in the fully degenerate case of the charged scalars, the expressions given by Eqs.~\eqref{eqs:C_coefficients_mumu} can be further simplified, leading to
\begin{subequations}
\begin{align}
    C^{(1)}_{\mu\mu,k,\alpha\beta} &\simeq -4m^4_hm^3_\mu\left(|\kappa_{k2}|^2R_{1\alpha}R_{1\beta} + |y_{2k}|^2R_{2\alpha}R_{2\beta}\right) \, ,\label{eq:approximation_for_C1_degenerate_scenario} \\
    C^{(4)}_{\mu\mu,k,\alpha\beta} &\simeq 0 \, , \label{eq:approximation_for_C4_degenerate_scenario} \\
    C^{(7)}_{\mu\mu,k,\alpha\beta} &\simeq -\frac{1}{2}\left\{(M^2_k - m^2_S)C^{(1)}_{\mu\mu,k,\alpha\beta} + 4m^6_hm^2_\mu M_k\kappa^\ast_{k2}y^{\ast}_{2k}R_{1\beta}R_{2\alpha}\right\} \nonumber \\
    &\quad \times C_0(m^2_h, m^2_\mu, m^2_\mu|\, m_S, m_S, M_k) \, , \label{eq:approximation_for_C7_degenerate_scenario}
\end{align}
\end{subequations}
where we have use the fact that $B_0(m^2_h, m_{S_\alpha}, m_{S_\beta}) \to -2$ when $m_{S_\alpha} \simeq m_{S_\beta} \gg m_h$. Furthermore, the coefficient $C^{(4)}_{\mu\mu,k,\alpha\beta}$ vanishes due to the logarithmic dependence. Under these conditions, deviations in $R_{\mu\mu}$ are dominated solely by the coefficient $C^{(1)}$.

From our previous considerations, points exhibiting sizable deviations while remaining compatible with all constraints are characterized by small mixing angles $\sin\theta \ll 1$. In this scenario we have $S^{\pm}_1 \simeq \phi^{\pm}$ and $S^{\pm}_2 \simeq \eta^{\pm}$, with corresponding masses $m_{S^\pm_1} \simeq \tilde{m}_\phi$ and $m_{S^\pm_2} \simeq \tilde{m}_\eta$. Additionally, this implies that $\mu \simeq 0$. Under these conditions, the expressions for the $C$-coefficients in Eqs.~\eqref{eqs:C_coefficients_mumu} simplify significantly when performing the summation over the indices $k$, $\alpha$, and $\beta$:
\begin{subequations}
\begin{align}
    C^{(1)}_{\mu\mu} &\simeq -\frac{v m_\mu}{16 \pi^2 m^2_h} \Biggl\{\lambda_{H\phi}\Bigl(|\kappa_{12}|^2 + |\kappa_{22}|^2 + |\kappa_{32}|^2\Bigr) |B_{11}| + \lambda_3\Bigl(|y_{21}|^2 + |y_{22}|^2 + |y_{23}|^2\Bigr) |B_{22}|\Biggr\} \, , \label{eq:coefficient_C1_small_angle} \\
    C^{(4)}_{\mu\mu} &\simeq 0 \, , \\
    C^{(7)}_{\mu\mu} &\simeq \frac{v m_\mu}{16 \pi^2 m^2_h} \sum^3_{k=1} \Biggl\{ \lambda_{H\phi}\Bigl(M^2_k - m^2_{S^\pm_1}\Bigr)|\kappa_{k2}|^2 |C_{11k}| + \lambda_3\Bigl(M^2_k - m^2_{S^\pm_2}\Bigr)|y_{2k}|^2 |C_{22k}| \Biggr\} \, .
\end{align}
\end{subequations}
Here $B_{\alpha\beta} := B_0(m^2_h, m_{S^{\pm}_\alpha}, m_{S^{\pm}_\beta})$ and $C_{\beta \alpha k} := C_0(m^2_h, m^2_\mu, m^2_\mu|\, m_{S^{\pm}_\beta}, m_{S^{\pm}_\alpha}, M_k)$. Also, we have used the fact that the coefficients $B_{\alpha\alpha}$ and $C_{\alpha \alpha k}$ are always negative. It is clear then that the sign of $C^{(1)}_{\mu\mu}$ will depend on the relation between $\lambda_{H\phi}$ and $\lambda_3$, while $C^{(7)}_{\mu\mu}$ can add either constructively or destructively, depending on the hierarchy between $m^2_{S^\pm_1}$, $m^2_{S^\pm_2}$, and $M^2_k$, making the additive balance between the two coefficients subtle. However, the factors $C_{11k}$ and $C_{22k}$ are typically of order $\mathcal{O}(10^{-6})$ or smaller so their contribution will be negligible compared to $C^{(1)}_{\mu\mu}$.

Taking these considerations into account, it is clear that the coefficient $C^{(1)}_{\mu\mu}$, given by Eq.~\eqref{eq:coefficient_C1_small_angle}, primarily determines the behavior of the deviations in $R_{\mu\mu}$ for most of the points in the open regions of the model's parameter space. The deviations are approximately given by
\begin{equation}
\label{eq:Rmu_approximation_small_angles}
    R_{\mu\mu} - 1 \simeq - \frac{1}{8 \pi^2} \frac{v m_\mu}{m^2_h} \frac{1}{\vartheta^{\text{SM}}_{\mu\mu}} \Biggl\{\lambda_{H\phi}\Bigl(|\kappa_{12}|^2 + |\kappa_{22}|^2 + |\kappa_{32}|^2\Bigr) |B_{11}| + \lambda_3\Bigl(|y_{21}|^2 + |y_{22}|^2 + |y_{23}|^2\Bigr) |B_{22}|\Biggr\} \, .
\end{equation}
The results of our numerical analysis show, however, that it is very difficult for both $\lambda_{H\phi}$ and $\lambda_3$ to be simultaneously negative. Moreover, in most cases, both coefficients are simultaneously positive. Therefore,  an important prediction of our model is that, after accounting for all current experimental bounds, the Higgs decay rate to dimuons, $\text{Br}(h \to \mu^+ \mu^-)$, is expected to be suppressed relatively to the Standard Model prediction, although an increase is also possible.

\section{Conclusions}
\label{section:conclusions}

In this paper, we explored a simple extension of the Scotogenic Model that features an additional complex singlet scalar. This extension is motivated by its potential to explain deviations in the Higgs decay $h \to \mu^+\mu^-$ while remaining consistent with the stringent constraints from lepton flavor violation, particularly $\mu^+ \to e^+\gamma$. The phenomenology of $h \to \mu^+\mu^-$ is intricately linked to the processes $\mu^+ \to e^+\gamma$ and $h \to \gamma\gamma$, as all three are governed by a common set of vertices. Our analysis reveals that deviations from the Standard Model prediction for the ratio $R_{\mu\mu} = \text{Br}(h \to \mu^+\mu^-)/\text{Br}(h \to \mu^+\mu^-)_{\text{SM}}$ are allowed. In particular, we found viable parameter points with deviations up to $|R_{\mu\mu} - 1| \lesssim 0.5$. This is the main result of our paper. In addition, we found that the model strongly favors values of $R_{\mu\mu}$ below one, reflecting the destructive interference of one-loop contributions involving the two new charged scalars and heavy right-handed neutrinos.

We also analyzed the analytical behavior of $R_{\mu\mu}$. Several limiting cases were explored, including the scenario in which the two charged scalars are mass degenerate. For most points in the viable parameter space, deviations in $R_{\mu\mu}$ exhibit a linear dependence on the parameter $\vartheta^L_{\mu\mu}$, which controls the loop effects in $h \to \mu^+\mu^-$. Interestingly, parameter points showing large deviations in $R_{\mu\mu}$ align precisely with this linear relation. When imposing the two most critical constraints, namely those from $\mu^+ \to e^+\gamma$ and $h \to \gamma\gamma$, the largest deviations in $h \to \mu^+\mu^-$ occur when the mixing between the inner charged scalars, controlled by the angle $\theta$, is close to zero. Consequently, sizable deviations in $R_{\mu\mu}$ are associated with $|\sin\theta| \simeq 0$. Notably, in this case the observable $R_{\mu\mu}$ takes a very simplified analytical form.

The model presented here represents a well-motivated and economical extension of the Standard Model, including a viable thermal dark matter candidate and a simple mechanism to induce radiative neutrino masses in agreement with oscillation data. While it remains a straightforward extension of the Scotogenic Model, it exhibits a very rich phenomenology, both at low energies and at collider experiments. In principle, the same statement should hold true for other low-energy neutrino mass models, although with quantitatively different conclusions. Future work should focus on further exploring collider signatures, particularly the Higgs decay $h \to \mu^+ \mu^-$, to uncover additional insights into the predictions and implications of other neutrino mass models.

\section*{Acknowledgments}

This work has been supported by the Spanish grants PID2023-147306NB-I00, CEX2023-001292-S (MCIU/AEI/10.13039/501100011033) and CIPROM/2021/054 (Generalitat Valenciana). V.M.L. acknowledges the financial support by Ministerio de Universidades and ``European Union-Next Generation EU/PRTR'' under the grant María Zambrano UP2021-044 (ZA2021-081) and Generalitat Valenciana under the grant CIPROM/2021/054. S.N.~acknowledges financial support from ANID Fellowship ANID BECAS/DOCTORADO NACIONAL 21200064 and Proyecto Milenio-ANID: ICN2019\_044. S.N.~would also like to thank Marco Aurelio Díaz for helpful observations on the analytical behavior of $R_{\mu\mu}$.


\begin{appendices}

\section{Mass eigenstate basis}
\label{appendix:mass_eigenstate_basis}

In this Section, we follow Ref.~\cite{Beniwal:2018hyi}. Note that in general, the rotation of a doublet of complex scalar weak eigenstates $\Xi^\dagger_\text{weak} = (\phi^+, \eta^+)$ to the respective mass eigenstates would demand the use of a general $U \in \text{SU}(2)$ (complex) rotation matrix,
\begin{equation}
    \Xi_\text{mass} = U^\dagger \Xi_\text{weak}
    \quad\Leftrightarrow\quad 
\begin{pmatrix}
    S^{-}_1 \\
    S^{-}_2
\end{pmatrix}
    =
\begin{pmatrix}
    \cos\theta e^{i(\alpha+\beta)} & -\sin\theta e^{-i(\alpha-\beta)} \\
    \sin\theta e^{i(\alpha-\beta)} & \cos\theta e^{-i(\alpha+\beta)}
\end{pmatrix}
\begin{pmatrix}
    \phi^{-} \\
    \eta^{-}
\end{pmatrix} \, ,
\end{equation}
where $\theta \in [0, \pi/2]$, $\alpha \in [0, 2\pi]$, and $\beta \in [0, \pi]$. The non-diagonal elements \eqref{eq:non_diagonal_elements} are then ``diagonalized'' as
\begin{equation}
    \Xi^\dagger\mathcal{M}^2\Xi = \Xi^\dagger(UU^\dagger)\mathcal{M}^2(UU^\dagger)\Xi = (\Xi^\dagger U)(U^\dagger\mathcal{M}^2U)(U^\dagger\Xi) \equiv \Xi^\dagger_\text{mass}\mathcal{M}^2_\text{diag}\Xi_\text{mass} \, ,
\end{equation}
where the (non-)diagonal mass matrix \eqref{eq:mass_matrix} satisfies
\begin{equation}\label{eq:diagonal_mass_matrix}
    \begin{pmatrix}
        m^2_{S^\pm_1} & 0 \\
        0 & m^2_{S^\pm_2} \\
    \end{pmatrix}
    =
    \mathcal{M}^2_\text{diag} = U^\dagger\mathcal{M}^2U \, .
\end{equation}
Now, because in our case, the entries of $\mathcal{M}^2$ are real, we are forced to set $\alpha = \beta \overset{!}{=} 0$. Then, by equating the left- and right-hand sides of Eq.~\eqref{eq:diagonal_mass_matrix}, we find the relations
\begin{subequations}
\begin{align}\label{eq:system_equations_M_and_masses}
    m^2_{S^\pm_1} &= \mathcal{M}^2_{11}\cos^2\theta + \mathcal{M}^2_{22}\sin^2\theta - \mathcal{M}^2_{12}\sin(2\theta) \, , \\
    m^2_{S^\pm_2} &= \mathcal{M}^2_{11}\sin^2\theta + \mathcal{M}^2_{22}\cos^2\theta + \mathcal{M}^2_{12}\sin(2\theta) \, , \\
    0 &= \mathcal{M}^2_{12}\cos(2\theta) - \frac{1}{2}(\mathcal{M}^2_{22} - \mathcal{M}^2_{11})\sin(2\theta) \, . \label{eq:third_constrain}
\end{align}
\end{subequations}
Equation \eqref{eq:third_constrain} implies the relation \eqref{eq:tan2theta} that defines the mixing angle $\theta$. The system of equations \eqref{eq:system_equations_M_and_masses} can be written as
\begin{equation}
    \begin{pmatrix}
        m^2_{S^\pm_1} \\
        m^2_{S^\pm_2} \\
        0
    \end{pmatrix}
    =
    \begin{pmatrix}
        \cos^2\theta & \sin^2\theta & -\sin(2\theta) \\
        \sin^2\theta & \cos^2\theta & \sin(2\theta) \\
        \frac{1}{2}\sin(2\theta) & -\frac{1}{2}\sin(2\theta) & \cos(2\theta)
    \end{pmatrix}
    \begin{pmatrix}
        \mathcal{M}^2_{11} \\
        \mathcal{M}^2_{22} \\
        \mathcal{M}^2_{12}
    \end{pmatrix}
\end{equation}
and easily inverted to obtain
\begin{equation}
\begin{pmatrix}
        \mathcal{M}^2_{11} \\
        \mathcal{M}^2_{22} \\
        \mathcal{M}^2_{12}
    \end{pmatrix}
    =
    \begin{pmatrix}
        \cos^2\theta & \sin^2\theta & \sin(2\theta) \\
        \sin^2\theta & \cos^2\theta & -\sin(2\theta) \\
        -\frac{1}{2}\sin(2\theta) & \frac{1}{2}\sin(2\theta) & \cos(2\theta)
    \end{pmatrix}
    \begin{pmatrix}
        m^2_{S^\pm_1} \\
        m^2_{S^\pm_2} \\
        0
    \end{pmatrix} \, .
\end{equation}
This allows one to obtain the inverse relations
\begin{align}
    \lambda_3 &= \frac{2(\mathcal{M}^2_{11} - m^2_\eta)}{v^2} = \frac{2}{v^2}\Bigl(m^2_{S^\pm_1}\cos^2\theta + m^2_{S^\pm_2}\sin^2\theta - m^2_\eta\Bigr) \, , \\
    \lambda_{H\phi} &= \frac{2(\mathcal{M}^2_{22} - m^2_\phi)}{v^2} = \frac{2}{v^2}\Bigl(m^2_{S^\pm_1}\sin^2\theta + m^2_{S^\pm_2}\cos^2\theta - m^2_\phi\Bigr) \, , \\
    \mu &= \frac{\sqrt{2}\mathcal{M}^2_{12}}{v} = \frac{\sqrt{2}}{2v}\left(m^2_{S^\pm_2} - m^2_{S^\pm_1}\right)\sin(2\theta) \, . \label{eq:mu_function}
\end{align}
In an (almost) degenerate scenario between charged scalars,
\begin{equation}
    \lambda_3 \simeq \frac{2}{v^2}\Bigl(m^2_{S^\pm} - m^2_\eta\Bigr) \, ,\quad \lambda_{H\phi} \simeq \frac{2}{v^2}\Bigl(m^2_{S^\pm} - m^2_\phi\Bigr) \, ,\quad 
    \mu \ll 1 \, .
\end{equation}

\section{Auxiliary Coefficients for \texorpdfstring{$\text{Br}(\boldsymbol{h \to \ell^{+}_i \ell^{-}_j)}$}{}}
\label{appendix:coefficients_hTOlilj}

The amplitude element of the vertex loop-diagram \eqref{fig:hTOll_loop_triangle} is
\begin{equation}\label{eq:Mloop_hTOll}
    -i\mathcal{M}_\text{loop} = -\sum^3_{k=1}\sum^2_{\alpha,\beta=1}\xi_{\alpha\beta} \int \frac{d^4q}{(2\pi)^4}\ \frac{\bar{u}(p')(-g^{R\ast}_{k\alpha i}P_L + g^{L\ast}_{k\alpha i}P_R)(\slashed{q} + M_k)(-g^{R}_{k\beta j}P_R + g^L_{k\beta j}P_L)v(p'')}{\Bigl[q^2 - M^2_k\Bigr]\Bigl[(p' + q)^2 - m^2_{S^{\pm}_\alpha}\Bigr]\Bigl[(q - p'')^2 - m^2_{S^{\pm}_\beta}\Bigr]} \, .
\end{equation}
The coefficients $i\xi_{\alpha\beta}$ correspond to the vertex factor between the Higgs, $h$, and the charged scalars $S^\pm_\alpha$ and $S^\pm_\beta$; they can be found in Appendix~\ref{appendix:vertex_factors_scalar_sector} (note that $\xi_{\beta\alpha} = \xi_{\alpha\beta}$). The auxiliary couplings $g^R_{k\beta i}$ and $g^L_{k\beta i}$ are defined in Eq.~\eqref{eq:auxiliary_coupling_definition}. On the other hand, the coefficient $\vartheta^L_{ij}$ used in calculation of Eq.~\eqref{eq:BR_hTOlili} can be identified from Eq.~\eqref{eq:Mloop_hTOll} (with the help of the \texttt{Mathematica} \cite{Mathematica} package \texttt{Package-X} \cite{Patel:2016fam}):
\begin{equation}
    \vartheta^L_{ij} := \sum^3_{k=1}\sum^2_{\alpha,\beta=1}\vartheta^L_{ij, k, \alpha\beta} \equiv C^{(1)}_{ij} + \cdots + C^{(7)}_{ij} \, ,
\end{equation}
where
\begin{equation}
    \vartheta^L_{ij, k, \alpha\beta} := -\frac{\xi_{\alpha\beta}}{32\pi^2m^2_hm_im_j\lambda(m^2_h, m^2_i, m^2_j)}\Bigl\{C^{(1)}_{ij,k,\alpha\beta} + \cdots + C^{(7)}_{ij,k,\alpha\beta}\Bigr\}
\end{equation}
and
\begingroup
\allowdisplaybreaks
\begin{align*}
    C^{(1)}_{ij,k,\alpha\beta} &:= 2m^2_hm_im_j\Bigl[m_i(m^2_h - m^2_i + m^2_j)g^L_{k\beta j}g^{L\ast}_{k\alpha i} + m_j(m^2_h + m^2_i - m^2_j)g^R_{k\beta j}g^{R\ast}_{k\alpha i}\Bigr]B_0(m^2_h, m_{S^{\pm}_\alpha}, m_{S^{\pm}_\beta}) \\
    &\simeq 2m^4_hm_im_j\Bigl[m_ig^L_{k\beta j}g^{L\ast}_{k\alpha i} + m_jg^R_{k\beta j}g^{R\ast}_{k\alpha i}\Bigr]B_0(m^2_h, m_{S^{\pm}_\alpha}, m_{S^{\pm}_\beta}) \, , \\
    C^{(2)}_{ij,k,\alpha\beta} &:= - 2m^2_hm^2_im_j\Bigl[2m_im_jg^R_{k\beta j}g^{R\ast}_{k\alpha i} + (m^2_h - m^2_i - m^2_j)g^L_{k\beta j}g^{L\ast}_{k\alpha i}\Bigr]B_0(m^2_i, m_{S^{\pm}_\alpha}, M_k) \\
    &\simeq - 2m^2_hm^2_im_j\Bigl[2m_im_jg^R_{k\beta j}g^{R\ast}_{k\alpha i} + m^2_hg^L_{k\beta j}g^{L\ast}_{k\alpha i}\Bigr]B_0(m^2_i, m_{S^{\pm}_\alpha}, M_k) \, , \\
    C^{(3)}_{ij,k,\alpha\beta} &:= - 2m^2_hm_im^2_j\Bigl[2m_im_jg^L_{k\beta j}g^{L\ast}_{k\alpha i} + (m^2_h - m^2_i - m^2_j)g^R_{k\beta j}g^{R\ast}_{k\alpha i}\Bigr]B_0(m^2_j, m_{S^{\pm}_\beta}, M_k) \\
    &\simeq - 2m^2_hm_im^2_j\Bigl[2m_im_jg^L_{k\beta j}g^{L\ast}_{k\alpha i} + m^2_hg^R_{k\beta j}g^{R\ast}_{k\alpha i}\Bigr]B_0(m^2_j, m_{S^{\pm}_\beta}, M_k) \, , \\
    C^{(4)}_{ij,k,\alpha\beta} &:= m_im_j\Bigl[-m_i(m^2_h + m^2_{S^{\pm}_\alpha} - m^2_{S_\beta})(m^2_h - m^2_i + m^2_j)g^L_{k\beta j}g^{L\ast}_{k\alpha i} \\
        &\quad + m_j(m^2_h - m^2_{S^{\pm}_\alpha} + m^2_{S^{\pm}_\beta})(m^2_h + m^2_i - m^2_j)g^R_{k\beta j}g^{R\ast}_{k\alpha i}\Bigr]\ln\frac{m^2_{S^{\pm}_\alpha}}{m^2_{S^{\pm}_\beta}} \nonumber \\
    &\simeq m^2_hm_im_j\Bigl[-m_i(m^2_h + m^2_{S^{\pm}_\alpha} - m^2_{S^{\pm}_\beta})g^L_{k\beta j}g^{L\ast}_{k\alpha i} + m_j(m^2_h - m^2_{S^{\pm}_\alpha} + m^2_{S^{\pm}_\beta})g^R_{k\beta j}g^{R\ast}_{k\alpha i}\Bigr]\ln\frac{m^2_{S^{\pm}_\alpha}}{m^2_{S^{\pm}_\beta}} \, , \nonumber \\
    C^{(5)}_{ij,k,\alpha\beta} &:= - m^2_hm_j\Bigl[2m_im_j(m^2_h - m^2_{S^{\pm}_\alpha} - m^2_j + M^2_k)g^R_{k\beta j}g^{R\ast}_{k\alpha i} \\
        &\quad - (m^2_h - m^2_i - m^2_j)(m^2_{S^{\pm}_\alpha} + m^2_i - M^2_k)g^L_{k\beta j}g^{L\ast}_{k\alpha i} \Bigr]\ln\frac{m^2_{S^{\pm}_\alpha}}{M^2_k} \\
    &\simeq - m^2_hm_j\Bigl[2m_im_j(m^2_h - m^2_{S^{\pm}_\alpha} + M^2_k)g^R_{k\beta j}g^{R\ast}_{k\alpha i} - m^2_h(m^2_{S^{\pm}_\alpha} - M^2_k)g^L_{k\beta j}g^{L\ast}_{k\alpha i} \Bigr]\ln\frac{m^2_{S^{\pm}_\alpha}}{M^2_k} \, , \\
    C^{(6)}_{ij,k,\alpha\beta} &:= - m^2_hm_i\Bigl[2m_im_j(m^2_h - m^2_{S^{\pm}_\beta} - m^2_i + M^2_k)g^L_{k\beta j}g^{L\ast}_{k\alpha i} \\
        &\quad - (m^2_h - m^2_i - m^2_j)(m^2_{S^{\pm}_\beta} + m^2_j - M^2_k)g^R_{k\beta j}g^{R\ast}_{k\alpha i}\Bigr]\ln\frac{m^2_{S^{\pm}_\beta}}{M^2_k} \\
    &\simeq - m^2_hm_i\Bigl[2m_im_j(m^2_h - m^2_{S^{\pm}_\beta} + M^2_k)g^L_{k\beta j}g^{L\ast}_{k\alpha i} - m^2_h(m^2_{S^{\pm}_\beta} - M^2_k)g^R_{k\beta j}g^{R\ast}_{k\alpha i}\Bigr]\ln\frac{m^2_{S^{\pm}_\beta}}{M^2_k} \, , \\
    C^{(7)}_{ij,k,\alpha\beta} &:= -2m^2_hm_im_j\Bigl\{-m_i[-2m^2_{S^{\pm}_\alpha}m^2_j + m^2_im^2_j - m^4_j + m^2_{S^{\pm}_\beta}(m^2_i + m^2_j) - M^2_k(m^2_i - m^2_j) \\
        &\quad + m^2_h(-m^2_{S^{\pm}_\beta} + m^2_j + M^2_k)]g^L_{k\beta j}g^{L\ast}_{k\alpha i} + M_k[m^4_h + (m^2_i - m^2_j)^2 - 2m^2_h(m^2_i + m^2_j)]g^L_{k\beta j}g^{R\ast}_{k\alpha i} \\
        &\quad - m_j[-2m^2_{S^{\pm}_\beta}m^2_i + m^2_{S^{\pm}_\alpha}(m^2_i + m^2_j) - m^4_i + m^2_im^2_j + (m^2_i - m^2_j)M^2_k \\
        &\quad + m^2_h(-m^2_{S^{\pm}_\alpha} + m^2_i + M^2_k)]g^R_{k\beta j}g^{R\ast}_{k\alpha i}\Bigr\}C_0(m^2_h, m^2_i, m^2_j|\, m_{S^{\pm}_\beta}, m_{S^{\pm}_\alpha}, M_k) \, .
\end{align*}
\endgroup
In each case, the approximations involve taking $m_i/m_h \ll 1$. The following identities might be useful:
\begin{equation}
\label{eq:relations_between_C_coefficients}
    \mathfrak{Re}[C^{(3)}_{ij,k,\alpha\beta}] = -\mathfrak{Re}[C^{(6)}_{ij,k,\alpha\beta}] \, ,\quad \mathfrak{Re}[C^{(2)}_{ij,k,\alpha\beta}] = -\mathfrak{Re}[C^{(5)}_{ij,k,\alpha\beta}] \, .
\end{equation}
The right-handed version of the coefficients is not independent, however, as they are related by
\begin{equation}
\label{eq:relation_between_varthetaLR_ijkalphabeta_coefficients}
    \vartheta^R_{ij, k, \alpha\beta} = (\vartheta^L_{ji, k, \beta\alpha})^\ast \, .
\end{equation}
To see this, we can first take the conjugate of the diagram in Figure \eqref{fig:hTOll_loop_triangle}, then perform a horizontal $180^\circ$ flip (i.e., a chirality flip), and finally relabel $i \leftrightarrow j$ and $\alpha \leftrightarrow \beta$. The resulting diagram (and corresponding vertex factor) must be equal to the original one, which leads to the relation given in Eq.~\eqref{eq:relation_between_varthetaLR_ijkalphabeta_coefficients}. It follows immediately that $\vartheta^R_{ij} = (\vartheta^L_{ji})^\ast$, i.e.
\begin{equation}
    \mathfrak{Re}[\vartheta^R_{ij}] = \mathfrak{Re}[\vartheta^L_{ji}] \quad \text{and} \quad  \mathfrak{Im}[\vartheta^R_{ij}] = -\mathfrak{Im}[\vartheta^L_{ji}] \, .
\end{equation}
For the Passarino-Veltman functions $B_0(x,y,z)$ and $C_0(a,b,c|x,y,z)$ found in the definition of the $C$-coefficients above, we use their definitions as given by \texttt{Package-X} \cite{Patel:2016fam}. The analytical expression for both functions can be straightforwardly found using the commands
\begin{align*}
    \texttt{DiscB[x,y,z] // DiscExpand} \\
    \texttt{ScalarC0[a,b,c,x,y,z] // C0Expand}
\end{align*}
after opening a new \texttt{Mathematica} session and initiating \texttt{Package-X}.

\section{Auxiliary Coefficients for \texorpdfstring{$\boldsymbol{\Delta a_{\ell_{i}}}$}{} and \texorpdfstring{$\boldsymbol{\text{Br}(\mu^{+} \to e^{+}\gamma)}$}{}}
\label{appendix:coefficients_a_and_muTOegamma}

The coefficients appearing in the calculations of the anomalous magnetic moments and $\text{Br}(\mu^{+} \to e^{+}\gamma)$ are
\begin{equation}\label{eq:coefficients_CRji}
    c^{\ell_j\ell_i}_R = \frac{e}{16\pi^2}\sum^3_{k=1} \sum^2_{\alpha=1} \frac{1}{m^2_{S^\pm_\alpha}} \Biggl[g^{L\ast}_{k\alpha j}g^R_{k\alpha i}M_k f_\Phi\left(\frac{M^2_k}{m^2_{S^\pm_\alpha}}\right) + \Bigl(m_{\ell_i}g^{L\ast}_{k\alpha j}g^L_{k\alpha i} + m_{\ell_j}g^{R\ast}_{k\alpha j}g^R_{k\alpha i}\Bigr)\tilde{f}_\Phi\left(\frac{M^2_k}{m^2_{S^\pm_\alpha}}\right)\Biggr] \, ,
\end{equation}
with
\begin{equation}
\label{eq:functions_f_and_ftilde}
    f_\Phi(x) = \frac{x^2 - 1 - 2x\ln x}{4(x - 1)^3} \, ,\quad \tilde{f}_\Phi(x) = \frac{2x^3 + 3x^2 - 6x + 1 - 6x^2\ln x}{24(x - 1)^4} \, .
\end{equation}
These functions have the following limits:
\begin{subequations}
\begin{align}
    \lim_{x \to 0} xf_\Phi\left(\frac{x^2}{a^2}\right) \, , &= 0 & \lim_{x \to a} xf_\Phi\left(\frac{x^2}{a^2}\right) &= \frac{a}{12} \, , & \lim_{x \to \infty} xf_\Phi\left(\frac{x^2}{a^2}\right) &= 0 \, , \\
    \lim_{x \to 0} \tilde{f}_\Phi\left(\frac{x^2}{a^2}\right) &= \frac{1}{24} \, , & \lim_{x \to a} \tilde{f}_\Phi\left(\frac{x^2}{a^2}\right) &= \frac{1}{48} \, , & \lim_{x \to \infty} \tilde{f}_\Phi\left(\frac{x^2}{a^2}\right) &= 0 \, .
\end{align}
\end{subequations}
Moreover, the function $xf_\Phi\left(x^2/a^2\right)$ reaches its maximum at $x = a$.

\section{Vacuum Stability}
\label{appendix:vacuum_stability}

We follow the co-positivity approach proposed in \cite{Kannike:2012pe, Kannike:2016fmd}. Let us parameterize the field bilinears as 
\begin{equation}
    H^\dagger H \equiv h^2_H \, ,\quad \eta^\dagger \eta \equiv h^2_\eta \, ,\quad \phi^\ast \phi \equiv h^2_\phi \, ,\quad H^\dagger \eta \equiv h_Hh_\eta\rho e^{i\phi} \, ,
\end{equation}
with $\rho \in [-1,1]$ and $\phi \in [0, 2\pi[$ free parameters. To study the vacuum stability of the potential, we ignore terms of dimension $d < 4$ in comparison to those quadratic in the fields because the former ones are negligible in the limit $|\text{fields}| \to \infty$. The quadratic part of the potential \eqref{eq:potential} has the structure of the quadratic form
\begin{align}
    V_\text{Quad}(h^2_H,h^2_\eta,h^2_\phi) &= \frac{\lambda_1}{2}(h^2_H)^2 + \frac{\lambda_2}{2}(h^2_\eta)^2 + \frac{\lambda_\phi}{2}(h^2_\phi)^2 + \lambda_3h^2_Hh^2_\eta + \lambda_4\rho^2h^2_Hh^2_\eta \nonumber \\
        &\quad + \frac{\lambda_5}{2}\rho^2h^2_Hh^2_\eta\cos(2\phi) + \lambda_{\eta\phi}h^2_\eta h^2_\phi + \lambda_{H\phi}h^2_Hh^2_\phi \nonumber \\
        &=
        \begin{pmatrix} h^2_H & h^2_\eta & h^2_\phi \end{pmatrix}
        \begin{pmatrix}
            a_{11} & a_{12} & a_{13} \\
            a_{12} & a_{22} & a_{23} \\
            a_{13} & a_{23} & a_{33}
        \end{pmatrix}
        \begin{pmatrix} h^2_H \\ h^2_\eta \\ h^2_\phi \end{pmatrix} \, ,
\end{align}
with
\begin{equation}
\begin{aligned}
    a_{11} = \lambda_1/2 \, ,\quad a_{22} &= \lambda_2/2 \, ,\quad a_{33} = \lambda_\phi/2 \, ,\quad a_{13} = \lambda_{H\phi}/2 \, ,\quad a_{23} = \lambda_{\eta\phi}/2 \, , \\
    a_{12} &= (1/2)[\lambda_3 + \rho^2(\lambda_4 + \lambda_5\cos2\phi)] \, .
\end{aligned}
\end{equation}
To guarantee the stability of the potential, we demand the matrix $(a_{ij})$ to be copositive. For a $3\times3$  matrix, the criteria reads
\begin{equation}
    \begin{aligned}
    0 &\leq \lambda_1, \lambda_2, \lambda_\phi \, , \\
    0 &\leq [\lambda_3 + \rho^2(\lambda_4 + \lambda_5\cos2\phi)] + \sqrt{\lambda_1\lambda_2} \equiv c_1 \, , \\
    0 &\leq \lambda_{H\phi} + \sqrt{\lambda_1\lambda_\phi} \equiv c_2 \, , \\
    0 &\leq \lambda_{\eta\phi} + \sqrt{\lambda_2\lambda_\phi} \equiv c_3 \, , \\
    0 &\leq \sqrt{\lambda_1\lambda_2\lambda_\phi} + [\lambda_3 + \rho^2(\lambda_4 + \lambda_5\cos2\phi)]\sqrt{\lambda_\phi} + \lambda_{H\phi}\sqrt{\lambda_2} + \lambda_{\eta\phi}\sqrt{\lambda_1} + \sqrt{2c_1c_2c_3} \, ,
\end{aligned}
\end{equation}
We can get rid of the free parameters $\rho$ and $\phi$ as follows:~\footnote{To see this, consider the functions $f(x,y) = x + y$ and $g(z^2,w) = z^2w$ with $z^2 \in [0,1]$. On the one hand, the function $f$ reaches its minimum when both arguments $x$ and $y$ reach their minimum simultaneously, i.e., $\text{min}[f(x,y)] = \text{min}(x) + \text{min}(y)$. On the other hand, it is not difficult to see that
\begin{equation}
    \text{min}[g(z^2,w)] = 
    \begin{cases}
        0 \, ,& \text{if $w > 0$} \, ,\\
        \text{min}(w) \, ,& \text{if $w < 0$}\, .
    \end{cases}
\end{equation}
} the term $\rho^2(\lambda_4 + \lambda_5\cos2\phi)$ gets its minimum, equal to 0, if $\lambda_4 + \lambda_5\cos2\phi > 0$ which in turn is true for every value of $\theta$ if $\lambda_4 - |\lambda_5| \geq 0$. On the other hand, if $\lambda_4 - |\lambda_5| < 0$, then the minimum of the term will be reached when $\rho^2 = 1$. From these considerations, we find the system of inequalities \eqref{eq:system_inequalities_vacuum_stability}. 

\section{Vertex Factors}
\label{appendix:vertex_factors}

\subsection{Scalar Sector}
\label{appendix:vertex_factors_scalar_sector}

The new interactions between scalars come from the potential \eqref{eq:potential}. The vertex factors involving three scalars, with all particles incoming, are
\begin{align*}
    \includegraphics[height=3cm, valign=c]{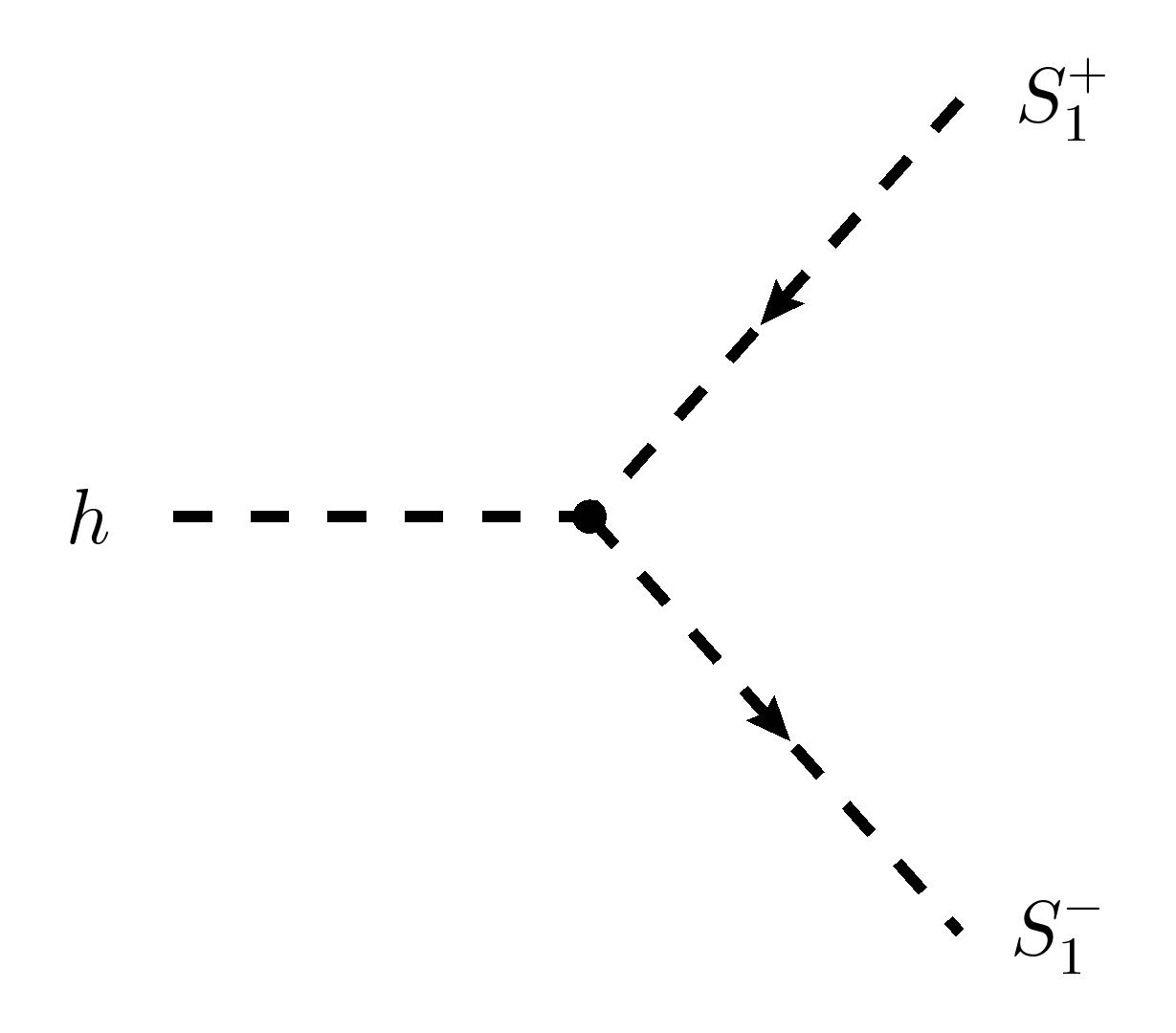} &\sim\ \begin{array}{l} {} \\ -i\Bigl[v \lambda_{H \phi} c^2_\theta - \sqrt{2} \mu c_\theta s_\theta \\ \quad + v \lambda_3 s^2_\theta\Bigr] \end{array} & \includegraphics[height=3cm, valign=c]{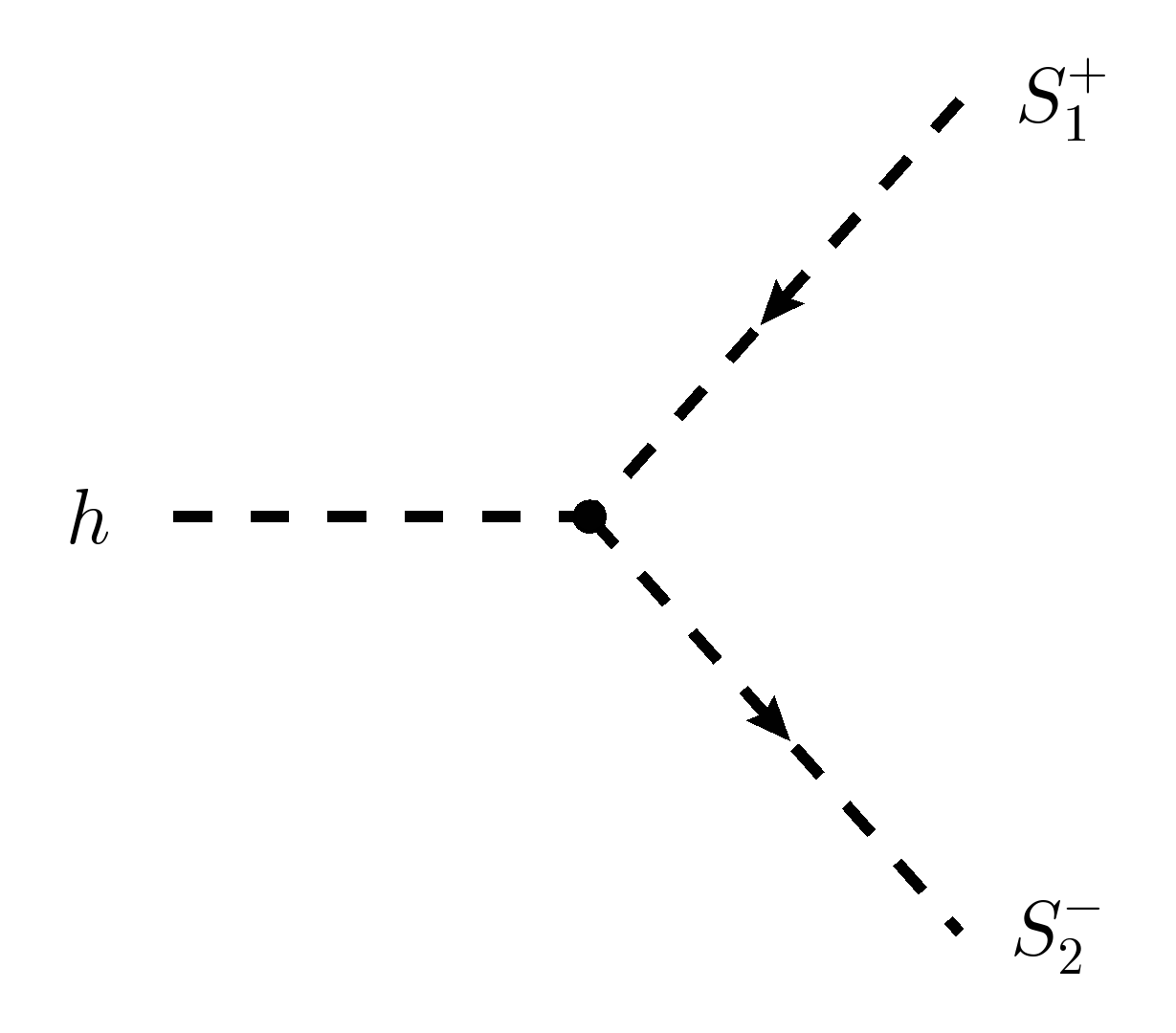} &\sim\ \begin{array}{l} {} \\ -\frac{i}{2}\Bigl[\sqrt{2} \mu c_{2\theta} + v (-\lambda_3  \\ + \lambda_{H\phi}) s_{2\theta}\Bigr] \end{array} \\
    \includegraphics[height=3cm, valign=c]{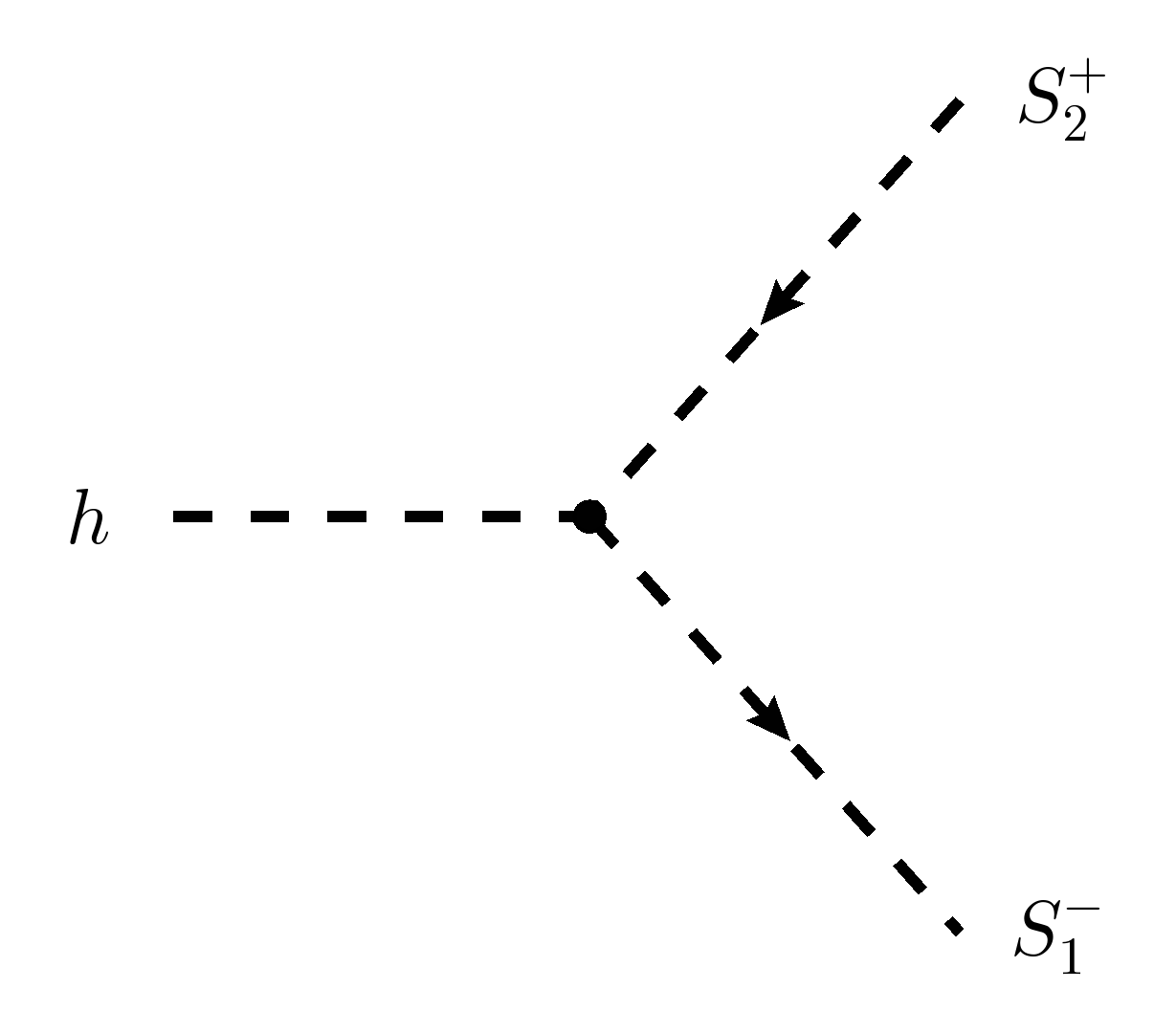} &\sim\ \begin{array}{l} {} \\ -\frac{i}{2}\Bigr[\sqrt{2} \mu c_{2\theta} + v (-\lambda_3 \\ \quad + \lambda_{H\phi}) s_{2\theta}\Bigl] \end{array} & \includegraphics[height=3cm, valign=c]{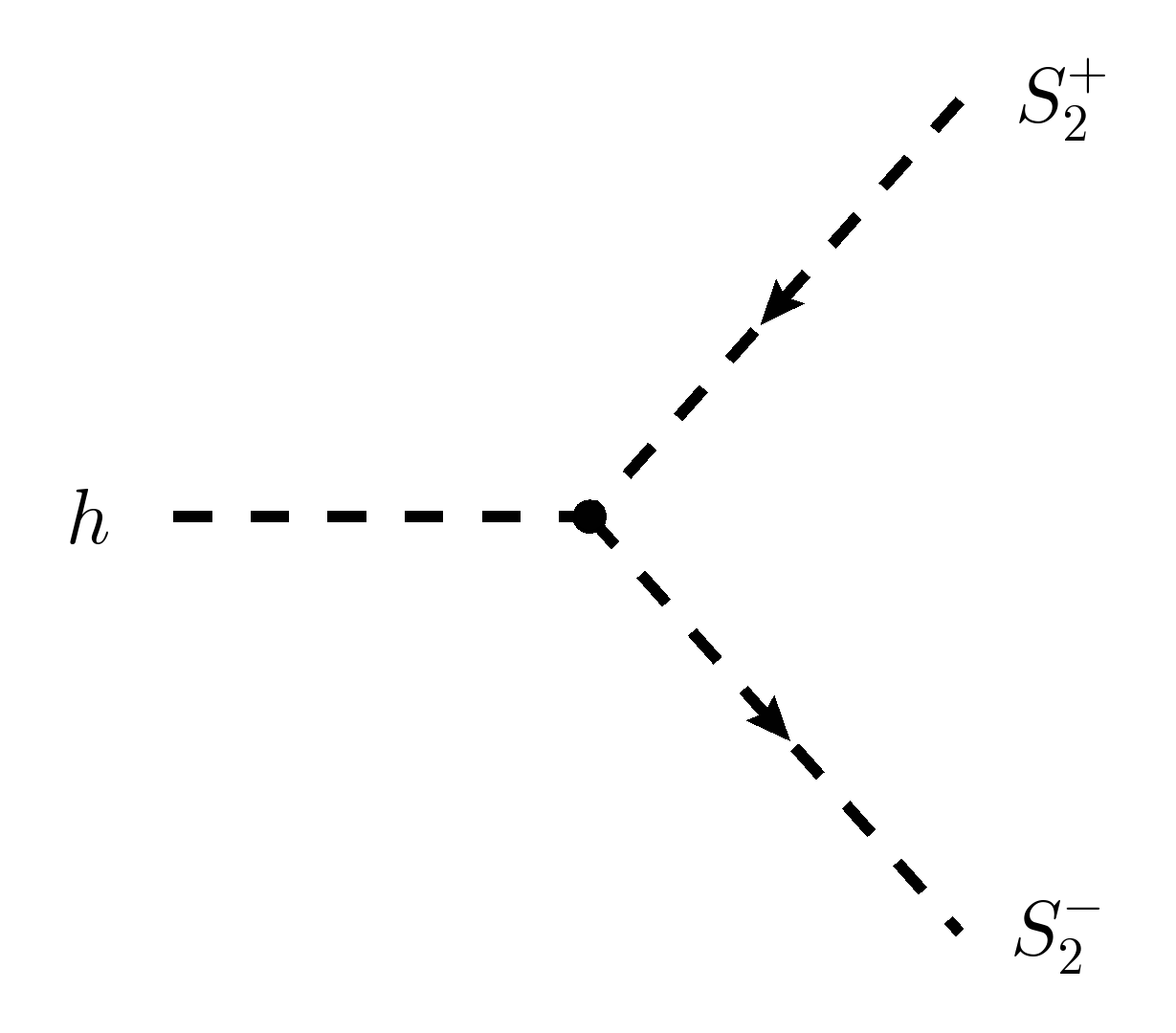} &\sim\ \begin{array}{l} {} \\ -i\Bigl[v \lambda_{H \phi}s^2_\theta + \sqrt{2} \mu c_\theta s_\theta \\ \quad + v \lambda_3 c^2_\theta\Bigr] \end{array} \\
    \includegraphics[height=3cm, valign=c]{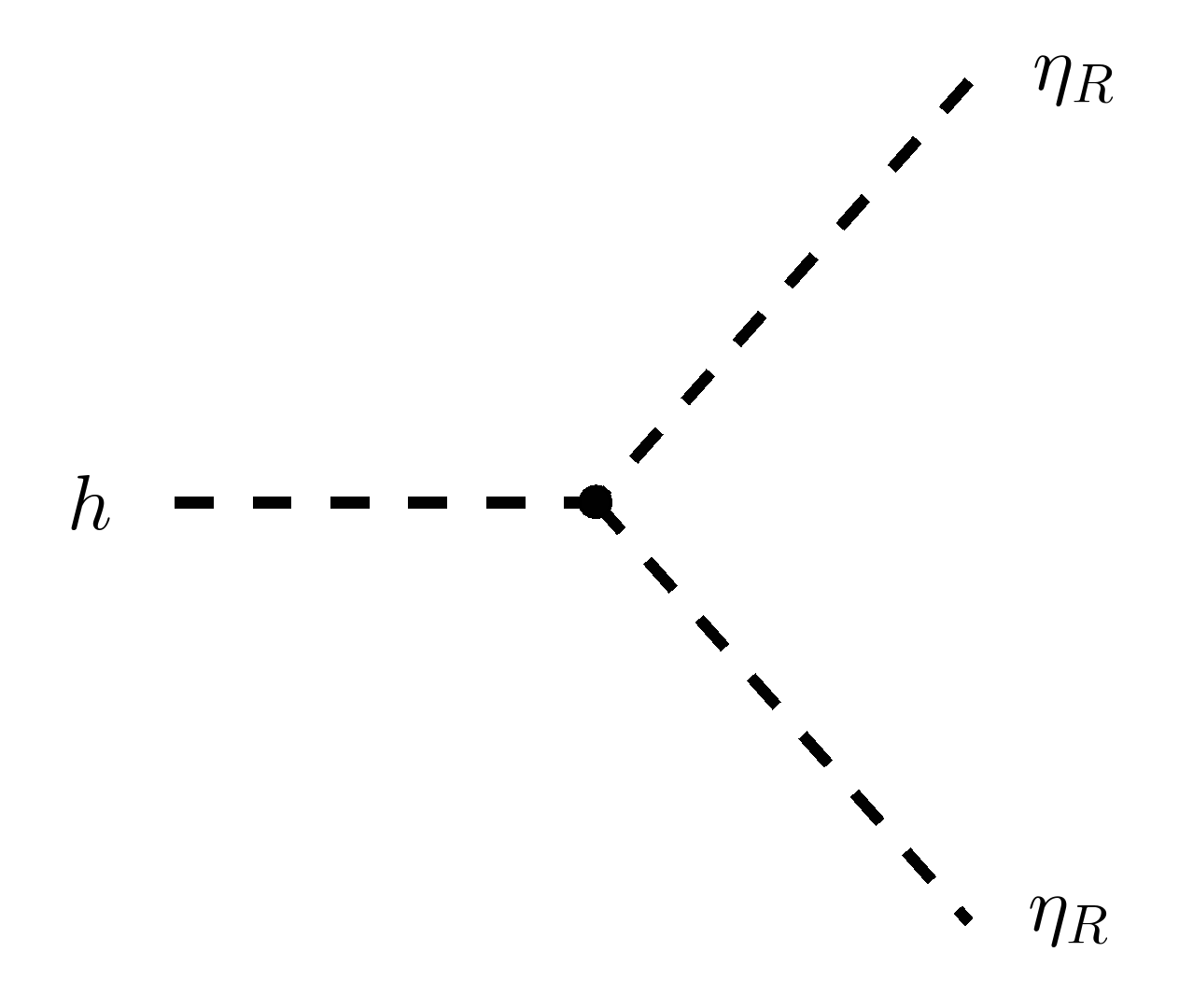} &\sim\ -2! \times i v (\lambda_3 + \lambda_4 + \lambda_5) & \includegraphics[height=3cm, valign=c]{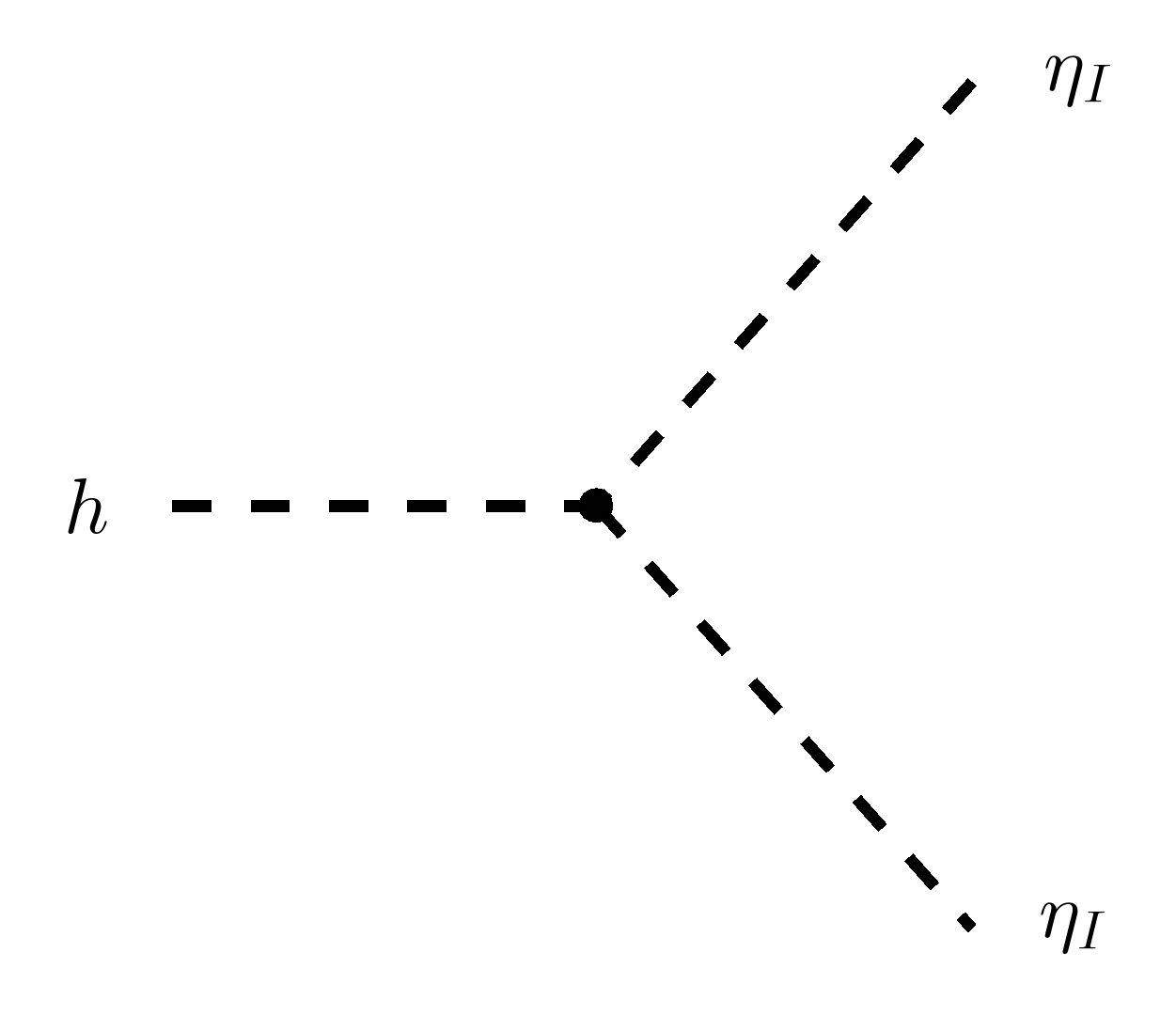} &\sim\ -2! \times i v (\lambda_3 + \lambda_4 - \lambda_5) \\
    \includegraphics[height=3cm, valign=c]{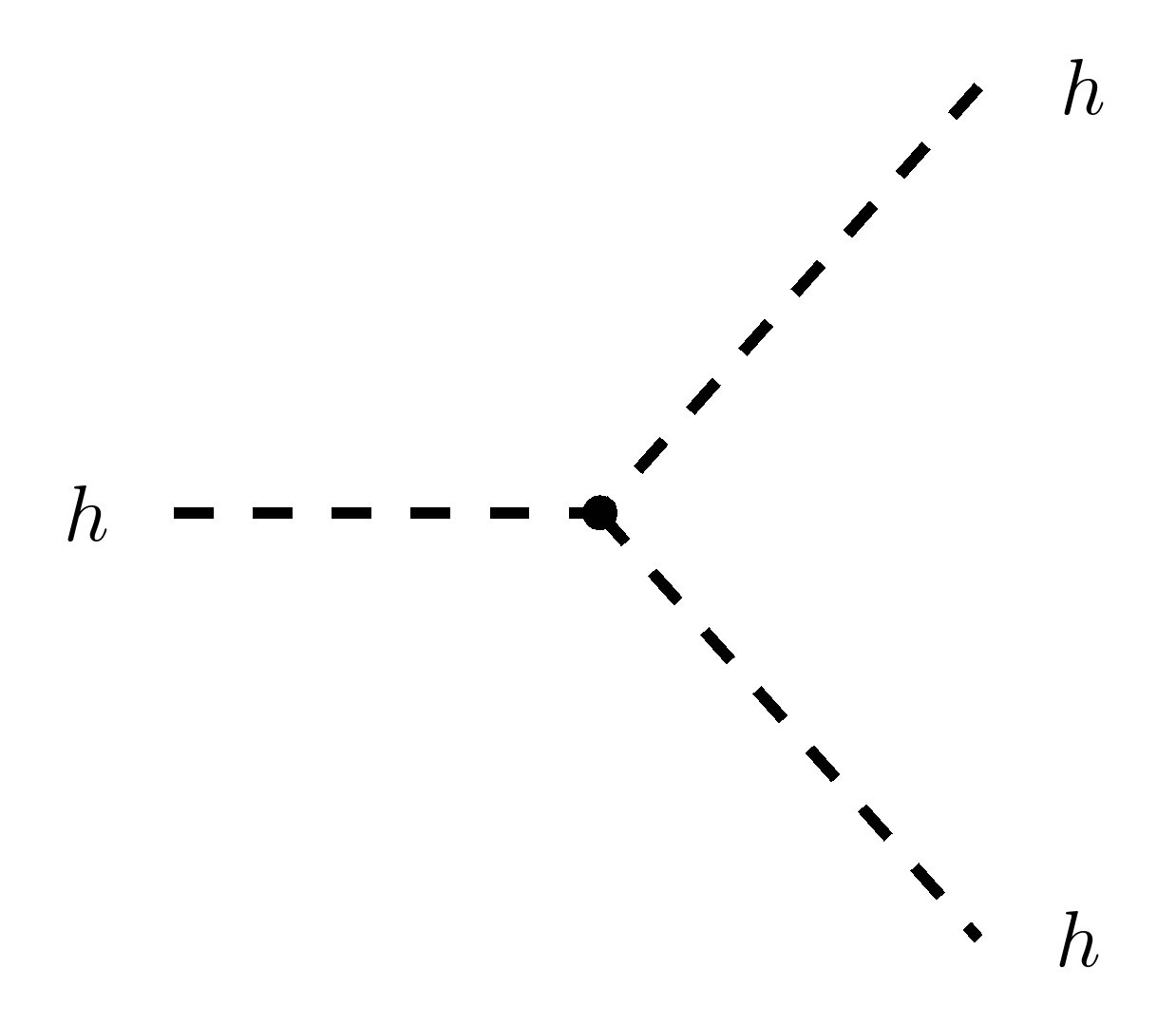} &\sim\ -3! \times 3 i v \lambda_1 & &
\end{align*}
The vertex factors involving four scalars, with all particles incoming, are
\begin{align*}
    \includegraphics[height=3cm, valign=c]{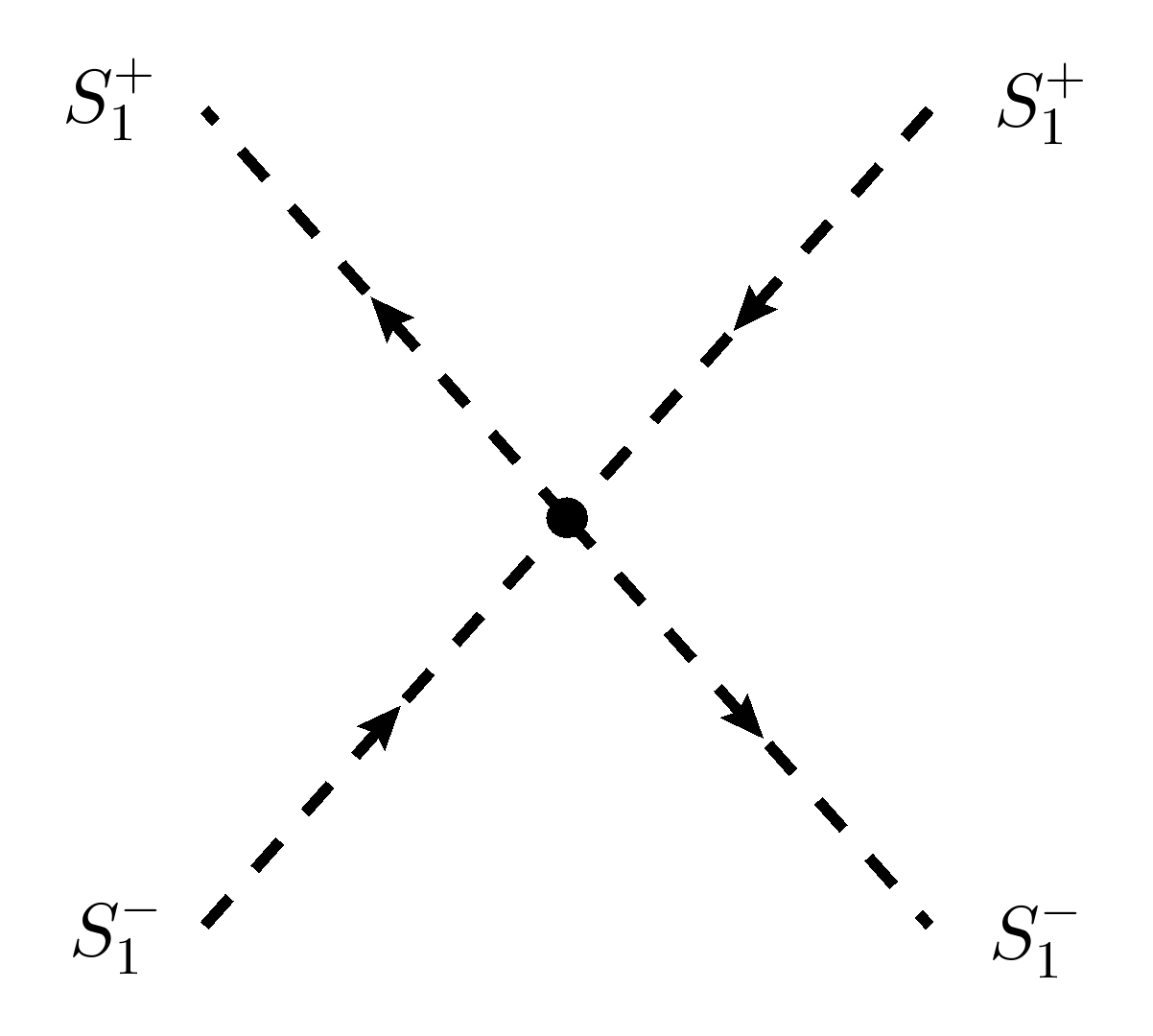} &\sim\ \begin{array}{l} {} \\ -2! \times 2! \times 2 i \Bigl[\lambda_{\Phi} c^4_\theta \\ \quad + 2 \lambda_{\eta\Phi} c^2_\theta s^2_\theta + \lambda_2 s^4_\theta\Bigr] \end{array} & \includegraphics[height=3cm, valign=c]{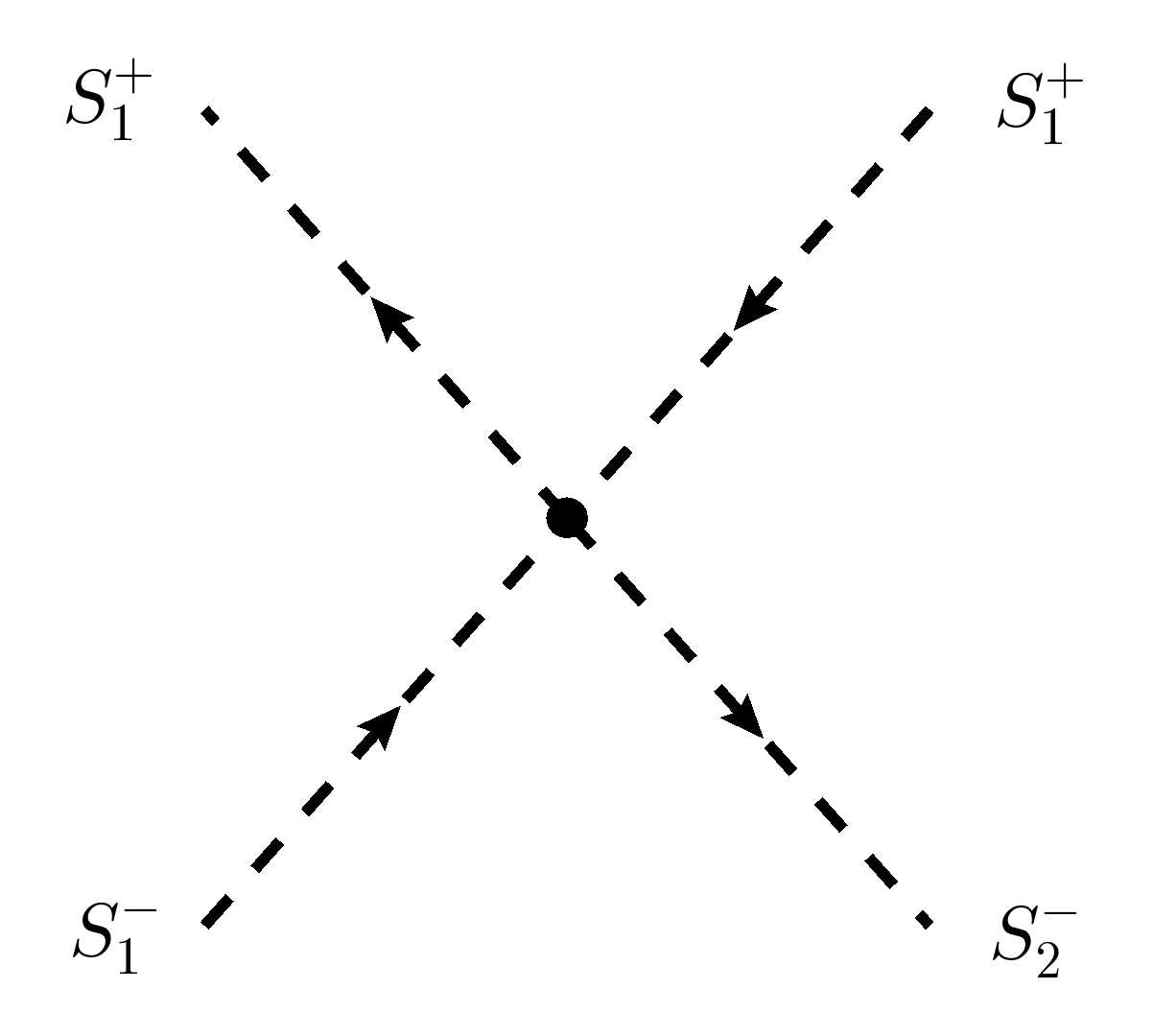} &\sim\ \begin{array}{l} {} \\ -2! \times \frac{i}{2} \Bigl[-\lambda_2 + \lambda_{\Phi} + (\lambda_2 \\ \quad - 2 \lambda_{\eta\Phi} + \lambda_{\Phi}) c_{2\theta}\Bigr] s_{2\theta} \end{array} \\
    \includegraphics[height=3cm, valign=c]{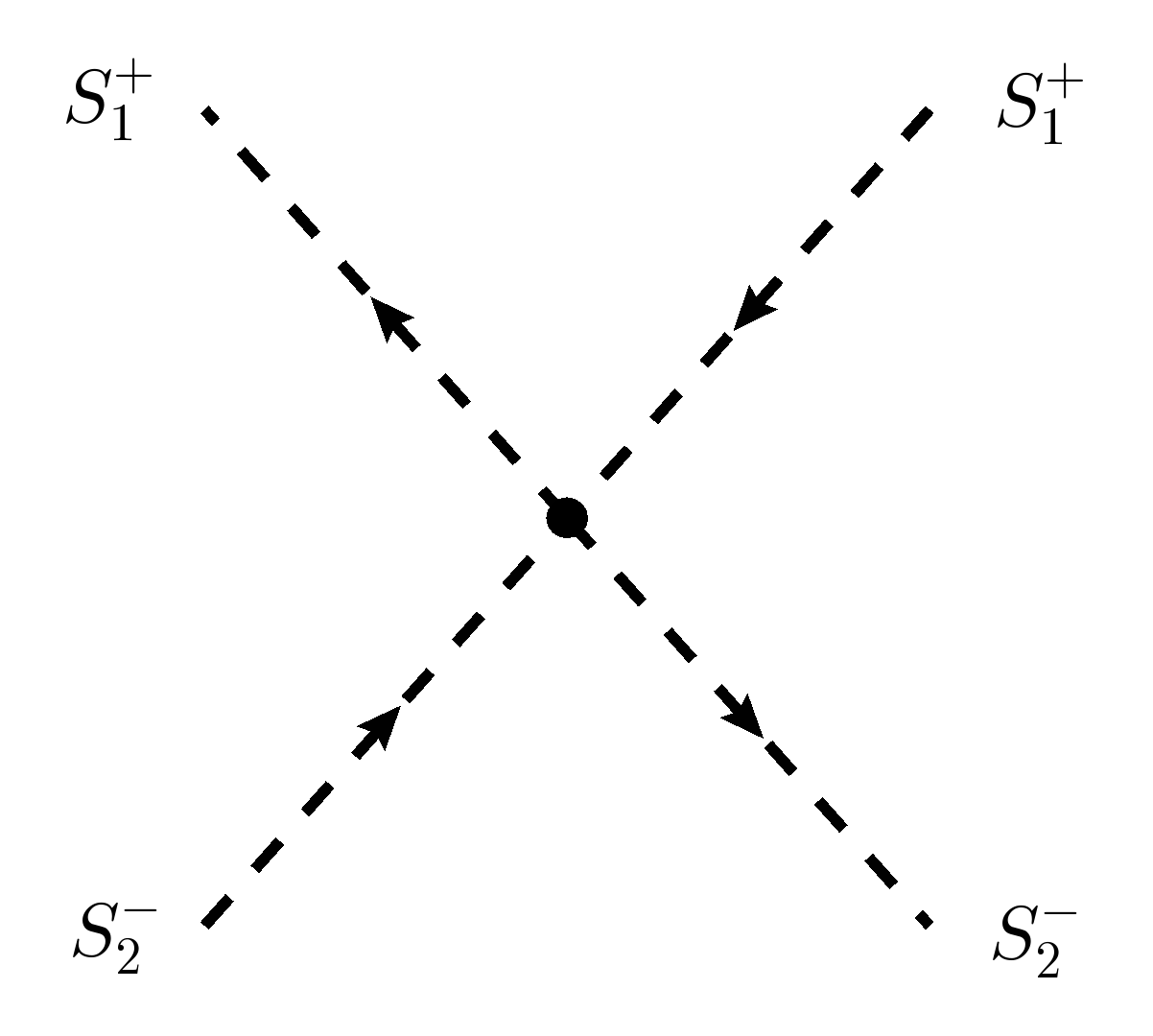} &\sim\ \begin{array}{l} {} \\ -2! \times 2! \times \frac{i}{2} \Bigl[\lambda_2 - 2 \lambda_{\eta\Phi} \\ \quad + \lambda_{\Phi}\Bigr] s^2_{2\theta} \end{array} & \includegraphics[height=3cm, valign=c]{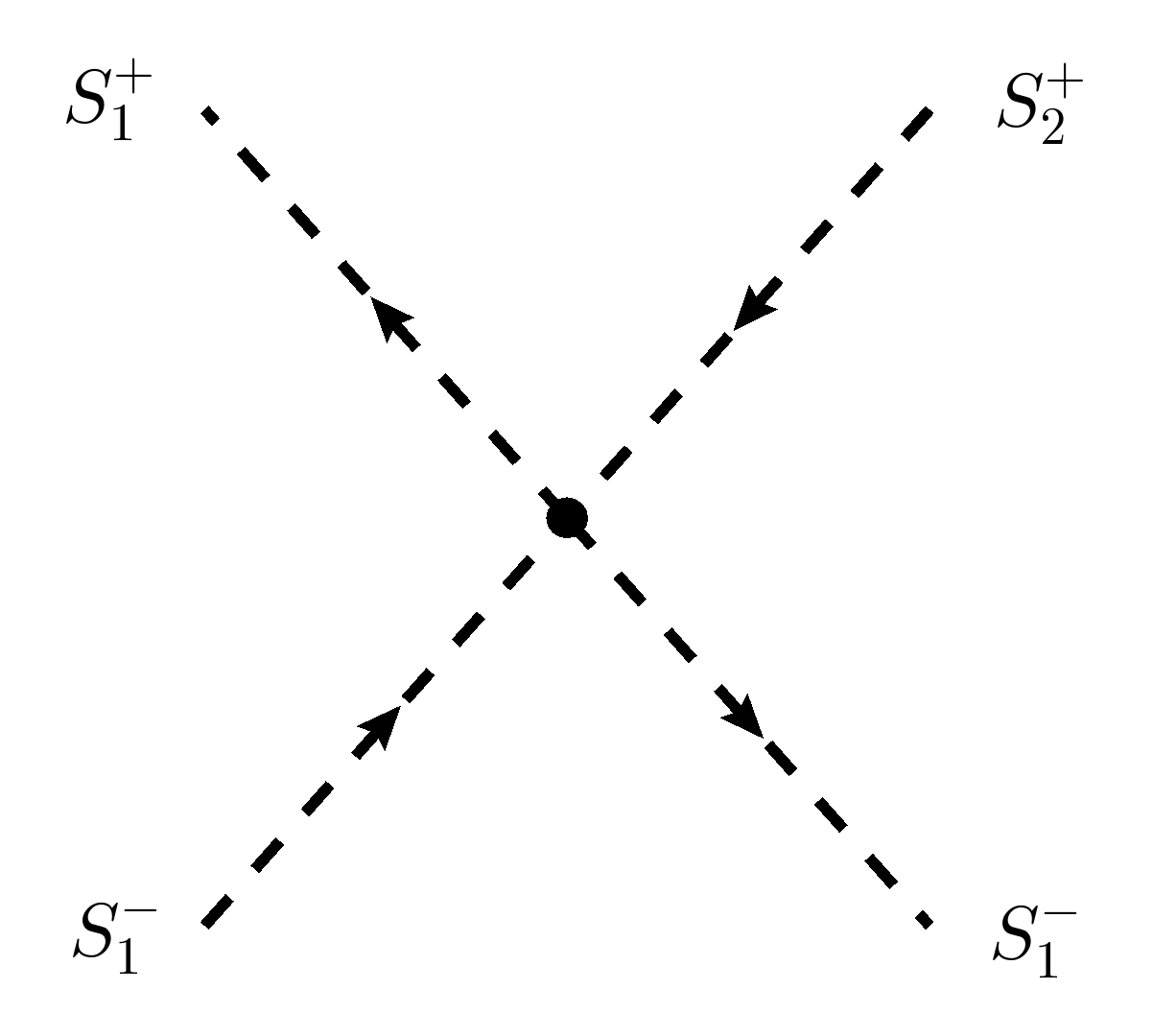} &\sim\ -2! \times \begin{array}{l} {} \\ \frac{i}{2} \Bigl[-\lambda_2 + \lambda_{\Phi} + (\lambda_2 \\ \quad - 2 \lambda_{\eta\Phi} + \lambda_{\Phi}) c_{2\theta}\Bigr] s_{2\theta} \end{array} \\
    \includegraphics[height=3cm, valign=c]{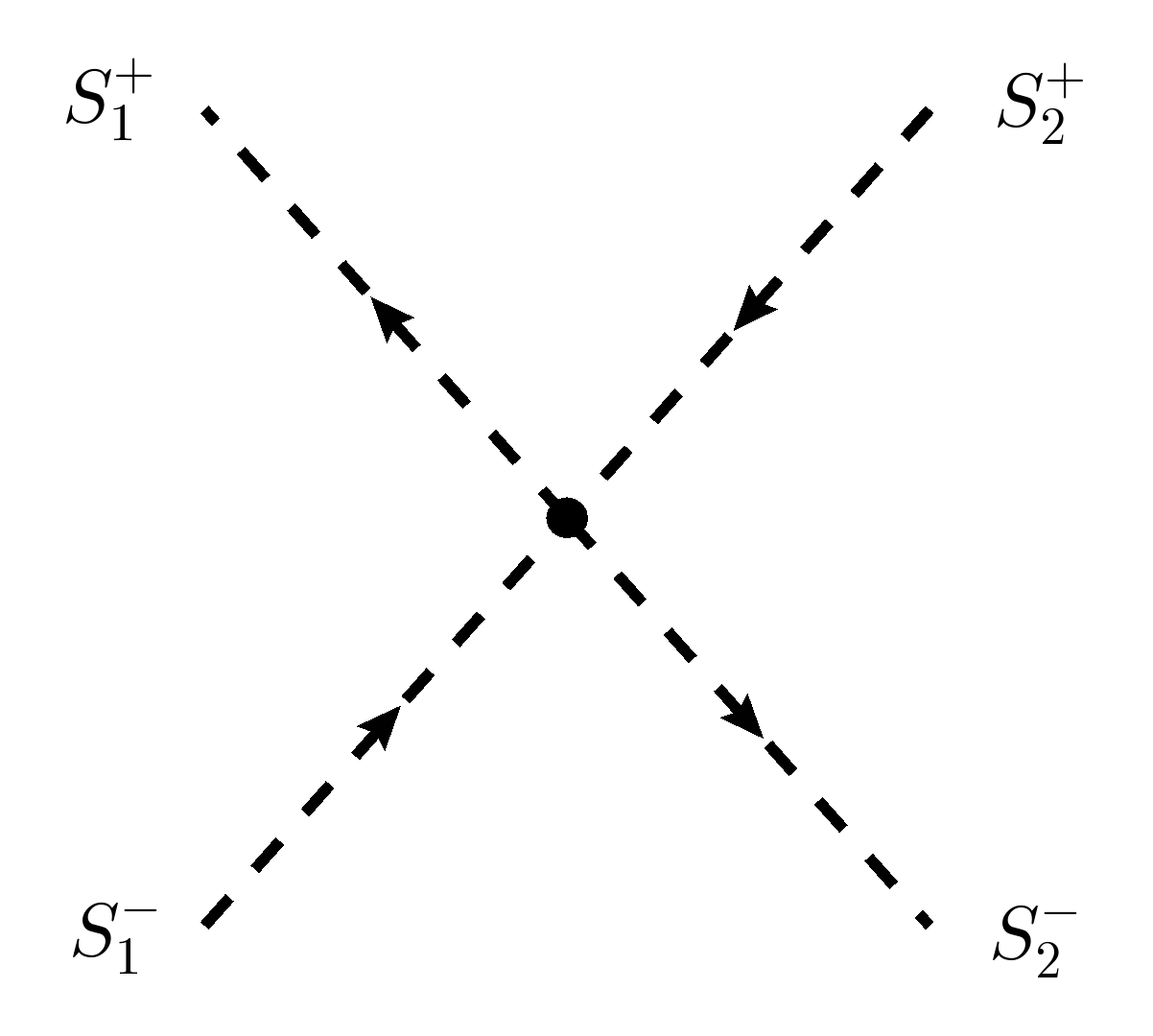} &\sim\ \begin{array}{l} {} \\ -\frac{i}{4} \Bigl[\lambda_2 + 2 \lambda_{\eta\Phi} + \lambda_{\Phi} - (\lambda_2 \\ \quad - 2 \lambda_{\eta\Phi} + \lambda_{\Phi}) c_{4\theta}\Bigr] \end{array} & \includegraphics[height=3cm, valign=c]{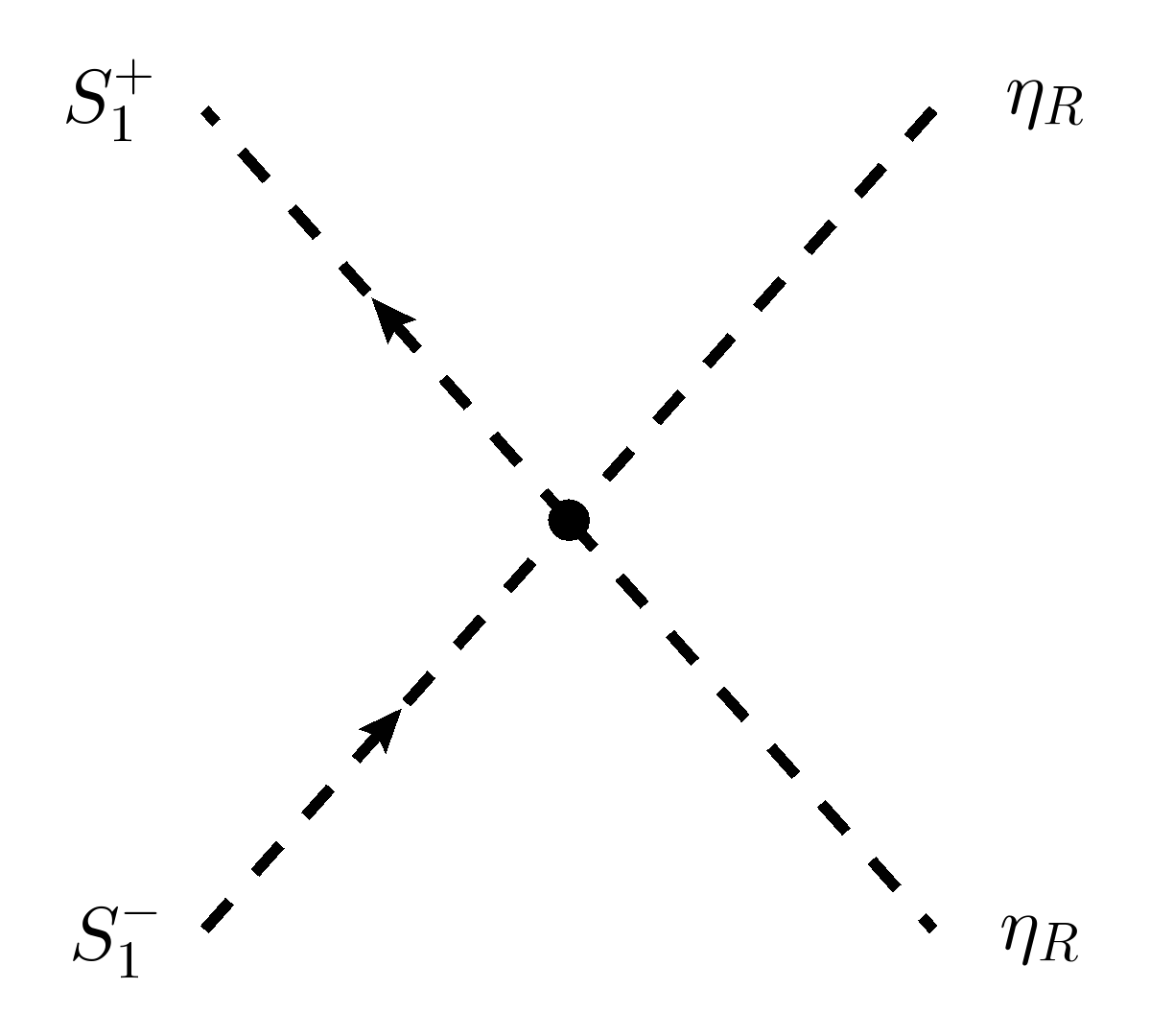} &\sim\ -2! \times i (\lambda_{\eta\Phi} c^2_\theta + \lambda_2 s^2_\theta) \\
    \includegraphics[height=3cm, valign=c]{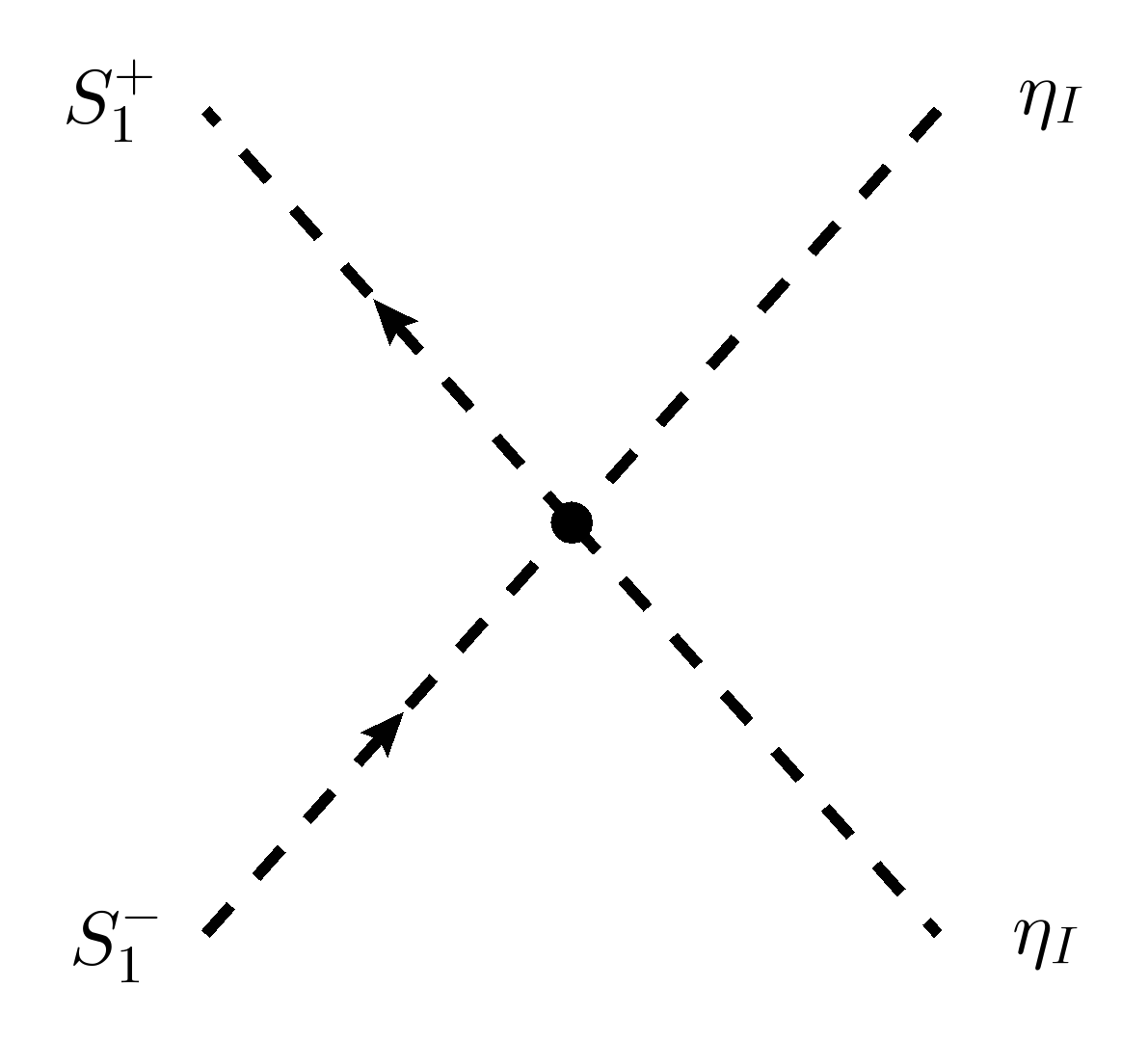} &\sim\ -2! \times i (\lambda_{\eta\Phi} c^2_\theta + \lambda_2 s^2_\theta) & \includegraphics[height=3cm, valign=c]{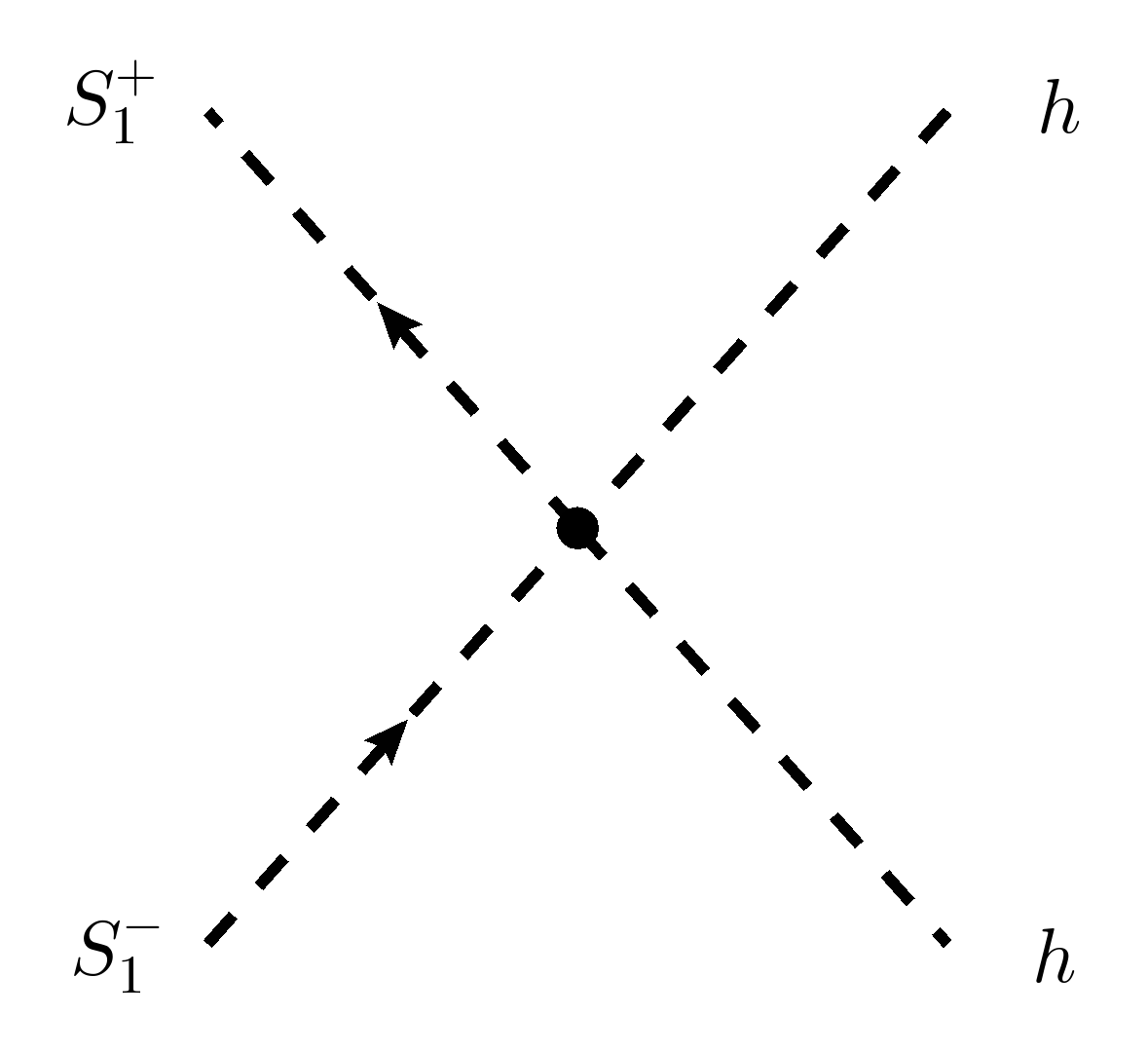} &\sim\ -2! \times i (\lambda_{H\phi} c^2_\theta + \lambda_3 s^2_\theta) \\
    \includegraphics[height=3cm, valign=c]{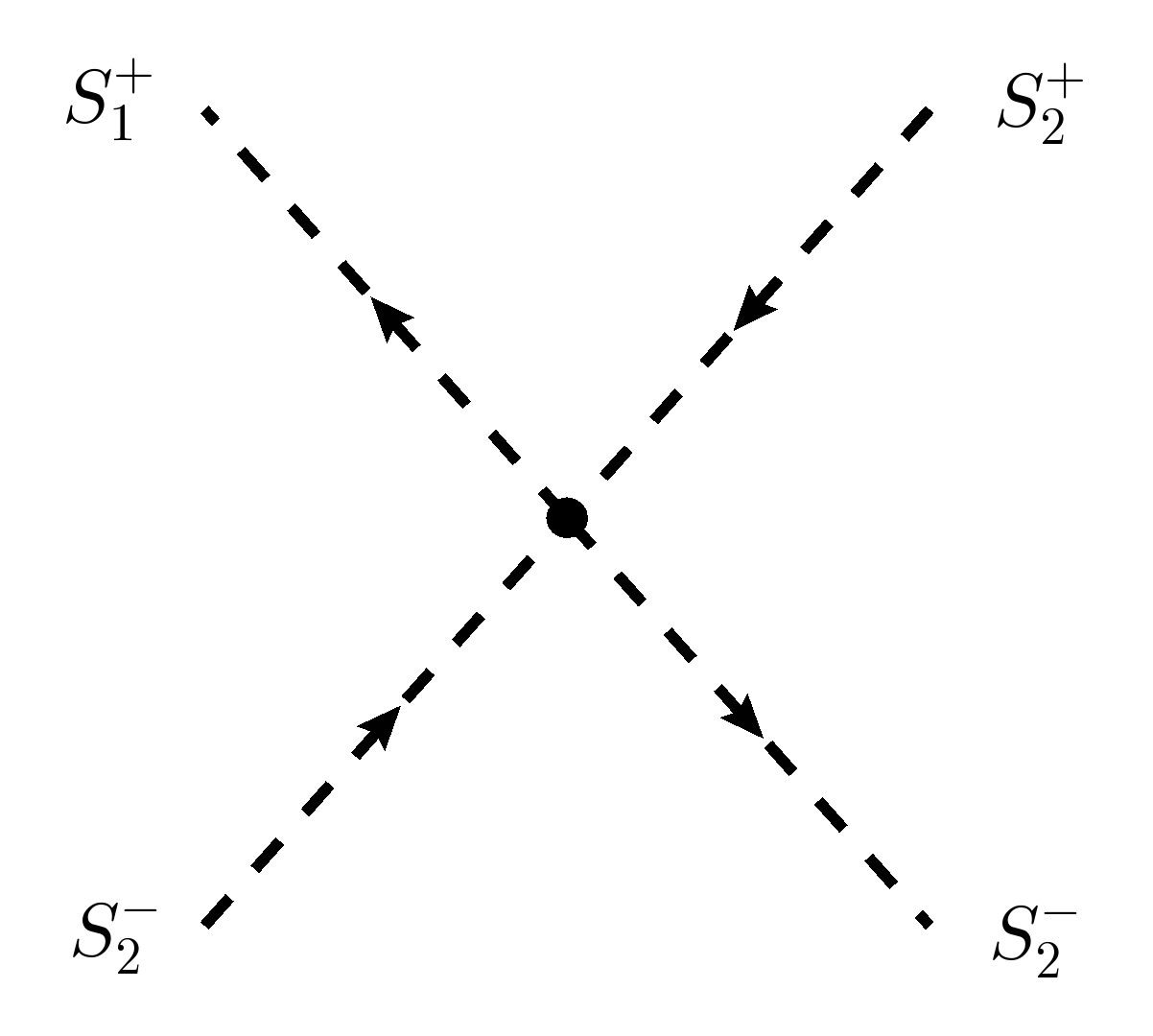} &\sim\ \begin{array}{l} {} \\ 2! \times \frac{i}{2} \Bigl[\lambda_2 - \lambda_{\Phi} + (\lambda_2  \\ \quad - 2 \lambda_{\eta\Phi} + \lambda_{\Phi}) c_{2\theta}\Bigr] s_{2\theta} \end{array} & \includegraphics[height=3cm, valign=c]{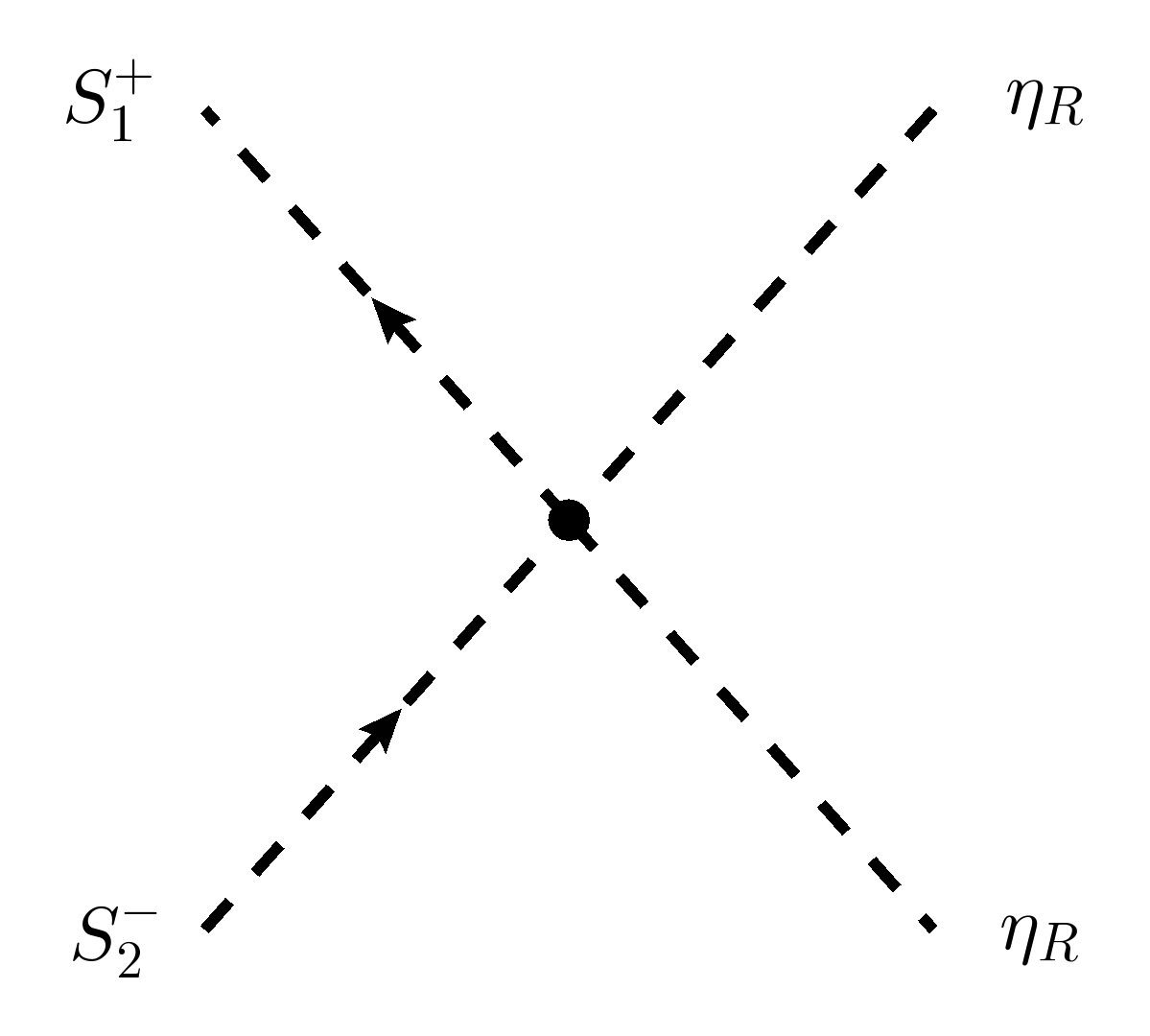} &\sim\ 2! \times i (\lambda_2 - \lambda_{\eta\Phi}) c_\theta s_\theta \\
    \includegraphics[height=3cm, valign=c]{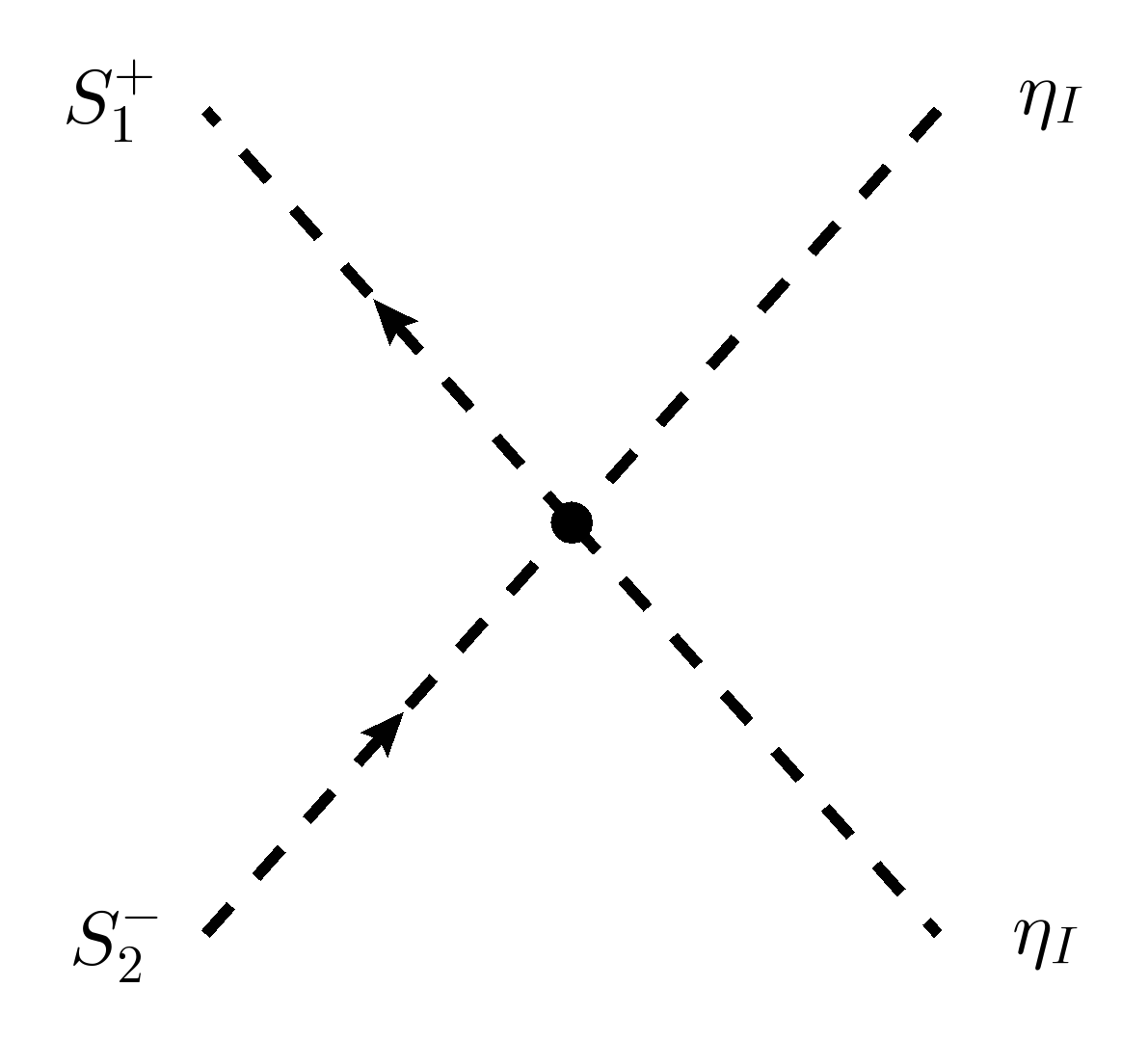} &\sim\ 2! \times i (\lambda_2 - \lambda_{\eta\Phi}) c_\theta s_\theta & \includegraphics[height=3cm, valign=c]{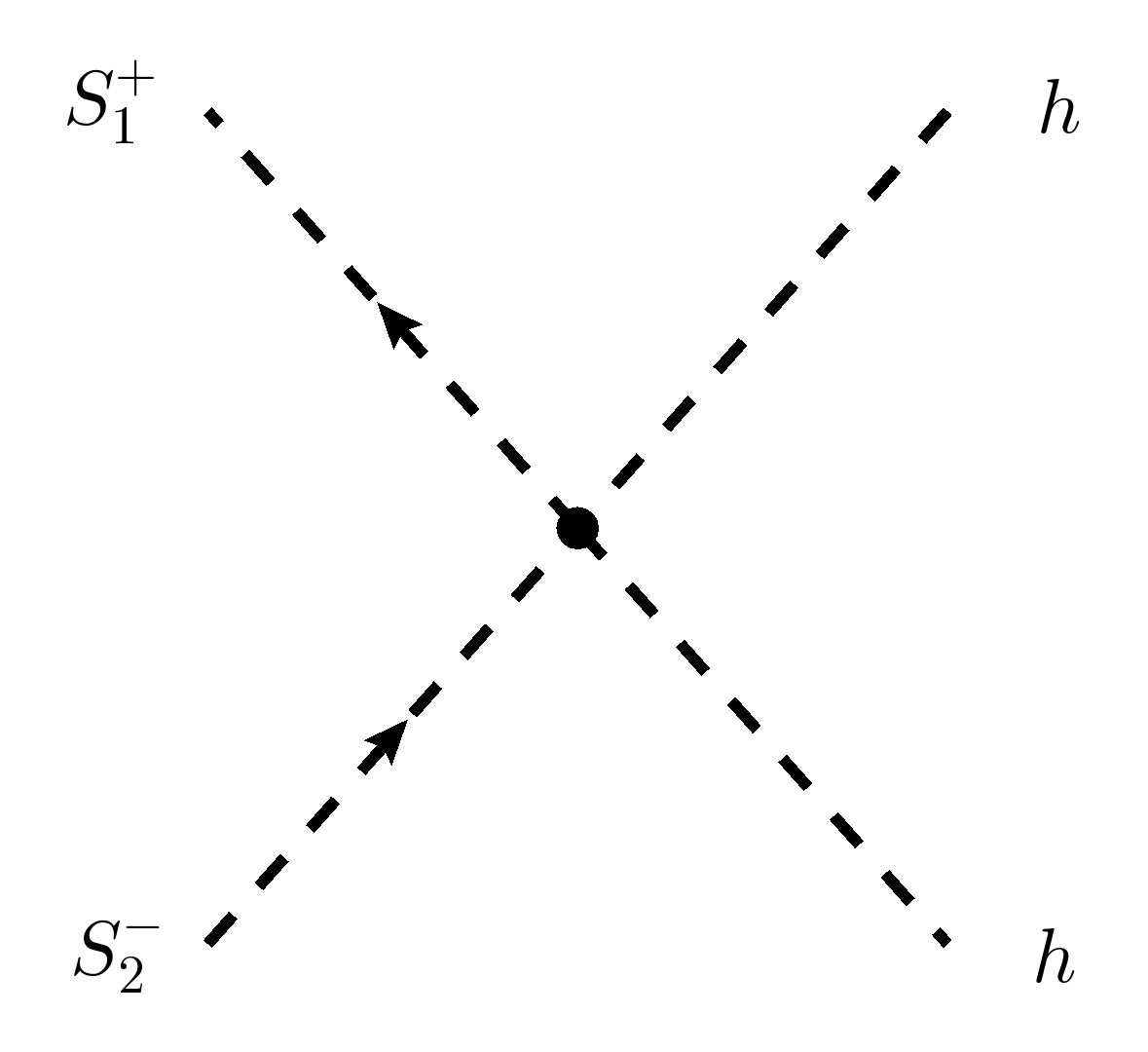} &\sim\ 2! \times i (\lambda_3 - \lambda_{H\phi}) c_\theta s_\theta \\
    \includegraphics[height=3cm, valign=c]{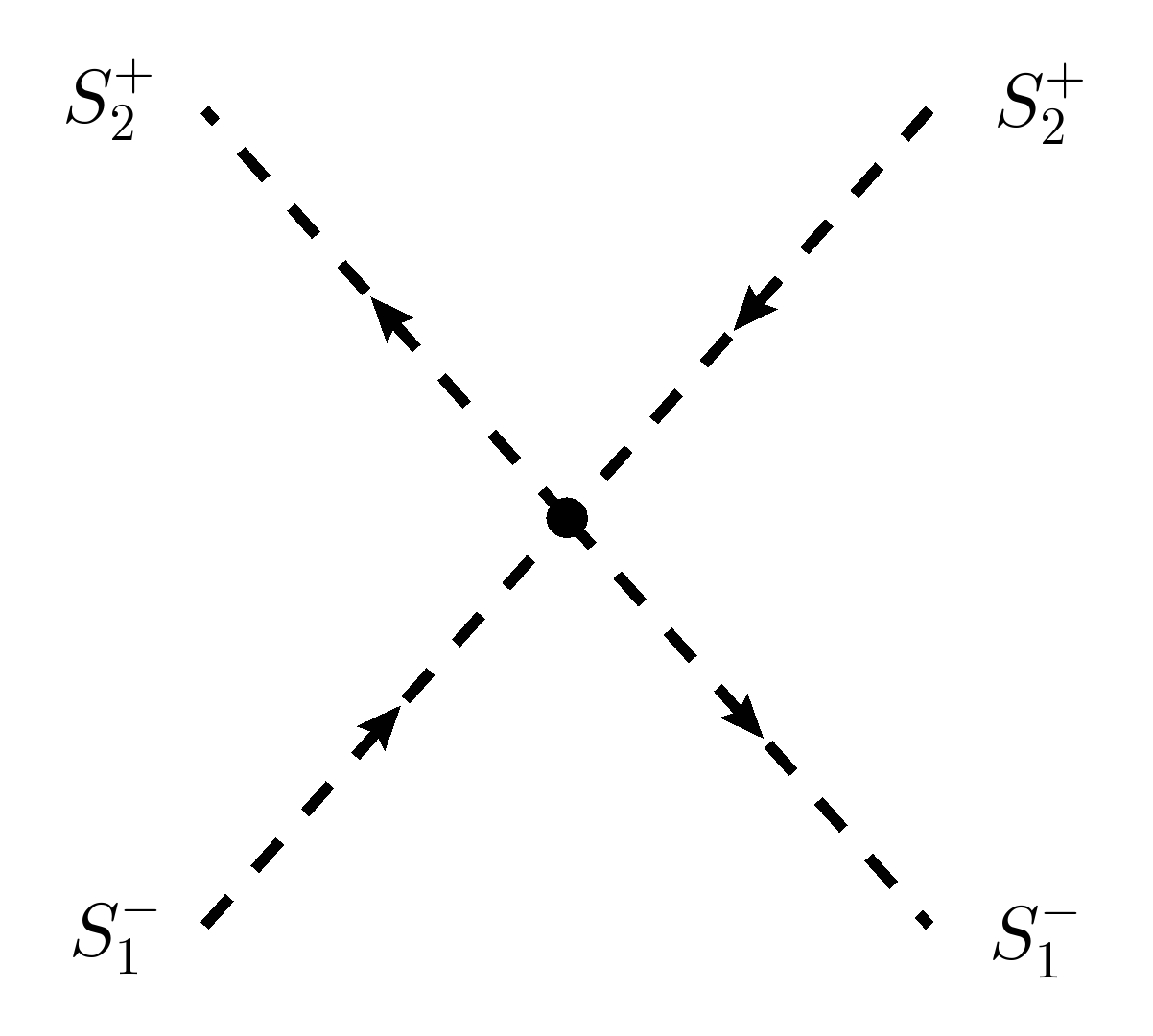} &\sim\ \begin{array}{l} {} \\ - 2! \times 2! \times \frac{i}{2} \Bigl[\lambda_2 - 2 \lambda_{\eta\Phi} \\ \quad + \lambda_{\Phi}\Bigr] s^2_{2\theta} \end{array} & \includegraphics[height=3cm, valign=c]{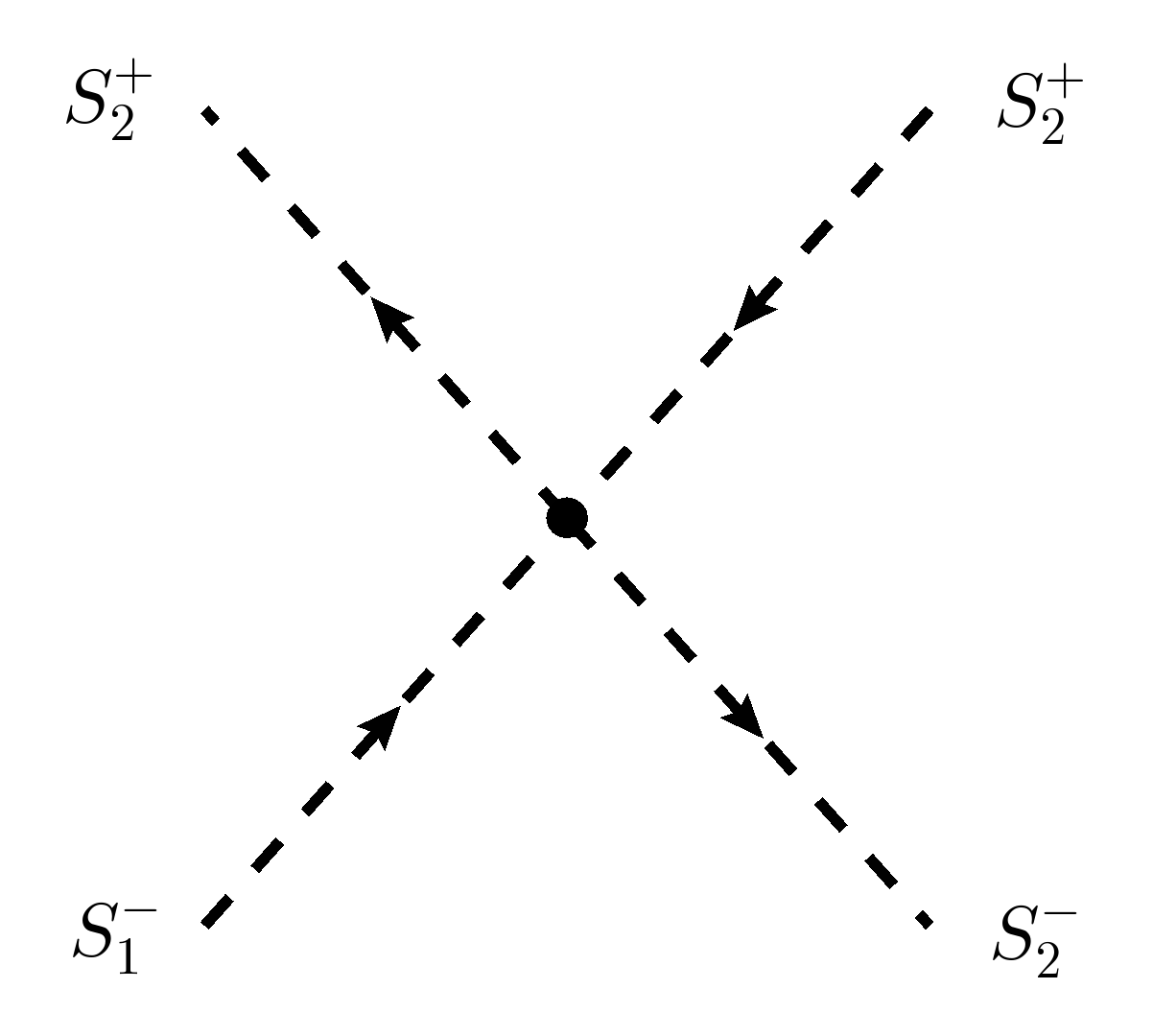} &\sim\ \begin{array}{l} {} \\ 2! \times \frac{i}{2} \Bigl[\lambda_2 - \lambda_{\Phi} + (\lambda_2 \\ \quad - 2 \lambda_{\eta\Phi} + \lambda_{\Phi}) c_{2\theta}\Bigr] s_{2\theta} \end{array} \\
    \includegraphics[height=3cm, valign=c]{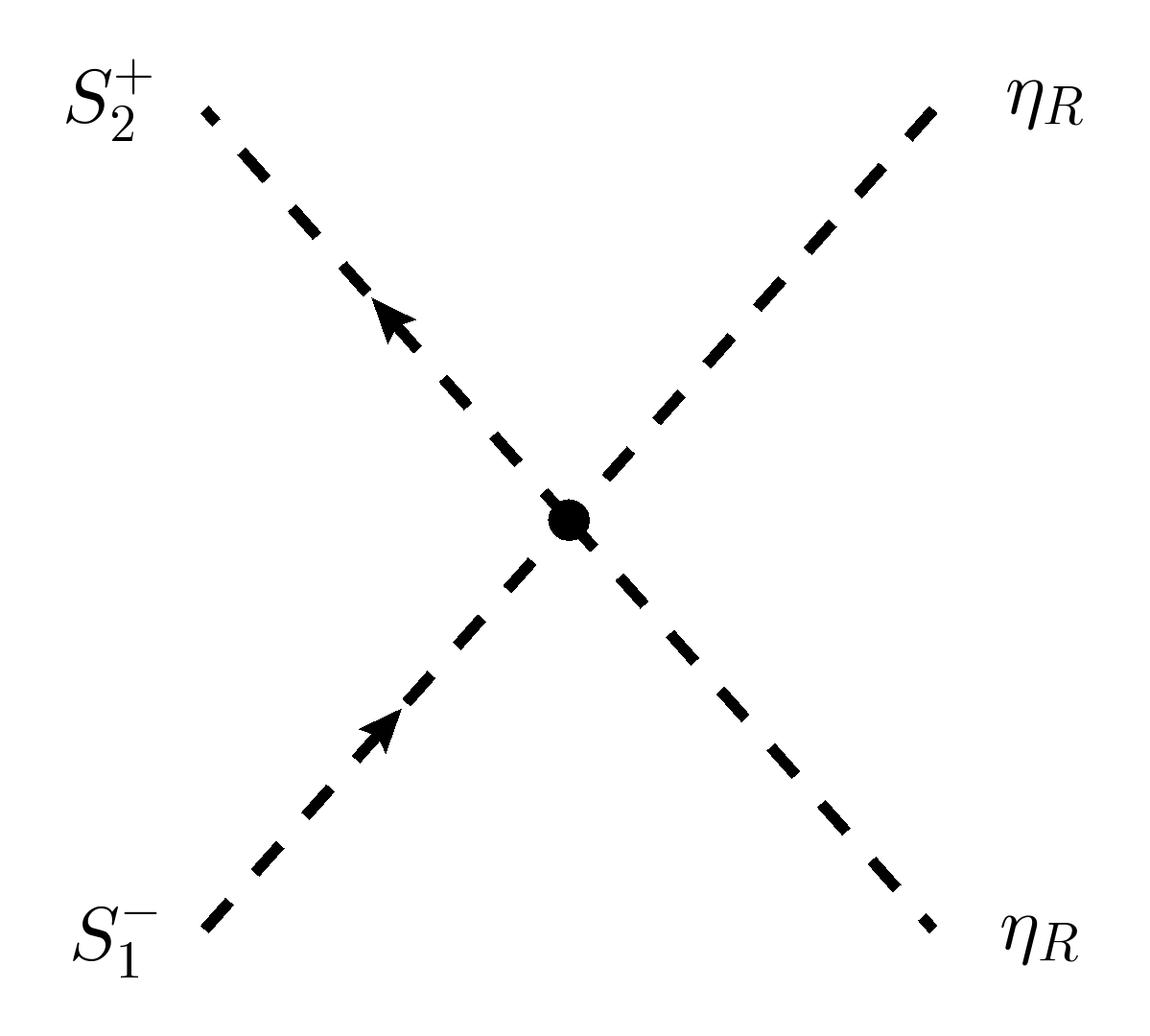} &\sim\ 2! \times i (\lambda_2 - \lambda_{\eta\Phi}) c_\theta s_\theta & \includegraphics[height=3cm, valign=c]{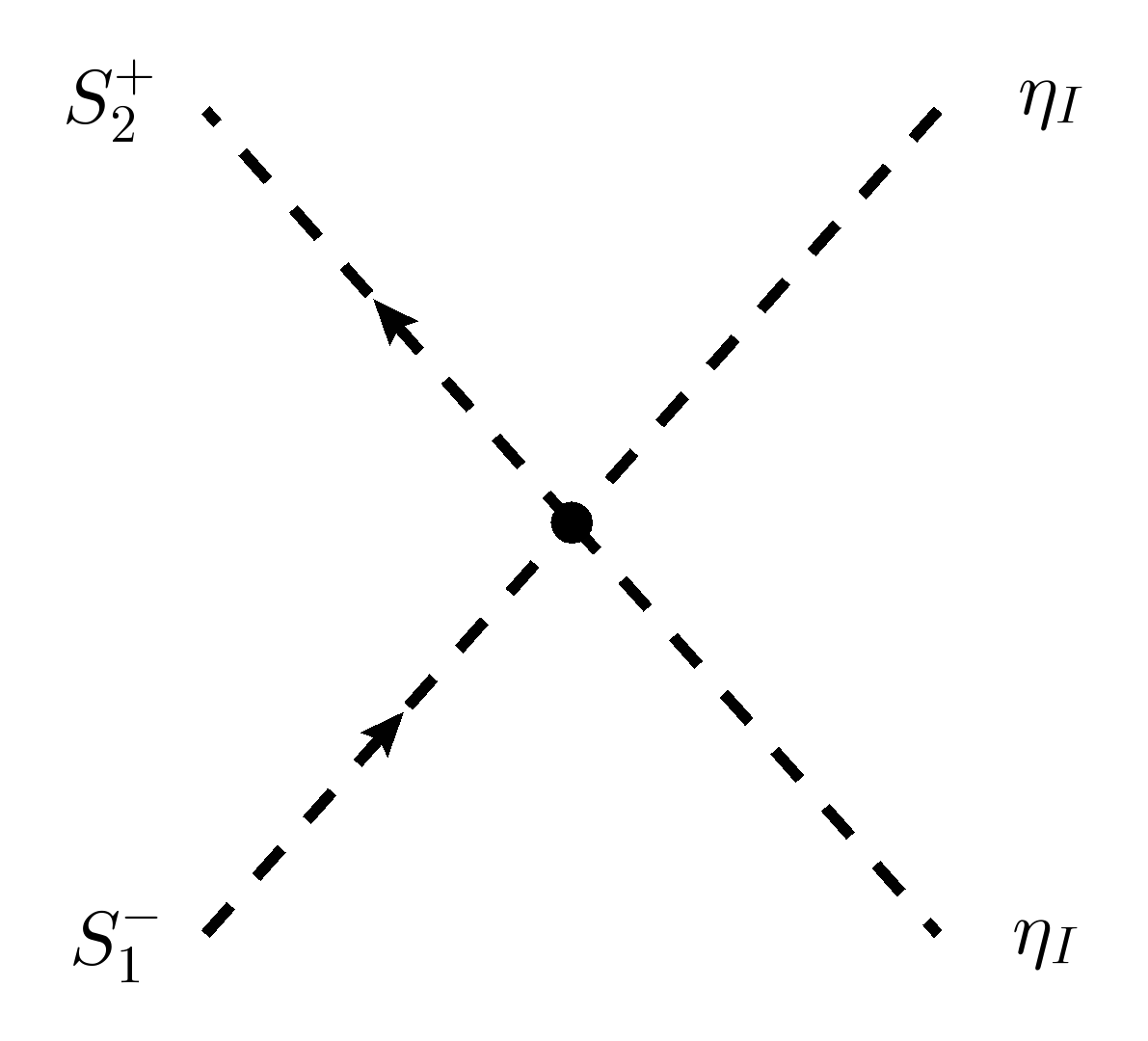} &\sim\ 2! \times i (\lambda_2 - \lambda_{\eta\Phi}) c_\theta s_\theta \\
    \includegraphics[height=3cm, valign=c]{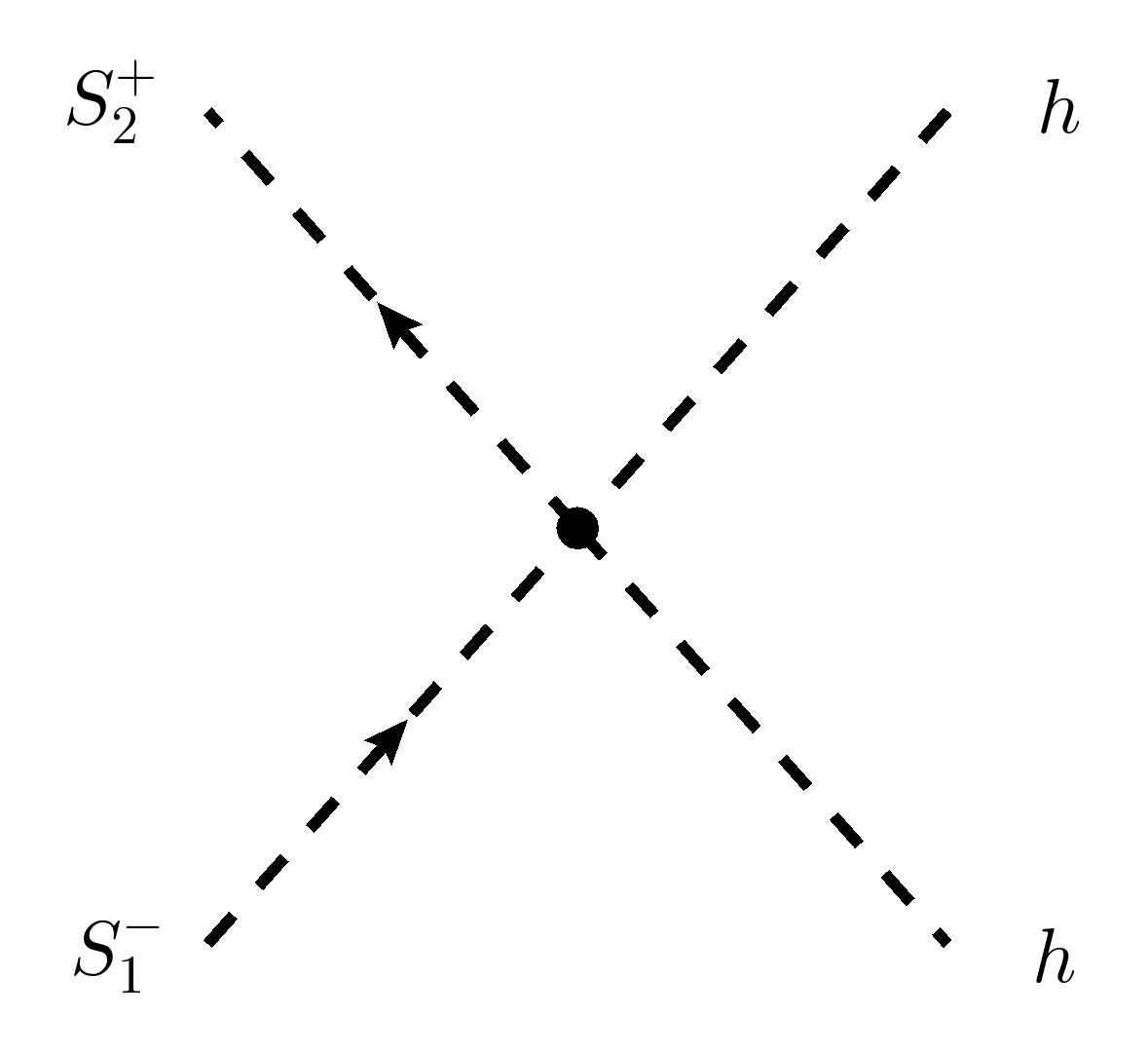} &\sim\ 2! \times i (\lambda_3 - \lambda_{H\phi}) c_\theta s_\theta & \includegraphics[height=3cm, valign=c]{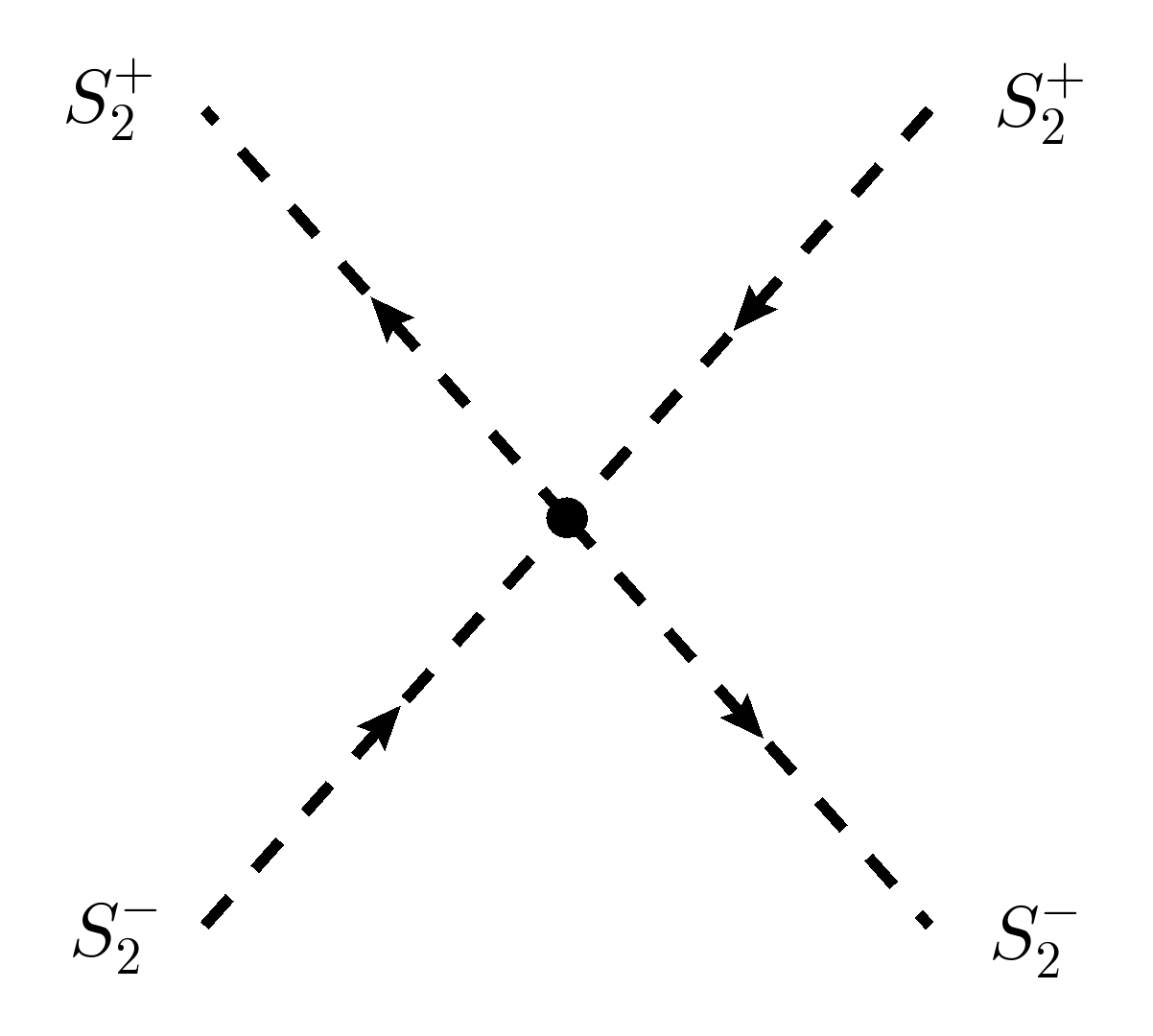} &\sim\ \begin{array}{l} {} \\ - 2! \times 2! \times 2 i \Bigl[\lambda_2 c^4_\theta \\ \quad + 2 \lambda_{\eta\Phi} c^2_\theta s^2_\theta + \lambda_{\Phi} s^4_\theta\Bigr] \end{array} \\
    \includegraphics[height=3cm, valign=c]{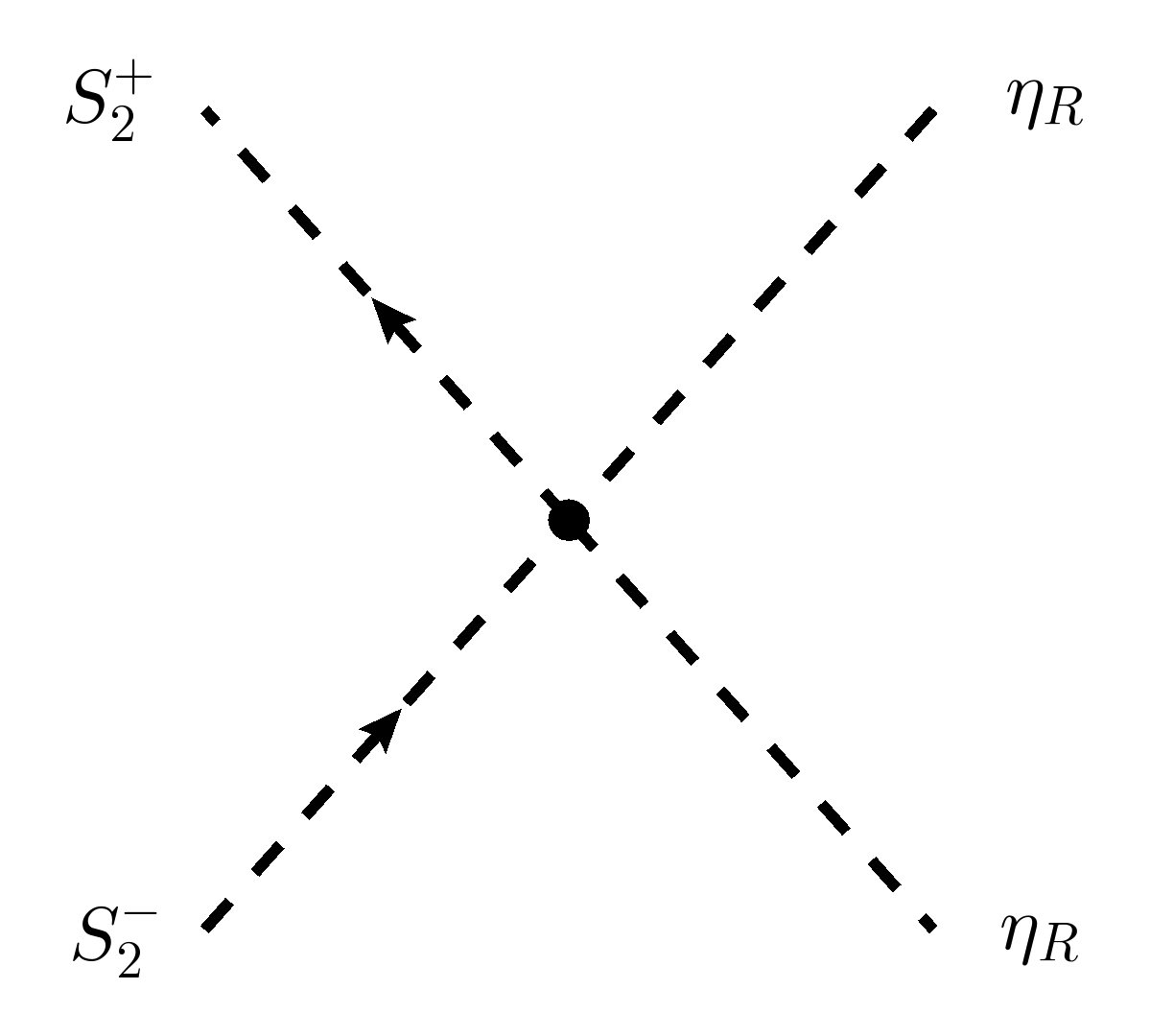} &\sim\ - 2! \times i (\lambda_2 c^2_\theta + \lambda_{\eta\Phi} s^2_\theta) & \includegraphics[height=3cm, valign=c]{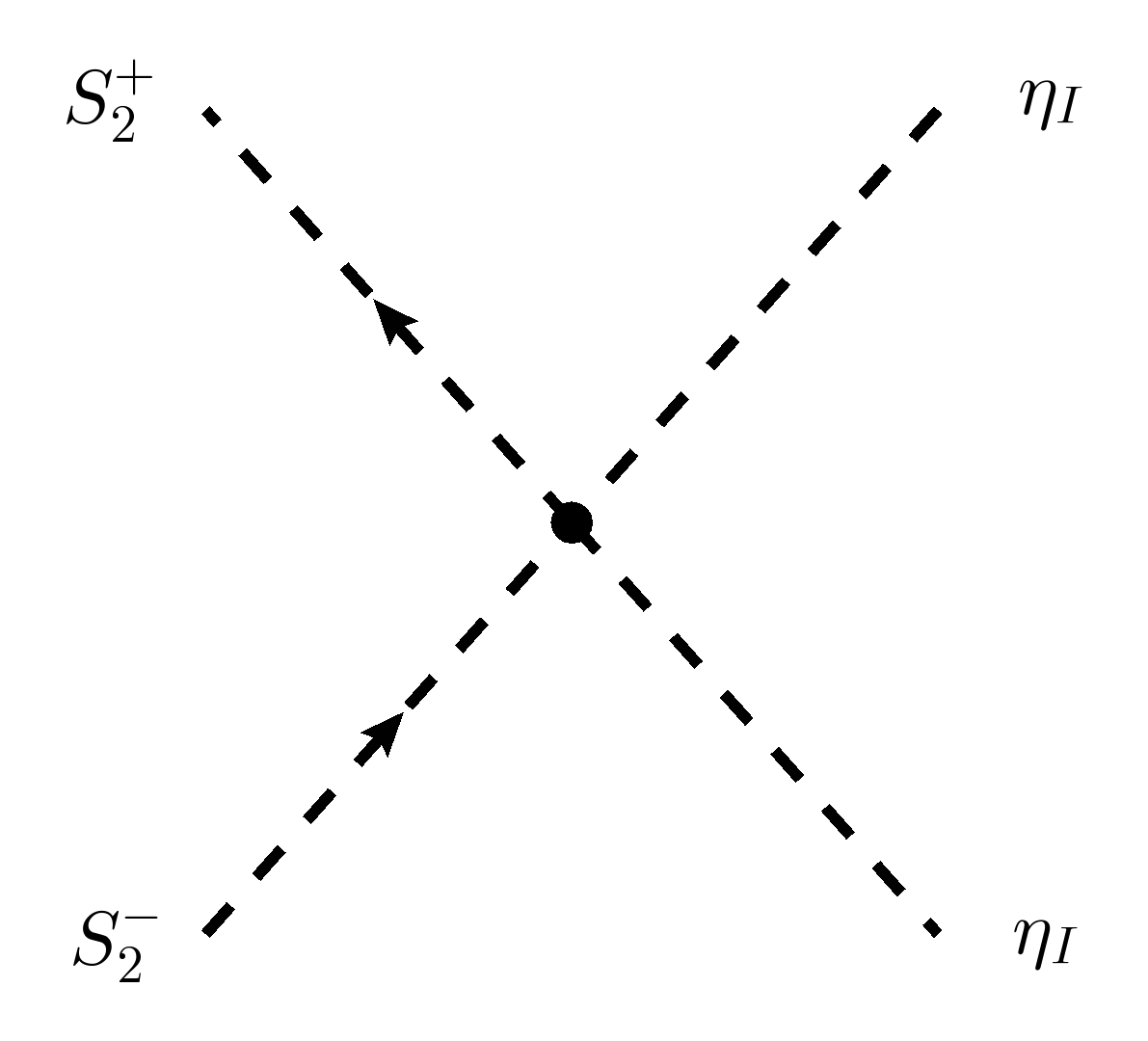} &\sim\ - 2! \times i (\lambda_2 c^2_\theta + \lambda_{\eta\Phi} s^2_\theta) \\
    \includegraphics[height=3cm, valign=c]{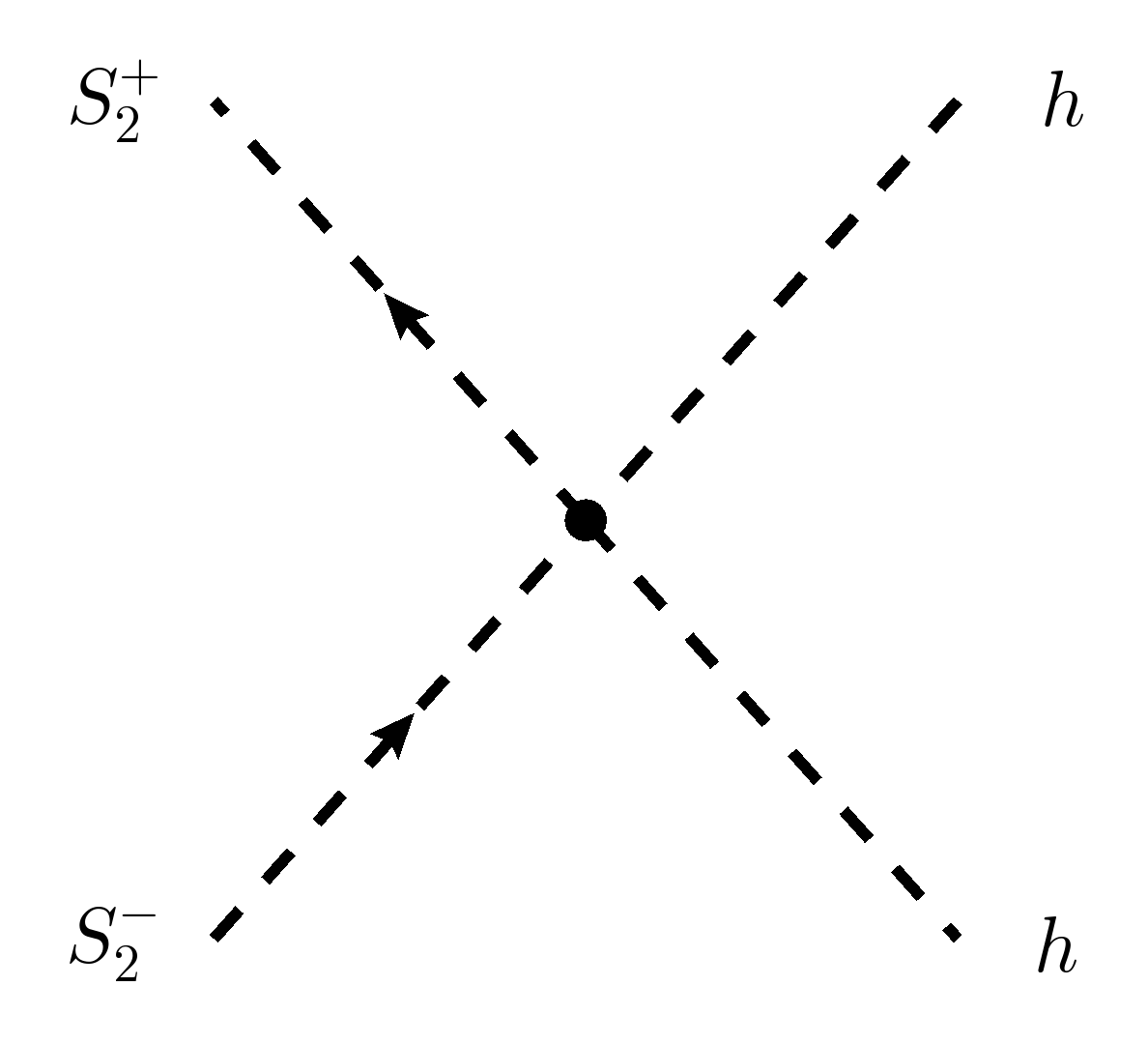} &\sim\ - 2! \times i (\lambda_3 c^2_\theta + \lambda_{H\phi} s^2_\theta) & \includegraphics[height=3cm, valign=c]{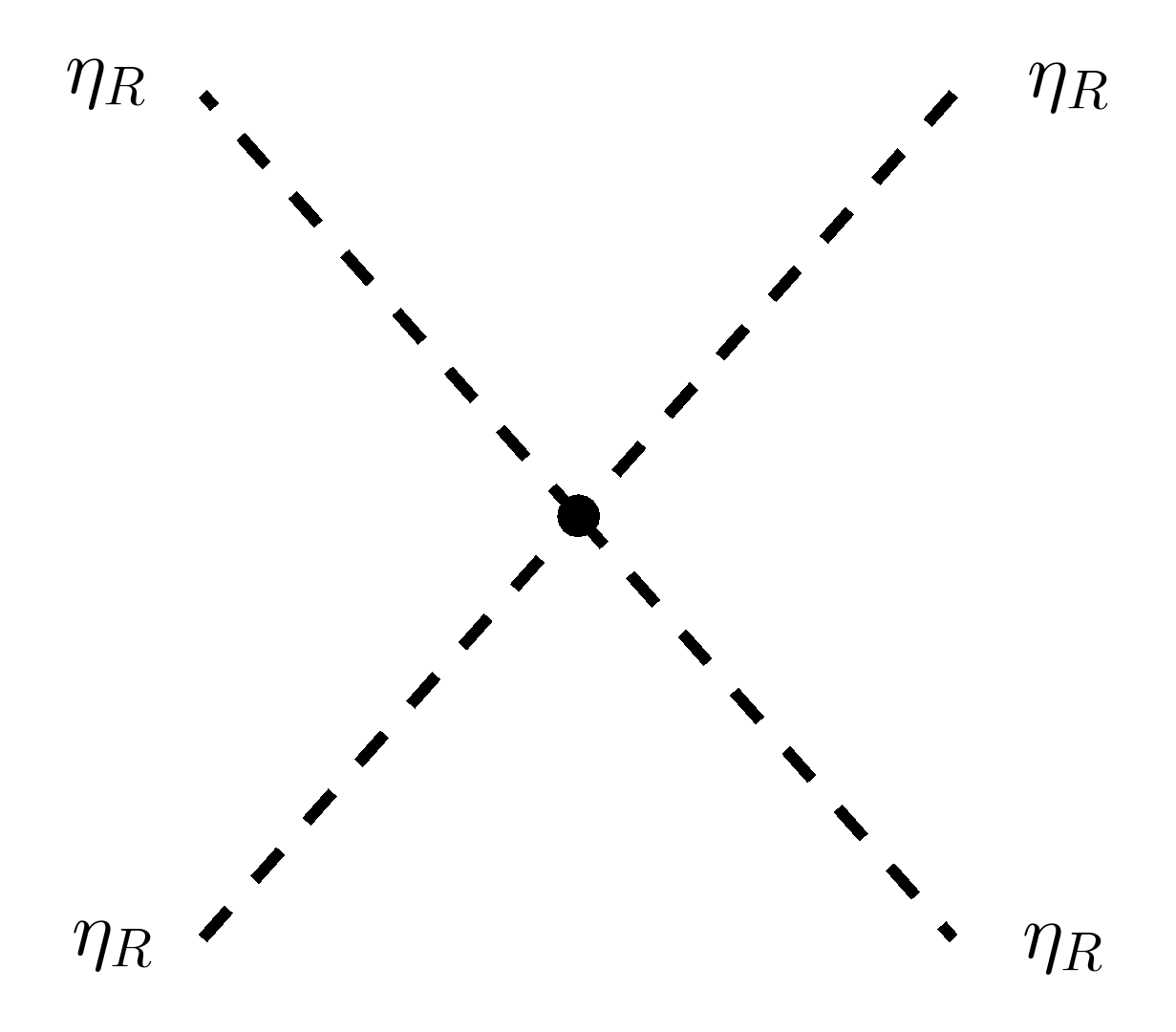} &\sim\ - 4! \times 3 i \lambda_2 \\
    \includegraphics[height=3cm, valign=c]{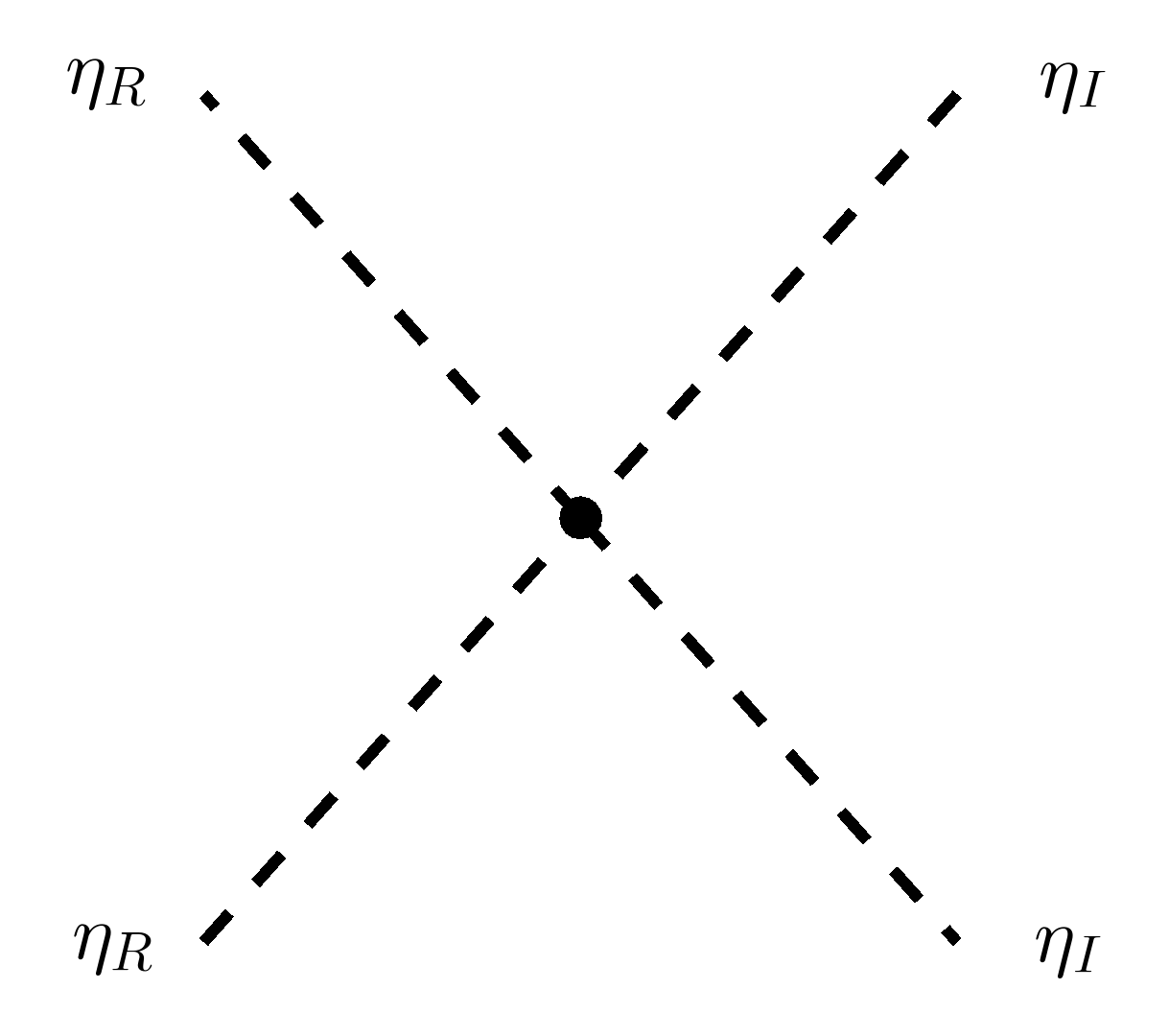} &\sim\ - 2! \times 2! \times i \lambda_2 & \includegraphics[height=3cm, valign=c]{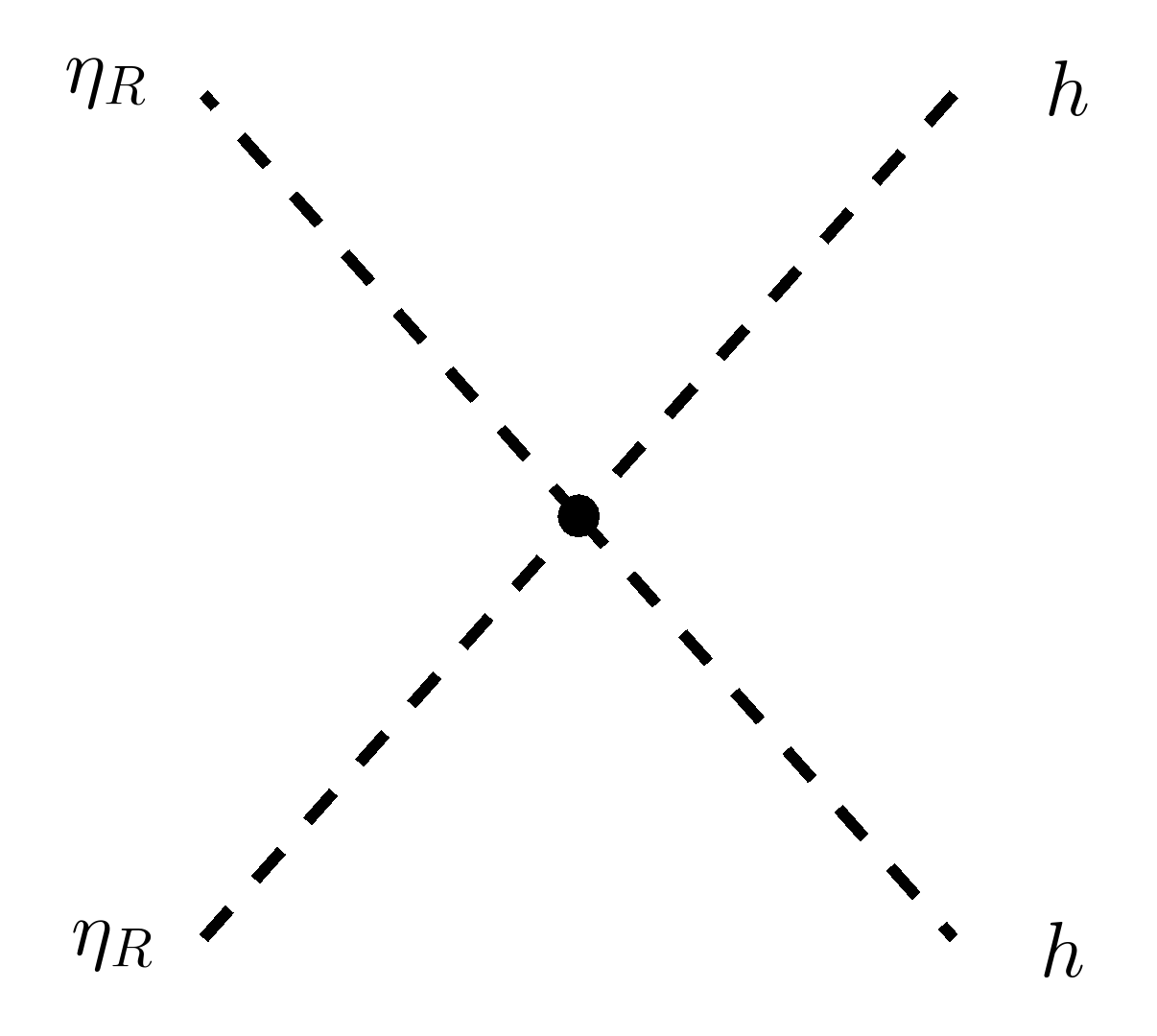} &\sim\ \begin{array}{l} {} \\ - 2! \times 2! \times i (\lambda_3 \\ \quad + \lambda_4 + \lambda_5) \end{array} \\
    \includegraphics[height=3cm, valign=c]{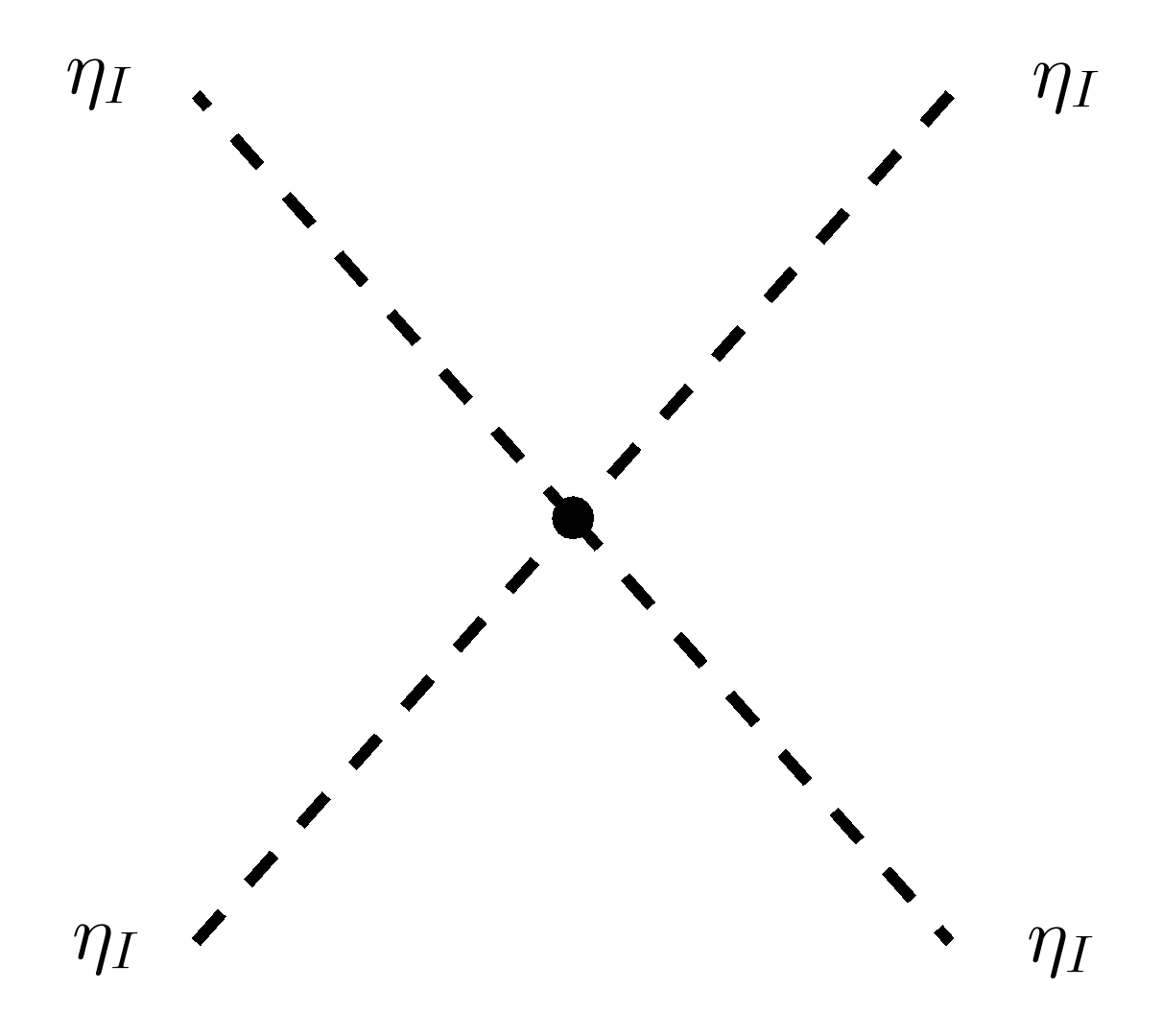} &\sim\ - 4! \times 3 i \lambda_2 & \includegraphics[height=3cm, valign=c]{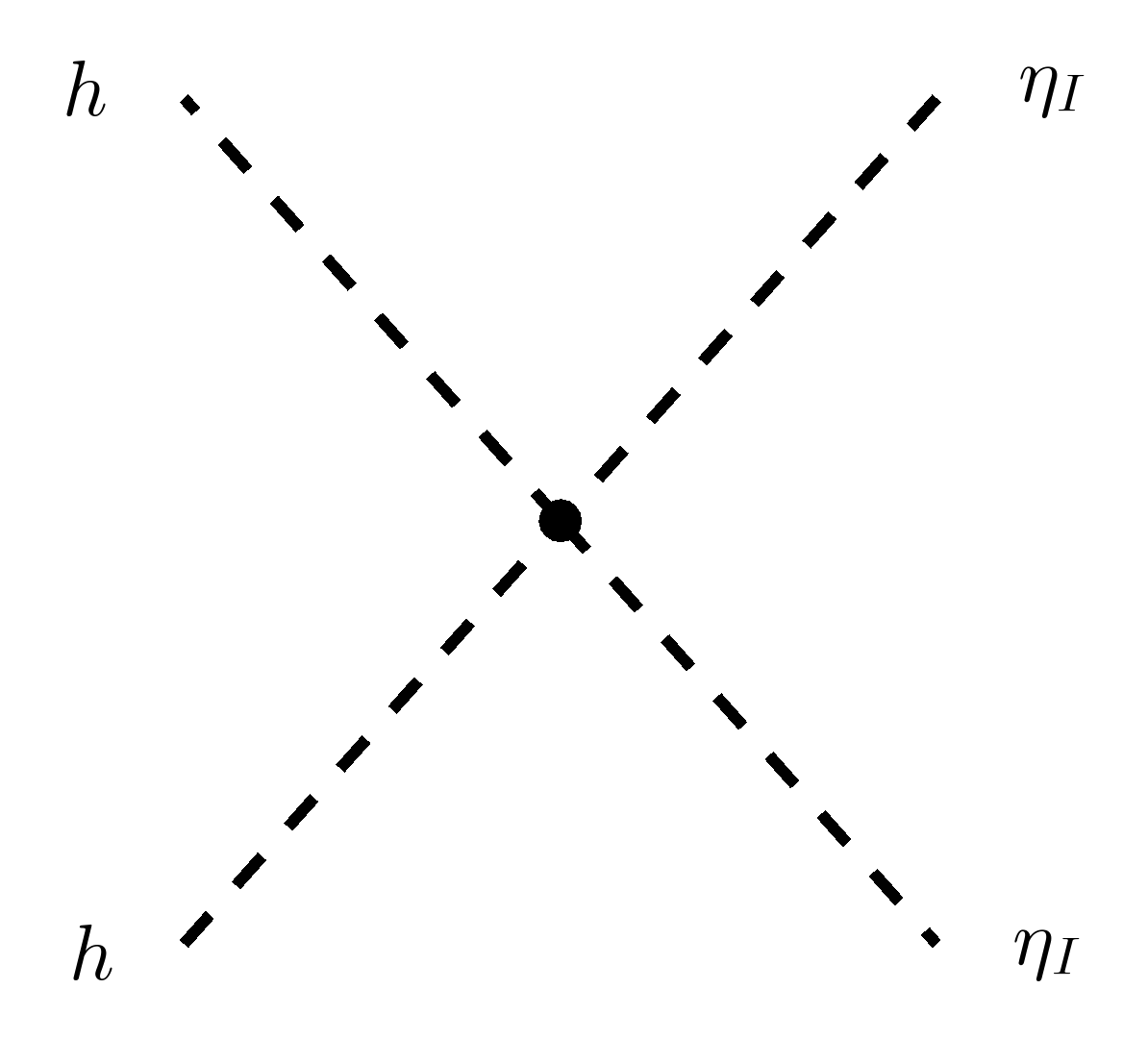} &\sim\ \begin{array}{l} {} \\ - 2! \times 2! \times i (\lambda_3 \\ \quad + \lambda_4 - \lambda_5) \end{array} \\
    \includegraphics[height=3cm, valign=c]{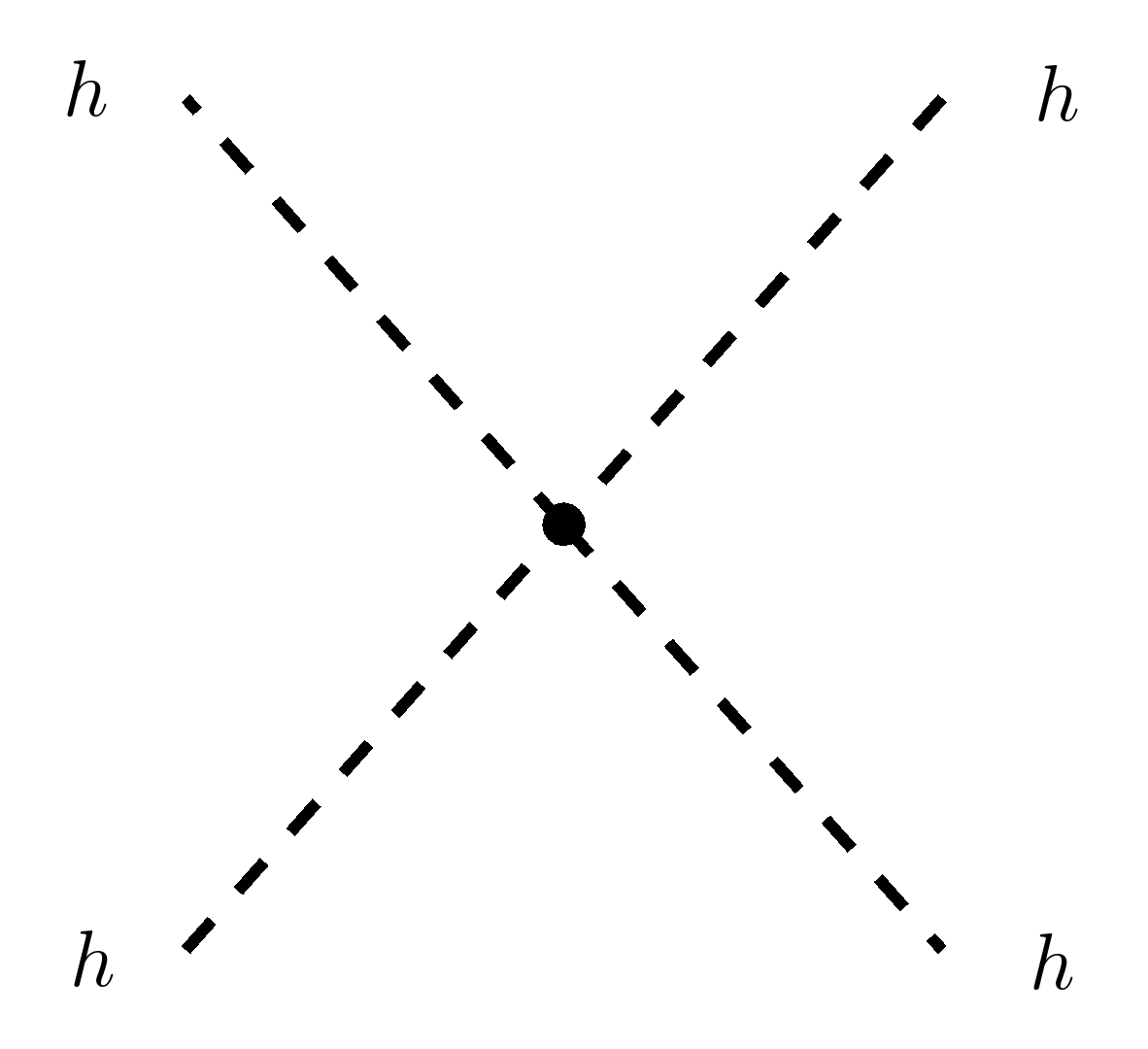} &\sim\ - 4! \times 3 i \lambda_1 & &
\end{align*}

\subsection{Gauge Sector}
\label{appendix:vertex_factors_gauge_sector}

The new interactions between gauge bosons and scalars come from~\footnote{Singlet fermions have the kinetic term $(D_\mu N_i)^\dagger D^\mu N_i = (\partial_\mu N_i)^\dagger\partial^\mu N_i$.}
\begin{equation}
    \mathcal{L} \supset (D_\mu\eta)^\dagger (D^\mu\eta) + (D_\mu\phi)^\dagger (D^\mu\phi) \, ,
\end{equation}
where the covariant derivatives are
\begin{subequations}
\begin{align}
    D_\mu\eta &= \Bigl(\partial_\mu - ig\frac{\sigma^j}{2}W^j_\mu - ig^\prime \frac{1}{2} B_\mu\Bigr)\eta \, ,\\
    D_\mu\phi &= \Bigr(\partial_\mu + ig^\prime B_\mu\Bigl)\phi \, .
\end{align}
\end{subequations}
The vertex factors involving one gauge boson and two scalars with incoming momenta $p$ and $p'$, with all particles incoming, are
\begin{align*}
    \includegraphics[height=3cm, valign=c]{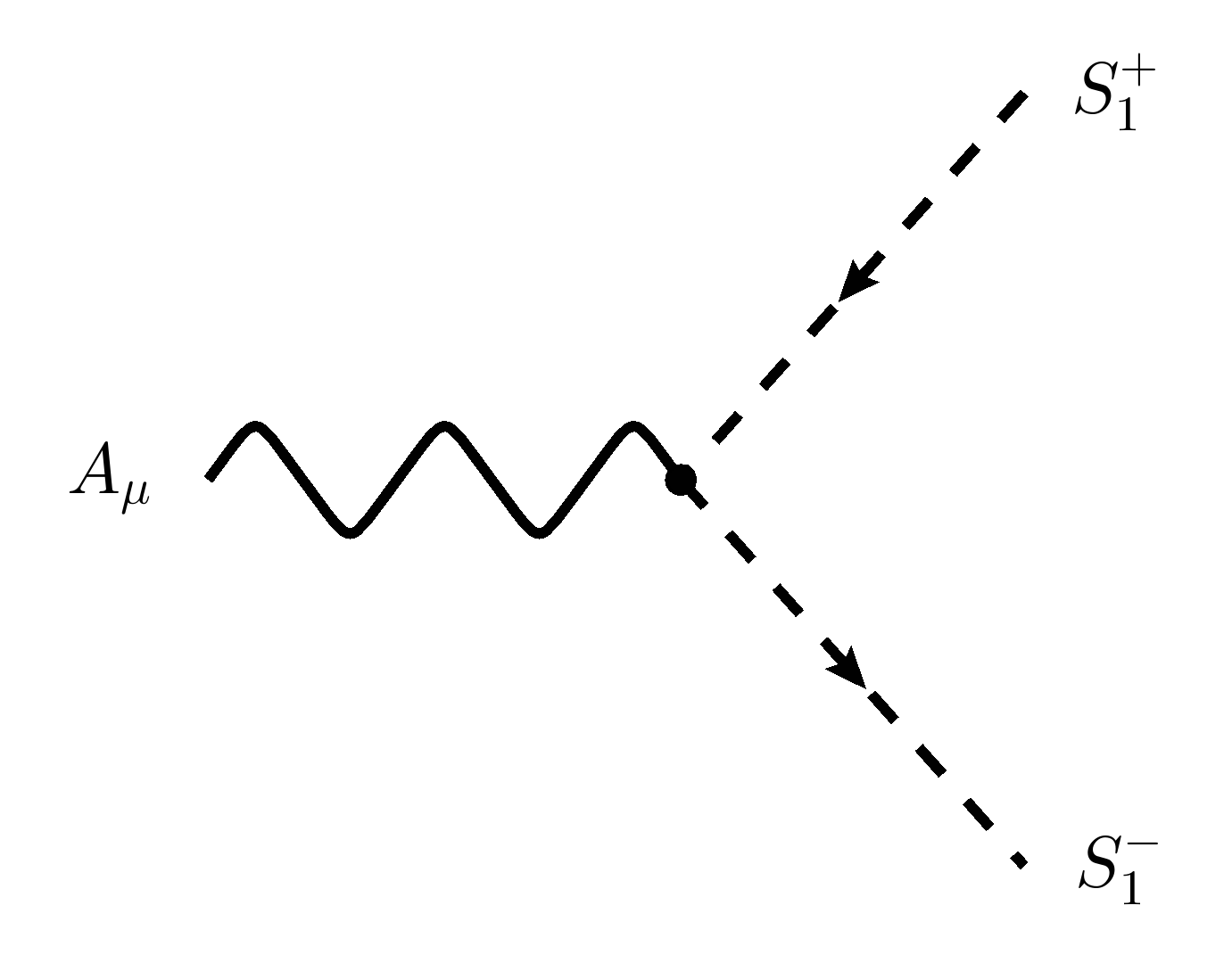} &\sim\ \begin{array}{l} {} \\ \frac{1}{2} i \Bigl[2 g' c_{\theta}^2 c_W + s_{\theta}^2 (g' c_W \\ \quad + g s_W)\Bigr](p - p')_\mu \end{array} & \includegraphics[height=3cm, valign=c]{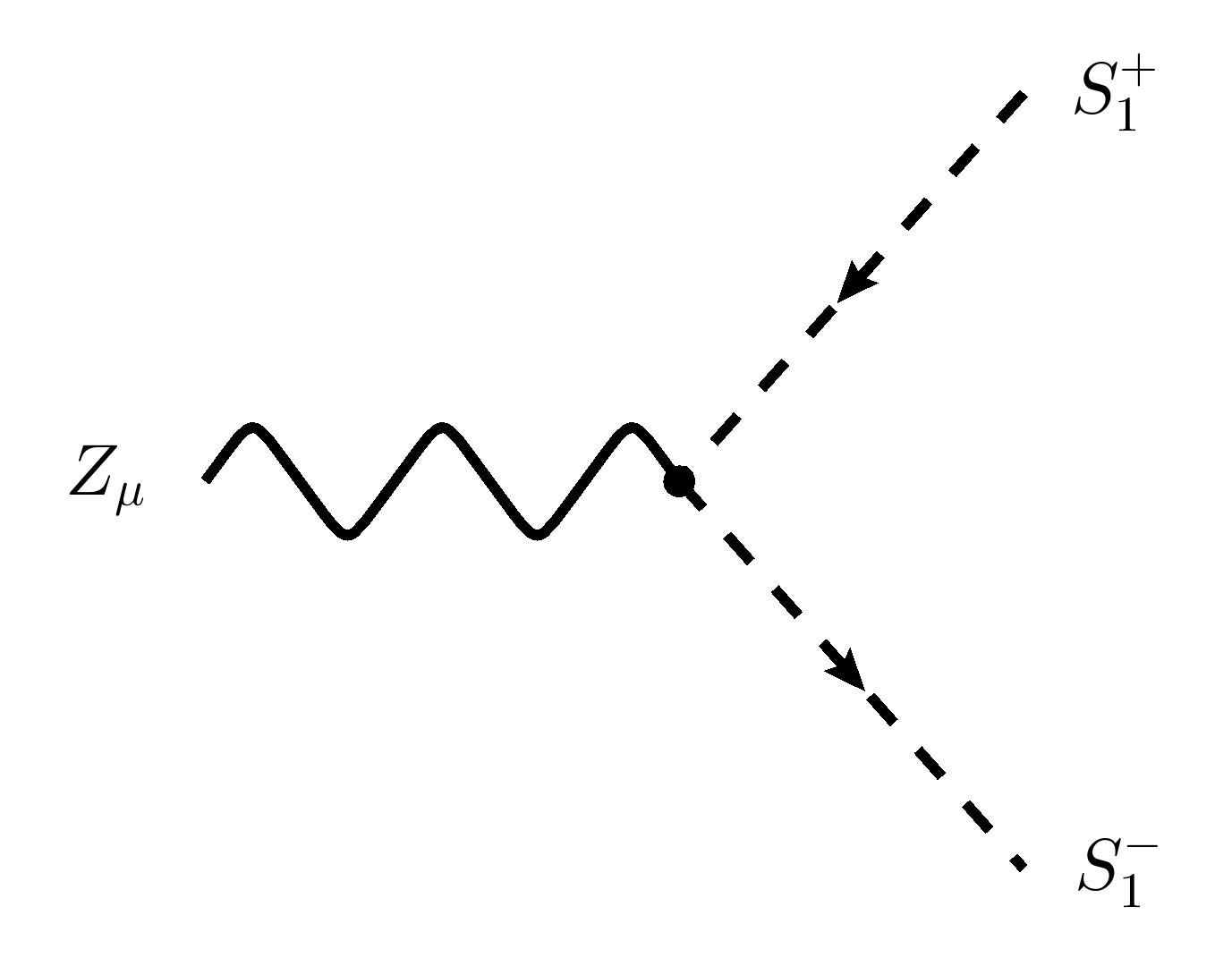} &\sim\ \begin{array}{l} {} \\ \frac{1}{2} i \Bigl[-g c_W s_{\theta}^2 + \frac{1}{2} g' (3 + \\ \quad \cos 2\theta) s_W\Bigr](p - p')_\mu \end{array} \\
    \includegraphics[height=3cm, valign=c]{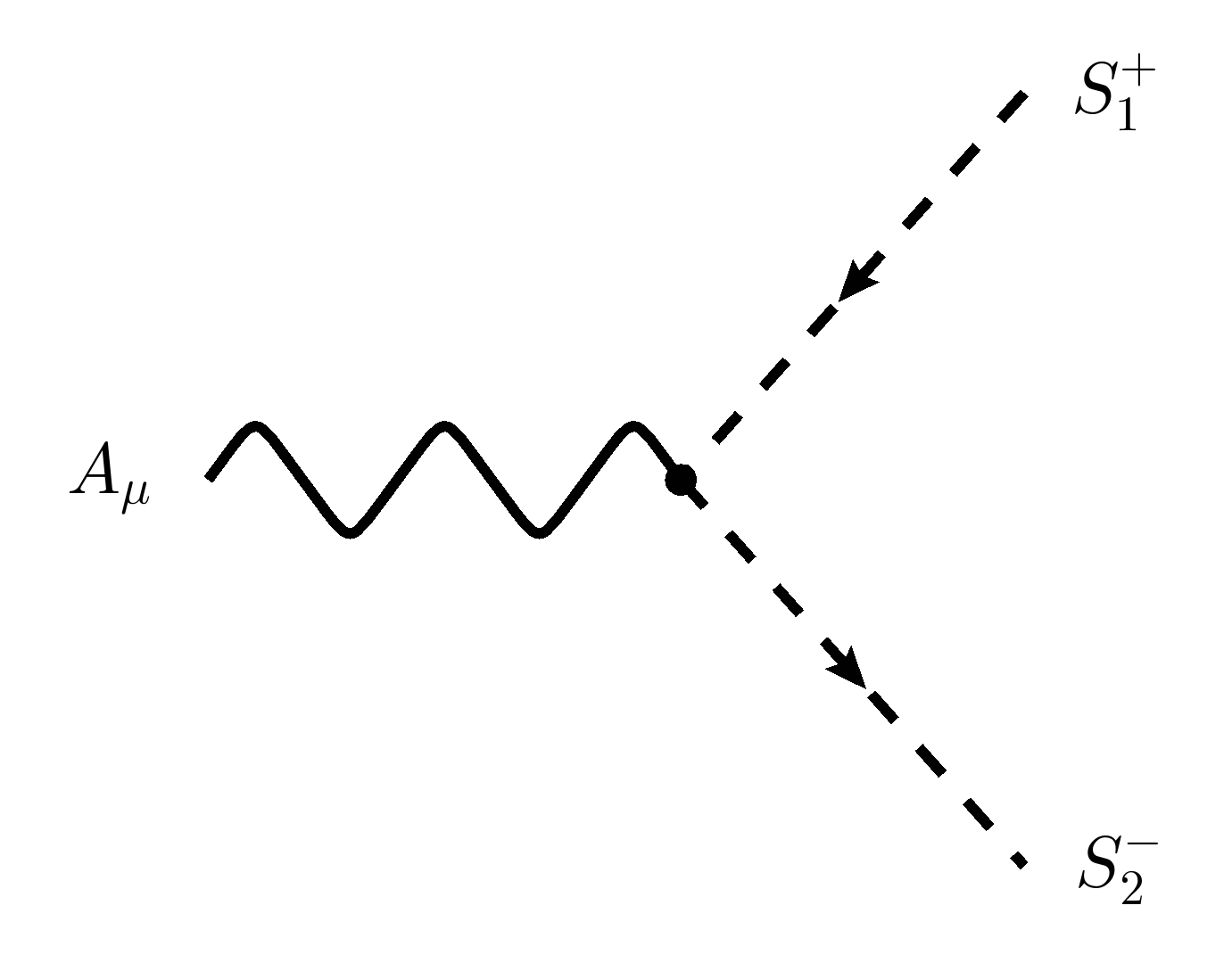} &\sim\ \begin{array}{l} {} \\ \frac{1}{4} i s_{2\theta} \left(g' c_W - g s_W\right) \\ \quad \times (p - p')_\mu \end{array} & \includegraphics[height=3cm, valign=c]{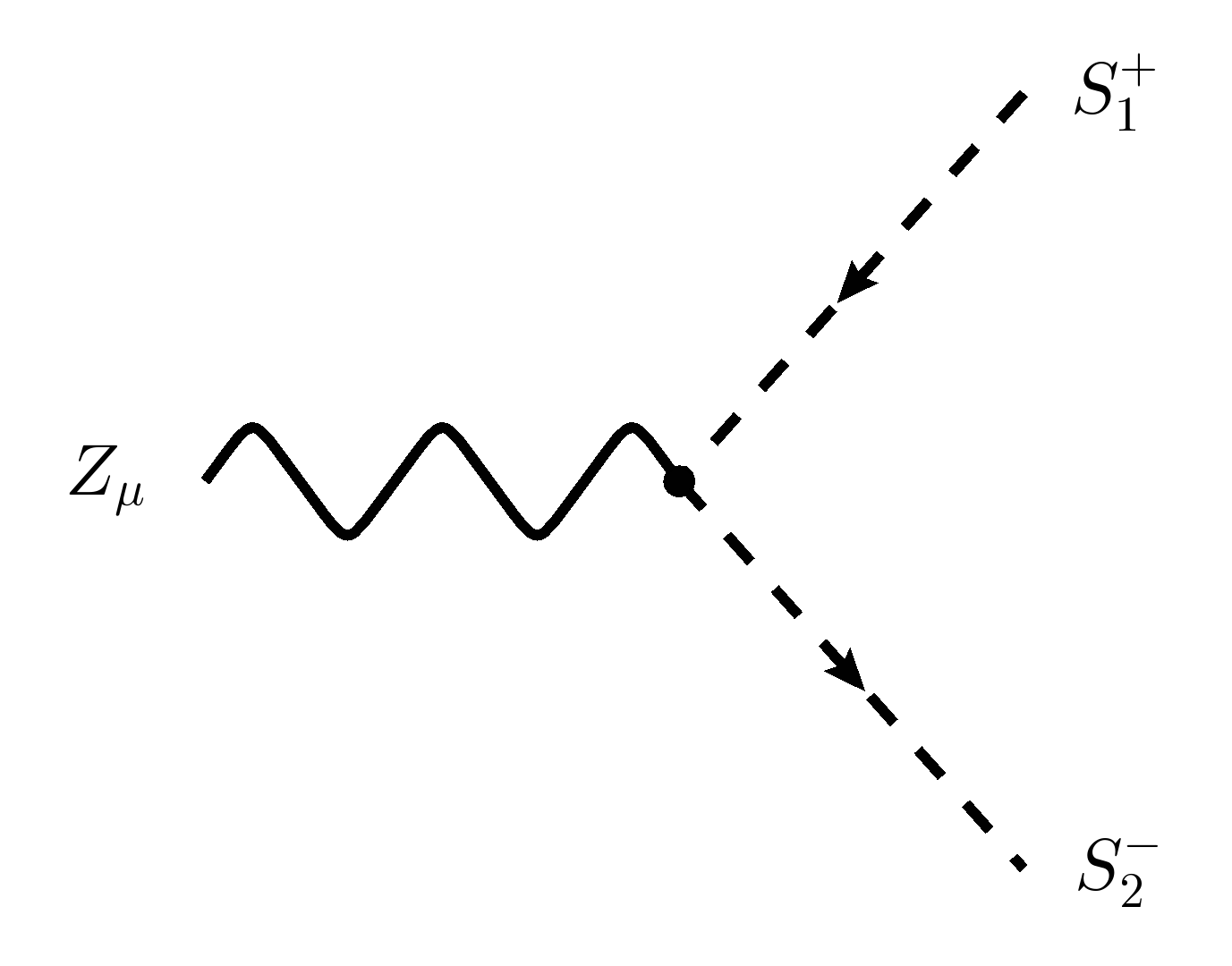} &\sim\ \begin{array}{l} {} \\ \frac{1}{4} i s_{2\theta} \left(g c_W + g' s_W\right) \\ \quad \times (p - p')_\mu \end{array}\\
    \includegraphics[height=3cm, valign=c]{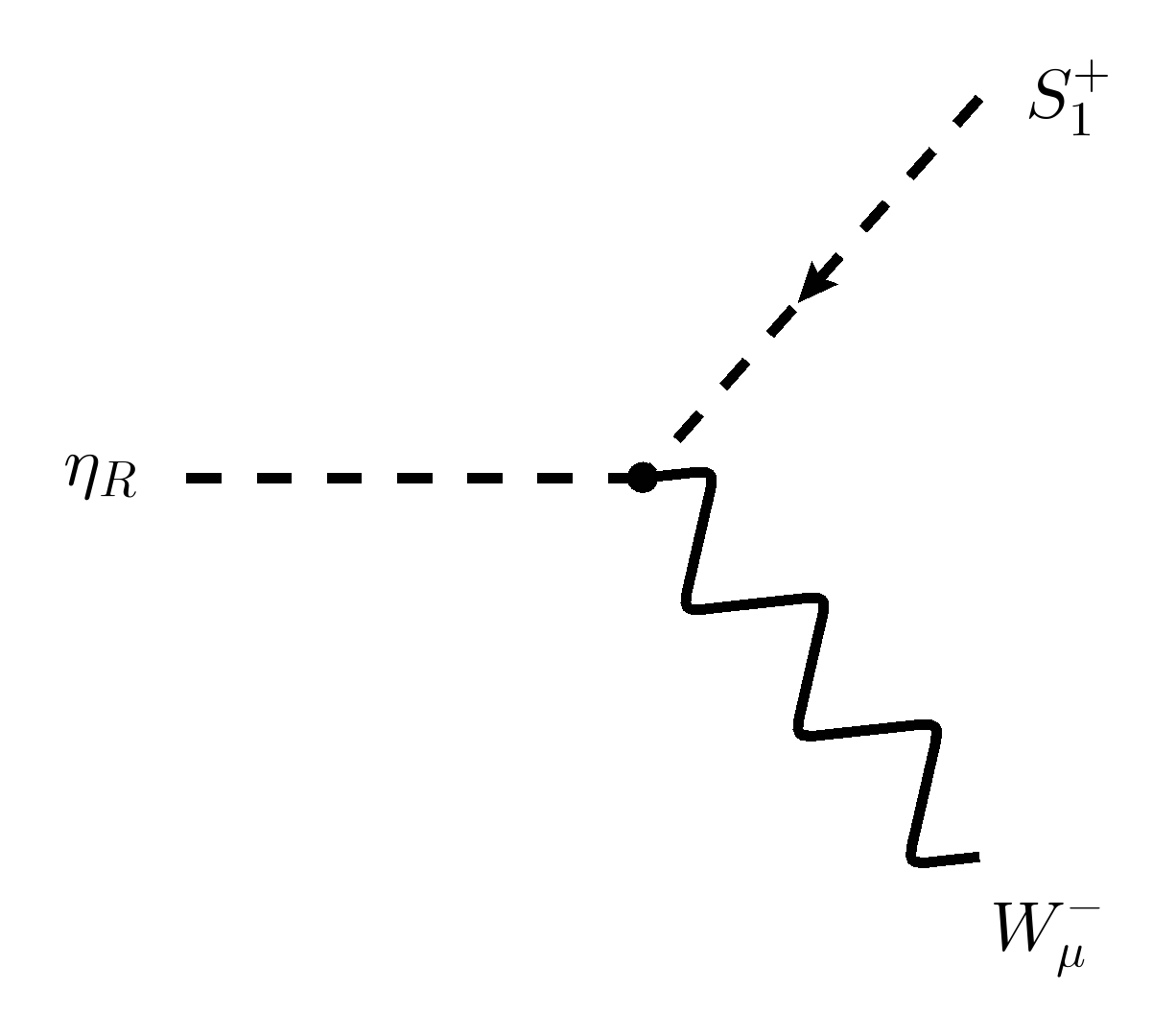} &\sim\ \frac{1}{2} i g s_{\theta}(p - p')_\mu & \includegraphics[height=3cm, valign=c]{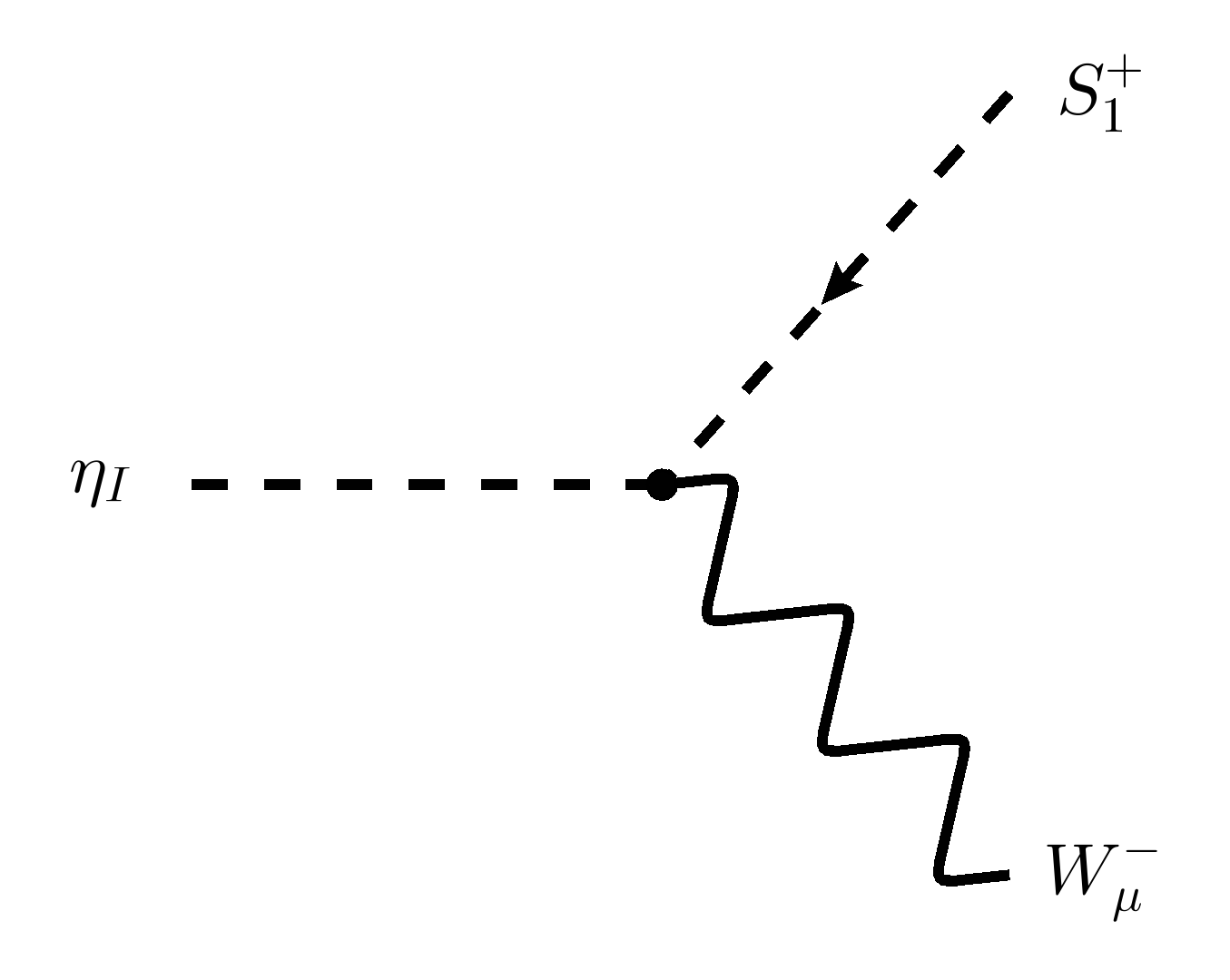} &\sim\ \frac{1}{2} g s_{\theta}(p - p')_\mu \\
    \includegraphics[height=3cm, valign=c]{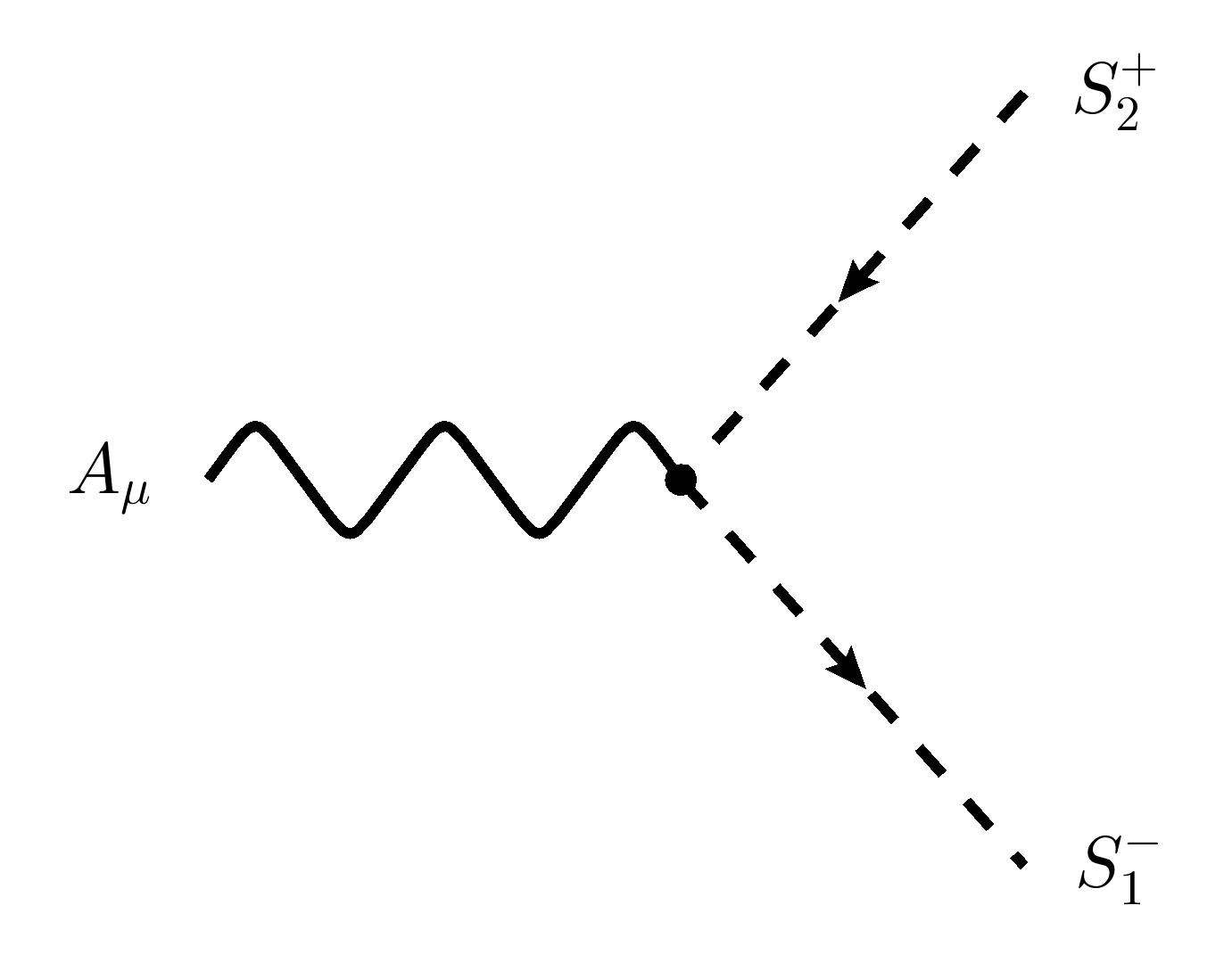} &\sim\ \begin{array}{l} {} \\ \frac{1}{4} i s_{2\theta} \left(g' c_W - g s_W\right) \\ \quad \times (p - p')_\mu \end{array} & \includegraphics[height=3cm, valign=c]{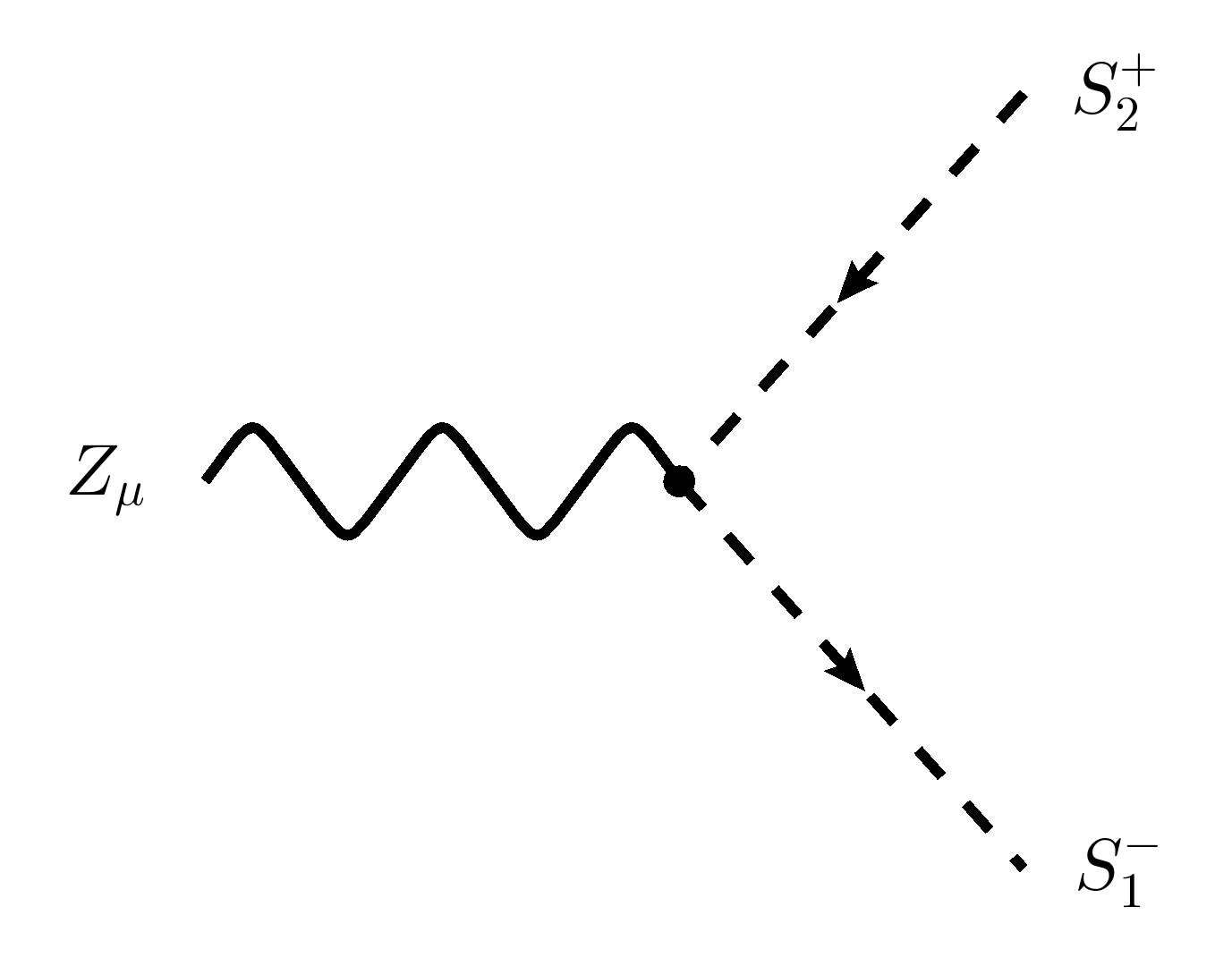} &\sim\ \begin{array}{l} {} \\ \frac{1}{4} i s_{2\theta} \left(g c_W + g' s_W\right) \\ \quad \times (p - p')_\mu \end{array} \\
    \includegraphics[height=3cm, valign=c]{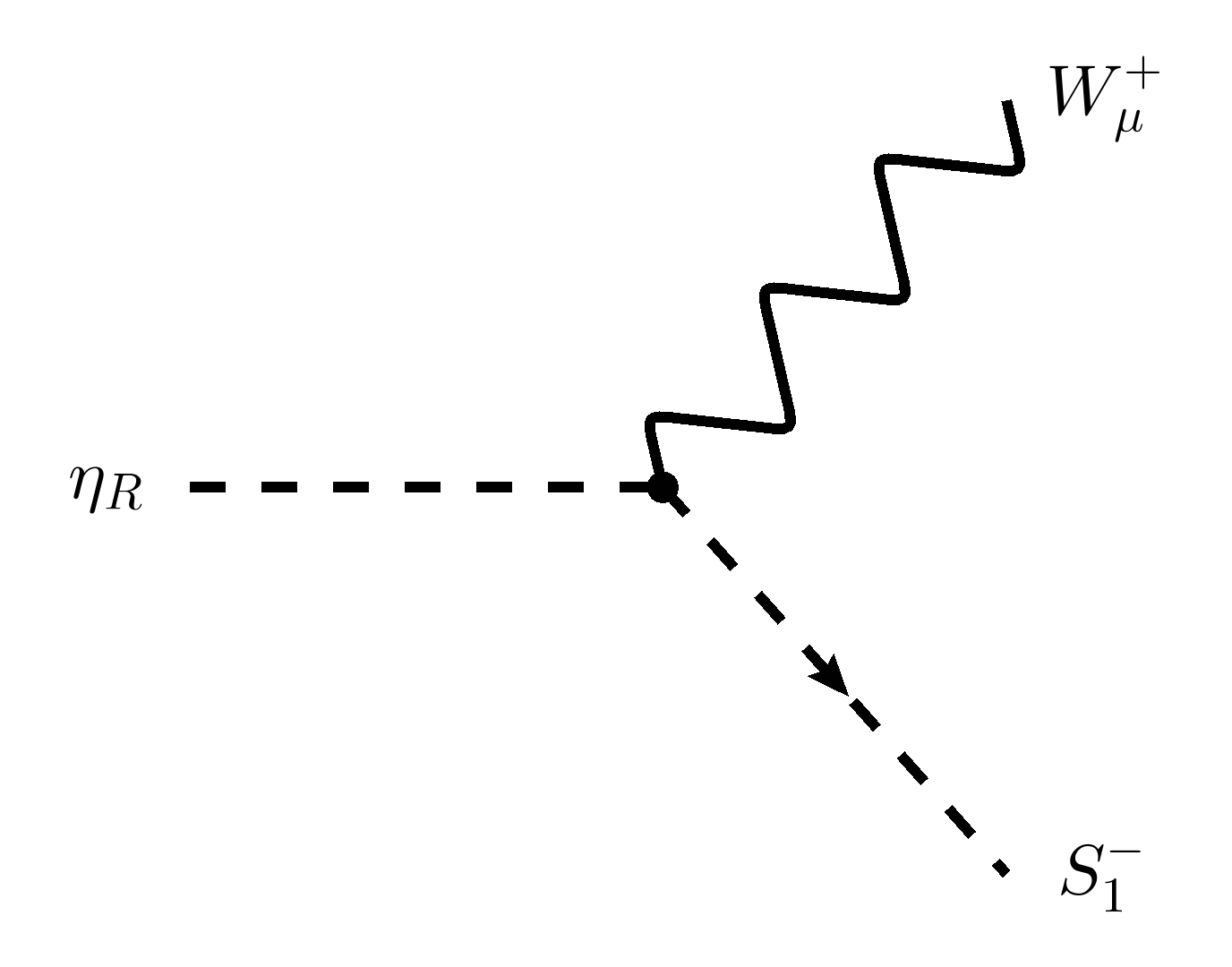} &\sim\ \frac{1}{2} i g s_{\theta}(p - p')_\mu & \includegraphics[height=3cm, valign=c]{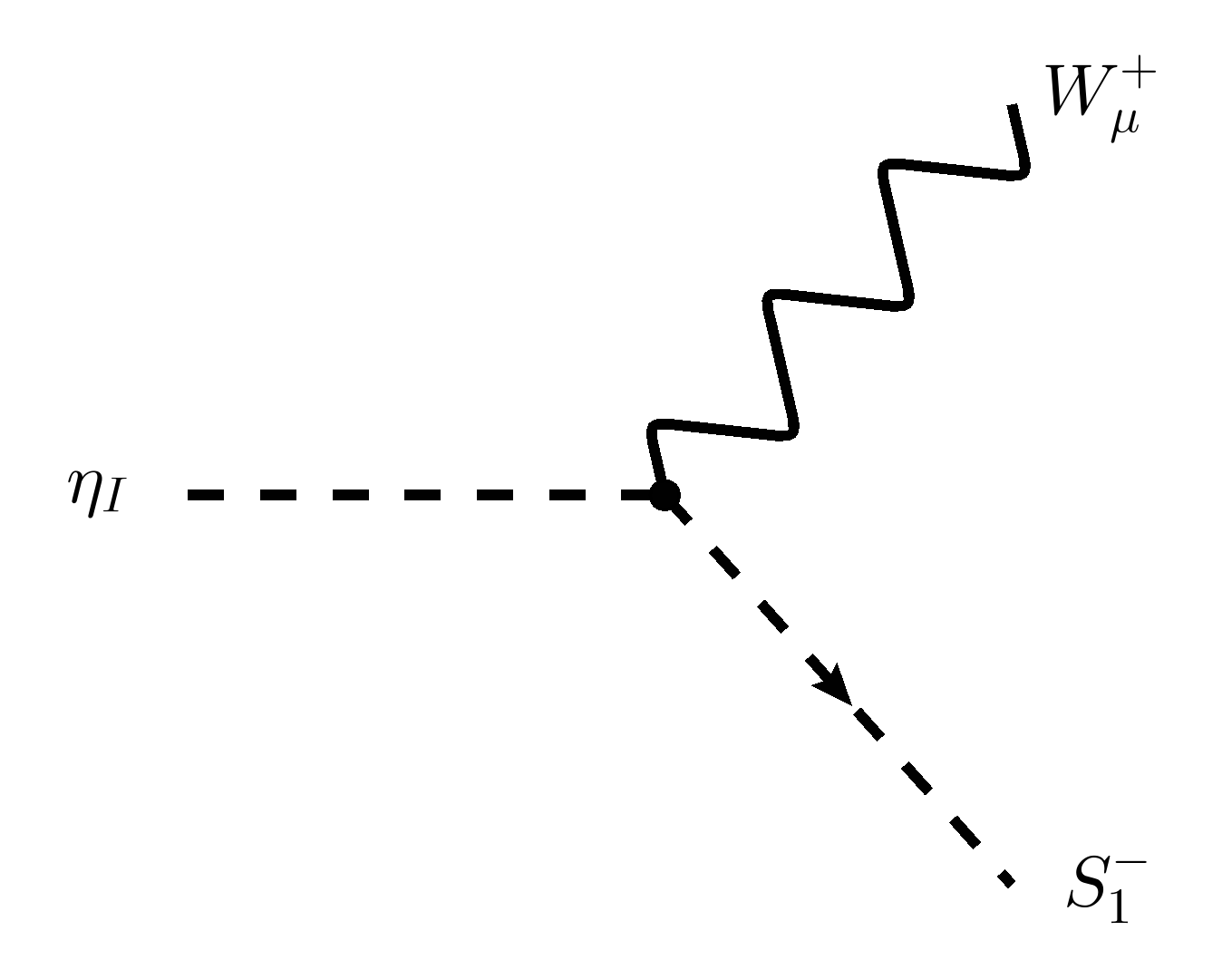} &\sim\ \frac{1}{2} g s_{\theta}(p - p')_\mu \\
    \includegraphics[height=3cm, valign=c]{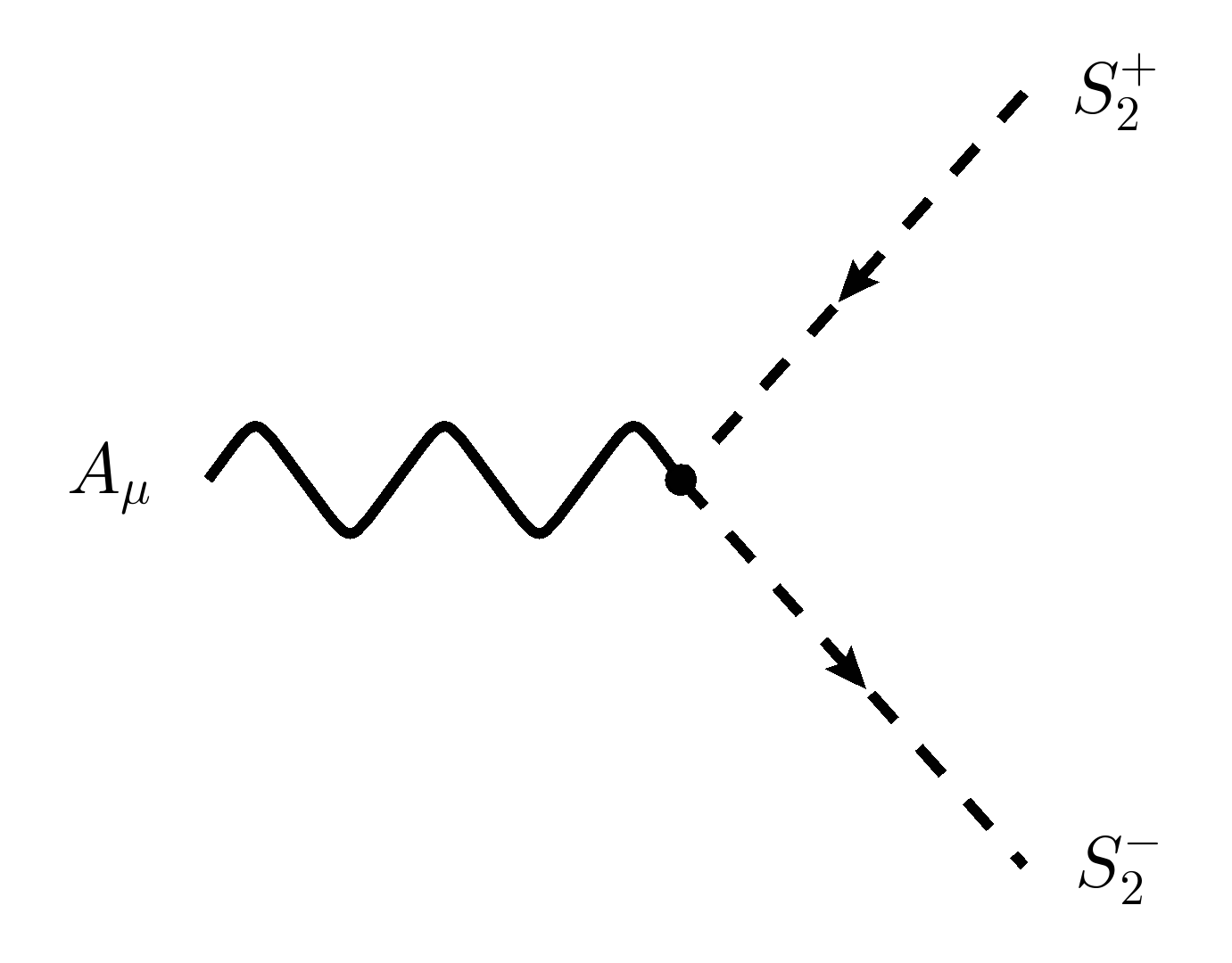} &\sim\ \begin{array}{l} {} \\ \frac{1}{2} i \Bigl[2 g' c_W s_{\theta}^2 + c_{\theta}^2 (g' c_W \\ \quad + g s_W)\Bigr](p - p')_\mu \end{array} & \includegraphics[height=3cm, valign=c]{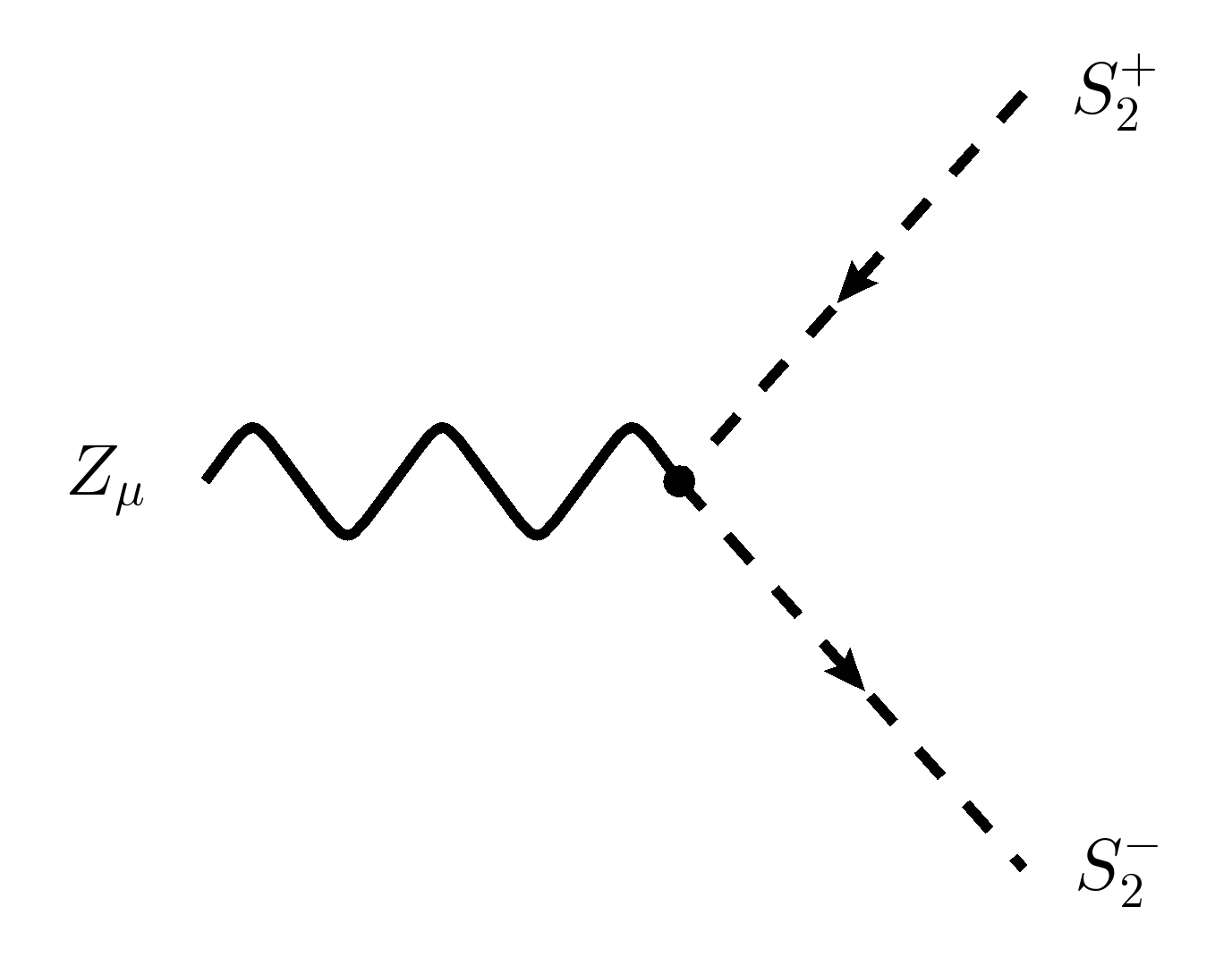} &\sim\ \begin{array}{l} {} \\ \frac{1}{2} i \Bigl[2 g' s_{\theta}^2 s_W + c_{\theta}^2 (-g c_W \\ \quad + g' s_W)\Bigr](p - p')_\mu \end{array} \\
    \includegraphics[height=3cm, valign=c]{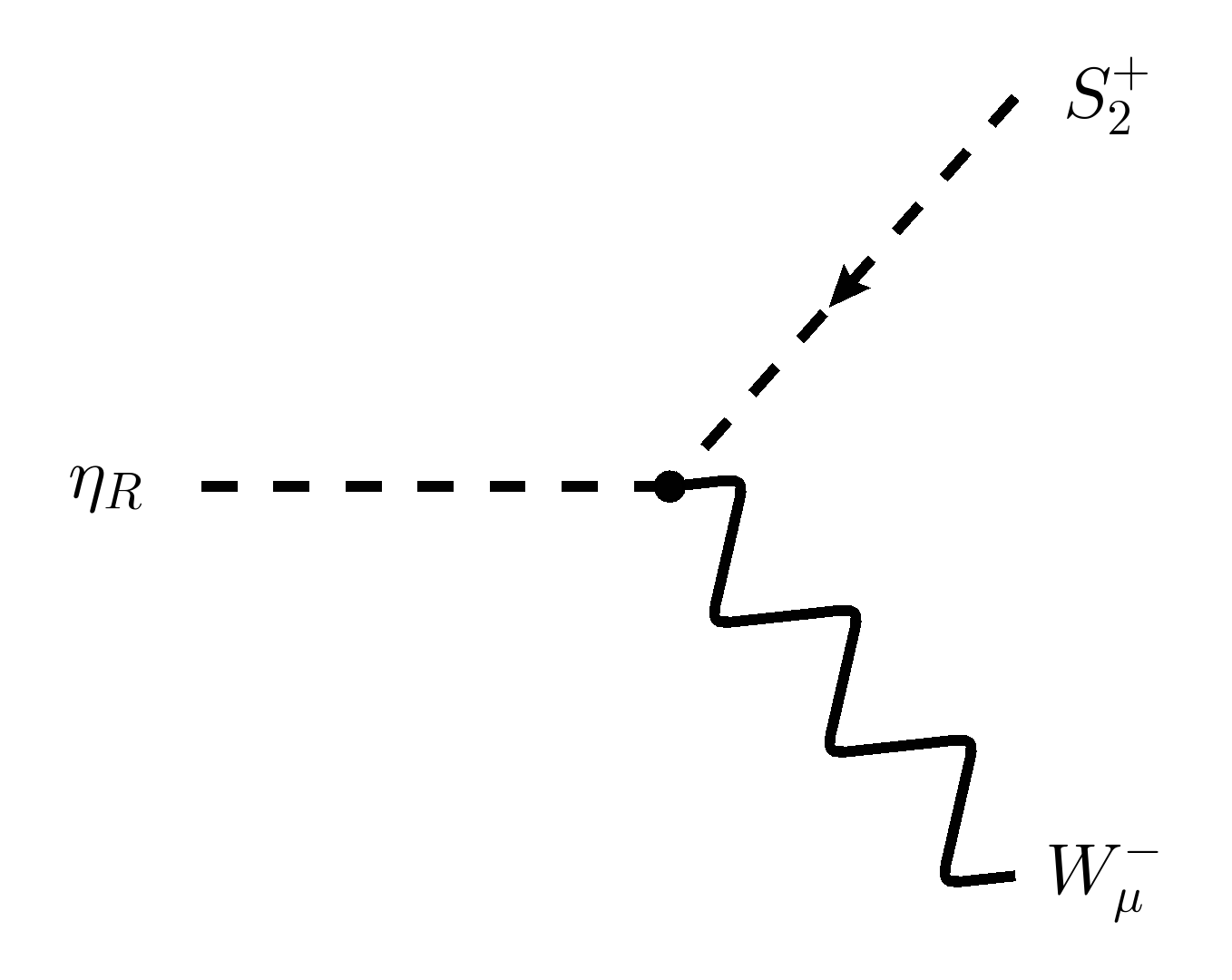} &\sim\ \frac{1}{2} i g c_{\theta}(p - p')_\mu & \includegraphics[height=3cm, valign=c]{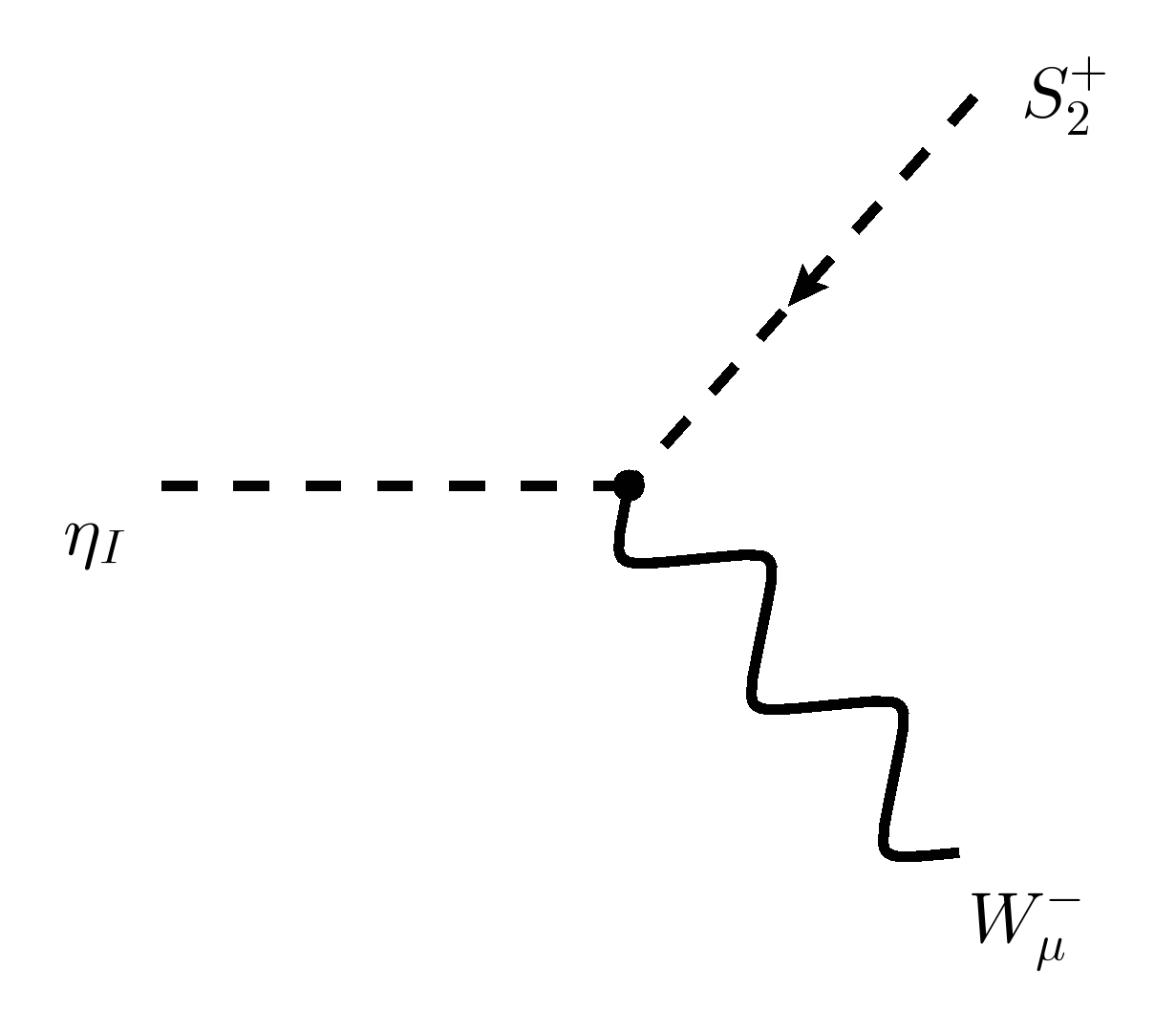} &\sim\ \frac{1}{2} g c_{\theta}(p - p')_\mu \\
    \includegraphics[height=3cm, valign=c]{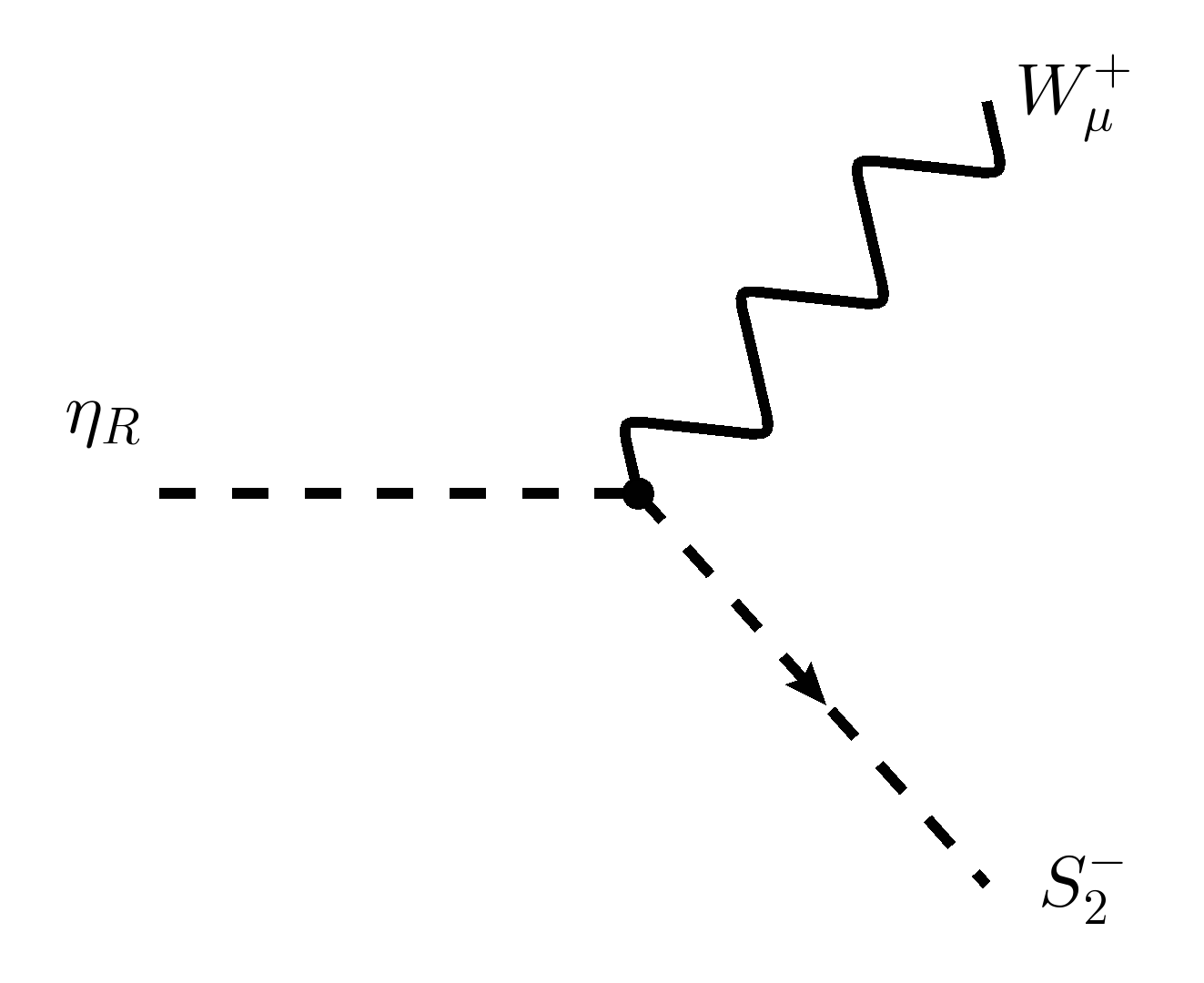} &\sim\ \frac{1}{2} i g c_{\theta}(p - p')_\mu & \includegraphics[height=3cm, valign=c]{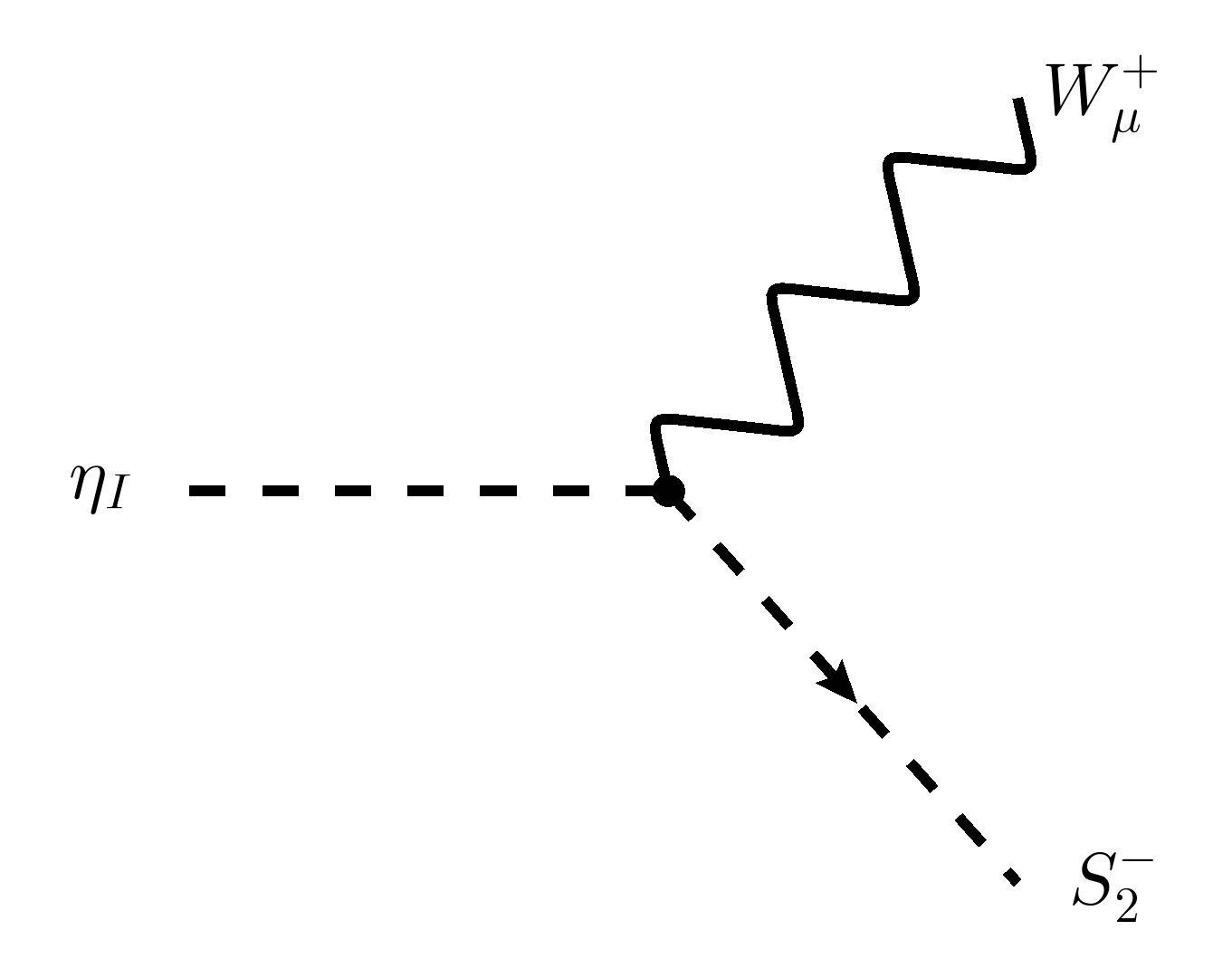} &\sim\ \frac{1}{2} g c_{\theta}(p - p')_\mu \\
    \includegraphics[height=3cm, valign=c]{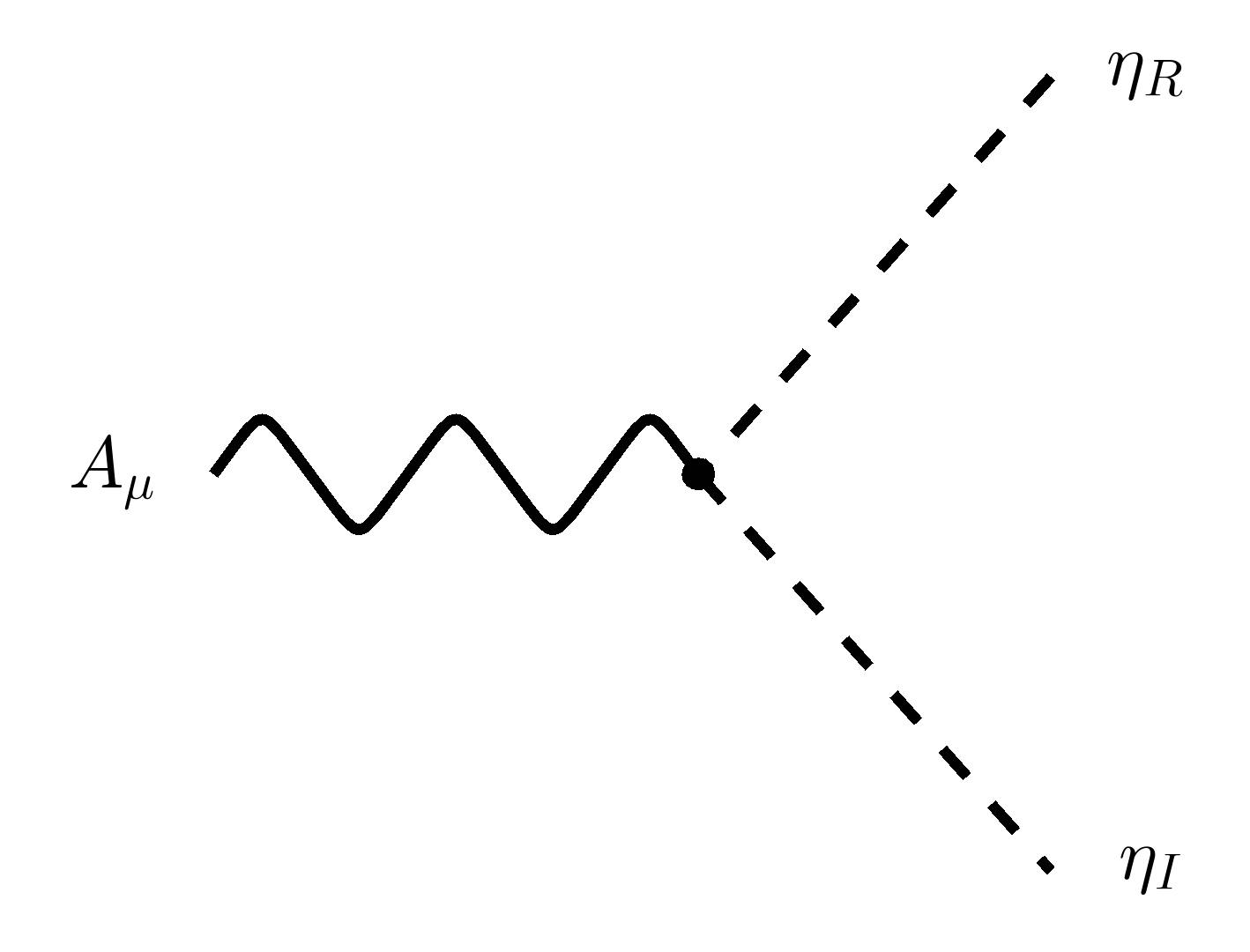} &\sim\ \begin{array}{l} {} \\ \frac{1}{2} \left(g' c_W - g s_W\right) \\ \quad \times (p - p')_\mu \end{array} & \includegraphics[height=3cm, valign=c]{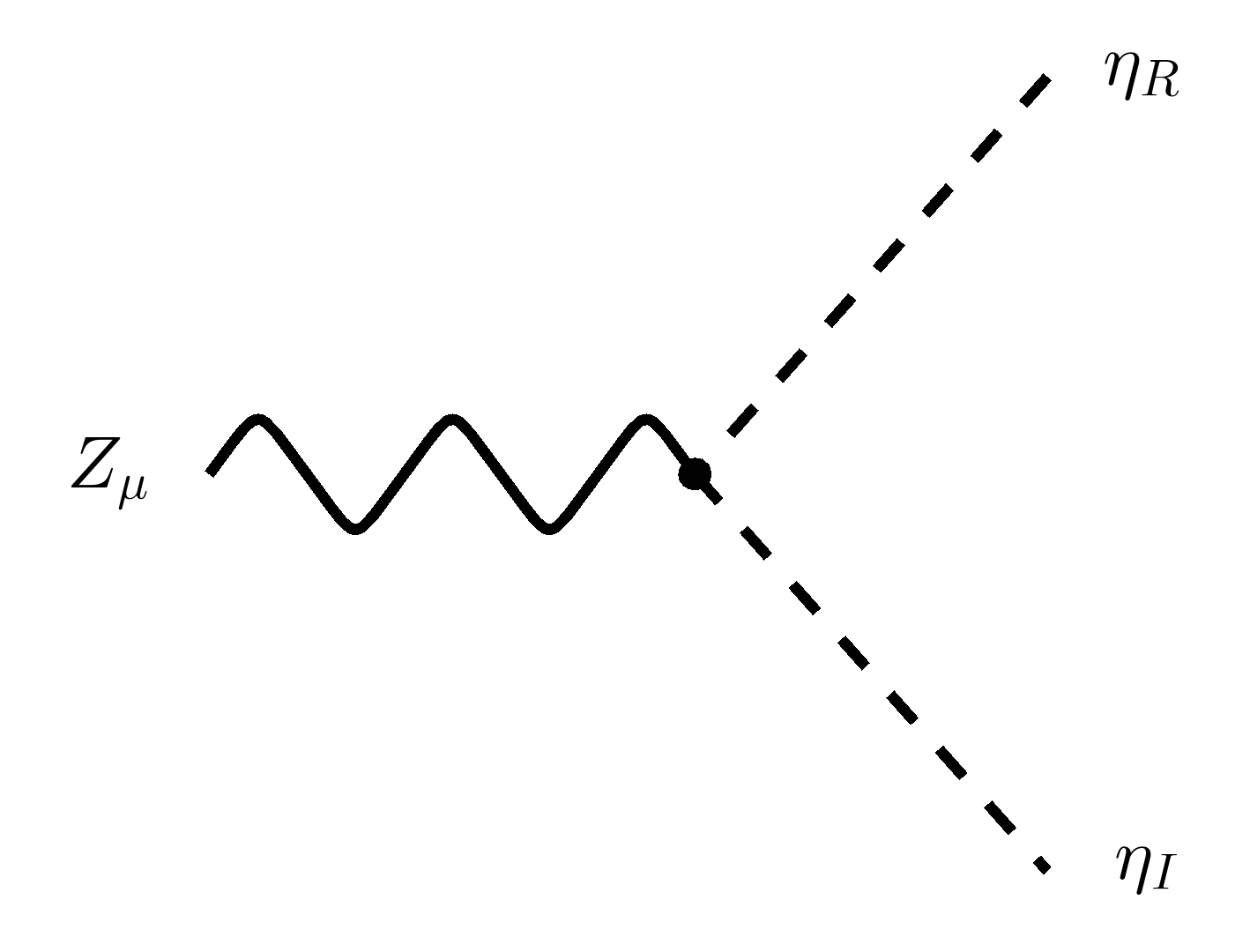} &\sim\ \begin{array}{l} {} \\ \frac{1}{2} \left(g c_W + g' s_W\right) \\ \quad \times (p - p')_\mu \end{array}
\end{align*}
The vertex factors involving two gauge bosons and two scalars, with all particles incoming, are
\begin{align*}
    \includegraphics[height=3cm, valign=c]{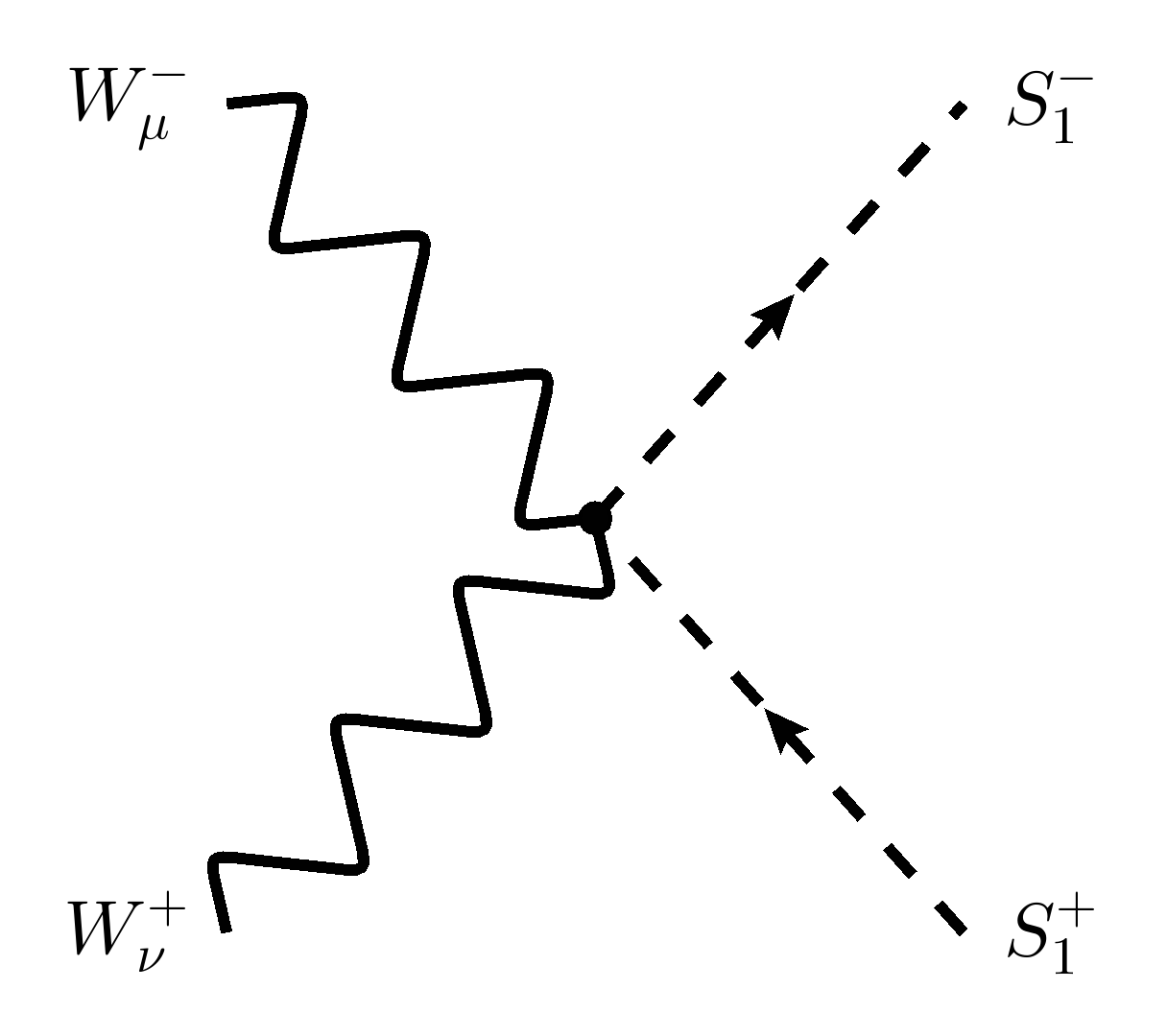} &\sim\ \frac{1}{2}ig^2s^2_\theta g_{\mu\nu} & \includegraphics[height=3cm, valign=c]{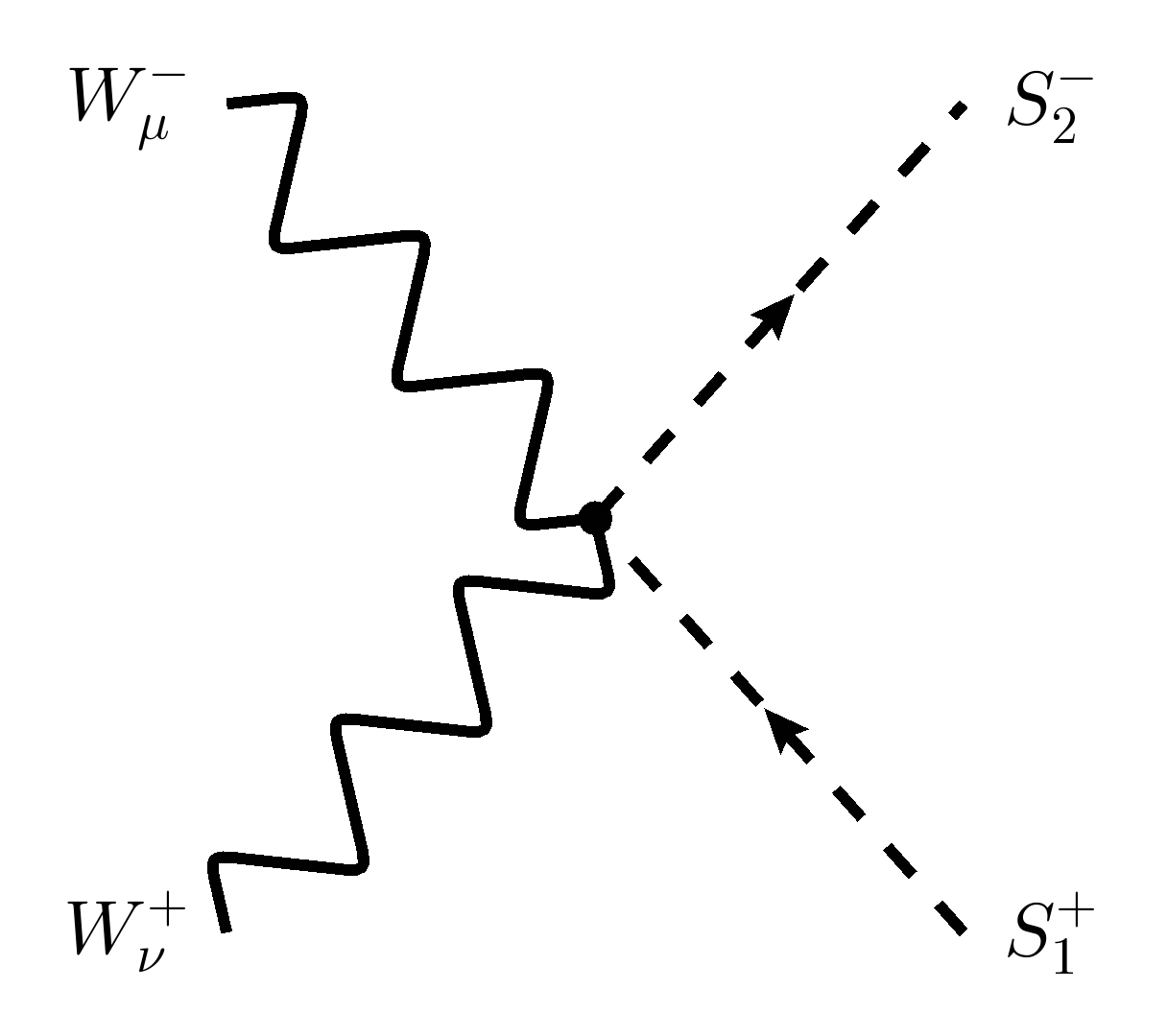} &\sim\ -\frac{1}{2}ig^2c_\theta s_\theta g_{\mu\nu} \\
    \includegraphics[height=3cm, valign=c]{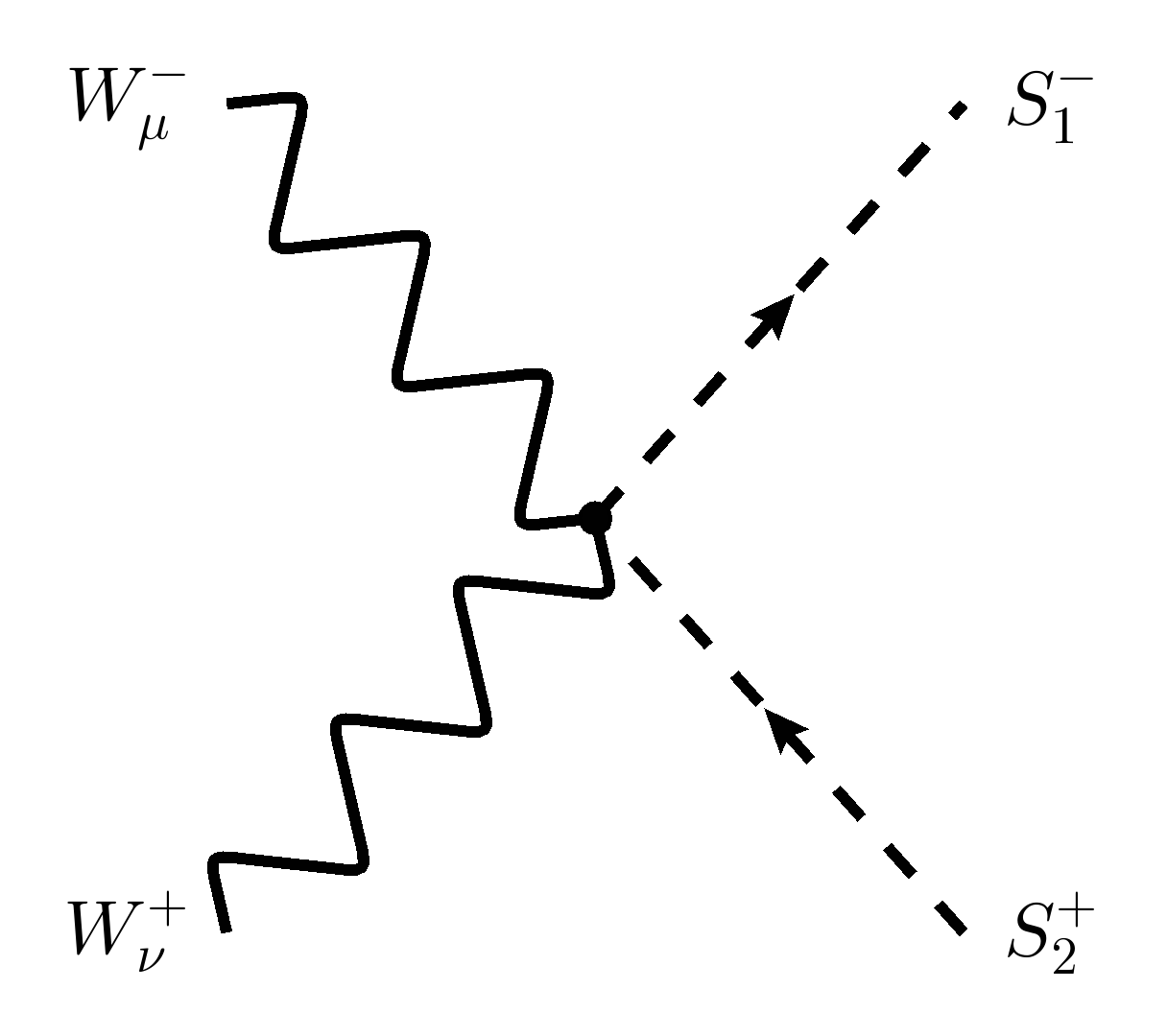} &\sim\ -\frac{1}{2}ig^2c_\theta s_\theta g_{\mu\nu} & \includegraphics[height=3cm, valign=c]{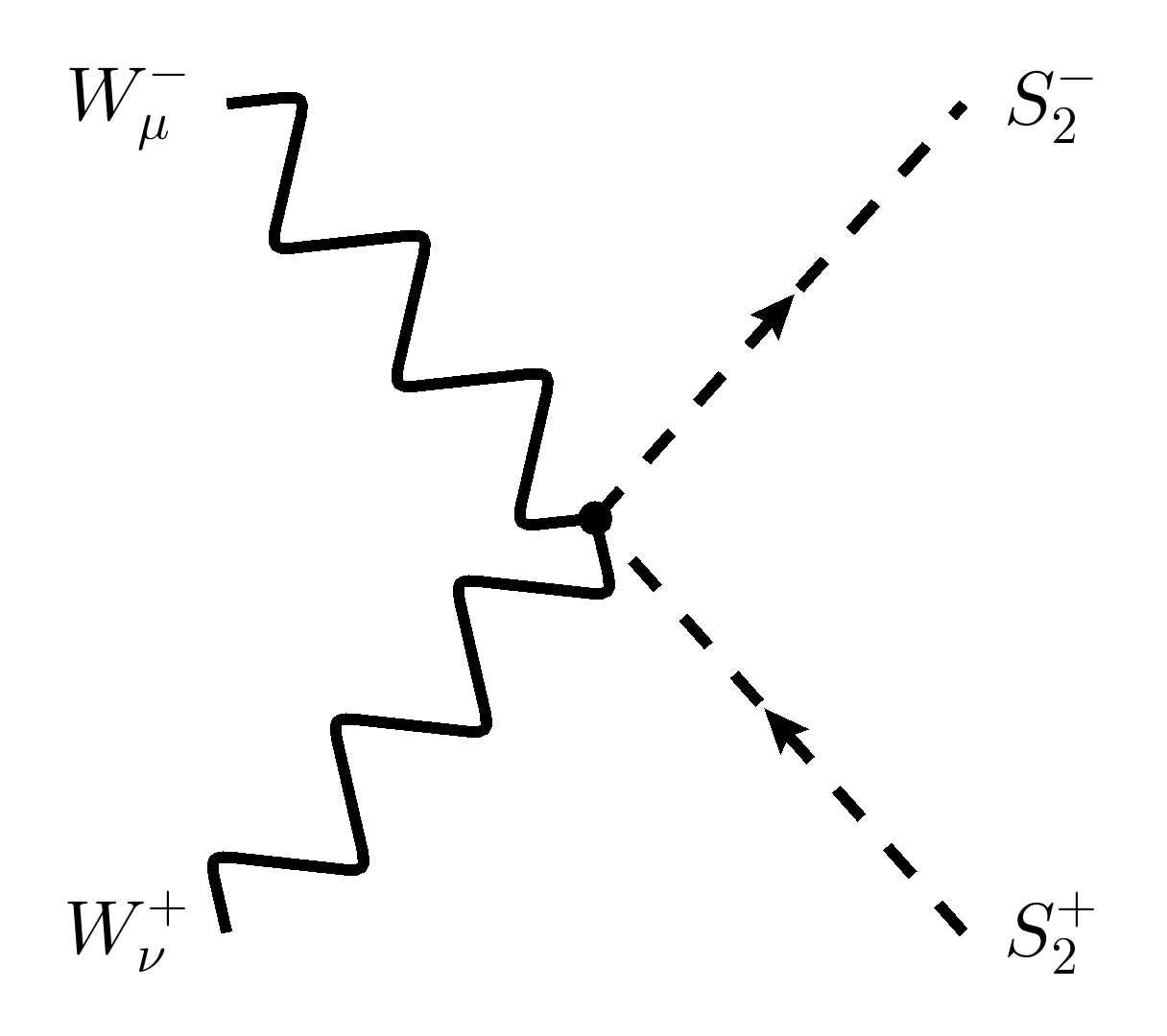} &\sim\ \frac{1}{2}ig^2c^2_\theta g_{\mu\nu} \\
    \includegraphics[height=3cm, valign=c]{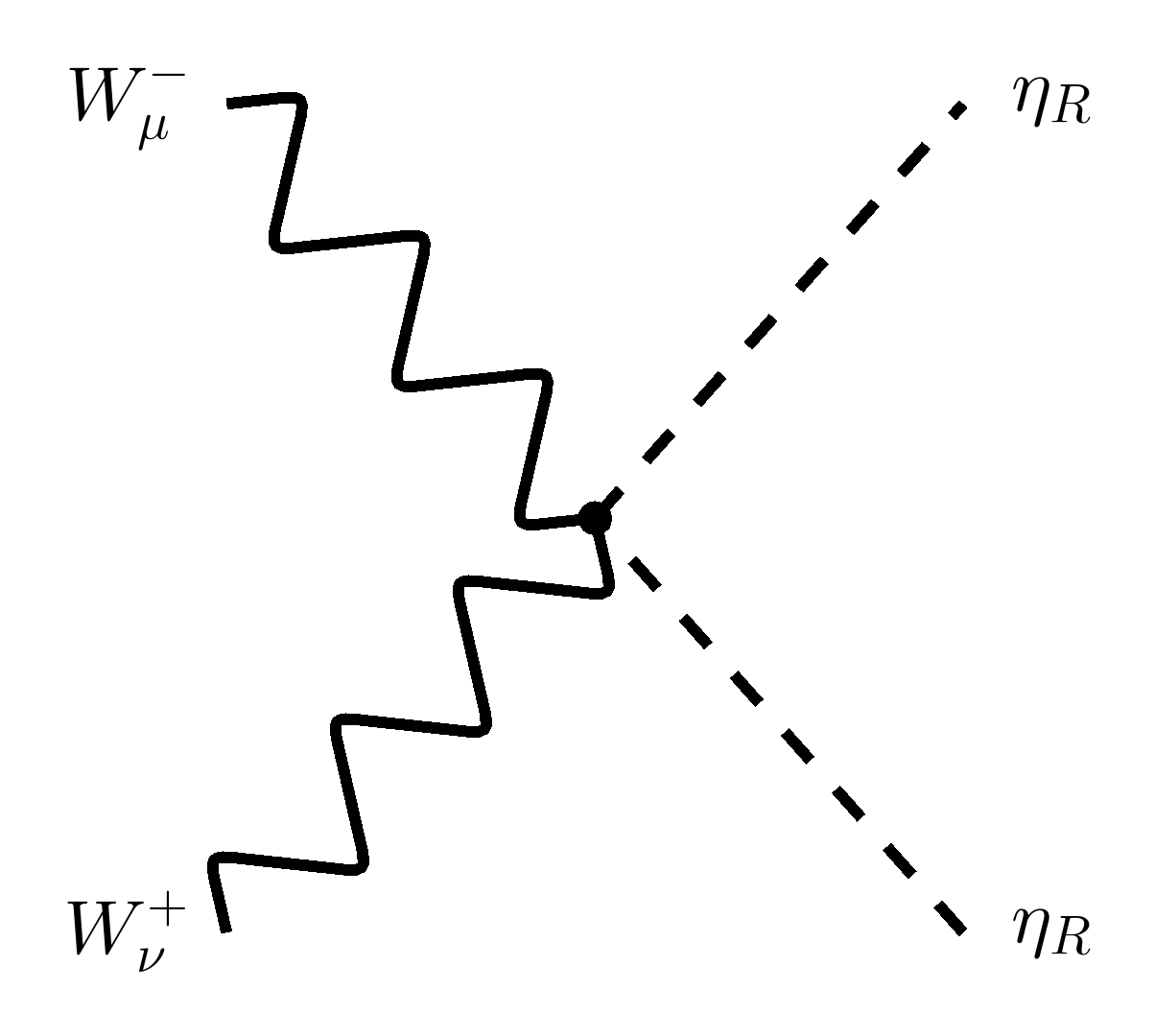} &\sim\ 2! \times \frac{1}{2}ig^2 g_{\mu\nu} & \includegraphics[height=3cm, valign=c]{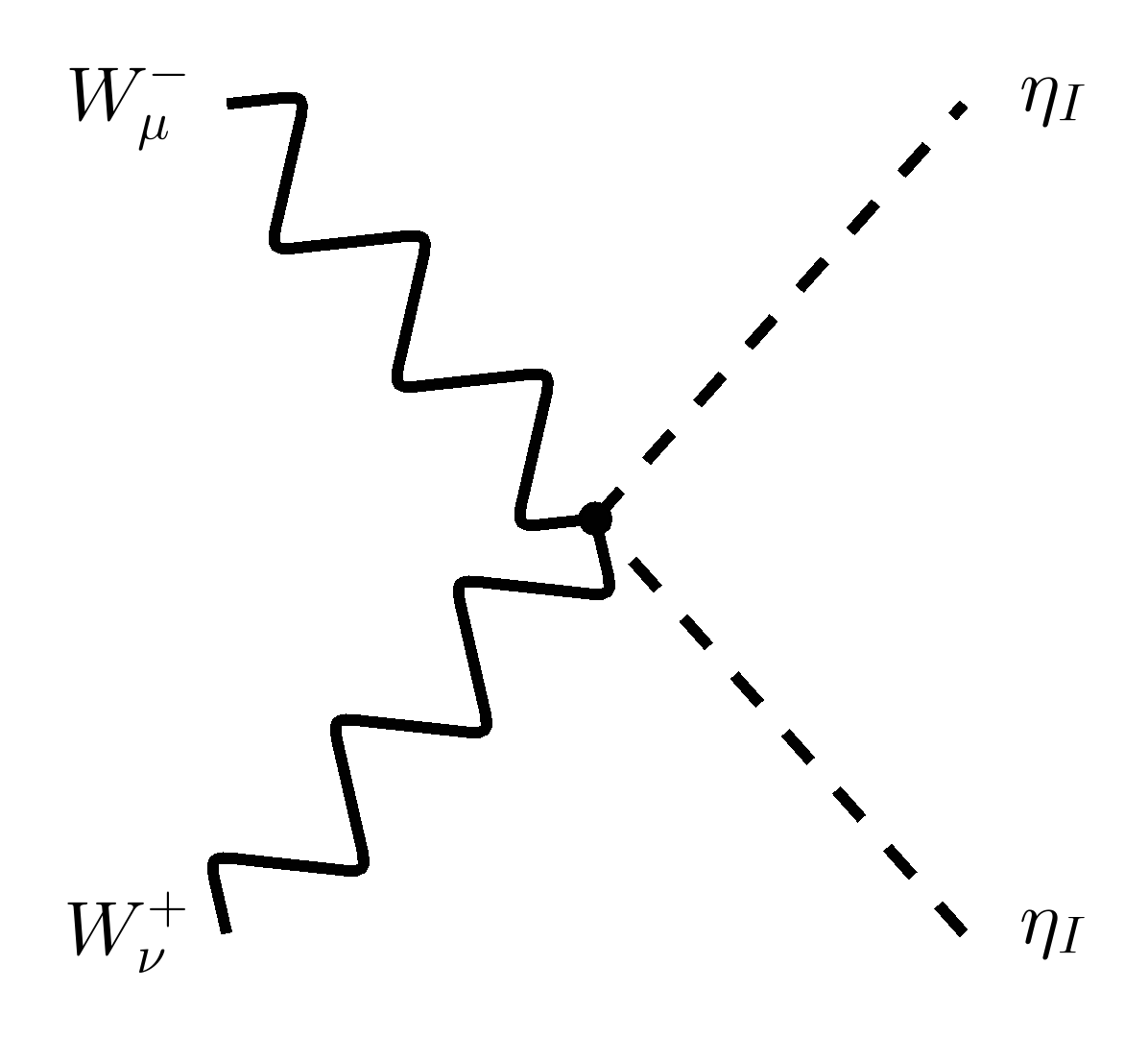} &\sim\ 2! \times \frac{1}{2}ig^2 g_{\mu\nu} \\
    \includegraphics[height=3cm, valign=c]{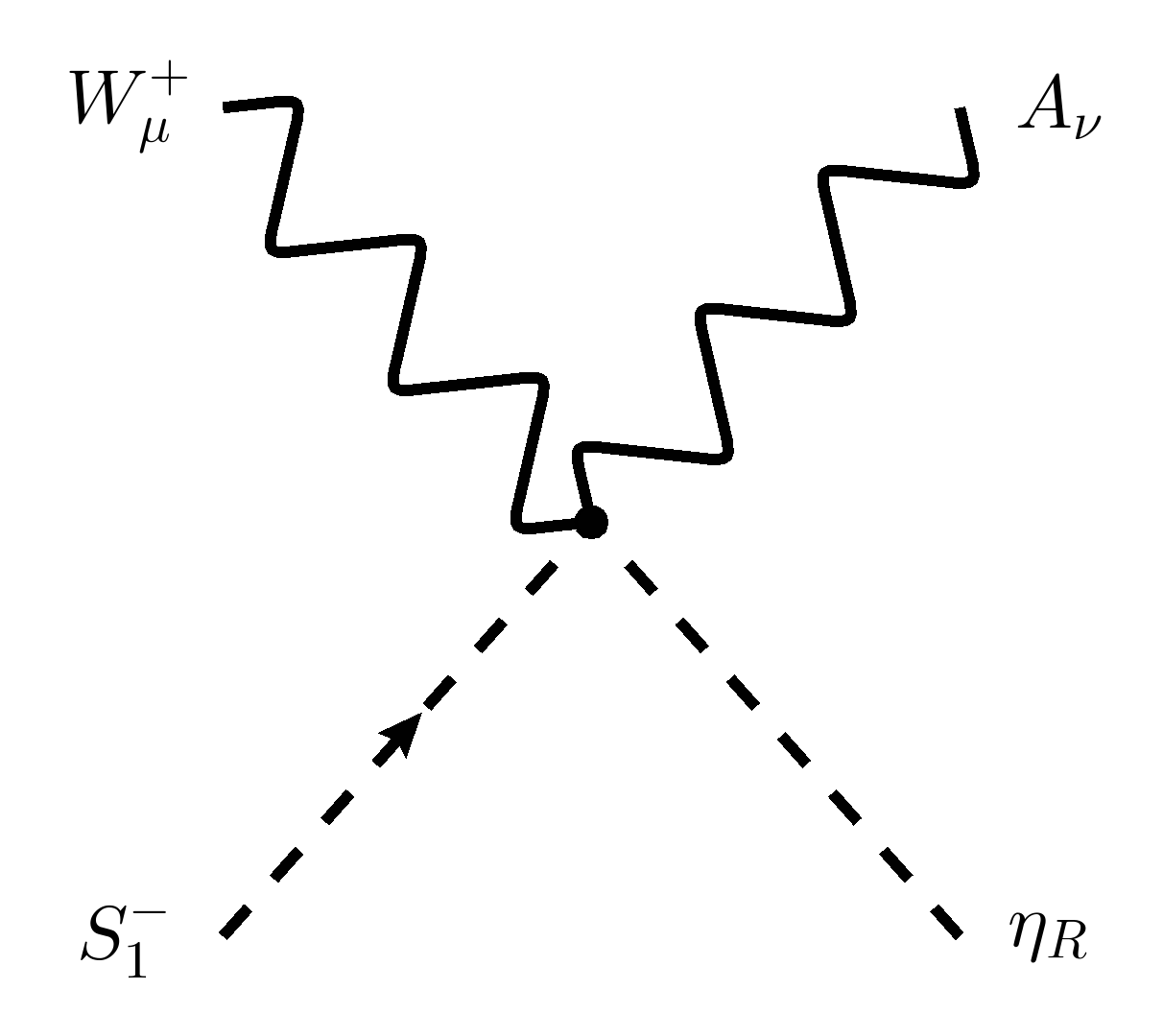} &\sim\ -\frac{1}{2}igg^\prime c_W s_\theta g_{\mu\nu} & \includegraphics[height=3cm, valign=c]{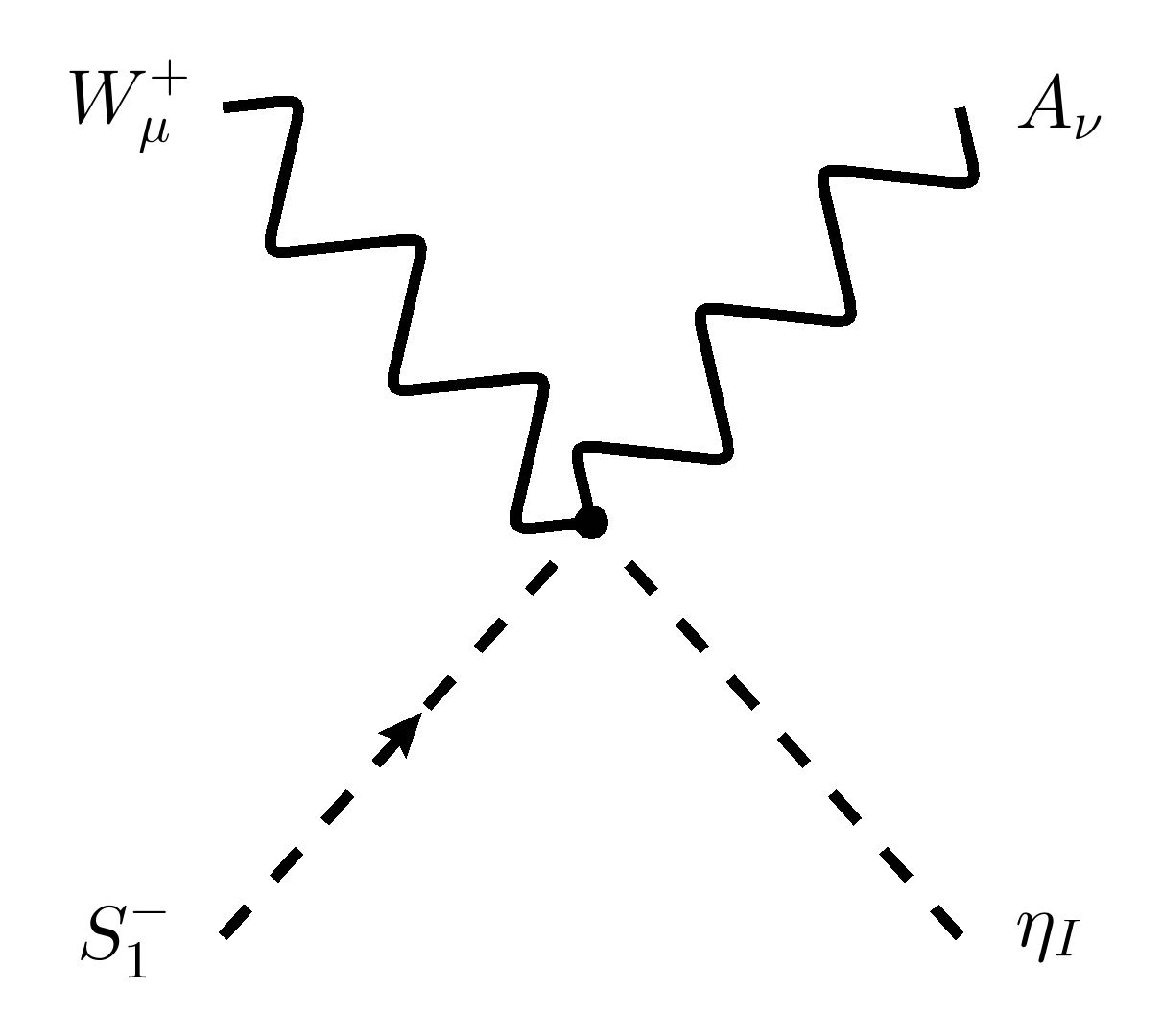} &\sim\ \frac{1}{2}gg^\prime c_W s_\theta g_{\mu\nu} \\
    \includegraphics[height=3cm, valign=c]{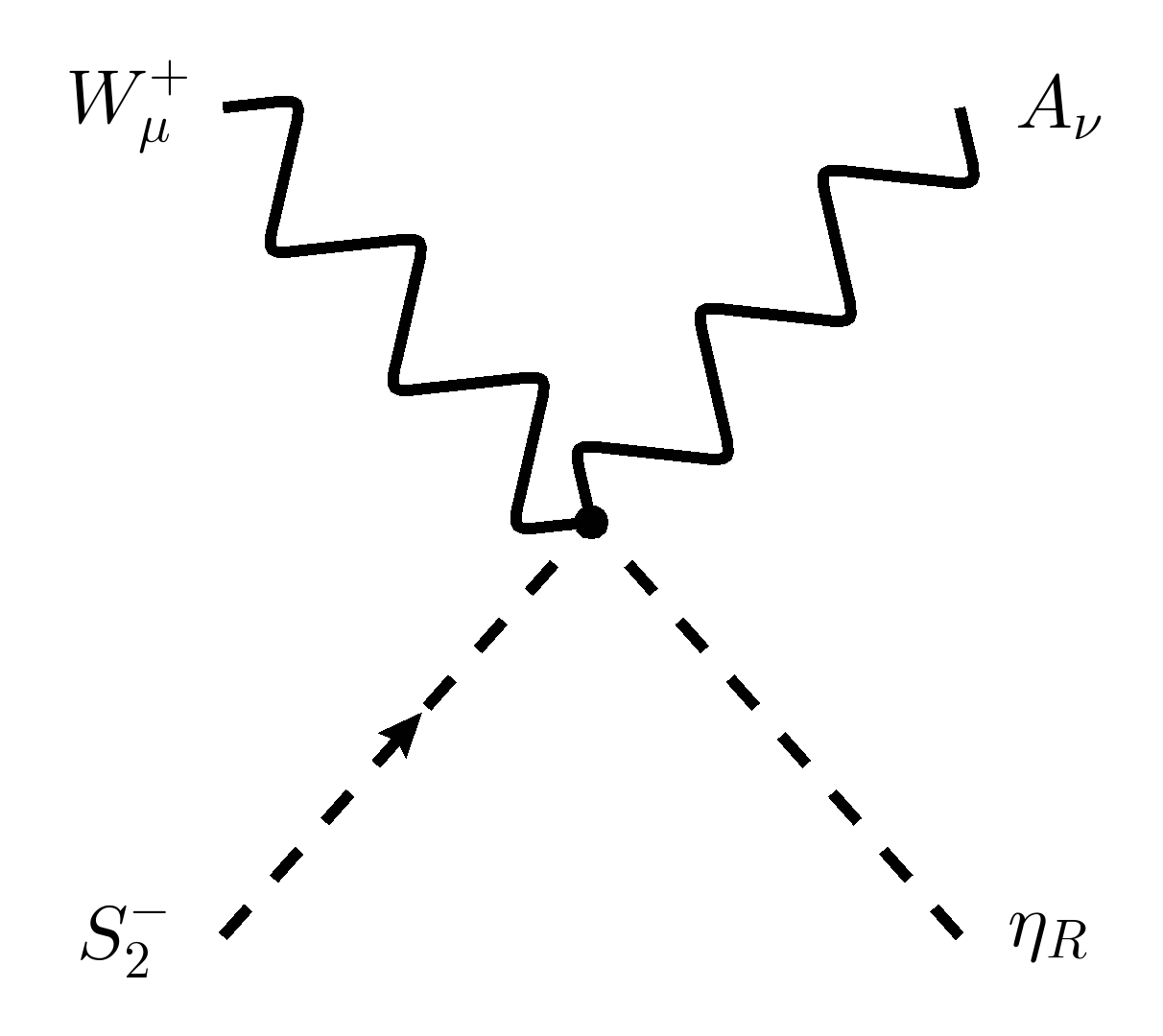} &\sim\ \frac{1}{2}igg^\prime c_\theta c_W g_{\mu\nu} & \includegraphics[height=3cm, valign=c]{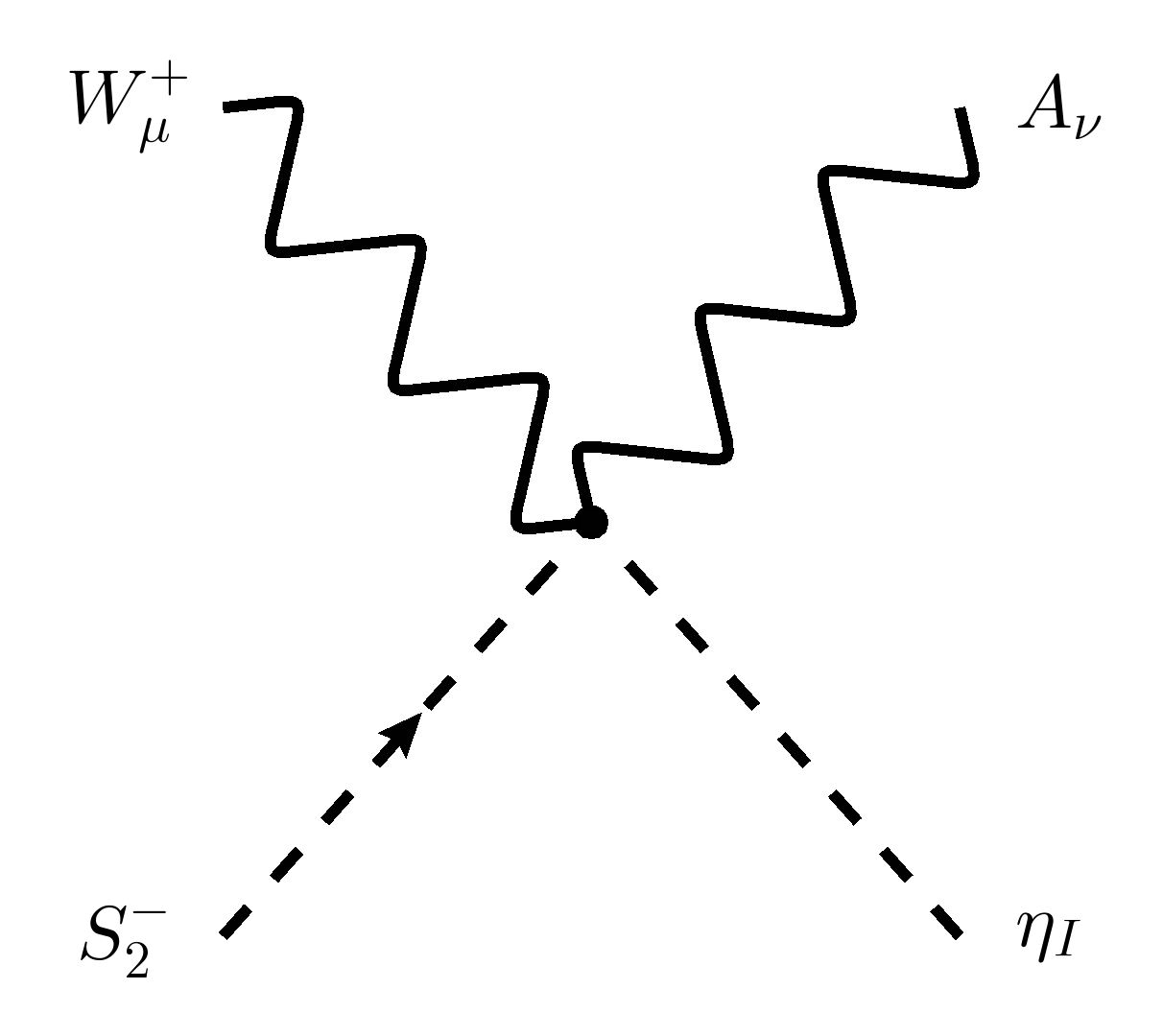} &\sim\ -\frac{1}{2}gg^\prime c_\theta c_W g_{\mu\nu}\\
    \includegraphics[height=3cm, valign=c]{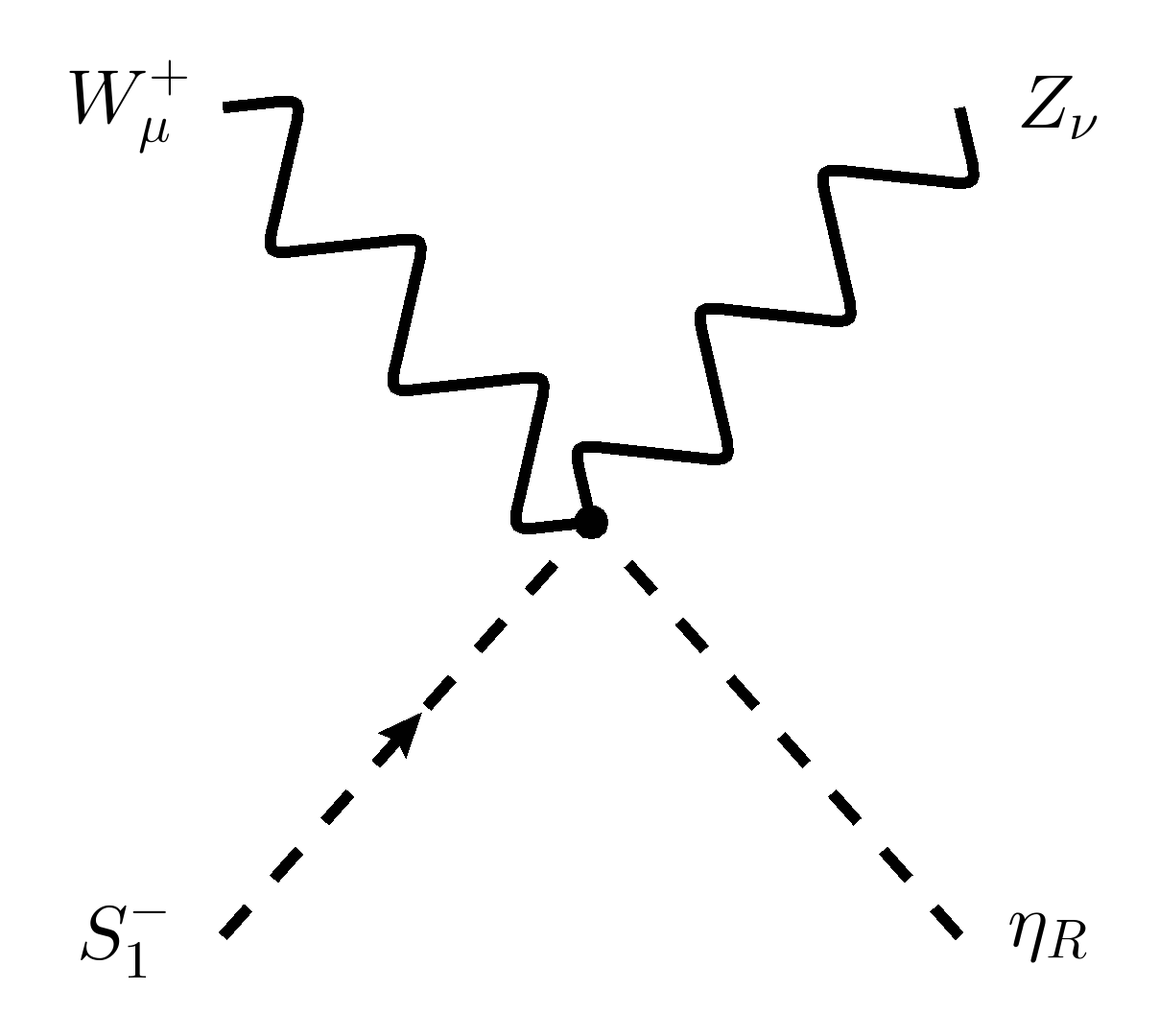} &\sim\ \frac{1}{2}igg^\prime s_\theta s_W g_{\mu\nu} & \includegraphics[height=3cm, valign=c]{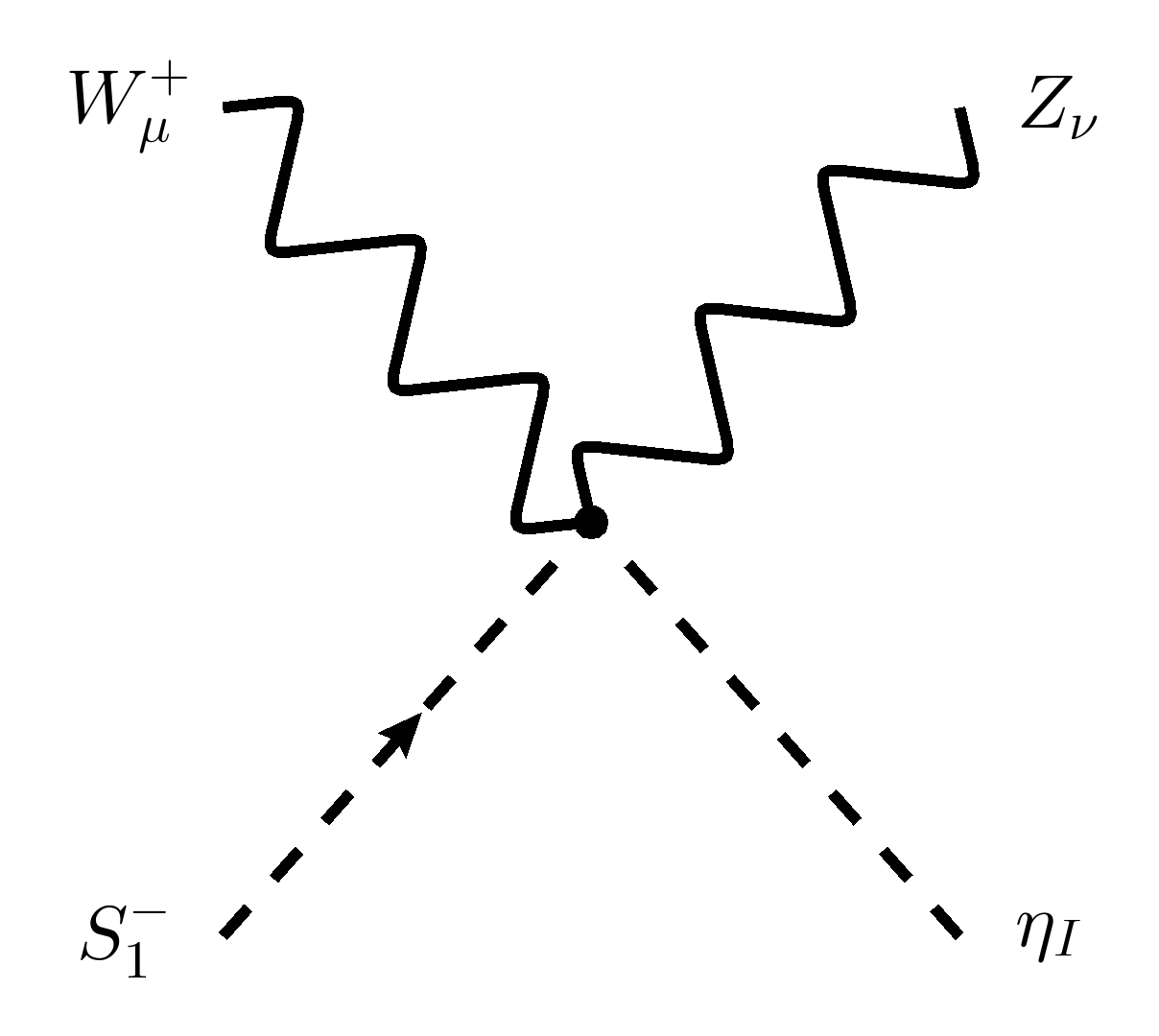} &\sim\ -\frac{1}{2}gg^\prime s_\theta s_W g_{\mu\nu} \\
    \includegraphics[height=3cm, valign=c]{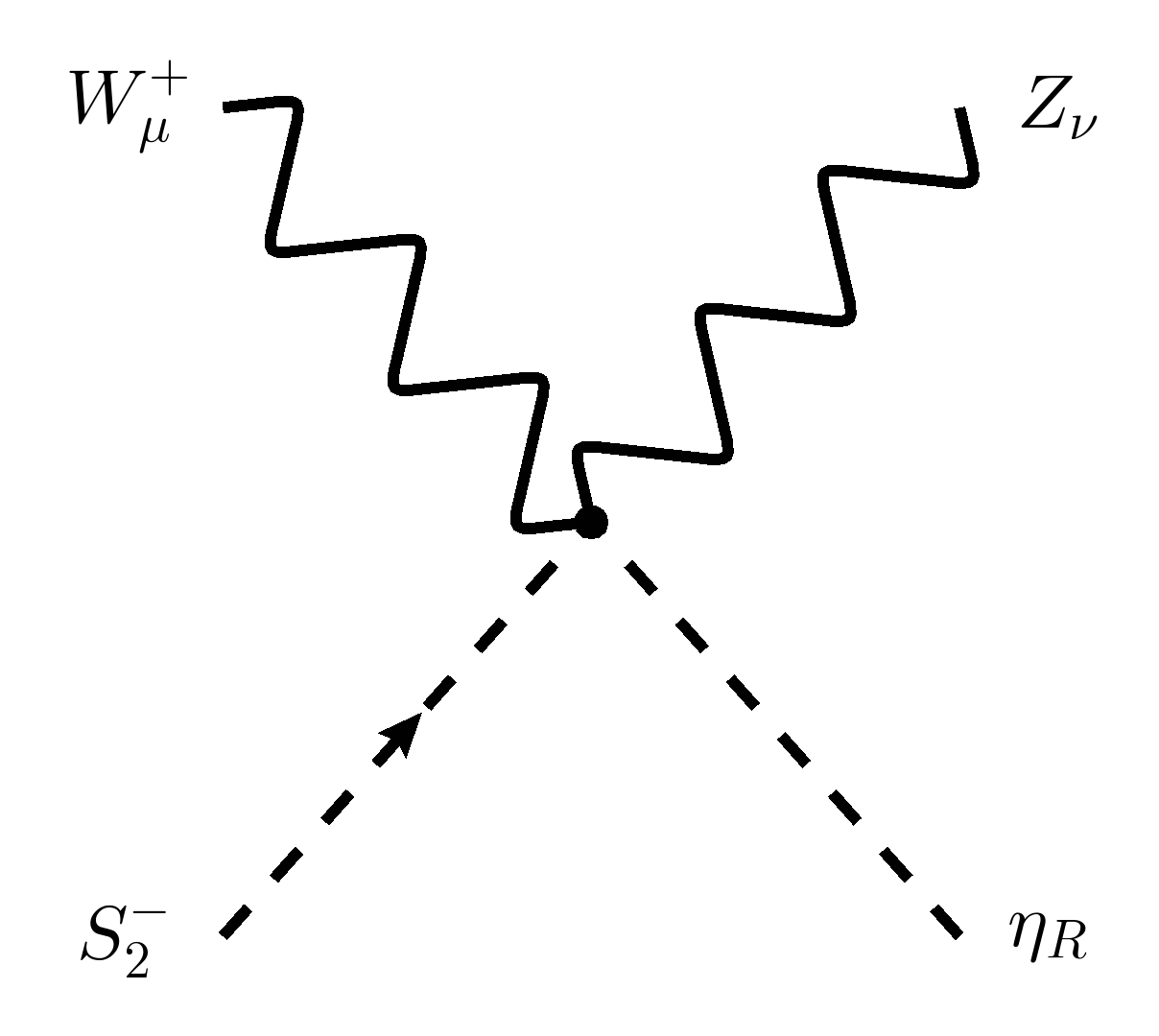} &\sim\ -\frac{1}{2}igg^\prime c_\theta s_W g_{\mu\nu} & \includegraphics[height=3cm, valign=c]{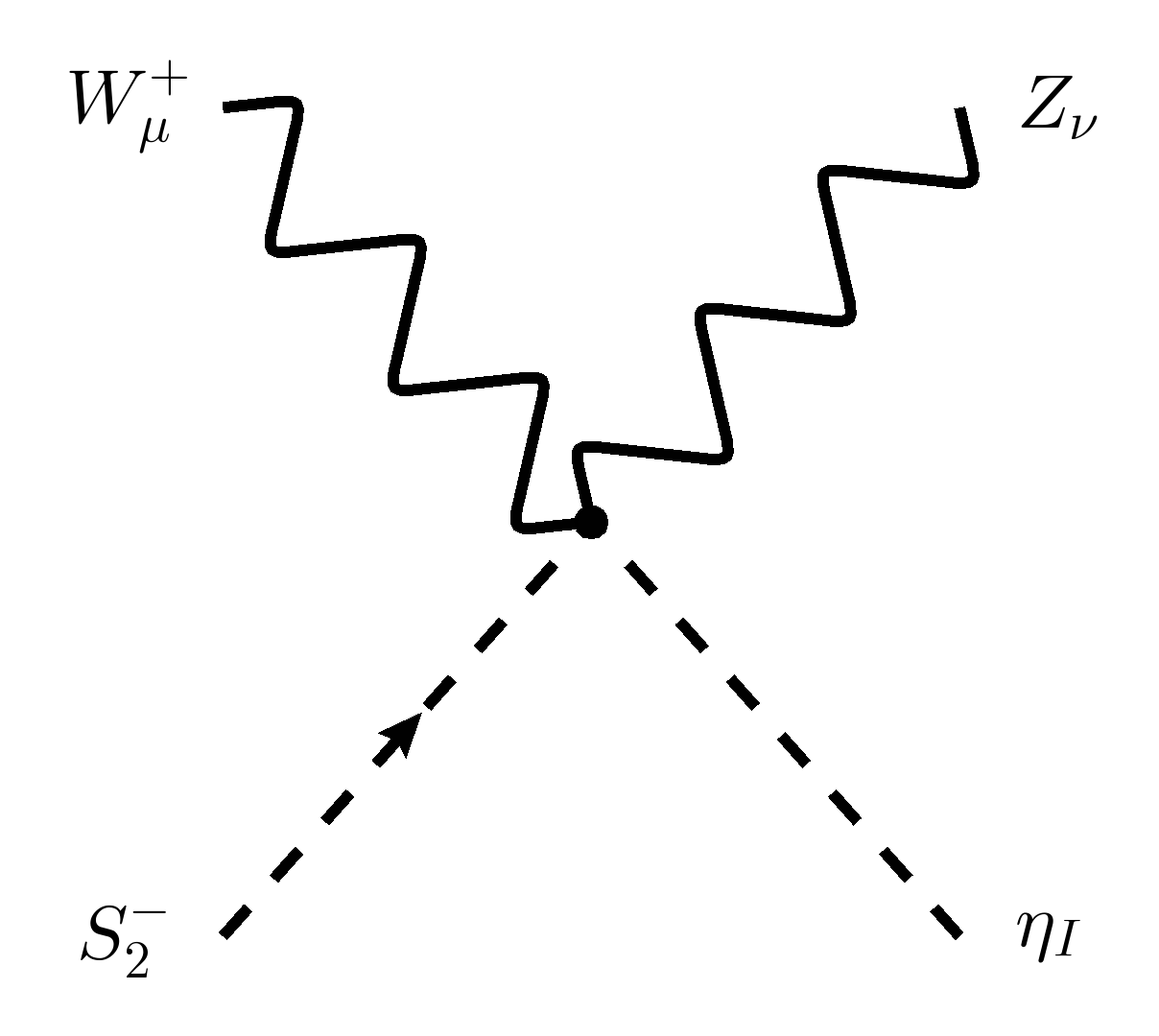} &\sim\ \frac{1}{2}gg^\prime c_\theta s_W g_{\mu\nu}\\
    \includegraphics[height=3cm, valign=c]{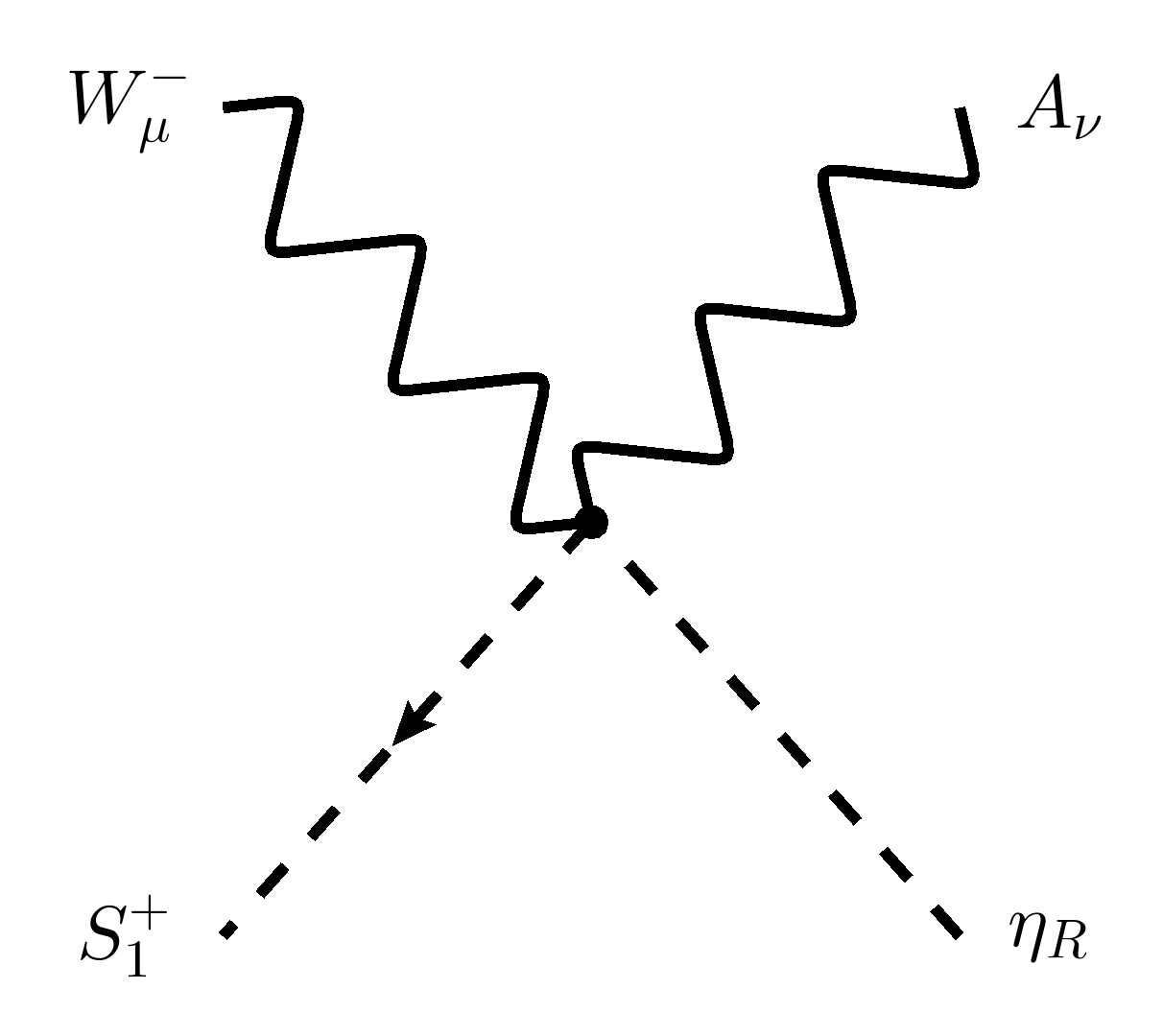} &\sim\ -\frac{1}{2}igg^\prime c_W s_\theta g_{\mu\nu} & \includegraphics[height=3cm, valign=c]{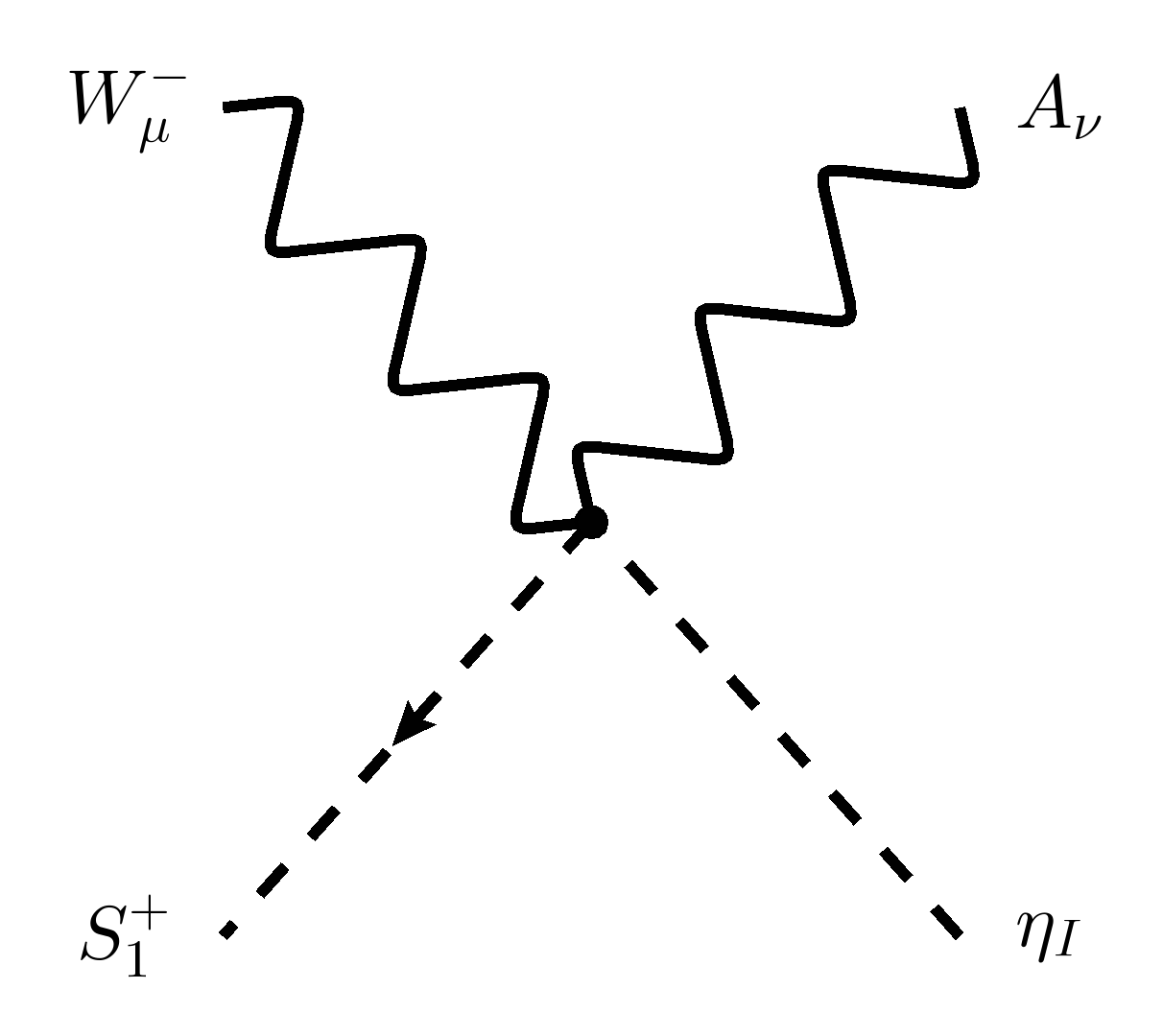} &\sim\ -\frac{1}{2}gg^\prime c_W s_\theta g_{\mu\nu} \\
    \includegraphics[height=3cm, valign=c]{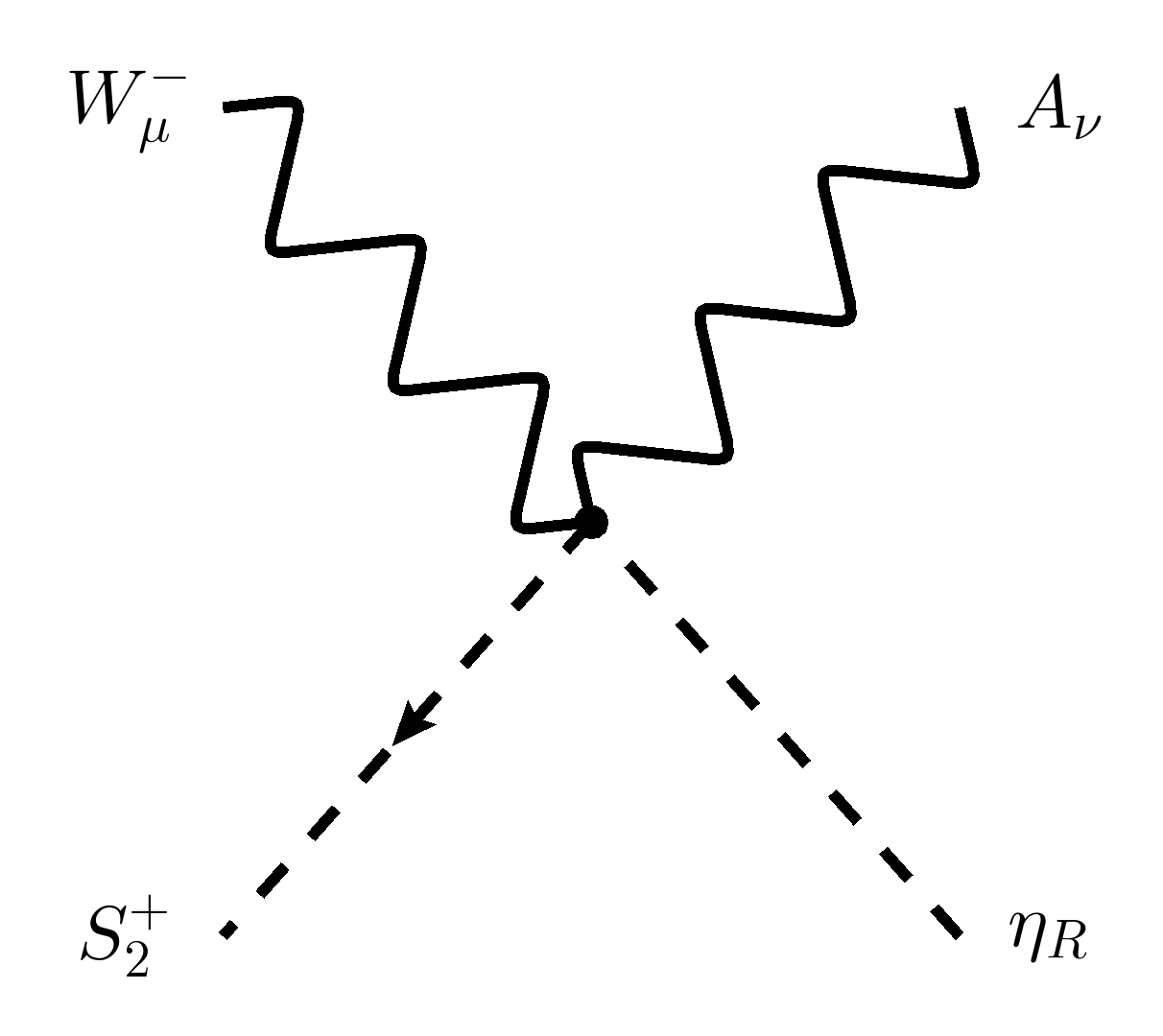} &\sim\ \frac{1}{2}igg^\prime c_\theta c_W g_{\mu\nu} & \includegraphics[height=3cm, valign=c]{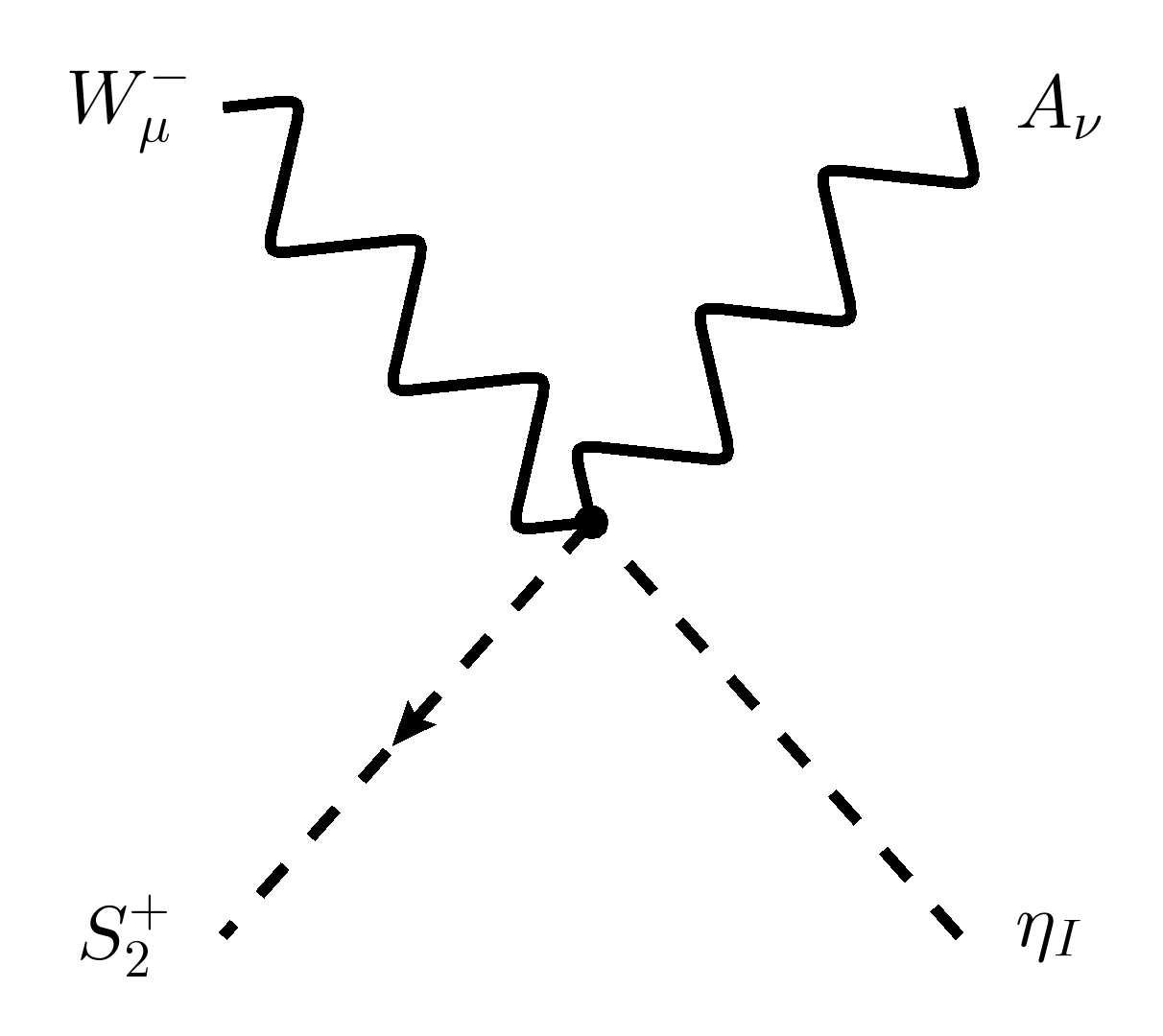} &\sim\ \frac{1}{2}gg^\prime c_\theta c_W g_{\mu\nu} \\
    \includegraphics[height=3cm, valign=c]{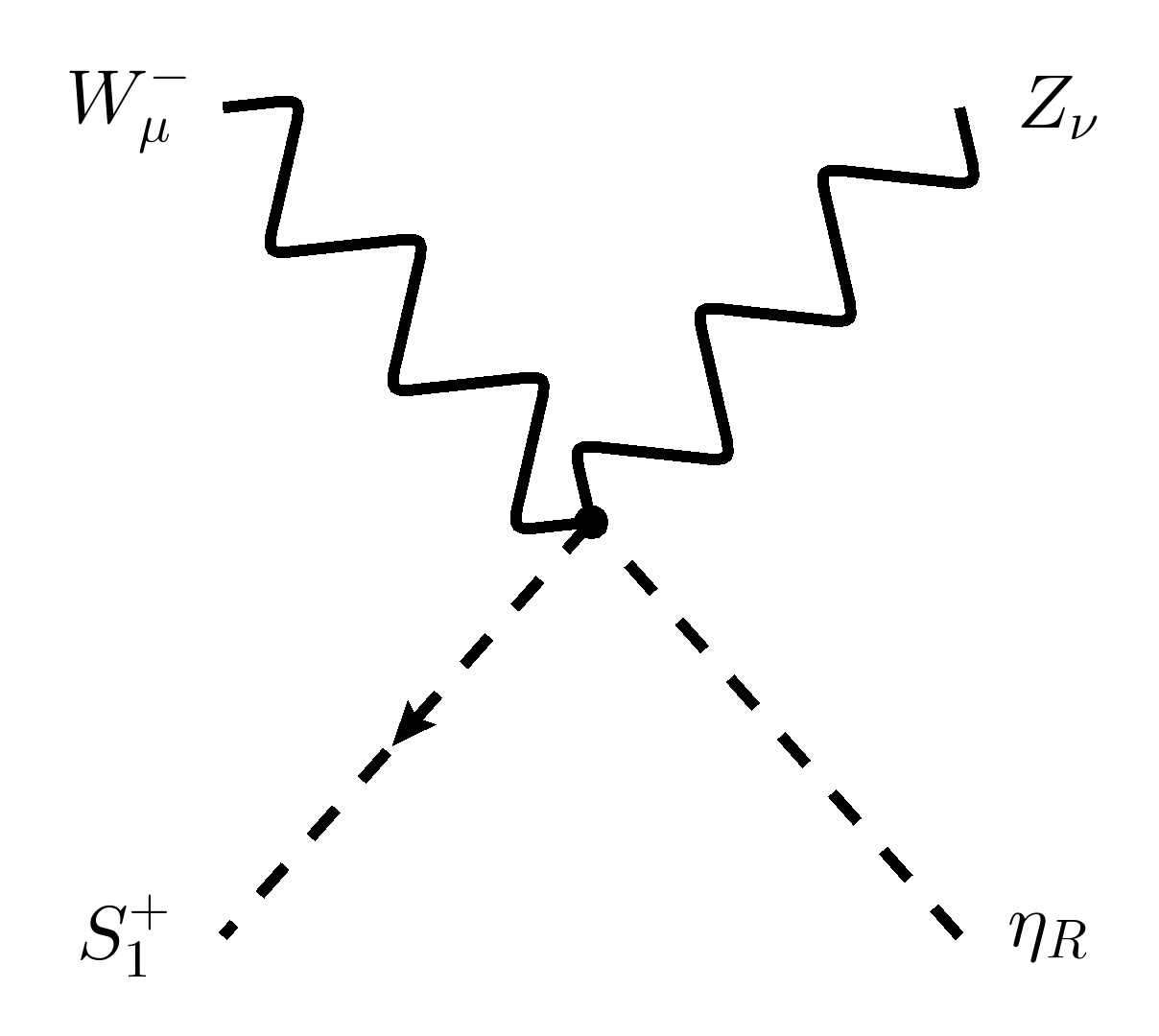} &\sim\ \frac{1}{2}igg^\prime s_\theta s_W g_{\mu\nu} & \includegraphics[height=3cm, valign=c]{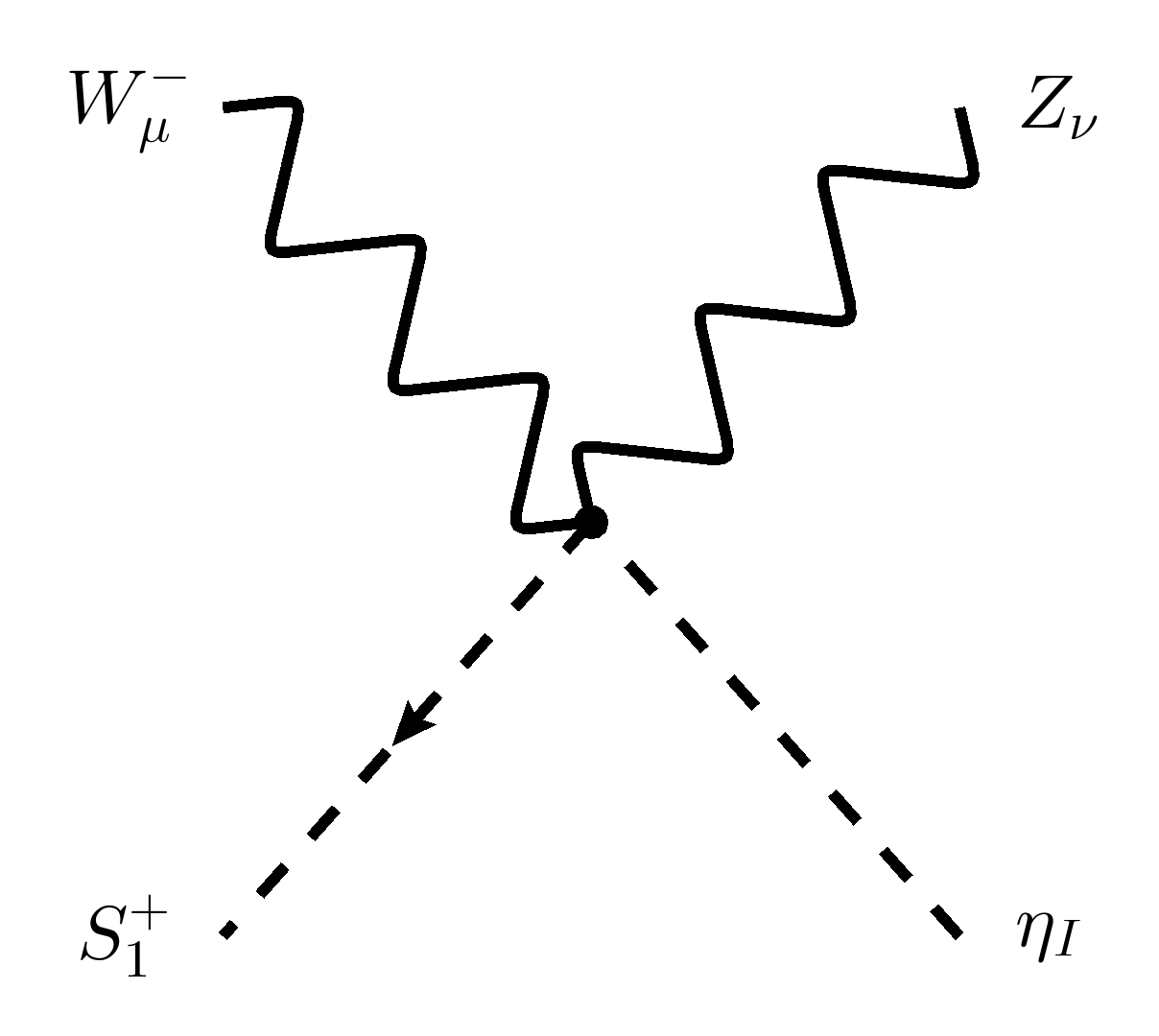} &\sim\ \frac{1}{2}gg^\prime s_\theta s_W g_{\mu\nu} \\
    \includegraphics[height=3cm, valign=c]{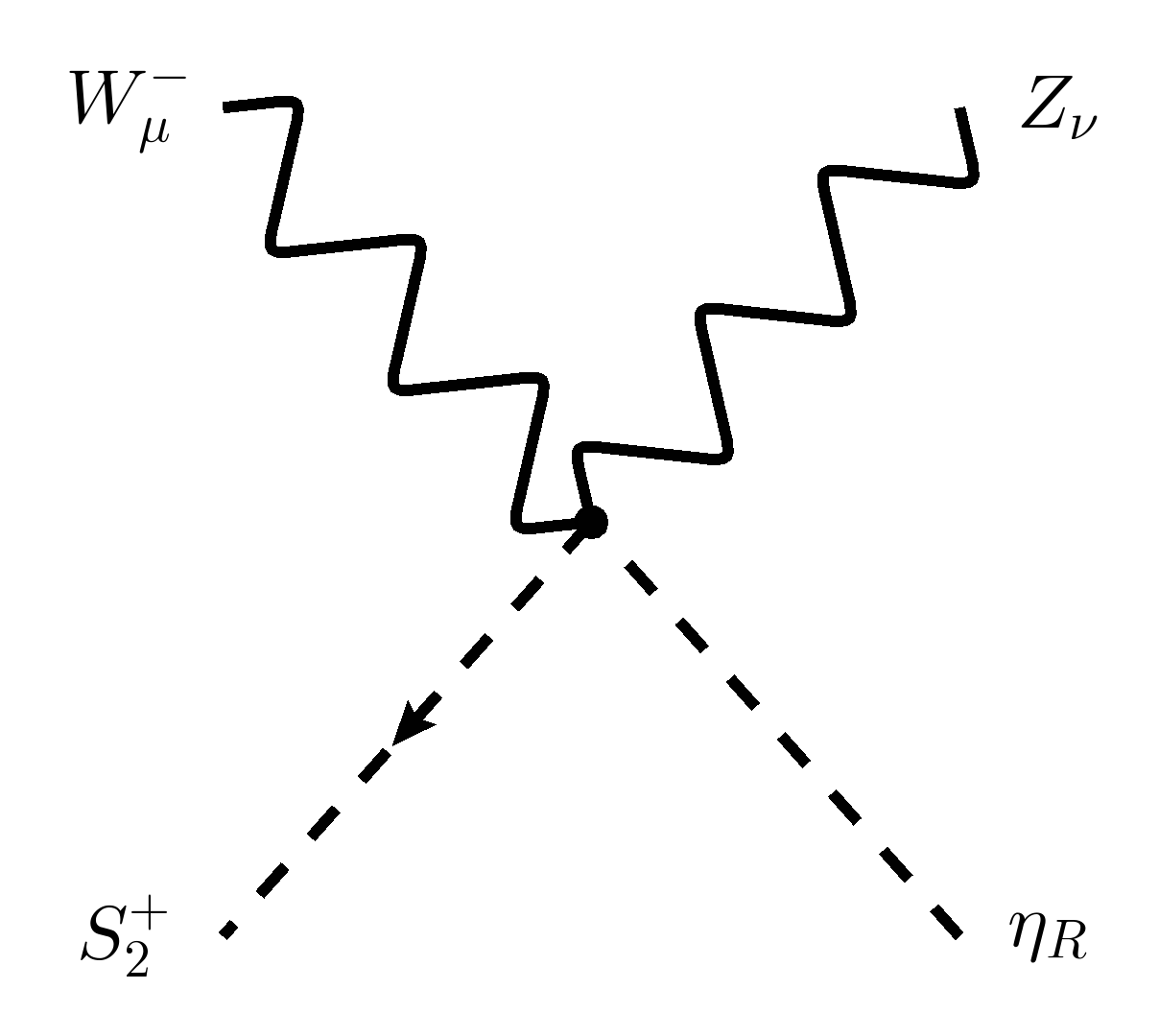} &\sim\ -\frac{1}{2}igg^\prime c_\theta s_W g_{\mu\nu} & \includegraphics[height=3cm, valign=c]{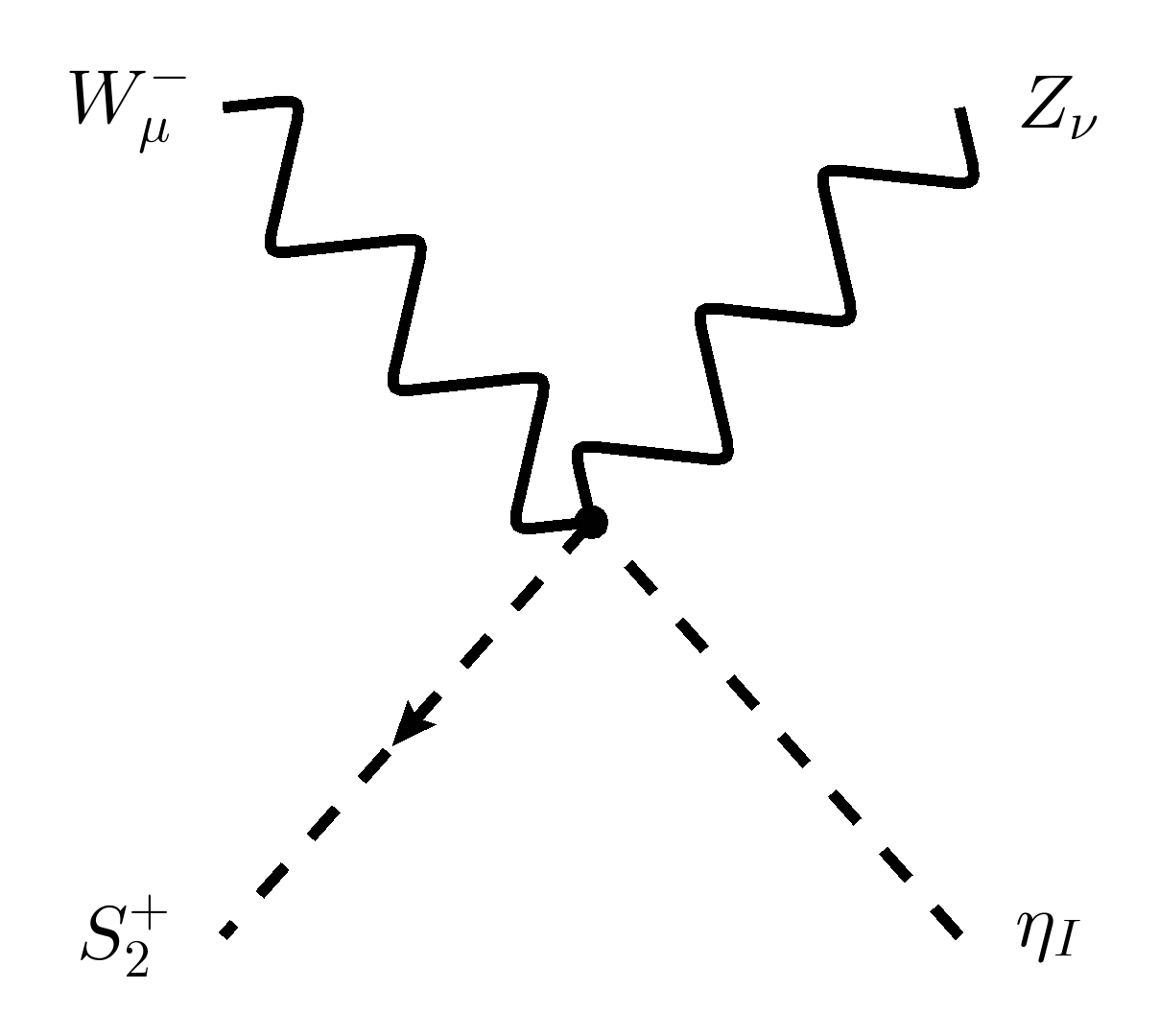} &\sim\ -\frac{1}{2}gg^\prime c_\theta s_W g_{\mu\nu} \\
    \includegraphics[height=3cm, valign=c]{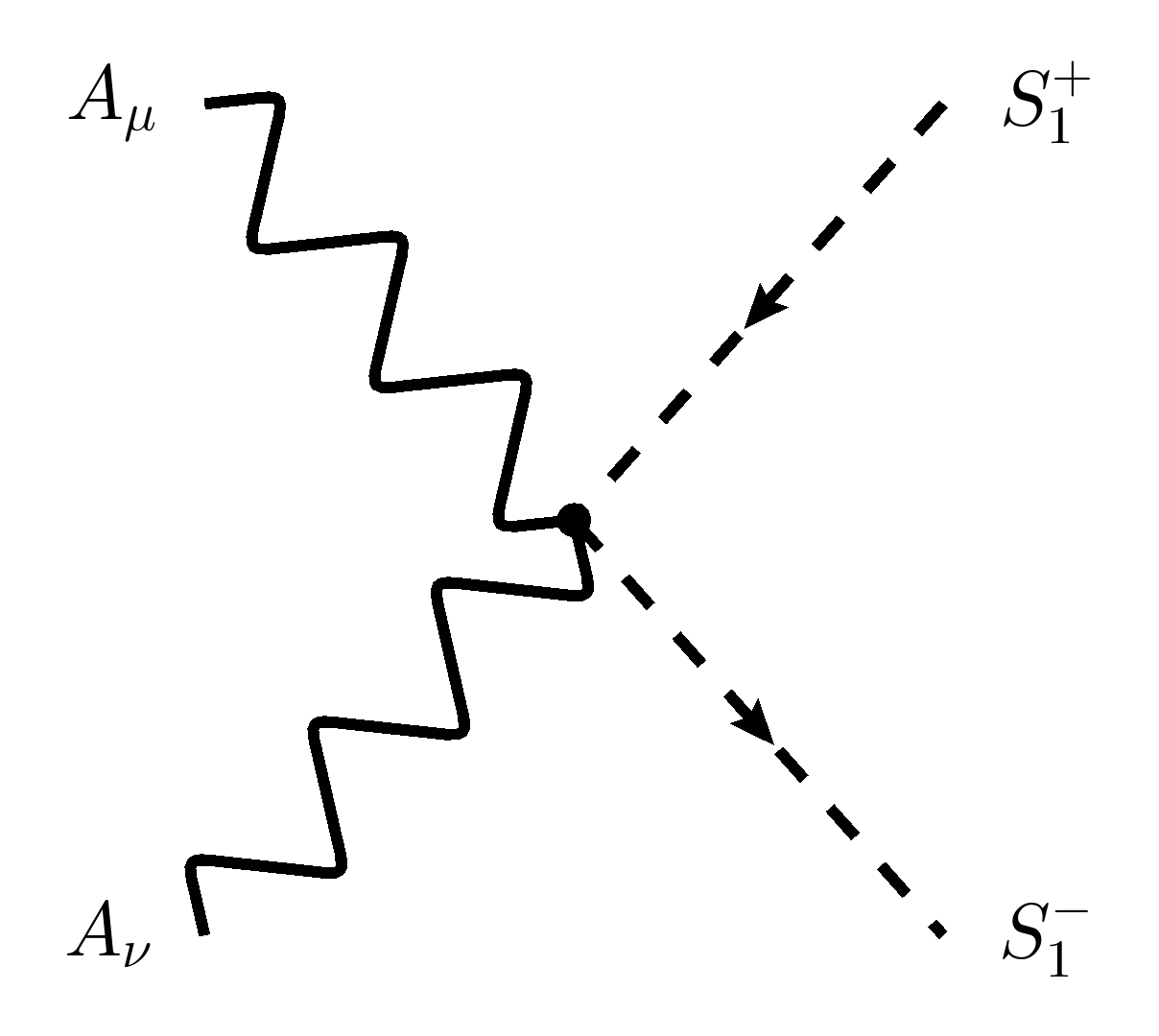} &\sim\ \begin{array}{l} {} \\ 2! \times \frac{1}{2}i\Bigl[4g^{\prime2}c^2_\theta c^2_W \\ \quad + s^2_\theta(g^\prime c_W + gs_W)^2\Bigr] g_{\mu\nu} \end{array} & \includegraphics[height=3cm, valign=c]{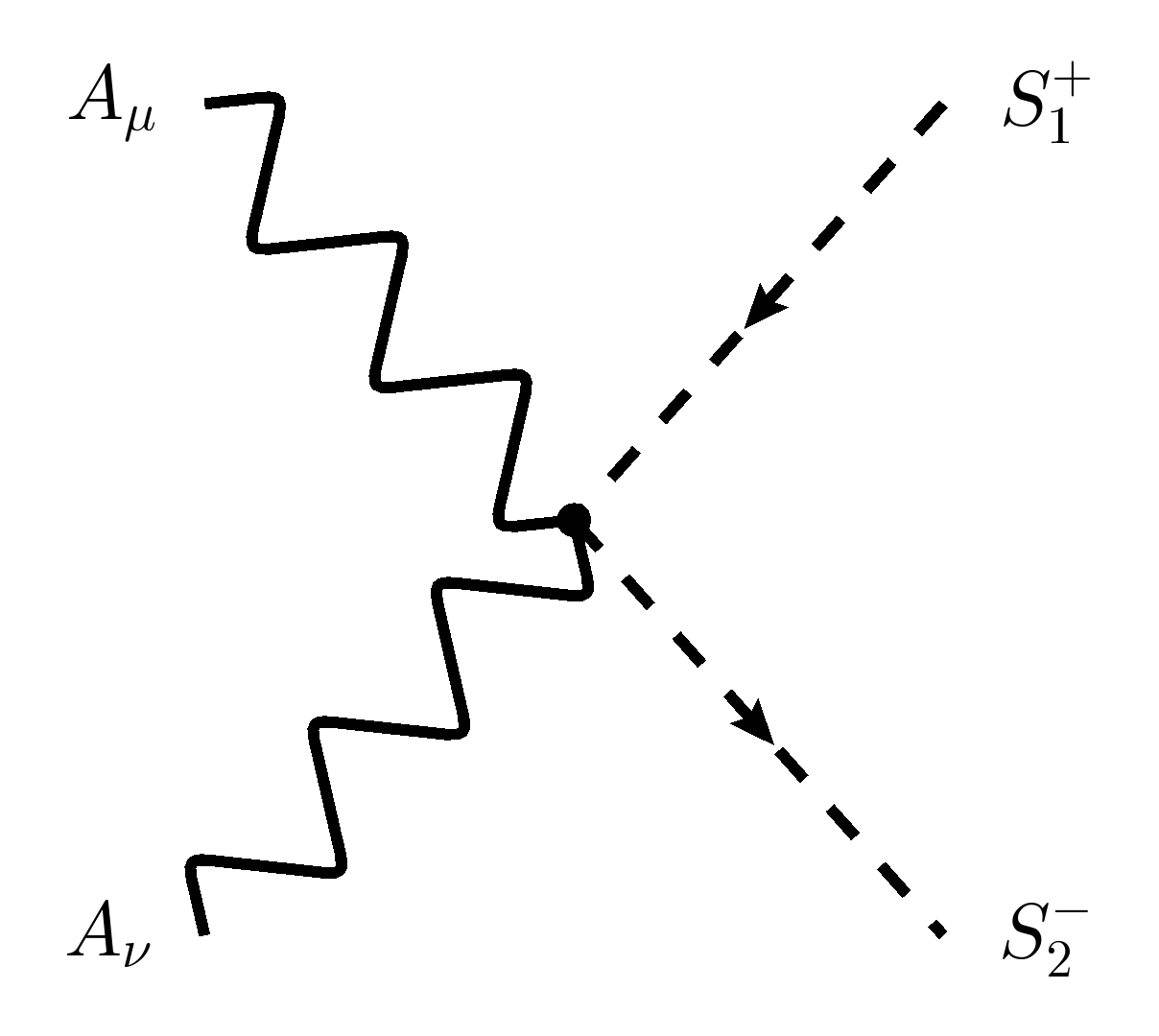} &\sim\ \begin{array}{l} {} \\ 2! \times \frac{1}{4}is_{2\theta}\Bigl[4g^{\prime2}c^2_W \\ \quad - (g^\prime c_W + gs_W)^2\Bigr] g_{\mu\nu} \end{array} \\
    \includegraphics[height=3cm, valign=c]{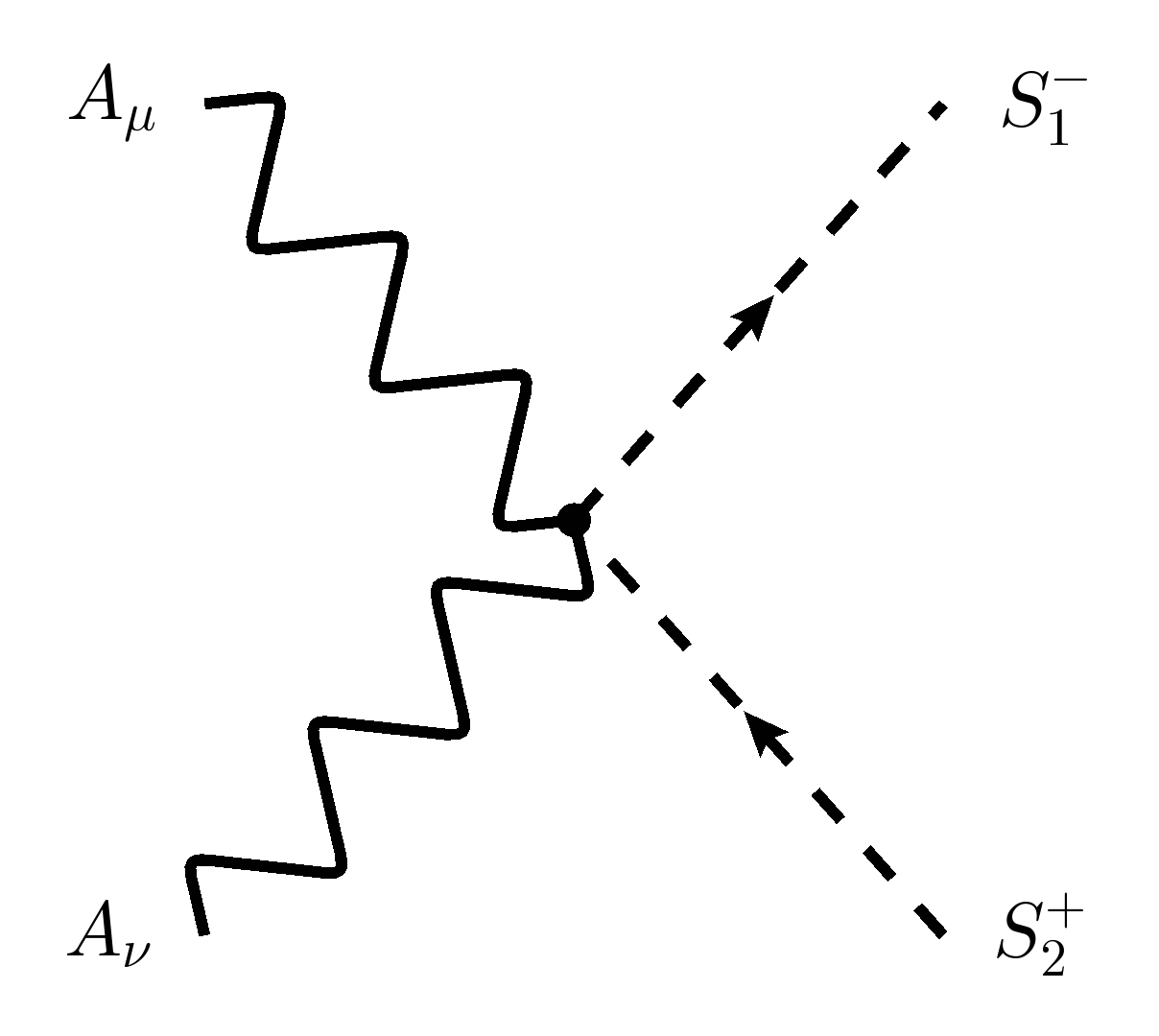} &\sim\ \begin{array}{l} 2! \times \frac{1}{4}is_{2\theta}\Bigl[4g^{\prime2}c^2_W \\ \quad - (g^\prime c_W + gs_W)^2\Bigr] g_{\mu\nu} \end{array} & \includegraphics[height=3cm, valign=c]{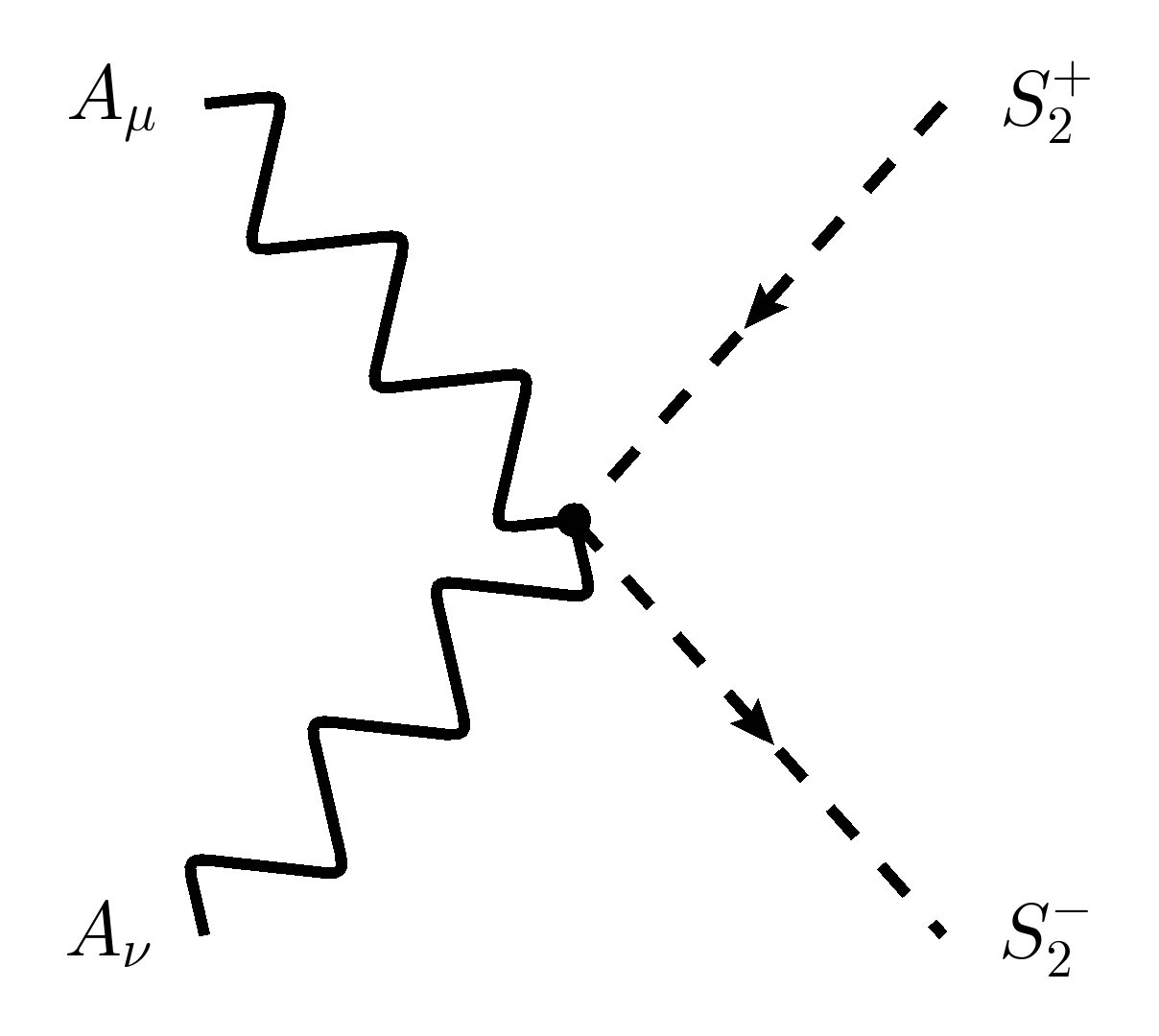} &\sim\ \begin{array}{l} {} \\ \frac{1}{2}i\Bigl[4g^{\prime2}c^2_W s^2_\theta \\ \quad + c^2_\theta(g^\prime c_W + gs_W)^2\Bigr] g_{\mu\nu} \end{array} \\
    \includegraphics[height=3cm, valign=c]{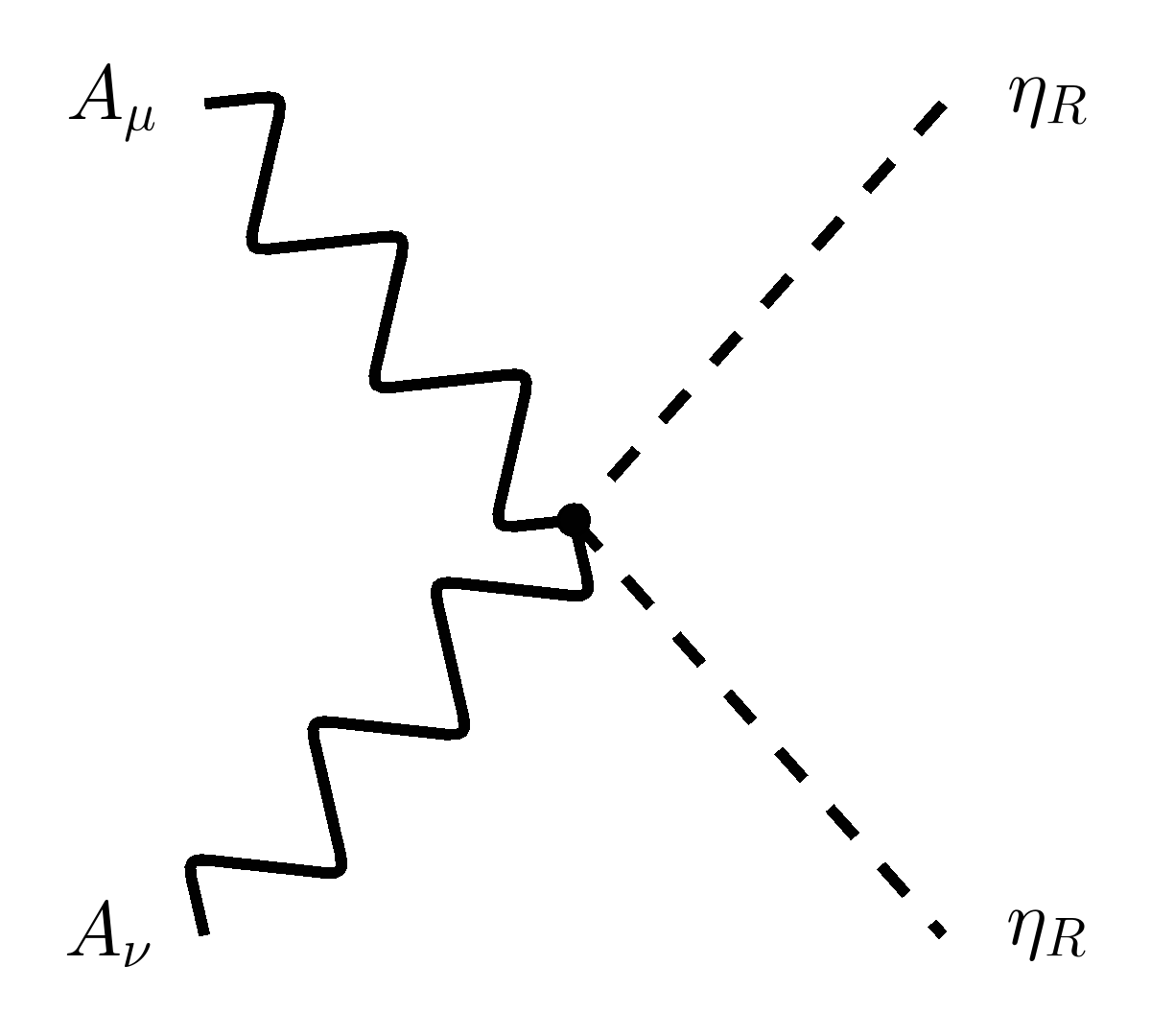} &\sim\ \begin{array}{l} {} \\ 2! \times 2! \\ \quad \times \frac{1}{2}i\Bigl[g^\prime c_W - gs_W\Bigr]^2 g_{\mu\nu} \end{array} & \includegraphics[height=3cm, valign=c]{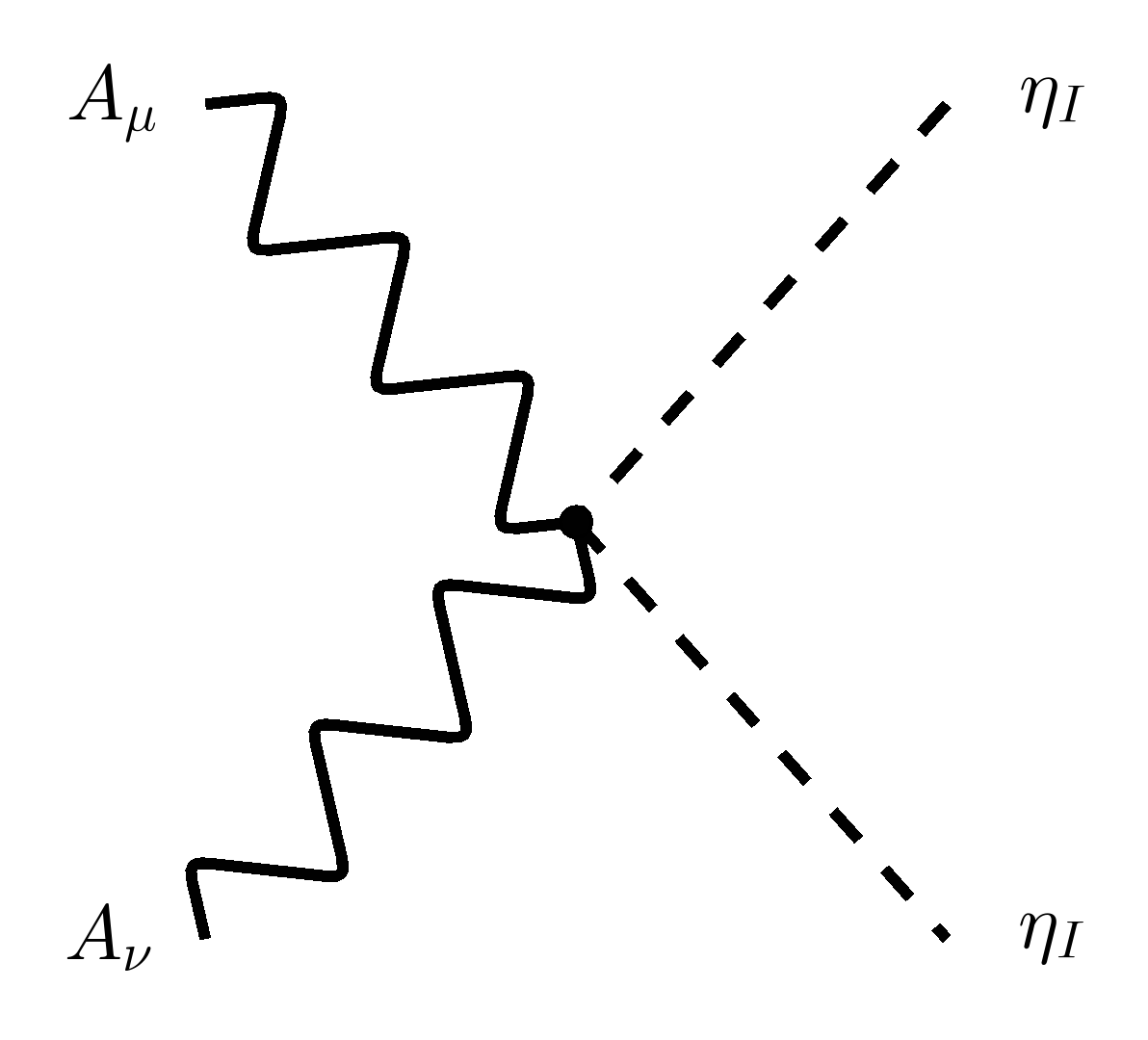} &\sim\ \begin{array}{l} {} \\ 2! \times 2! \\ \quad \times \frac{1}{2}i\Bigl[g^\prime c_W - gs_W\Bigr]^2 g_{\mu\nu} \end{array} \\
    \includegraphics[height=3cm, valign=c]{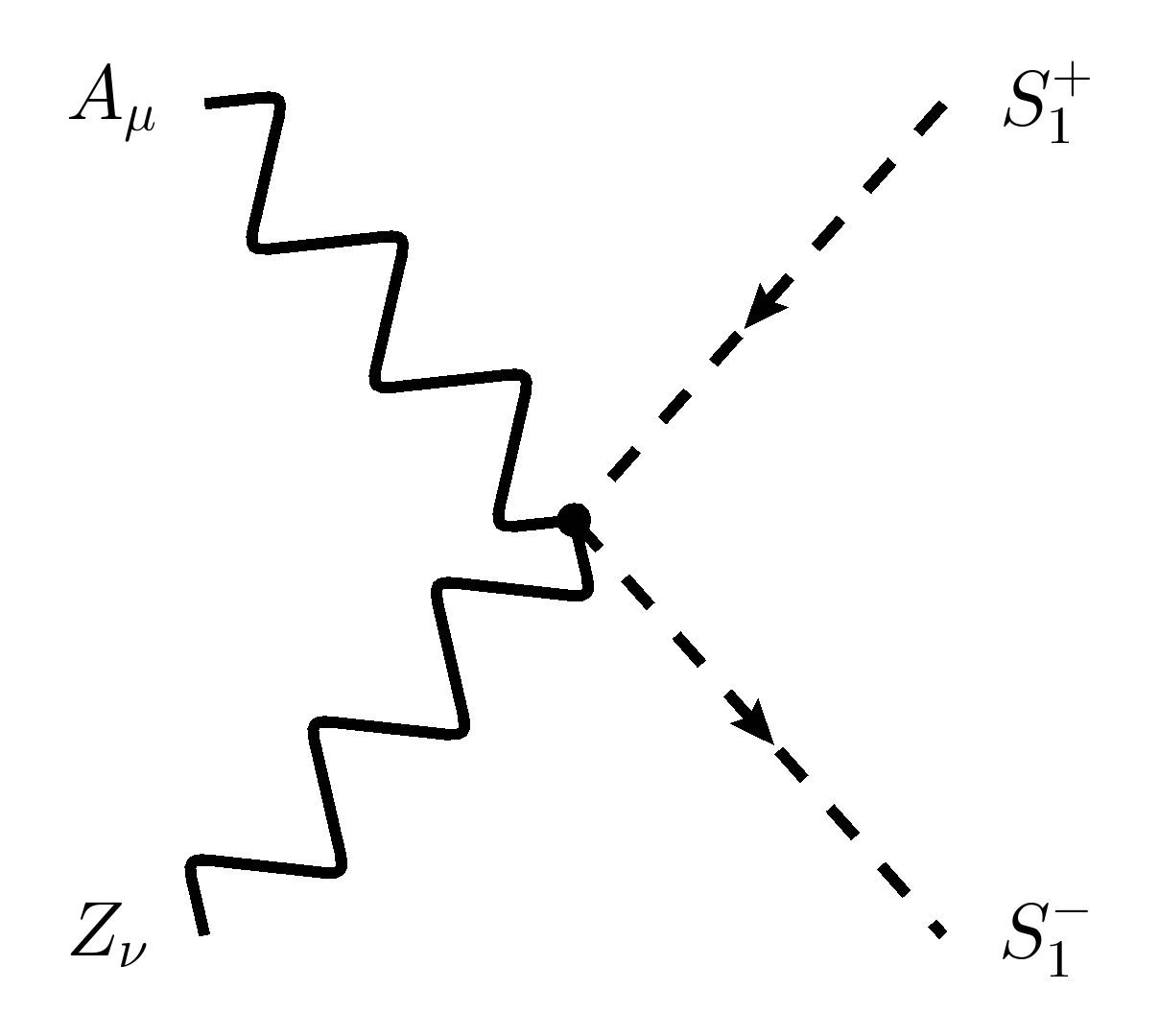} &\sim\ 
    \begin{array}{l} {} \\ i\Bigl[-2g^{\prime2}c^2_\theta c_W s_W \\ \quad + \frac{1}{4}s^2_\theta(2gg^\prime c_{2W} \\ \quad + (g^2 - g^{\prime2})s_{2W})\Bigr] g_{\mu\nu} \end{array} & \includegraphics[height=3cm, valign=c]{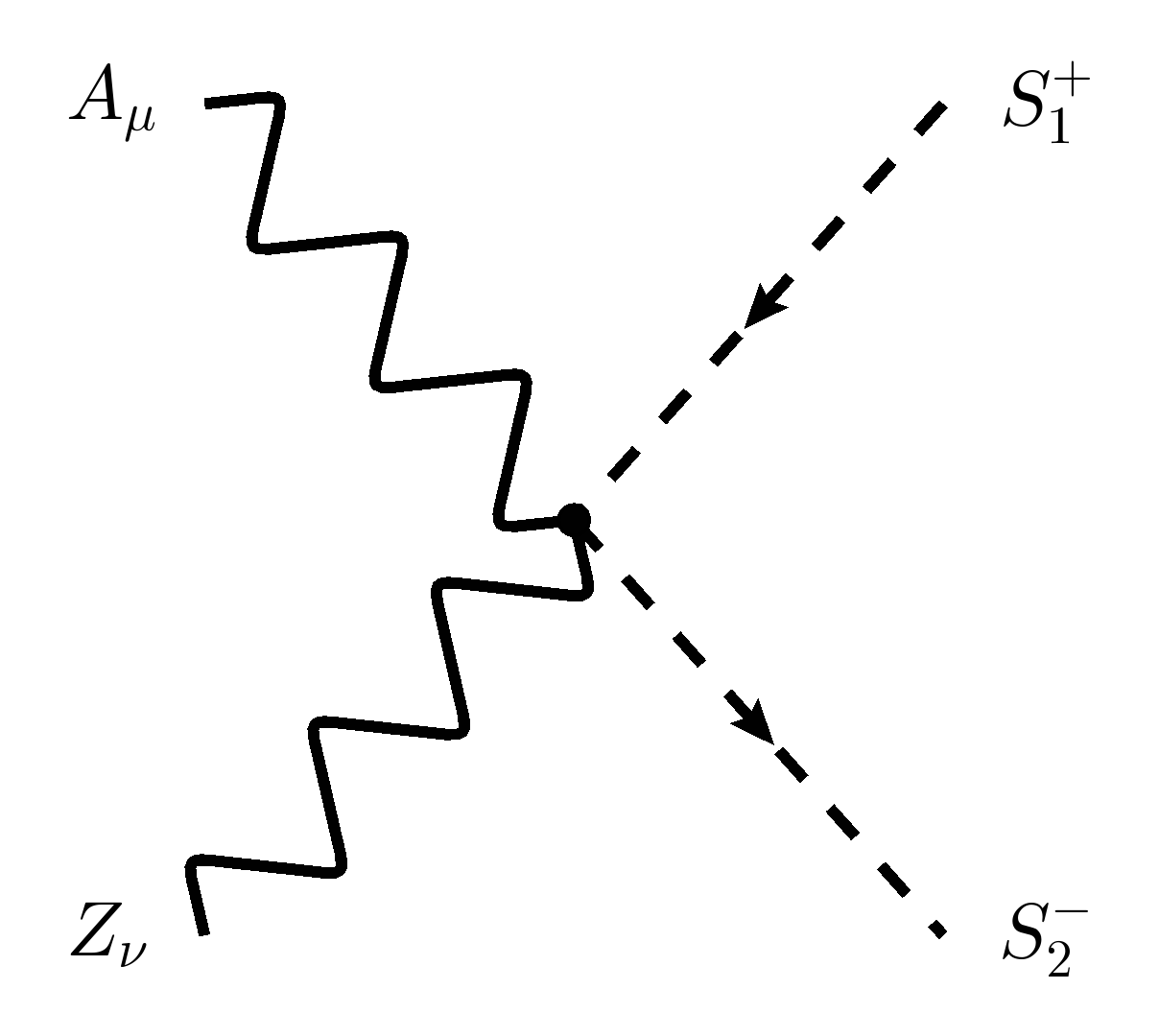} &\sim\ \begin{array}{l} {} \\ -\frac{1}{8}is_{2\theta}\Bigl[2gg^\prime c_{2W} \\ \quad + (g^2 + 3g^{\prime2})s_{2W}\Bigr] g_{\mu\nu} \end{array} \\
    \includegraphics[height=3cm, valign=c]{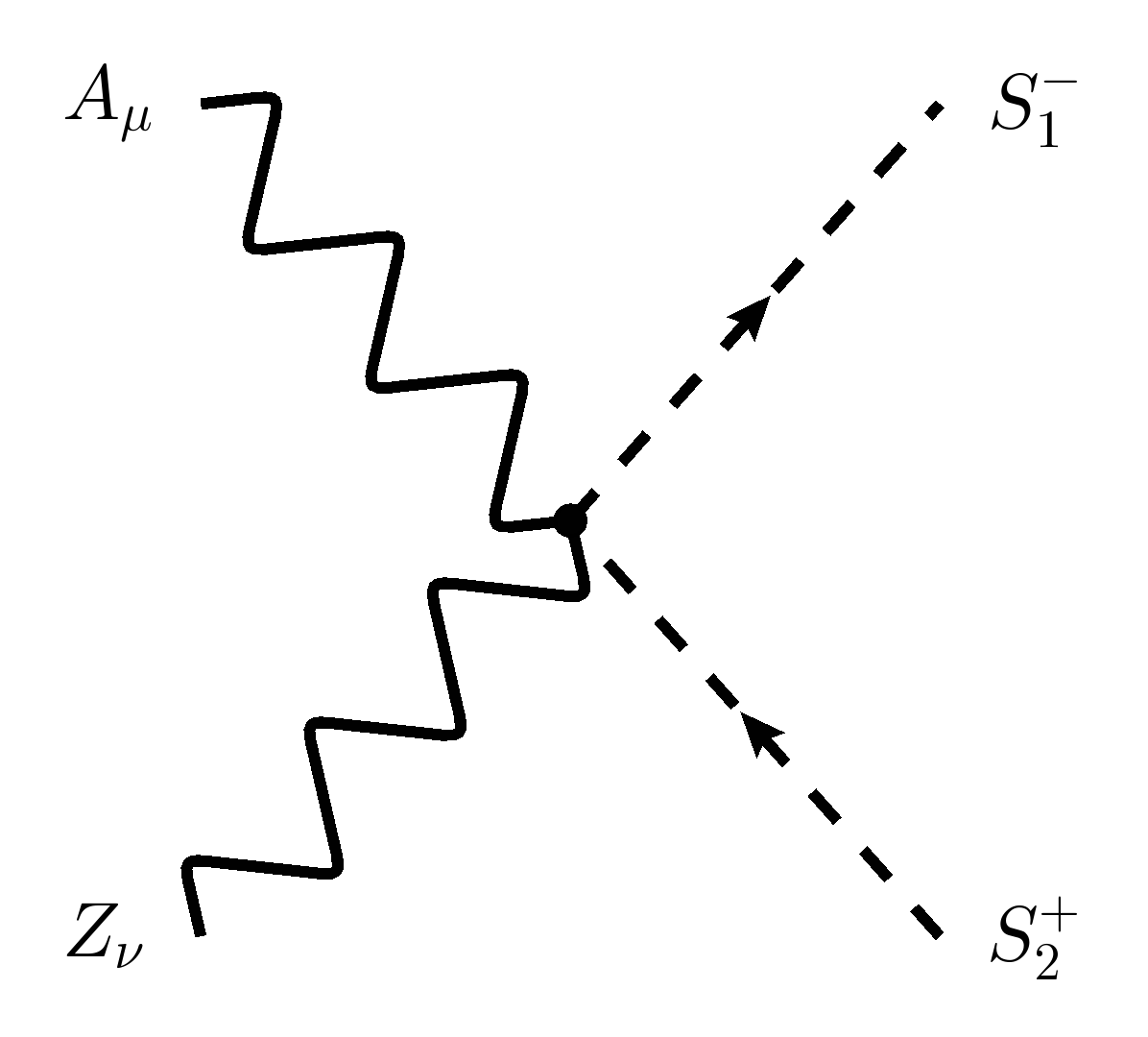} &\sim\ \begin{array}{l} {} \\ -\frac{1}{8}is_{2\theta}\Bigl[2gg^\prime c_{2W} \\ \quad + (g^2 + 3g^{\prime2})s_{2W}\Bigr] g_{\mu\nu} \end{array} & \includegraphics[height=3cm, valign=c]{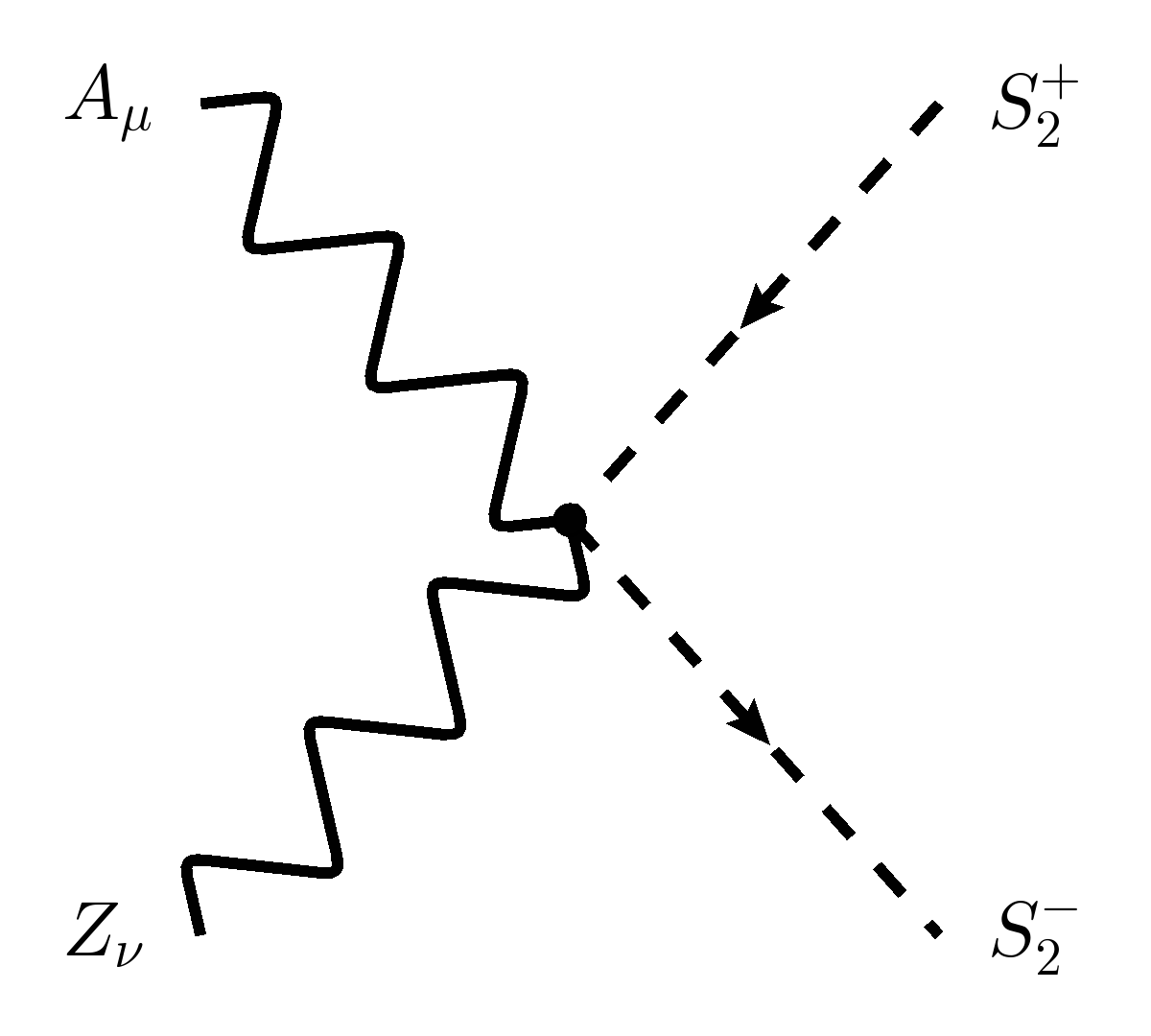} &\sim\ \begin{array}{l} {} \\ \frac{1}{4}i\Bigl[-8g^{\prime2} c_W s^2_\theta s_W \\ \quad + c^2_\theta(2gg^\prime c_{2W} \\ \quad + (g^2 - g^{\prime2})s_{2W})\Bigr] g_{\mu\nu} \end{array} \\
    \includegraphics[height=3cm, valign=c]{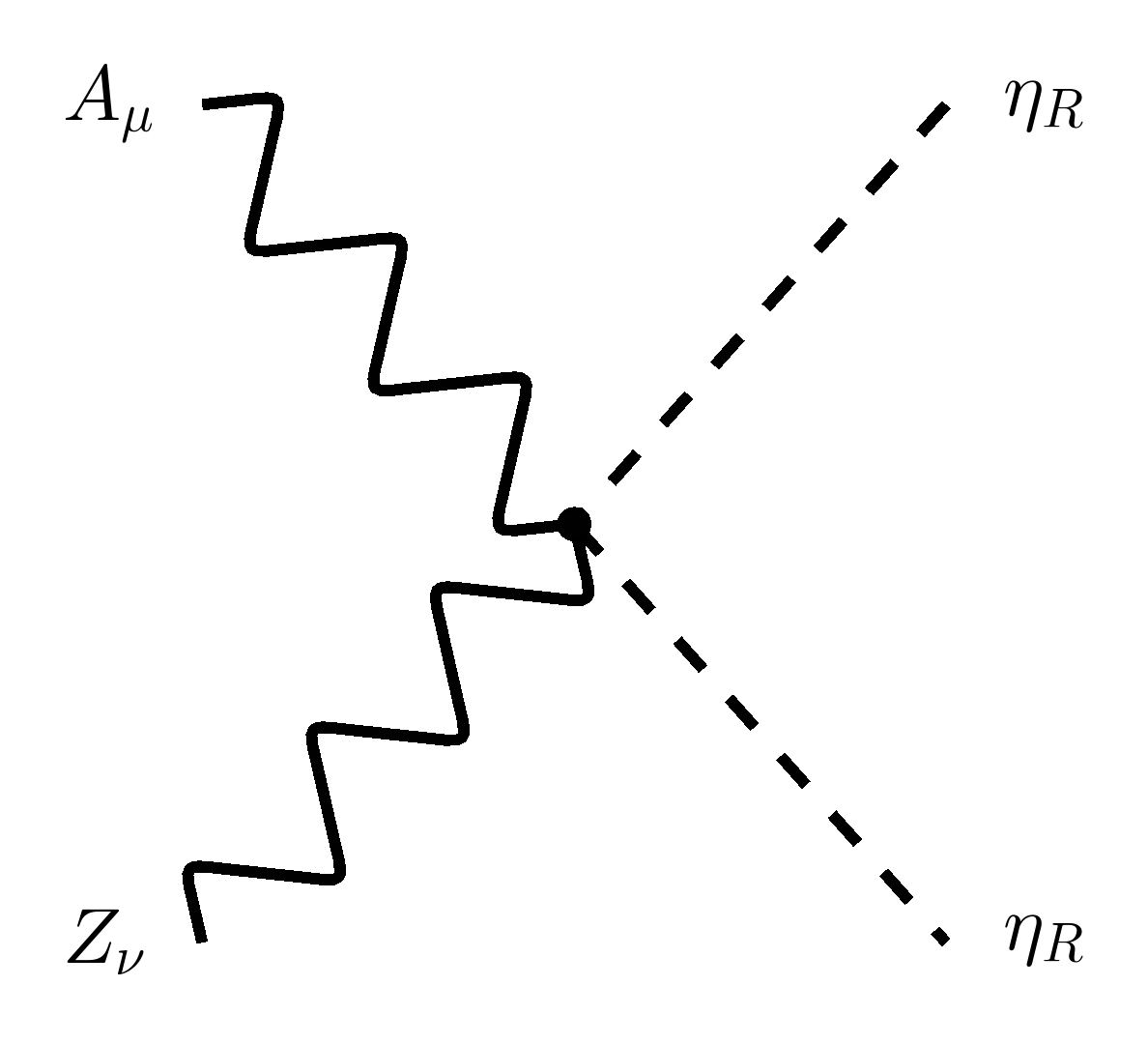} &\sim\ \begin{array}{l} {} \\ 2! \times \frac{1}{4}i\Bigl[-2gg^\prime c_{2W} \\ \quad + (g^2 - g^{\prime2})s_{2W}\Bigr] g_{\mu\nu} \end{array} & \includegraphics[height=3cm, valign=c]{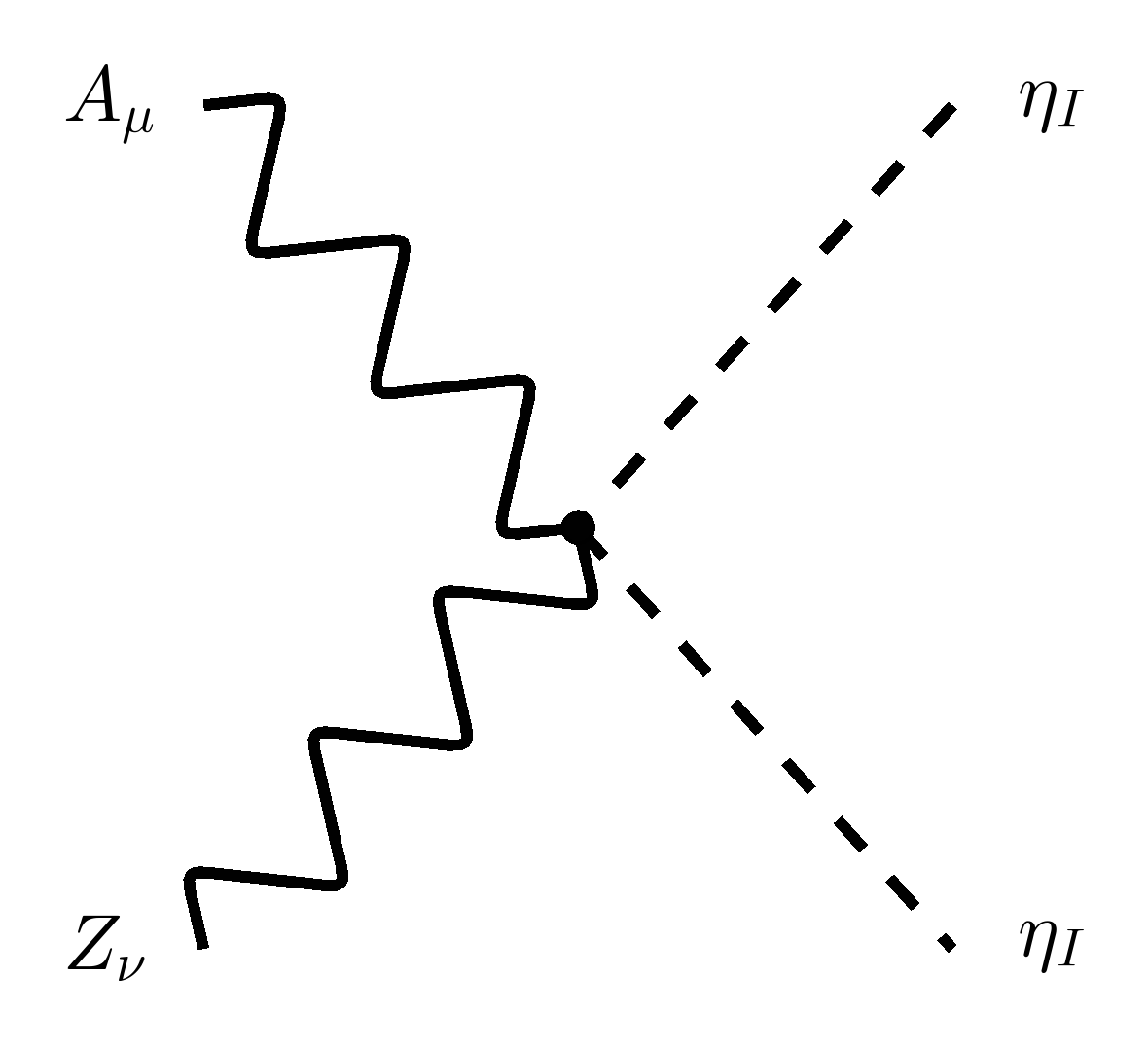} &\sim\ \begin{array}{l} {} \\ 2! \times \frac{1}{4}i\Bigl[-2gg^\prime c_{2W} \\ + (g^2 - g^{\prime2})s_{2W}\Bigr] g_{\mu\nu} \end{array} \\
    \includegraphics[height=3cm, valign=c]{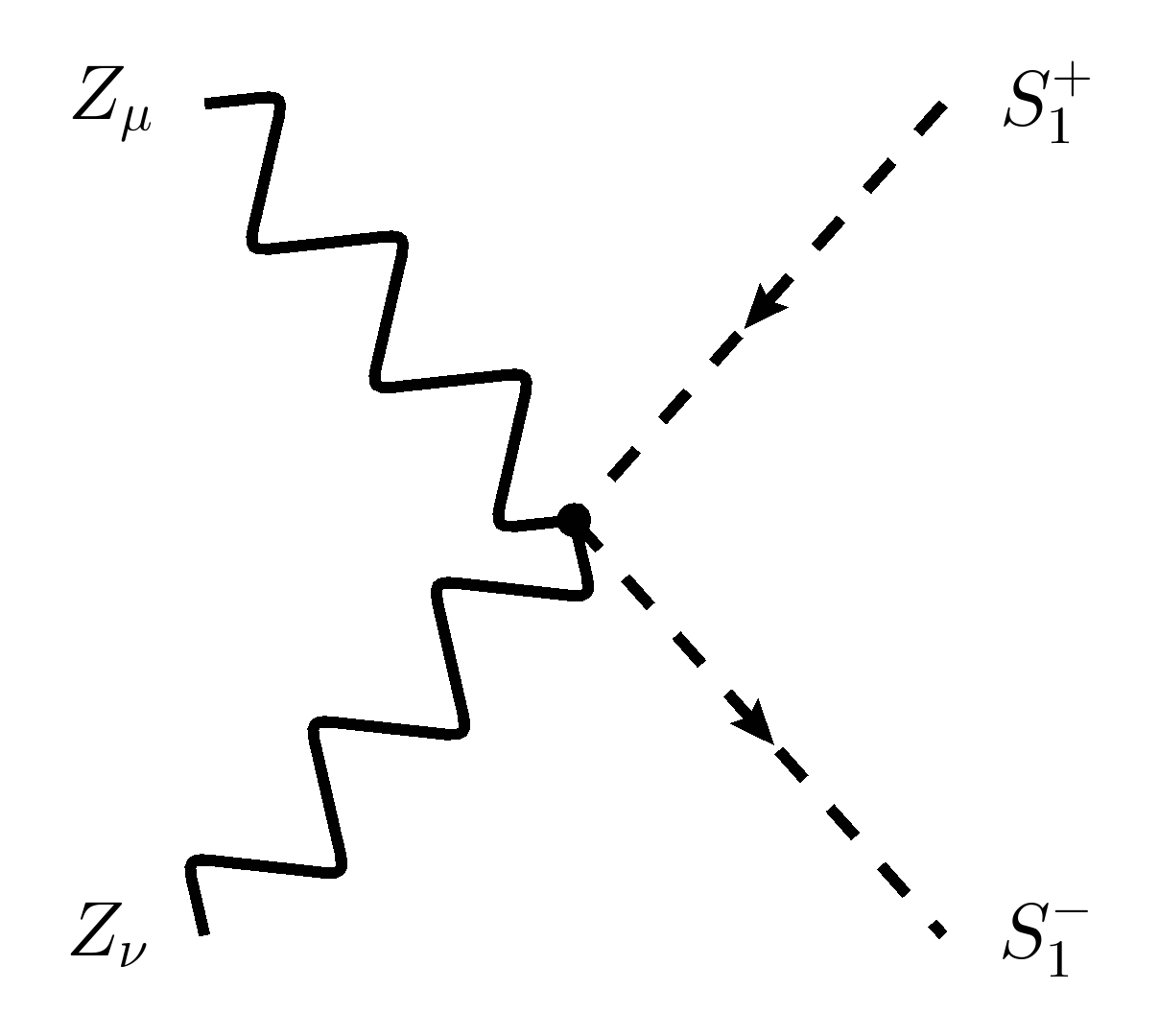} &\sim\ \begin{array}{l} {} \\ 2! \times i\Bigl[2g^{\prime2}c^2_\theta s^2_W \\ \quad + \frac{1}{2}s^2_\theta(gc_W - g^\prime s_W)\Bigr] g_{\mu\nu} \end{array} & \includegraphics[height=3cm, valign=c]{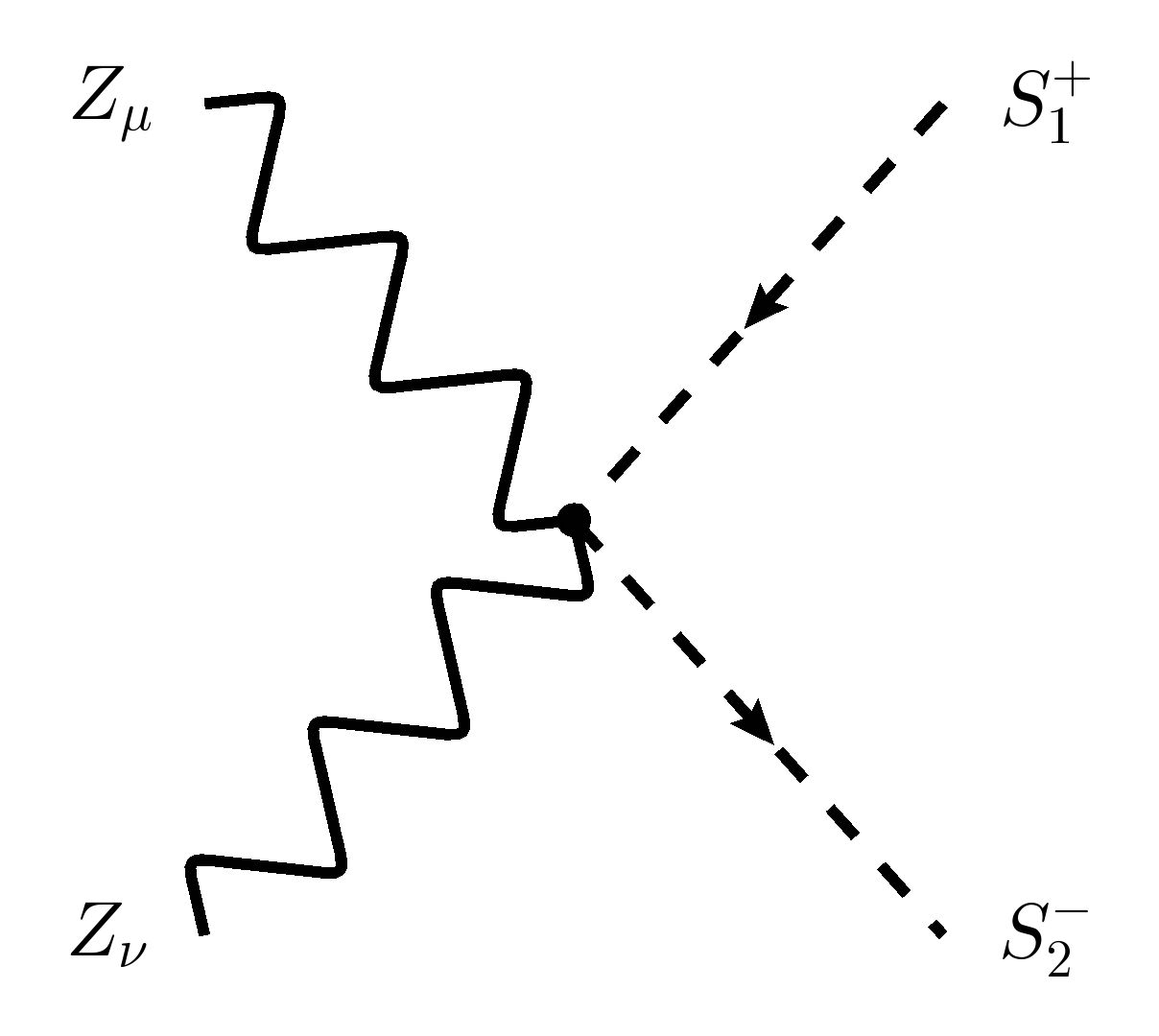} &\sim\ \begin{array}{l} {} \\ 2! \times \frac{1}{4}is_{2\theta}\Bigl[4g^{\prime2}s^2_W \\ \quad - (gc_W - g^\prime s_W)^2\Bigr] g_{\mu\nu} \end{array} \\
    \includegraphics[height=3cm, valign=c]{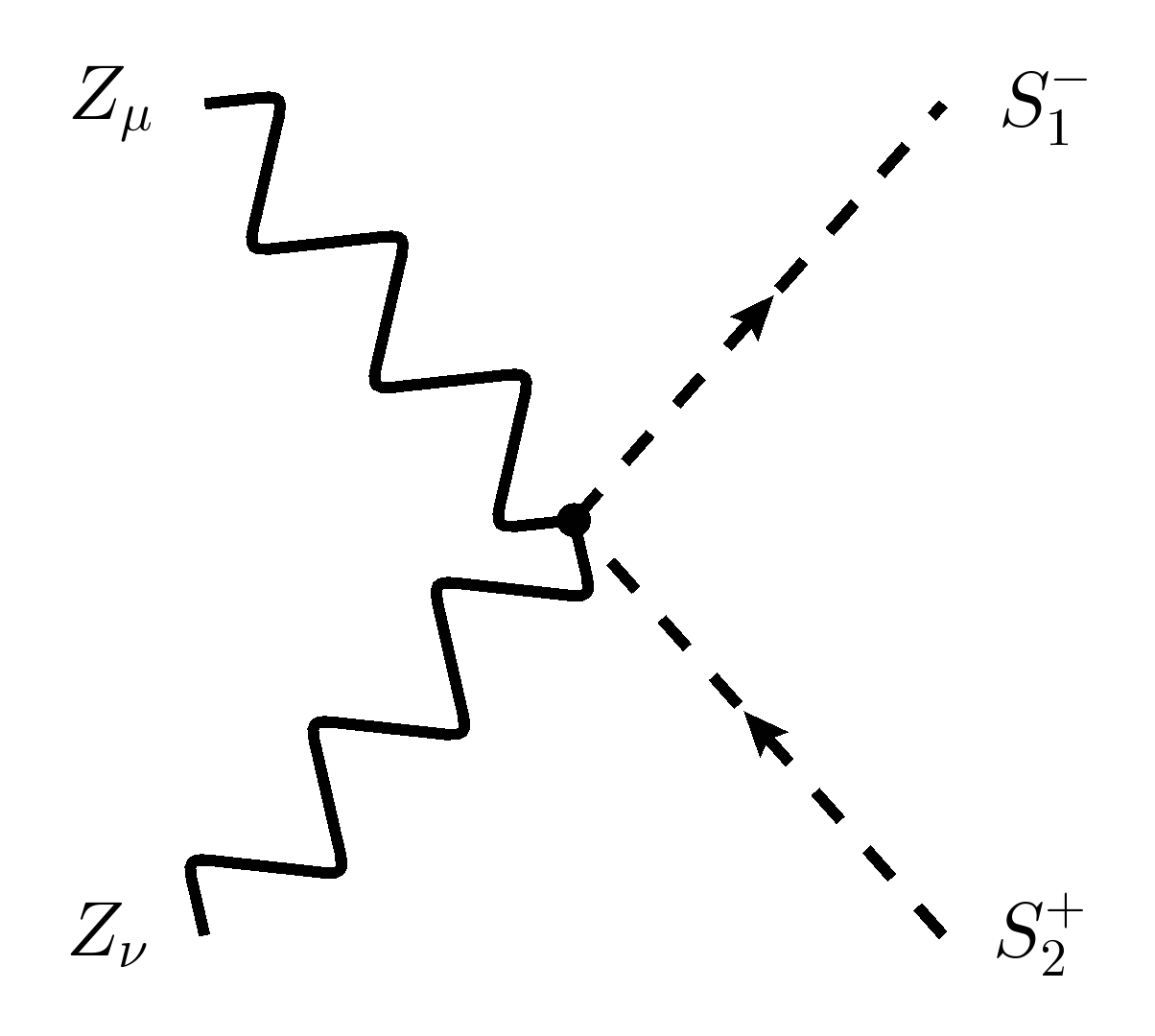} &\sim\ \begin{array}{l} {} \\ 2! \times \frac{1}{4}is_{2\theta}\Bigl[4g^{\prime2}s^2_W \\ \quad - (gc_W - g^\prime s_W)^2\Bigr] g_{\mu\nu} \end{array} & \includegraphics[height=3cm, valign=c]{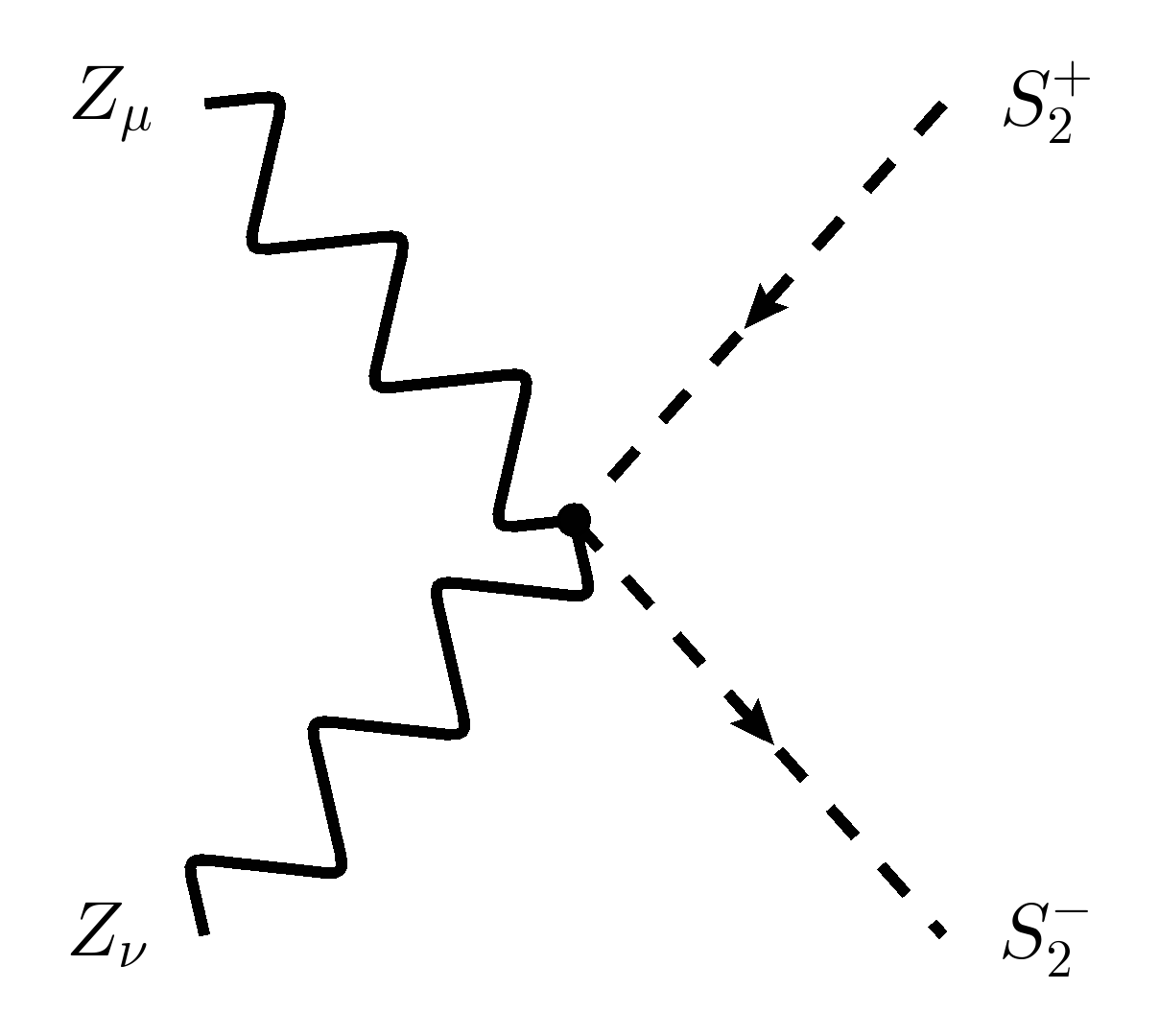} &\sim\ \begin{array}{l} {} \\ 2! \times \frac{1}{2}i\Bigl[4g^{\prime2}s^2_W \\ + c^2_\theta(gc_W - g^\prime s_W)^2\Bigr] g_{\mu\nu} \end{array} \\
    \includegraphics[height=3cm, valign=c]{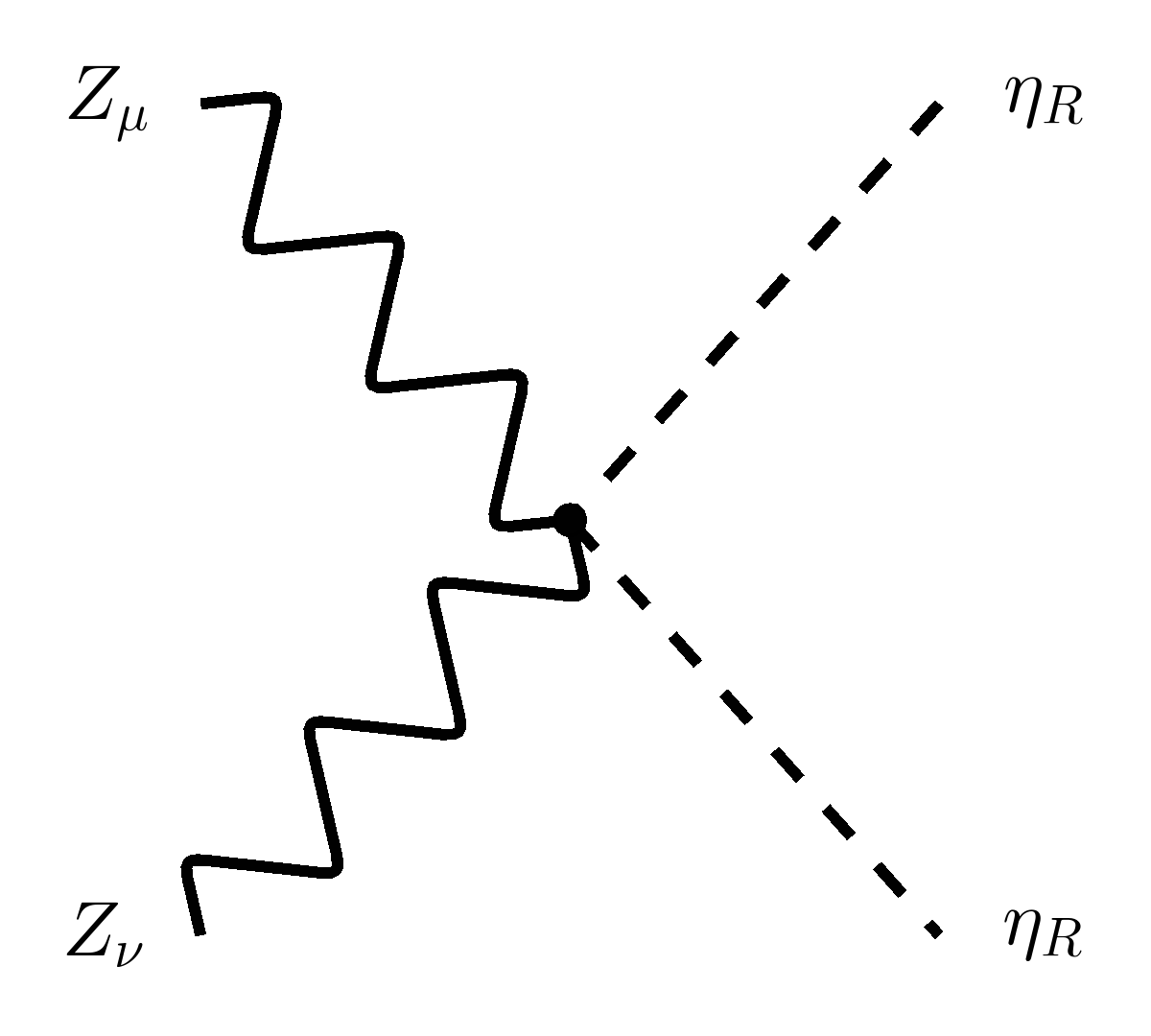} &\sim\ \begin{array}{l} {} \\ 2! \times 2! \\ \quad \times \frac{1}{2}i\Bigl[gc_W + g^\prime s_W\Bigr]^2 g_{\mu\nu} \end{array} & \includegraphics[height=3cm, valign=c]{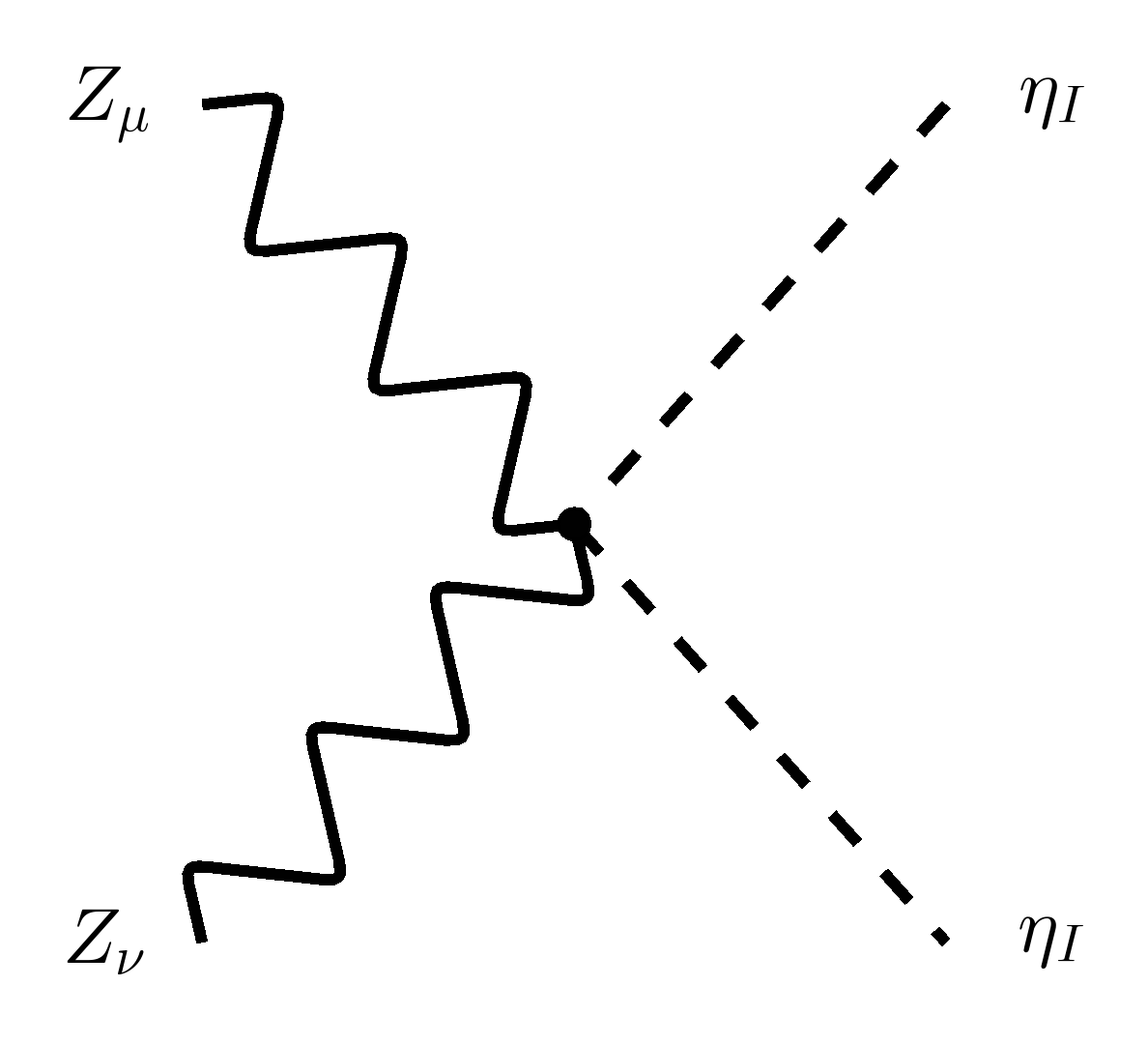} &\sim\ \begin{array}{l} {} \\ 2! \times 2! \\ \quad \times \frac{1}{2}i\Bigl[gc_W + g^\prime s_W\Bigr]^2 g_{\mu\nu} \end{array}
\end{align*}

\subsection{Yukawa Sector}
\label{appendix:vertex_factors_yukawa_sector}

From the new Yukawa sector
\begin{align}
    -\mathcal{L} &\supset \frac{y_{ik}}{\sqrt{2}}U^\dagger_{ni}\eta_R\overline{\nu_{nL}}N_{kR} - \frac{iy_{ik}}{\sqrt{2}}U^\dagger_{ni}\eta_I\overline{\nu_{nL}}N_{kR} + y_{ik}s_\theta S^{-}_1\overline{e_{iL}}N_{kR} - y_{ik}c_\theta S^{-}_2\overline{e_{iL}}N_{kR} \nonumber \\
    &\quad + \kappa_{ki}c_\theta S^{+}_1\overline{N^c_{kR}}e_{iR} + \kappa_{ki}s_\theta S^{+}_2\overline{N^c_{kR}}e_{iR} + \frac{y^\ast_{ik}}{\sqrt{2}}U_{in}\eta_R\overline{N_{kr}}\nu_{nL} + \frac{iy^\ast_{ik}}{\sqrt{2}}U_{in}\eta_I\overline{N_{kR}}\nu_{nL} \nonumber \\
    &\quad + y_{ik}s_\theta S^{+}_1\overline{N_{kR}}e_{iL} - y^\ast_{ik}c_\theta S^{+}_2\overline{N_{kR}}e_{iL} + \kappa^\ast_{ki}c_\theta S^{-}_1\overline{e_{iR}}N^c_{kR} + \kappa^\ast_{ki}s_\theta S^{-}_2\overline{e_{iR}}N^c_{kR}
\end{align}
we get the following vertices:
\begin{align*}
    \includegraphics[height=3cm, valign=c]{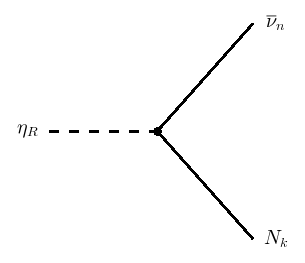} &\sim\ -i\frac{y_{ik}}{\sqrt{2}} U^\dagger_{ni} P_R \label{eq:vertex_NebarHm} &
    \includegraphics[height=3cm, valign=c]{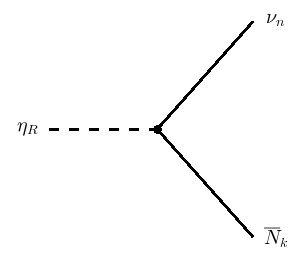} &\sim\ -i\frac{y^\ast_{ik}}{\sqrt{2}} U_{in} P_L \\
    \includegraphics[height=3cm, valign=c]{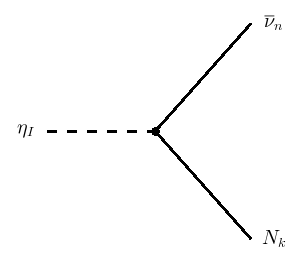} &\sim\ -\frac{y_{ik}}{\sqrt{2}} U^\dagger_{ni} P_R &
    \includegraphics[height=3cm, valign=c]{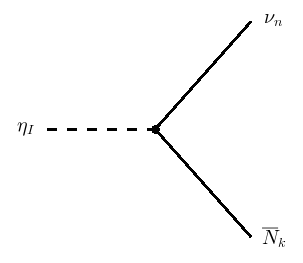} &\sim\ \frac{y^\ast_{ik}}{\sqrt{2}} U_{in} P_L \\
    \includegraphics[height=3cm, valign=c]{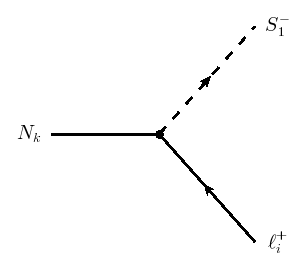} &\sim\ -i(y_{ik} s_\theta P_R + \kappa^\ast_{ki} c_\theta P_L) &
    \includegraphics[height=3cm, valign=c]{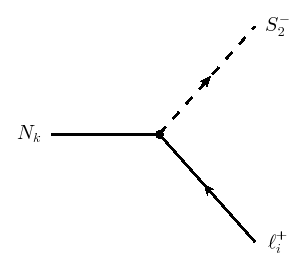} &\sim\ -i(-y_{ijk} c_\theta P_R + \kappa^\ast_{ki} s_\theta P_L) \\
    \includegraphics[height=3cm, valign=c]{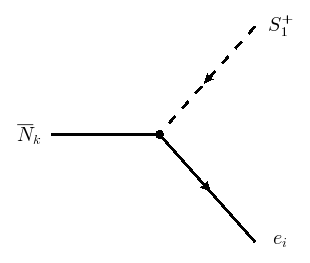} &\sim\ -i(y^\ast_{ik} s_\theta P_L + \kappa_{ki} c_\theta P_R) &
    \includegraphics[height=3cm, valign=c]{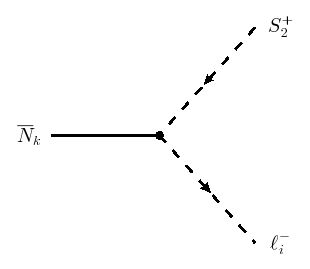} &\sim\ -i(-y^\ast_{ik} c_\theta P_L + \kappa_{ki} s_\theta P_R)
\end{align*}
Here $U = U_\text{PMNS}$, and $P_L = (1 - \gamma^5)/2$ and $P_R = (1 + \gamma^5)/2$ are the left and right chiral projectors, respectively. In this work, we summarize and use the last four diagrams as
\begin{align*}
    \includegraphics[height=3cm, valign=c]{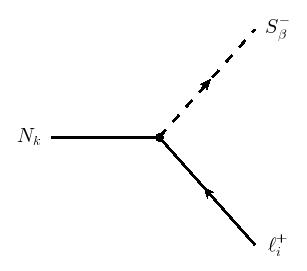} &\sim\ -i(-g^R_{k\beta i} P_R + g^L_{k\beta i} P_L) &
    \includegraphics[height=3cm, valign=c]{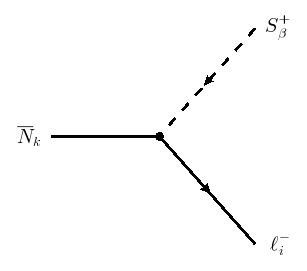} &\sim\ -i(-g^{R\ast}_{k\beta i} P_L + g^{L\ast}_{k\beta i} P_R)
\end{align*}
where we have defined the auxiliary couplings
\begin{equation}\label{eq:auxiliary_coupling_definition}
    g^R_{k\beta i} = y_{ik}R_{2\beta} \quad \text{and} \quad g^L_{k\beta i} = \kappa^\ast_{ki}R_{1\beta} \, ,
\end{equation}
with
\begin{equation}
    R
    =
    \begin{pmatrix}
        R_{11} & R_{12} \\
        R_{21} & R_{22}
    \end{pmatrix}
    =
    \begin{pmatrix}
        \cos\theta & \sin\theta \\
        -\sin\theta & \cos\theta
    \end{pmatrix} \, .
\end{equation}

\end{appendices}


\newpage

\providecommand{\href}[2]{#2}
\begingroup
\raggedright

\endgroup


\end{document}